\documentclass[twosides,a4paper,12pt,openright]{book}
\usepackage[a4paper,top=2cm,bottom=2cm,outer=2cm]{geometry}

\usepackage[utf8]{inputenc}
\usepackage[T1]{fontenc}
\usepackage[polish,main=english]{babel}
\usepackage[toc,page]{appendix}
\usepackage{amsmath}
\usepackage{booktabs}
\usepackage{multirow}
\usepackage{longtable}
\usepackage{xltabular}
\usepackage{makecell}
\usepackage{tabularx,booktabs}
\usepackage{graphicx}
\usepackage{wrapfig}
\usepackage{lscape}
\usepackage[figuresleft]{rotating}
\usepackage{floatflt}
\usepackage{subcaption}
\usepackage{float}
\usepackage{xspace}
\usepackage[outercaption]{sidecap}
\usepackage{pdflscape}
\usepackage{xurl}
\usepackage{footnote}
\usepackage{footmisc}
\usepackage{tablefootnote}
\usepackage{enumitem}
\usepackage{siunitx}
\usepackage[hidelinks]{hyperref}
\usepackage[toc,acronym]{glossaries}
\usepackage{threeparttable}
\usepackage{afterpage}
\usepackage{csquotes}
\usepackage[version=4]{mhchem}
\usepackage[nobottomtitles*]{titlesec}
\usepackage[capitalise]{cleveref}

\crefformat{app}{Appendix #2#1#3}
\crefrangeformat{app}{Appendices #3#1#4 to #5#2#6}
\crefmultiformat{app}{Appendices #2#1#3}{ and #2#1#3}{, #2#1#3}{ and #2#1#3}
\crefrangemultiformat{app}{Appendices #3#1#4 to #5#2#6}{ and #3#1#4 to #5#2#6}{, #3#1#4 to #5#2#6}{ and #3#1#4 to #5#2#6}

\Crefformat{app}{Appendix #2#1#3}
\Crefrangeformat{app}{Appendices #3#1#4 to #5#2#6}
\Crefmultiformat{app}{Appendices #2#1#3}{ and #2#1#3}{, #2#1#3}{ and #2#1#3}
\Crefrangemultiformat{app}{Appendices #3#1#4 to #5#2#6}{ and #3#1#4 to #5#2#6}{, #3#1#4 to #5#2#6}{ and #3#1#4 to #5#2#6}

\defglsentryfmt[acronym]{\glsgenacfmt}
\makeglossaries
\glsenableentrycount

\usepackage{fancyhdr}
\fancypagestyle{plain}{%
   \fancyhead{}
}

\newcommand{\blankpage}{\null \thispagestyle{empty} \newpage} 
\renewcommand{\deg}{\ensuremath{^\circ}\xspace} 
\newcommand{\eg}{\emph{e.g.}\xspace} 
\newcommand{\ie}{\emph{i.e.}\xspace} 
\newcommand{\etal}{\mbox{\emph{et al.}}\xspace}	
\newcommand{\etc}{\mbox{\emph{etc.}}\xspace} 
\newcommand{\anhpeak}{\SI{511}{\kilo\electronvolt}\xspace} 
\newcommand{\peakpos}{$\mu_{\textrm{511~keV}}$\xspace} 
\newcommand{\Na}{\ce{^{22}Na}\xspace} 
\newcommand{\chiNDF}{$\chi^2/ndf$\xspace} 
\newcommand{\MLR}{$M_{\textrm{LR}}$\xspace} 
\newcommand{\MLRx}{$M_{\textrm{LR}}$(x)\xspace} 
\newcommand{\MLRstar}{$M_{\textrm{LR}}^*$\xspace} 
\newcommand{\Zeff}{$Z_{\textrm{eff}}$\xspace} 
\newcommand{\tzero}{$T_{\textrm{0}}$\xspace} 
\newcommand{\baso}{\ce{BaSO_4}\xspace} 
\newcommand{\Qavg}{$Q_{\mathrm{avg}}$\xspace}
\newcommand{\Qavgstar}{$Q_{\mathrm{avg}}^{*}$\xspace}

\DeclareSIUnit{\photon}{ph}
\DeclareSIUnit{\cps}{cps}
\DeclareSIUnit{\pe}{PE}
\DeclareSIUnit{\revPE}{PE^{-1}}
\DeclareSIUnit{\adc}{ADC}
\DeclareSIUnit{\au}{a.u.}
\DeclareSIUnit{\year}{y}

\newacronym{gl:PG}{PG}{Prompt gamma}
\newacronym{gl:ELA}{ELA}{Exponential light attenuation} 
\newacronym{gl:ELAR}{ELAR}{Exponential light attenuation with light reflection}
\newacronym{gl:BL}{BL}{Base line}
\newacronym{gl:PE}{PE}{Photoelectron}
\newacronym{gl:PMT}{PMT}{Photomultiplier tube}
\newacronym{gl:SPAD}{SPAD}{Single-photon avalanche apotodiode}
\newacronym{gl:SiPM}{SiPM}{Silicon photomultiplier}
\newacronym{gl:dSiPM}{dSiPM}{Digital silicon photomultiplier}
\newacronym{gl:HEP}{HEP}{High energy physics}
\newacronym{gl:WLS}{WLS}{Wavelength shifter}
\newacronym{gl:PDE}{PDE}{Photodetection efficiency}
\newacronym{gl:DCR}{DCR}{Dark count rate}
\newacronym{gl:ADC}{ADC}{Analog-to-digital converter}
\newacronym{gl:SNR}{SNR}{Signal-to-noise ratio}
\newacronym{gl:TDC}{TDC}{Time-to-digital converter}
\newacronym{gl:CT}{CT}{Computed tomography}
\newacronym{gl:SPECT}{SPECT}{Single-photon emission computed tomography}
\newacronym{gl:PET}{PET}{Positron emission tomography}
\newacronym{gl:LED}{LED}{Leading edge discriminator}
\newacronym{gl:LLT}{LLT}{Low level threshold}
\newacronym{gl:DAQ}{DAQ}{Data acquisition system}
\newacronym{gl:SiFi-CC}{SiFi-CC}{Silicon Photomultipiers and Scintillating Fibers based Compton Camera}
\newacronym{gl:TOT}{TOT}{Time over threshold}
\newacronym{gl:SSP}{SSP}{Small scale prototype}
\newacronym{gl:JU}{JU}{Jagiellonian University}
\newacronym{gl:PMI}{PMI}{Physics of Molecular Imaging Systems Department}
\newacronym{gl:CM}{CM}{Coded mask}
\newacronym{gl:CC}{CC}{Compton camera}
\newacronym{gl:PCB}{PCB}{Printed circuit board}
\newacronym{gl:FFC}{FFC}{Flexible flat cable}
\newacronym{gl:FBC}{FBC}{Fan-beam collimator}
\newacronym{gl:FEE}{FEE}{Front-end electronics}
\newacronym{gl:FWHM}{FWHM}{Full width at half maximum}
\newacronym{gl:HIT}{HIT}{Heidelberger Ionenstrahl-Therapiezentrum}
\newacronym{gl:PMMA}{PMMA}{Poly(methyl methacrylate)}
\newacronym{gl:MRI}{MRI}{Magnetic resonance imaging}
\newacronym{gl:PDPC}{PDPC}{Philips Digital Photon Counting}
\newacronym{gl:DPC}{DPC}{Digital photon counter}
\newacronym{gl:FPGA}{FPGA}{Field-programmable gate array}
\newacronym{gl:DAPS}{DAPS}{Data acquisition and processing server}
\newacronym{gl:SPU}{SPU}{Singles processing unit}
\newacronym{gl:ROI}{ROI}{Region of interest}
\newacronym{gl:COG}{COG}{Center of gravity}
\newacronym{gl:HVD}{HVD}{Horizontal vertical diagonal}
\newacronym{gl:LYSO:Ce}{\ce{LYSO}:\ce{Ce}}{Lutetium Yttrium Orthosilicate (\ce{Lu3Al5O12}:\ce{Ce})}
\newacronym{gl:LuAG:Ce}{\ce{LuAG}:\ce{Ce}}{Lutetium aluminum garnet doped with cerium (\ce{Lu2Al5O12}:\ce{Ce})}
\newacronym{gl:GAGG:Ce}{\ce{GAGG}:\ce{Ce}}{Gadolinium aluminum gallium garnet doped with cerium (\ce{Gd3Al2Ga3O12}:\ce{Ce})}
\newacronym{gl:GAGG:Ce,Mg}{\ce{GAGG}:\ce{Ce},\ce{Mg}}{Gadolinium aluminum gallium garnet doped with cerium and magnesium (\ce{Gd3Al2Ga3O12}:\ce{Ce},\ce{Mg})}
\newacronym{gl:MLEM}{MLEM}{Maximum-Likelihood Expectation-Maximization}
\newacronym{gl:CCB}{CCB}{Centrum Cyklotronowe Bronowice (Bronowice Cyclotron Center)}
\newacronym{gl:FOV}{FOV}{Field of view}
\newacronym{gl:MURA}{MURA}{Modified uniform redundant array}
\newacronym{gl:BP}{BP}{Bragg peak}
\newacronym{gl:SOBP}{SOBP}{Spread-out Bragg peak}
\newacronym{gl:RBE}{RBE}{Relative biological effectiveness}
\newacronym{gl:LET}{LET}{Linear energy transfer}
\newacronym{gl:IMXT}{IMXT}{Intensity modulated X-ray therapy}
\newacronym{gl:EBRT}{EBRT}{External beam radiotherapy}
\newacronym{gl:INSIDE}{INSIDE}{Innovative Solution for In-beam Dosimetry}
\newacronym{gl:ASIC}{ASIC}{Application-specific integrated circuit}

\newglossaryentry{gl:peakpos}{
name=\peakpos,
description={Position of the \anhpeak peak}}

\newglossaryentry{gl:Zeff}{
name=\Zeff,
description={Effective atomic number}}

\newglossaryentry{gl:n}{
name=$n$,
description={Refractive index}}

\newglossaryentry{gl:MLR}{
name=\MLR,
description={Quantity combining the correlated events recorded at both ends of the investigated scintillating fiber, according to \cref{eq:MLR}}
}

\newglossaryentry{gl:MLRstar}{
name=\MLRstar,
description={Quantity combining the primary signal components of the correlated signals recorded at both ends of the investigated fiber, calculated based on the ELAR parameterization of the experimental data, defined as \cref{eq:MLR-corrected}}
}

\newglossaryentry{gl:Qavg}{
name=\Qavg,
description={Geometric mean of the charges of the correlated events recorded at both ends of the investigated fiber, defined as \cref{eq:qavg}}
}

\newglossaryentry{gl:Qavgstar}{
name=\Qavgstar,
description={Geometric mean of the primary signal components of the correlated events recorded at both ends of the investigated fiber, calculated using the ELAR parameterization of the experimental data, defined as \cref{eq:qavg-corrected}}
}

\newglossaryentry{gl:tzero}{
name=\tzero,
description={Time at which signal crosses set threshold value. It defines the beginning of the signal duration}
}

\newglossaryentry{gl:Xreal}{
name=$X_{\textrm{real}}$,
description={Known position of the \Na source along the investigated fiber or the prototype, determined by the electronic collimation system}
}

\newglossaryentry{gl:Xreco}{
name=$X_{\textrm{reco}}$,
description={Reconstructed position of the interaction along the investigated fiber or the prototype}
}

\usepackage[
 backend=biber,
 style=science,
 citestyle=numeric-comp,
 defernumbers,
 doi=true,
 sorting=none,
 maxbibnames=10,
 url=true,
 isbn=false
]{biblatex}
\begin{document}

{
\thispagestyle{empty}
\newgeometry{%
  top=1in,%
  bottom=1in,%
  left=1.0in,%
  right=1.0in,%
  hmarginratio=2:1%
}


\begin{titlepage}

\begin{center}
\selectlanguage{polish}
\Large{\bf Jagiellonian University in Krak\'ow}

\Large{Faculty of Physics, Astronomy and Applied Computer Science}

\begin{figure}[htbp]
\centering
\includegraphics[width=.2\textwidth]{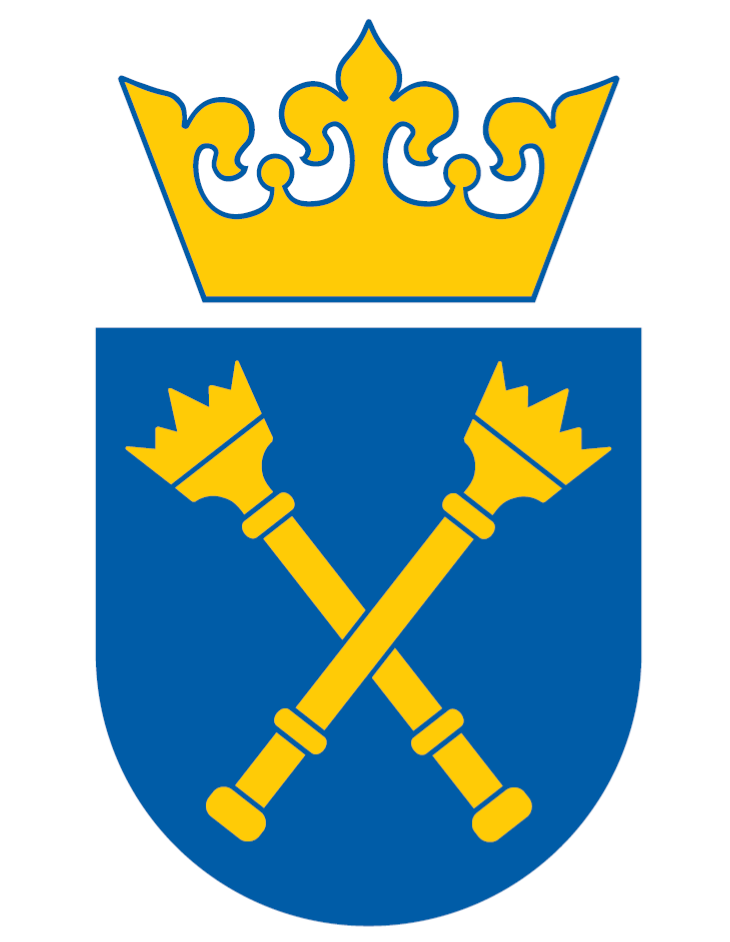}
\end{figure}

\large{Doctoral dissertation}

\vspace{1,0cm}

{\large \textbf{\textsc{The SiFi-CC detector for beam range monitoring in proton therapy - characterization of components and a prototype detector module}}}\\

\vspace{1,0cm}

\large{mgr Katarzyna Rusiecka} \\

\vspace{2cm}

\large{
Supervisor: prof. dr hab. Andrzej Magiera \\
Auxiliary supervisor: dr Aleksandra Wro\'nska}

\vspace{2cm}

\large{
Marian Smoluchowski Institute of Physics \\
Department of Hadron Physics
}

\vspace{3cm}

\selectlanguage{polish}
Krak\'ow, 2023
\end{center}

\end{titlepage}

\blankpage 

}


\begin{titlepage}

\begin{center}
\selectlanguage{polish}
\Large{\bf Uniwersytet Jagiello\'nski w Krakowie}

\Large{Wydzia\l{} Fizyki, Astronomii i Informatyki Stosowanej}

\begin{figure}[htbp]
\centering
\includegraphics[width=.2\textwidth]{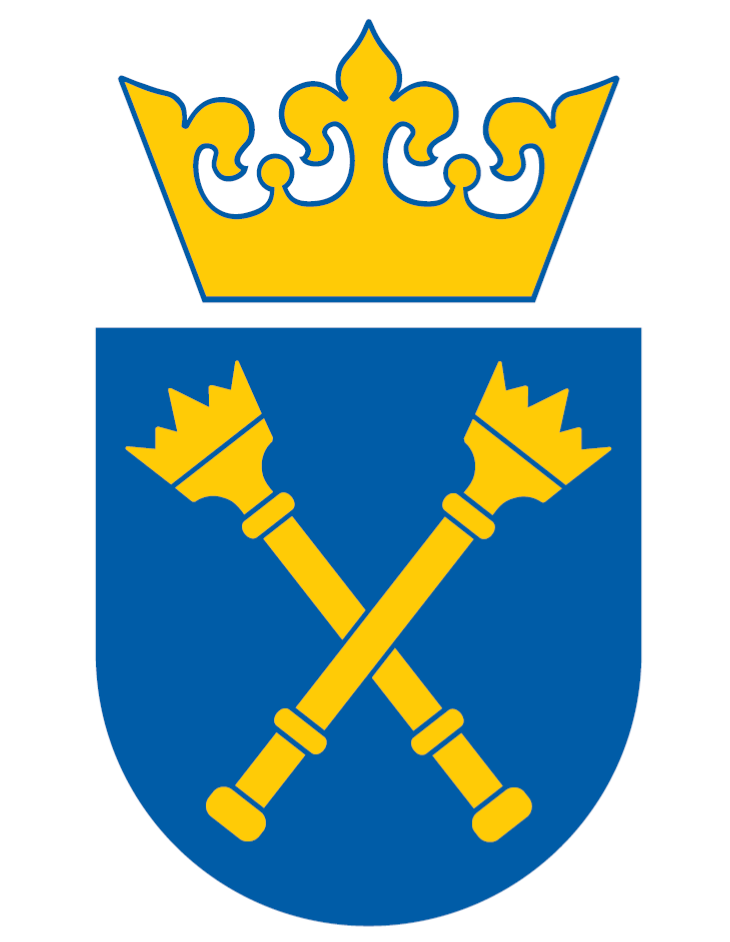}
\end{figure}

\large{Rozprawa doktorska}

\vspace{1,0cm}

{\large \textbf{\textsc{Detektor SiFi-CC do monitorowania zasi\k{e}gu wi\k{a}zki w terapii protonowej - charakterystyka komponent\'ow i prototypu modu\l{}u detektora}}}\\

\vspace{1,0cm}

\large{mgr Katarzyna Rusiecka} \\

\vspace{2cm}

\large{
Promotor: prof. dr hab. Andrzej Magiera \\
Promotor pomocniczy: dr Aleksandra Wro\'nska}

\vspace{2cm}

\large{
Instytut Fizyki im. Mariana Smoluchowskiego \\
Zak\l{}ad Fizyki Hadron\'ow
}

\vspace{3cm}

\selectlanguage{polish}
Krak\'ow, 2023
\end{center}

\end{titlepage}

\blankpage 


\frontmatter


\thispagestyle{empty}
\selectlanguage{polish}

\begin{center}
Wydzia\l{} Fizyki, Astronomii i Informatyki Stosowanej \\
Uniwersytet Jagiello\'nski

\vspace{2cm}

{\large\bf{O\'swiadczenie}}
\end{center}

Ja ni\.zej podpisana Katarzyna Rusiecka (nr indeksu: 1078159) doktorantka Wydzia\l{}u Fizyki, Astronomii i Informatyki Stosowanej Uniwersytetu Jagiello\'nskiego o\'swiadczam, \.ze przed\l{}o\.zona przeze mnie rozprawa doktorska pt. ''The SiFi-CC detector for beam range monitoring in proton therapy - characterization of the components and a prototype detector module'' (''Detektor SiFi-CC do monitorowania zasi\k{e}gu wi\k{a}zki w terapii protonowej - charakterystyka komponent\'ow i prototypu modu\l{}u detektora'') jest oryginalna i przedstawia wyniki bada\'n wykonanych przeze mnie osobi\'scie, pod kierunkiem prof.~dr~hab. Andrzeja Magiery oraz dr Aleksandry Wro\'nskiej. Prac\k{e} napisa\l{}am samodzielnie.

O\'swiadczam, \.ze moja rozprawa doktorska zosta\l{}a opracowana zgodnie z Ustaw\k{a} o prawie autorskim i prawach pokrewnych z dnia 4 lutego 1994 r. (Dziennik Ustaw 1994 nr 24 poz. 83 wraz z p\'o\'zniejszymi zmianami).

Jestem \'swiadoma, \.ze niezgodno\'s\'c niniejszego o\'swiadczenia z prawd\k{a} ujawniona w dowolnym czasie, niezale\.znie od skutk\'ow prawnych wynikaj\k{a}cych z ww. ustawy, mo\.ze spowodowa\'c uniewa\.znienie stopnia nabytego na podstawie tej rozprawy.

\vspace{1cm}

\begin{table}[htbp]
\centering
\begin{tabularx}{1.0\textwidth} { 
      >{\raggedright\arraybackslash}X 
      >{\raggedleft\arraybackslash}X }
Krak\'ow, dnia ................................... & ............................................................... \\
{} & \emph{podpis doktorantki}
\end{tabularx}
\end{table}

\newpage

\blankpage

\selectlanguage{english}

\chapter*{\centering{Abstract}}

The following thesis presents research which constitutes the first steps towards the construction of a novel \acrshort{gl:SiFi-CC} detector intended for real-time monitoring of proton therapy. The detector construction will be based on inorganic scintillating fibers and silicon photomultipliers. The scope of the presented thesis includes the design optimization of the components of the proposed detector, as well as the construction, characterization, and tests of a prototype. 

The design optimization comprised an extensive systematic comparison of chosen inorganic scintillating materials, different types of scintillator surface modifications (wrappings and coatings), and different types of interface materials ensuring optical contact between the scintillators and the photodetector. The propagation of scintillating light in all investigated samples was described using two models: the exponential light attenuation model (\acrshort{gl:ELA}), and the exponential light attenuation model with light reflection (\acrshort{gl:ELAR}). The two models yielded the corresponding methods for energy and position reconstruction. Furthermore, the samples were investigated for energy and position resolution, light collection, and timing properties.  

Based on the results obtained from the optimization study, the detector prototype was constructed. Prototype tests were performed with two different photodetectors and data acquisition systems. The performance of the prototype was evaluated using the same metrics as in the case of single-fiber measurements. The best results were obtained in measurements with Philips Digital Photon Counting photosensor and the Hyperion platform, yielding a position resolution of \SI{33.38}{\milli\meter} and an energy resolution of \SI{7.73}{\percent}. The results obtained are satisfactory and sufficient for the successful operation of the proposed \acrshort{gl:SiFi-CC} detector. 

\vspace{1cm}
\noindent
\textbf{Keywords:} real-time monitoring of proton therapy, range verification, scintillators, silicon photomultipliers, scintillating detectors, detector construction and optimization.


\blankpage

\selectlanguage{polish}

\chapter*{\centering{Streszczenie}}

Przedstawiona praca doktorska prezentuje badania stanowiące pierwsze kroki w kierunku zbudowania nowatorskiego detektora \acrshort{gl:SiFi-CC} do monitorowania terapii protonowej w czasie rzeczywistym. Konstrukcja proponowanego detektora ma opierać się na nieorganicznych włóknach scyntylacyjnych oraz fotopowielaczach krzemowych. Zakres prezentowanej pracy zawiera optymalizację komponentów detektora oraz budowę i testy prototypu. 

Optymalizacja detektora obejmowała obszerne systematyczne porównanie wybranych materiałów scyntylacyjnych, różnych typów modyfikacji powierzchni scyntylatora oraz różnych materiałów zapewniających kontakt optyczny pomiędzy scyntylatorem i fotosensorem. Do opisania propagacji światła scyntylacyjnego w badanych próbkach wykorzystano dwa modele: eksponencjalny model tłumienia światła (\acrshort{gl:ELA}) oraz eksponencjalny model tłumienia światła z uwzględnieniem odbicia (\acrshort{gl:ELAR}). Z powyższych modeli wynikają odpowiadające metody rekonstrukcji pozycji interakcji i depozytu energii w scyntylatorze.  Badane próbki były ponadto scharakteryzowane pod kątem pozycyjnej i energetycznej zdolności rozdzielczej, uzysku światła oraz własności czasowych. 

W oparciu o uzyskane wyniki optymalizacji zbudowano prototyp detektora. Prototyp ten został przetestowany z dwoma różnymi fotosensorami i systemami akwizycji danych. Charakteryzację przeprowadzono w sposób analogiczny jak w przypadku pomiarów z pojedynczymi próbkami włókien scyntylacyjnych, uwzględniając te same własności. Najlepsze wyniki uzyskanow w pomiarach przeprawdzaonych a fotodetektorem Philips Digital Photon Counting oraz platformą Hyperion, t.j. pozycyjną zdolność rozdzielczą \SI{33.38}{\milli\meter} oraz energetyczną zdolność rozdzielczą \SI{7.73}{\percent}. Otrzymane wyniki są satysfakcjonujące i wystarczające do działania przyszłego detektora \acrshort{gl:SiFi-CC}.

\vspace{1cm}
\noindent
\textbf{S\l{}owa kluczowe:} monitorowanie terapii protonowej w czasie rzeczywistym, weryfikacja zasięgu, scyntylatory, fotopowielacze krzemowe, detektory scyntylacyjne, budowa i optymalizacja detektora.


\blankpage
\selectlanguage{english}
\tableofcontents

\clearpage

\mainmatter


\newpage

\cleardoublepage


\chapter{Introduction}
\label{chap:intro}

In this chapter, the motivation for the work presented in this thesis is explained. An overview of proton therapy is presented, including the physical and biological rationale. Subsequently, the need to develop a method for real-time monitoring of the spatial distribution of the dose administered to the patient during therapy is discussed. A brief description of the currently explored solutions is given. Finally, the \acrshort{gl:SiFi-CC} project is introduced, with its proposed approach to develop the real-time monitoring method for proton therapy.

\section{Particle therapy}
\label{sec:proton-therapy}

\subsection{Context}
\label{ssec:broad context}

Approximately 18.1 million people worldwide are estimated to have suffered various types of cancer in 2020 alone. Due to the worsening pollution in the environment and the unhealthy lifestyle, an additional increase in cancer occurrence is predicted, with 28 million new cases each year by 2040 \cite{cancer-statistics, who-cancer}. With these grim statistics, cancer remains one of the most frequently occurring diseases in the human population \cite{who-cancer}. Therefore, constant effort is being made to perfect existing treatment methods and develop new and more effective ones, such as immunotherapy, which was awarded the Nobel Prize in 2018 \cite{Smyth2018}. The most commonly used treatment methods include surgery, chemotherapy, hormone therapy, immunotherapy, targeted therapy, and radiotherapy \cite{american-cancer-soc}. During the surgery, the tumors are removed from the patient's body. It is the most straightforward treatment method, however, it is often not possible to operate on the patient or remove abnormal tissues completely. In chemotherapy, patients are administered the anti-cancer drug. This treatment method can cause serious side effects and is devastating to the human body. The growth of some cancers is driven by hormones. In that case, it is beneficial to inhibit the production of those hormones or alter their operation (hormone therapy). In immunotherapy, the patient's immune system is enhanced or redirected to fight cancer cells. During targeted therapy, patients are given medications that are directed against specific compounds produced by cancer cells which stimulate their division. Finally, during radiation therapy, cancer tissues are irradiated with ionizing radiation to kill abnormal cells. This type of therapy also carries a risk, with the increased probability of secondary cancer later in the patient's life. The chosen type of therapy depends on the type and stage of the disease; often different types of therapy are combined to increase the chances of full recovery \cite{cancer-org}. 

Different types of radiation therapy can be distinguished considering the placement of the used ionizing radiation source relative to the patient. In brachytherapy, the radioactive source is inserted into the patient's body, in the vicinity of the treated abnormal tissues. On the contrary, during external beam radiotherapy (\acrshort{gl:EBRT}) the source of radiation is placed outside of the patient's body. In this case, the radiation can originate from the radioactive source (\eg ${}^{60}$\ce{Co}, ${}^{137}$\ce{Cs}, ${}^{226}$\ce{Ra}), or from the particle accelerator \cite{Podgorsak}. 

Another criterion for distinguishing different types of radiation therapy is the type of ionizing radiation used. Historically, $\gamma$ radiation and X-rays were employed first in cancer treatment. The methods of photon-based radiotherapy range from very straightforward methods, such as using ${}^{60}$\ce{Co} sources for irradiation, to state-of-the-art sophisticated methods, such as intensity modulated X-ray therapy \acrshort{gl:IMXT}. In \acrshort{gl:IMXT} the photon beam
produced with the use of a linear accelerator and shaped by a dedicated leaf-collimator to fit the tumor shape. This helps reduce irradiation of the neighboring healthy tissues. Electrons can also be used for irradiation of cancer tissues. Finally, there is particle therapy, sometimes also called hadron therapy \cite{Podgorsak}, which will be discussed in the next sections.

\subsection{Physical aspects of particle therapy}
\label{ssec:}

Particle therapy uses accelerated beams of ions (\eg \ce{{}^4He}, \ce{{}^{12}C}, \ce{{}^{16}O}) for the irradiation of cancer tissues.  In particular, proton therapy which uses \ce{{}^1H} ions is distinguished. Among all of the ions used in particle therapy, protons are utilized the most frequently. The success of hadron therapy in cancer treatment can be attributed to the specific energy deposition pattern of heavy charged particles in matter. A charged particle penetrating through matter interacts with the medium atoms via electromagnetic and strong interactions. In this section, the interactions will be explained on the example of a proton. 

Matter can be considered as a mixture of free electrons and atomic nuclei. The accelerated proton penetrating through experiences electromagnetic interaction with both of them. However, the effects of the interaction with electrons and nuclei are very different. The mass of most types of nuclei is significantly larger than the mass of a proton. Therefore, if a proton collides with a nucleus, it loses only a small fraction of its kinetic energy. However, as a result of such a collision, the proton trajectory can change significantly. In case of interactions with electrons, the effect is opposite, \ie the energy transfer is large, but the change in proton direction is small. Therefore, interactions with electrons are mostly responsible for the energy loss of the proton, and interactions with nuclei are responsible for the change in its trajectory. On its way in the medium, a charged particle causes excitations and ionizations. Some of the electrons which acquired sufficient energy in the interaction with the proton can travel a small distance in the matter on their own ($\delta$-electrons). They can also cause excitations and ionizations \cite{Tavernier}.  

\begin{figure}[!ht]
    \centering
    \includegraphics{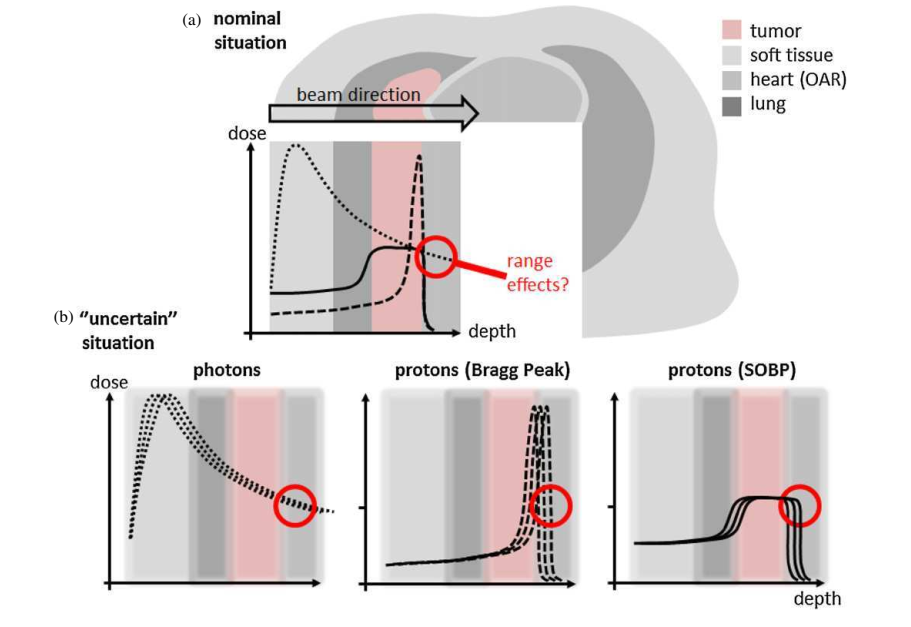}
    \caption{A: dose deposition profiles for different irradiation methods. The dotted line is for photons, the dashed line is for mono-energetic protons and the solid line represents the spread-out Bragg peak (\acrshort{gl:SOBP}). B: influence of uncertainties for the above irradiation methods. Picture reprinted from \cite{Knopf2013}.}
    \label{fig:bragg-peak}
\end{figure}

The mean energy loss of charged, heavy particles due to electromagnetic interactions with electrons in matter per unit length of the track is given by the Bethe-Bloch formula \cite{henley}: 
\begin{equation}
- \bigg \langle \frac{dE}{dx} \bigg \rangle = Kz^2 \frac{Z}{A}\frac{1}{\beta^2} \Bigg[ \frac{1}{2} \ln \frac{2m_\textrm{e} c^2 \beta^2 \gamma^2 T_{\textrm{max}}}{I^2} - \beta^2 - \frac{\delta(\beta \gamma)}{2} \Bigg]\ .
\label{eq:bethe}
\end{equation}
All symbols from the equation are explained in \cref{tab:variables}. From the Bethe-Bloch equation it can be seen that the energy loss increases with larger charge of incident particle $z$, and decreases with the relative velocity of incident particle $\beta$. Therefore, as the proton penetrates through matter, the energy loss per unit length changes. For fast, high-energetic protons, the energy loss is relatively constant and starts increasing with decreasing kinetic energy. When the proton reaches the velocity comparable with the velocity of electrons in atoms, the energy loss reaches a sharp maximum called the Bragg peak (\acrshort{gl:BP}). At this point, the Bethe-Bloch formula is no longer valid. After reaching the Bragg peak, the energy loss drops almost immediately to zero, which means that the particle stops completely. This point defines the range of the particle in the medium \cite{Tavernier}. The dependence of the mean energy loss per unit track as a function of the penetration depth for monoenergetic protons  is called the Bragg curve and is illustrated in \cref{fig:bragg-peak} (a) with the dashed line.   

\begin{table}[!ht]
\centering
\caption{Description of variables used in \cref{eq:bethe} \cite{henley}.}
\begin{tabular}{p{1.5cm}|p{9cm}|p{5cm}}
Variable & Definition & Value or units \\ \hline
$K$ & $4 \pi N_\textrm{A} r_\textrm{e}^2 m_\textrm{e} c^2$ & \SI{0.307075}{\mega\electronvolt\per\mol\centi\meter\squared} \\
$r_\textrm{e} $ & classical electron radius & \SI{2.81794}{\femto\meter} \\ 
$c$ & speed of light & \SI{299.792}{\kilo\meter\per\second} \\
$N_\textrm{A}$ & Avogadro's number & \SI{6.022e23}{\per\mole} \\
$z$ & charge number of incident particle & {} \\
$Z$ & atomic number of medium & {} \\
$A$ & atomic mass of medium & \si{\gram\per\mol} \\
$\beta$ & speed relative to $c$ & {} \\
$m_\textrm{e} c^2$ & electron mass $\times$ $c^2$ & \SI{511}{\kilo\electronvolt} \\
$\gamma$ & Lorentz factor & {} \\
$T_{\textrm{max}}$ & maximum energy transfer to a free electron in a single collision &  \si{\electronvolt} \\
$I$ & mean excitation energy & \si{\electronvolt} \\
$\delta(\beta \gamma)$ & density effect correction to ionization energy loss & {} \\
\end{tabular}
\label{tab:variables}
\end{table}

The presence of the sharp maximum in the Bragg curve is favorable for irradiation of abnormal tissues. The position of the Bragg peak in the medium can be tuned with the initial energy of the impinging particles. Therefore, it is possible to target specific areas in the human body. The energy deposition pattern ensures the deposition of maximum dose in the desired area. At the same time, healthy tissues upstream of the targeted area receive relatively small radiation dose and tissues located deeper than the targeted area receive no dose at all. Conformal irradiation of the full volume of the tumor can be achieved by the modulation of the broadened proton beam with custom collimators and range compensators, which results in the creation of the so-called spread-out Bragg peak (\acrshort{gl:SOBP}), as shown in \cref{fig:bragg-peak} (b). Alternatively, the tumor volume can be scanned point-by-point using monoenergetic pencil proton beams (active scanning) \cite{Knopf2013}.

\Cref{fig:bragg-peak}~(a) shows the comparison of the energy deposition curves for protons and photons, which are also used in radiotherapy. Photons interact with matter via completely different mechanisms, mostly in the photoelectric effect, Compton effect, and pair creation. Therefore, the energy deposition pattern for photons is also completely different compared to protons. The curve for photons reaches its maximum relatively shallow in matter, which is followed by an exponential decrease. To irradiate the abnormal tissues with the sufficient dose the patient is often irradiated with multiple photon beams from different directions. Even in the most advanced and conformal photon radiotherapy technique \acrshort{gl:IMXT}, the irradiation results in a significant dose delivered to surrounding healthy tissues. On the other hand, as shown in \cref{fig:bragg-peak}~(b), the energy deposition pattern of protons with its sharp Bragg peak makes proton therapy particularly sensitive to range uncertainties. Even small range shifts in the order of millimeters can cause complications in treatment. At the same time, photon-based radiotherapy is insensitive to range shifts during treatment \cite{Knopf2013}.

Other electromagnetic interactions of charged particles with matter are the Cherenkov effect, emission of transition radiation, and Bremsstrahlung; however, their significance in proton therapy is negligible \cite{Tavernier}. 

At higher energies of impinging particles, in the order of several \si{\mega\electronvolt}, the strong interactions also start to play a significant role. 
If a particle has sufficient energy, nuclear reactions can occur. 
For low-energy protons (several hundred \si{\kilo\electronvolt}), the cross section for nuclear reactions is very small, as the electrostatic repulsion between the positively charged proton and the atomic nucleus prevents the two to get close enough for the strong interaction to become dominant. Therefore, nuclear reactions occur along the particle path before the particle has a chance to reach the Bragg peak. As a result of a nuclear reaction, a new nucleus is created, usually with the emission of a light fragment (\eg proton, neutron or $\alpha$ particle) or emission of $\gamma$ quanta \cite{Tavernier}.



\subsection{Radiobiological aspects of particle therapy}
\label{ssec:radiobiology}

As previously stated, the accelerated charged particles penetrating the matter cause ionizations of atoms and excitations of atoms and nuclei along their tracks. As a result, radicals are created that are chemically active and can induce a chain of chemical reactions. The radicals can damage the DNA molecules leading to cell necrosis or stopping its further division. The ionization and excitation can also occur directly in the DNA molecules, leading to single-strand breaks and double-strand breaks. To some extent, cells are capable of repairing their damaged DNA. However, if the number of strand impairments is large and concentrated, the cell undergoes necrosis or is prevented from proliferation. Double-strand breaks are particularly difficult to fix, and they are the most damaging to cells. The aim of radiotherapy is to damage the DNA of cancer cells and thus cause a decrease in their number while sparing normal cells \cite{encyclopedia-can}.  

For radiobiological studies, a linear energy transfer (\acrshort{gl:LET}) was introduced. It is defined as the energy loss per unit track length of the impinging particle in the closest vicinity of the trajectory. Therefore, the $\delta$-electrons created along the way do not contribute to the \acrshort{gl:LET} \cite{radiology-radiobiol, Tavernier}. The \acrshort{gl:LET} is higher for lower energies of impinging particles, \ie for charged particles, in the region of the Bragg peak. Consequently, a high \acrshort{gl:LET} implies a greater number of occurring interactions that damage the DNA of cells. The typical \acrshort{gl:LET} of $\gamma$ radiation emitted from ${}^{60}$\ce{Co} radioactive source is \SI{0.2}{\kilo\electronvolt\per\micro\meter}. The typical \acrshort{gl:LET} for protons ranges from \SI{0.5}{\kilo\electronvolt\per\micro\meter} for \SI{150}{\mega\electronvolt} (beginning of the Bragg curve) to \SI{4.7}{\kilo\electronvolt\per\micro\meter } for \SI{10}{\mega\electronvolt} (close to the Bragg peak) \cite{radiology-radiobiol}. The values listed indicate that protons are more effective in radiation therapy than photons. 

Another parameter introduced to quantify the effects of irradiation in radiotherapy is the relative biological effectiveness (\acrshort{gl:RBE}). It helps to compare the biological effects of two types of radiation. It is calculated as the ratio of the reference radiation dose and dose of the radiation of interest required to produce the same biological effects \cite{radiology-radiobiol}. Usually, the $\gamma$ radiation of ${}^{60}$\ce{Co} source is used as a reference. Typically, the value of \acrshort{gl:RBE} increases with growing \acrshort{gl:LET}. However, the \acrshort{gl:RBE} is a more complex parameter, as it also takes into account many additional factors, such as dose rate, number of radiotherapy fractions and dose administered per fraction, the oxygen concentration in cells, and cell-cycle phase. Therefore, two different types of radiation of the same \acrshort{gl:LET} can result in different \acrshort{gl:RBE} values. The \acrshort{gl:RBE} for protons determined in \emph{in vitro} and \emph{in vivo} experiments is 1 -- 1.1 \cite{Jiang2012}.  

The \acrshort{gl:RBE} of carbon ions, the second most common ion species used in particle therapy, differs strongly depending on the depth in tissue. At small depths, it ranges from \num{1.06} to \num{1.32}, and in the Bragg peak region, it ranges from \num{2.0} to \num{3.01} \cite{Karger2017}. Therefore, based on radiobiological premises, the effectiveness of carbon ion therapy is expected to be superior to that of proton therapy \cite{Jiang2012}. However, to date there is no strong evidence for a significant advantage of carbon versus proton therapy in clinical practice. Comparison is difficult to make, due to different dose and fractionation regimes and different methods for \acrshort{gl:RBE} calculation. Furthermore, data on carbon ion treatment are limited, due to the small number of facilities offering this type of treatment \cite{Balosso2022, Weber2009, Uhl2014}. The lack of clear evidence for the superiority of carbon ion therapy over proton therapy and the significantly higher cost of construction and maintenance of adequate facilities are the main reasons why proton therapy remains the most popular type of particle therapy. 

\subsection{History and current status}
\label{ssec:history}

The use of protons for cancer treatment was first proposed by Robert Wilson in 1946 \cite{Wilson1946}. This idea, combined with the earlier invention of the cyclotron by Ernest O. Lawrence, led to the first treatment of a patient in 1954 at the University of California (Berkeley, USA) \cite{Lawrence1958}. Within three years, the facility was adapted to accelerate helium ions. Concurrently, another proton therapy center was being prepared in Uppsala, Sweden, where the first patient was treated in 1957. In the following several years, ten more proton therapy facilities were brought into operation worldwide. All of them were operating in physics research centers where cyclotrons were present. However, they did not have the medical infrastructure for patient care \cite{Giap2012}.  

The main advancements made in the 1970s and 1980s included the construction of early synchrotrons, allowing the acceleration of ions of larger masses, as well as the development of computed tomography (\acrshort{gl:CT}) and magnetic resonance imaging (\acrshort{gl:MRI}). The latter two improved not only the diagnostic capabilities but also the planning of proton therapy. At the same time, intensive research was conducted on different types of cancer and radiobiological aspects of irradiation \cite{Giap2012}.  

In 1990, the first proton therapy facility located at the hospital was launched in Loma Linda (California, USA). It helped to establish proton therapy as one of the standard radiotherapy methods rather than an experimental procedure. At that time, private companies had begun production of off-the-shelf commercial solutions for proton therapy facilities. This triggered the rapid growth of a number of new clinical proton therapy centers around the world \cite{Giap2012}. 

Today, proton therapy is a well-established and widespread radiotherapy modality with 123 particle therapy facilities currently in operation and another 38 under construction (status in May 2023). Until the end of 2022, over \num{360000} patients worldwide were treated with particle therapy, out of which approximately \num{312000} with proton therapy \cite{PTCOG}. 

\section{Monitoring of proton therapy}
\label{sec:pt-monitoring}

Before the first irradiation of a patient, a treatment plan must be prepared. It is a simulation of the dose distribution delivered to the tumor and surrounding tissues as a result of the irradiation. It helps to optimize the procedure taking into account local control of the disease and toxicity of treatment \cite{nupecc}. Since proton therapy is particularly sensitive to range shifts, as explained in the previous section, it is necessary to impose so-called safety margins in treatment plans, \ie additional volume around the tumor that is added to the target volume. The safety margins ensure that the entire volume of the abnormal tissue will be irradiated. This increases the chances of eliminating all cancer cells even if small unintentional shifts of the \acrshort{gl:BP} occur during irradiation. Possible sources of beam range uncertainties and errors comprise an uncertainties in the translation of the \acrshort{gl:CT} images to the maps of stopping power for protons, incorrect data transfer from the treatment planning system to the delivery device, constant anatomical changes occurring in the patient's body, limited precision of patient positioning, and human errors \cite{wronska-hal, Knopf2013}. The size of the safety margins depends on the type and location of the tumor. Each particle therapy facility has its own procedure for determination of safety margins, which is based on their research and experience. Safety margins range from a few millimeters up to over a centimeter for deep-located lesions. Additional irradiated volume increases the possibility of long-term side effects and short-term toxicity. The possibility of monitoring the distribution of the dose administered to the patient during treatment in real time would allow for a reduction of the safety margins. Consequently, the quality of treatment would improve with a reduced probability of side effects and more conformal irradiation \cite{wronska-hal}.  

The need for such a real-time method of particle therapy monitoring was emphasized in the report of the Nuclear Physics European Collaboration Committee in 2014 \cite{nupecc}. To date, several approaches have been proposed by different research groups. All utilize one of the by-products of irradiation, such as secondary ions, prompt $\gamma$ radiation or $\beta^+$ emitters \cite{nupecc}.  

Secondary ions, such as protons, $\alpha$ particles, or other light fragments, are emitted as a result of nuclear reactions that occur during irradiation. They are produced in all types of particle therapy; however, only in heavy-ion therapy secondary fragments produced have sufficient energy to penetrate outside the patient tissues. These high-energy particles come from projectile fragmentation. Furthermore, in the case of heavy-ion therapy, the efficiency of other currently developed monitoring methods is decreased due to a significant background of neutrons and uncorrelated $\gamma$ radiation. A scintillating tracker is currently being tested for the monitoring of particle therapy using secondary ions at the Centro Nazionale per l'Adroterapia Oncologica (CNAO, Pavia, Italy) as part of the monitoring system \acrshort{gl:INSIDE} (Innovative Solution for In-beam Dosimetry) \cite{Traini2019}.

As described in the previous section, other important by-products of particle therapy are $\beta^+$ emitters. The distribution of produced $\beta^+$ emitters can be determined by position emission tomography (\acrshort{gl:PET}). For many years now, \acrshort{gl:PET} has been a well-established and widely used imaging technique in nuclear medicine. Until recently, it was only considered for post-irradiation treatment control. It was related to the spatial incompatibility of \acrshort{gl:PET} scanners and sophisticated gantries for particle therapy, both of which require full solid-angle access to the patient. Moreover, a relatively long half-life (from several \si{seconds} to several \si{minutes}) of nuclei of interest results in long acquisition times required to collect a sufficient number of events for image reconstruction. During the elongated acquisition time, the biological washout effects become significant and cause deterioration of the reconstructed images \cite{wronska-hal}. Therefore, it was proposed to focus on short-lived $\beta+$ emitters, with half-lives in the order of several \si{\milli\second} to a few \si{\second} \cite{Dendooven2015}. The mentioned above \acrshort{gl:INSIDE} system, besides the secondary ion tracker, also contains a high-acceptance and high-efficiency in-beam \acrshort{gl:PET} scanner. The first clinical tests were performed in 2018 and showed very promising results with a range control sensitivity of \SI{1}{\milli\meter} \cite{Ferrero2018}.  

Another important by-product of patient irradiation is prompt-gamma (\acrshort{gl:PG}) radiation. The $\gamma$ spectrum registered during particle therapy has two components: a continuum component and discrete peaks. \acrshort{gl:PG} radiation is emitted on the \si{ps} time scale after nuclear interaction. The energy reaches up to \SIrange{7}{8}{\mega\electronvolt}, which is sufficient for $\gamma$ radiation to escape the tissues of the patient without much disturbance. The continuum component of \acrshort{gl:PG} radiation is difficult to resolve or discriminate from the inevitable neutron background. Therefore, the attention is mainly focused on the discrete peaks. They result from deexcitations of nuclei excited by impinging particles \cite{wronska-hal}. In the case of proton therapy, there are two main $\gamma$ lines recognized as potentially useful for real-time monitoring: at \SI{4.44}{\mega\electronvolt} and \SI{6.13}{\mega\electronvolt}. They originate from the following nuclear interactions: ${}^{12}$C($p$, $p'$$\gamma_{\SI{4.44}{\mega\electronvolt}}$)${}^{12}$C, ${}^{16}$O($p$, $X\gamma_{\SI{4.44}{\mega\electronvolt}}$)${}^{12}$C and ${}^{16}$O($p$, $p'\gamma_{\SI{6.13}{\mega\electronvolt}}$)${}^{16}$O. Due to large cross sections for these processes, reaching their maxima at small proton energies, \ie, close to the Bragg peak, as well as large abundance of ${}^{12}$C and ${}^{16}$O nuclei in human tissues, the resulting $\gamma$ lines are dominant in the \acrshort{gl:PG} energy spectrum \cite{Kelleter2017}. 

In 2006, it was experimentally shown for the first time that there is a correlation between the \acrshort{gl:PG} radiation yield and the position of the Bragg peak in matter \cite{Min2006}. This ultimately proved that \acrshort{gl:PG} radiation has the potential to be used for real-time monitoring of particle therapy. It triggered intensive research to characterize the timing, energetic and spatial features of the \acrshort{gl:PG} radiation \cite{Polf2013, Pinto2015, Kelleter2017}. Currently, many research groups around the world are developing methods and devices that utilize \acrshort{gl:PG} radiation with future applications in particle therapy facilities in mind. The variety of the proposed methods can be divided into three main categories: prompt gamma imaging, prompt gamma timing and prompt gamma spectroscopy \cite{wronska-hal}. In prompt gamma imaging, the spatial distribution of \acrshort{gl:PG} vertices is reconstructed. The devices operating on that principle include knife-edge slit cameras \cite{Richter2016, Ready2016, Xie2017}, multi-slit cameras \cite{Smeets2016, Park2019} or Compton cameras (see \cref{sec:sifi-cc}). The prompt gamma timing approach is based on the idea that the particles' transit time is dependent on their range in matter. It should be noted that the \acrshort{gl:PG} radiation can only be produced before the particles come to a stop, and the lifetimes of the excited states created in collisions are negligible compared with the transition times of the particles. The reference for the time measurements is provided by a  beam tagging detector placed before the entrance of the beam into the patient, necessary in this method. The registered time-of-flight distribution can also be correlated with the \acrshort{gl:BP} position \cite{Golnik2014, Testa2014, Hueso-Gonzalez2015, Werner2019, Marcatili2020}.
The timing information can be additionally enriched with spectroscopic information when only events corresponding to specific discrete transitions are taken into account (prompt gamma integrals method) \cite{Krimmer2017}. Finally, in prompt gamma spectroscopy, it is required to register a spectroscopic-quality \acrshort{gl:PG} energy spectrum from a small volume of irradiated matter preceding the expected \acrshort{gl:BP} position. Then the spectral analysis focuses on chosen discrete transitions. The cross sections for various interactions resulting in \acrshort{gl:PG} emission are energy dependent. Moreover, the \acrshort{gl:PG} yields of different transitions are different. Therefore, knowing the ratios of yields corresponding to chosen transitions it is possible to calculate the residual energy of impinging particles and thus deduce their residual range in the material \cite{Verburg2014, Hueso-Gonzalez2018, DalBello2020}.

Various alternative approaches are also investigated, including the detection of secondary electron Bremsstrahlung \cite{Yamaguchi2016_2}, secondary neutrons \cite{Marafini2017, Lerendegui-Marco2022} or acoustic waves \cite{Parodi2015}. However, those research directions are less mainstream in the field of particle therapy monitoring. 

\section{The SiFi-CC project}
\label{sec:sifi-cc}

The \acrshort{gl:SiFi-CC} collaboration was established in 2016. The name stands for Scintillating Fiber and Silicon Photomultiplier-based Compton Camera. Currently, it associates scientists from the Jagiellonian University (Krak\'ow, Poland), the RWTH Aachen University (Aachen, Germany), and the University of L\"ubeck (L\"ubeck, Germany). The group evolved from the \emph{gammaCCB} collaboration (2012--2016), which was investigating \acrshort{gl:PG} emission in proton therapy and its possible usefulness for real-time range monitoring. The group performed several measurements with proton beams and tissue-like phantoms at \acrshort{gl:CCB} (Krak\'ow, Poland) and \acrshort{gl:HIT} (Heidelberg, Germany) therapy centers. The obtained results showed a clear correlation between the expected position of the Bragg peak in the target and the yield of the \acrshort{gl:PG} radiation. The conclusions of those experiments were published in scientific journals \cite{Wronska2015, Rusiecka2016, Kelleter2017, Wronska2017, Rusiecka2018, Wronska2021}\footnote{The author of this thesis has been a member of the \emph{gammaCCB} and \acrshort{gl:SiFi-CC} collaborations since 2014.}.

As a natural next step, the group went on to explore the possibility of using \acrshort{gl:PG} radiation for real-time monitoring of the dose distribution delivered during proton therapy. Therefore, the aim of the collaboration is to develop appropriate methods and algorithms and construct a dedicated detection setup for that purpose. Since the group consists of not only medical physicists but also experimental nuclear physicists, we apply the technologies which are commonly used and well understood in nuclear and high energy physics (\acrshort{gl:HEP}) in the design of a medical imaging device. The proposed detection setup for real-time monitoring of proton therapy will operate in two modes: as a Compton camera (\acrshort{gl:CC}) and as a coded mask (\acrshort{gl:CM}) \cite{Kasper2020}.

\subsubsection*{Compton camera}

\begin{figure}[ht]
\centering
\includegraphics[width=0.8\textwidth]{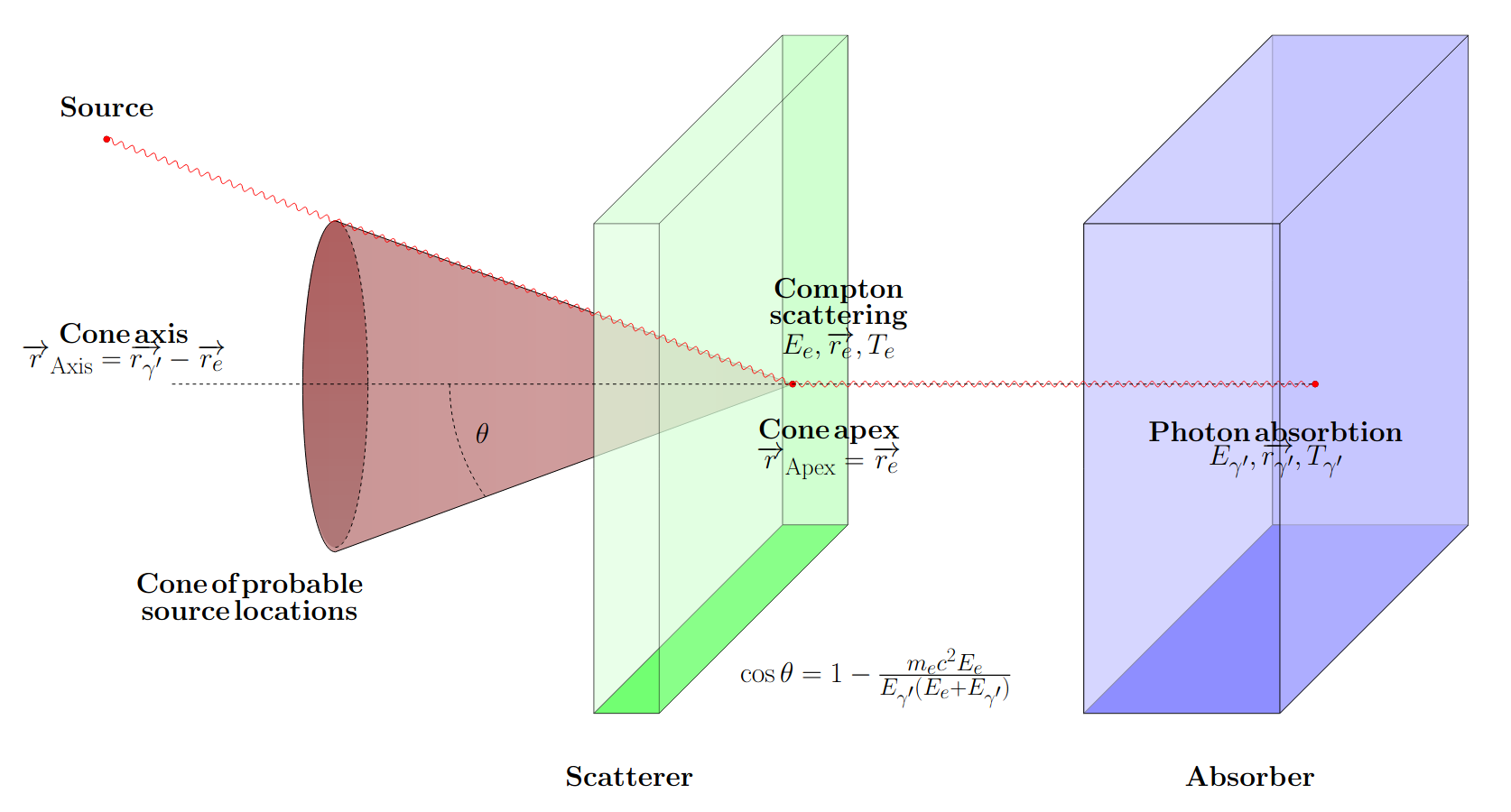}
\caption{Scheme showing the principle of operation of a Compton camera. Reprinted from \cite{Kasper2022}.}
\label{fig:cc-operation}
\end{figure}

Compton camera takes advantage of the kinematics of the Compton effect to reconstruct a spatial source distribution of $\gamma$ radiation. It usually consists of two modules: a scatterer and an absorber, as shown in \cref{fig:cc-operation}. In the ideal scenario, the incoming $\gamma$ radiation first reaches the scatterer where it undergoes Compton scattering. The scattered $\gamma$ subsequently reaches the absorber where it is completely absorbed. Assuming this ideal Compton event topology, the two energy deposits in the two detector parts ($E_{\textrm{e}}$ and $E_{\gamma'}$, as shown in \cref{fig:cc-operation}) sum up to the energy $E$ of the incoming $\gamma$:
\begin{equation}
\label{eq:gamma-sum}
E = E_{\textrm{e}} + E_{\gamma'} \ .
\end{equation}
Knowing the energy deposits and positions of interactions in both detector modules it is possible to calculate the Compton scattering angle $\theta$ as follows \cite{Parajuli2022}:
\begin{equation}
\label{eq:Compton-angle}
\cos(\theta) = 1 - \frac{m_{\textrm{e}} c^2 E_{\textrm{e}}}{E_{\gamma'} (E_{\textrm{e}} + E_{\gamma'})} \ ,
\end{equation}
where $ m_{\textrm{e}}c^2$ is the electron rest mass. Having the Compton scattering angle, the origin of the incoming $\gamma$ can be restricted to the surface of a so-called Compton cone. The apex of the Compton cone is the interaction point in the scatterer ($\vec{r}_{\textrm{e}} = \vec{r}_{\textrm{Apex}}$) and the cone axis is calculated as the line connecting the interaction points in both modules ($\vec{r}_{\textrm{Axis}} = \vec{r}_{\gamma'} - \vec{r}_{\textrm{e}}$). Reconstruction and superposition of many Compton cones results in a 3-dimensional image of the radiation source distribution, which is a big advantage of this detector type. Moreover, \acrshort{gl:CC} does not require additional collimation. However, for satisfactory performance, an excellent energy and position resolutions are required \cite{Parajuli2022}. 

Alternative designs of a Compton camera were proposed with a multi-stage scatterer made of many thin strips of semiconductor detectors. They allow tracking of the Compton electron and thus provide additional kinematic information. This leads to the restriction of the radiation origin to the surface of the cone section, rather than full cone surface \cite{McCleskey2015}. 

The Compton camera concept was first proposed in 1973 for measurements of \SIrange{1}{10}{\mega\electronvolt} $\gamma$ radiation of extraterrestrial origin \cite{Schonfelder1973}. Therefore, Compton cameras were initially used mostly in astrophysics \cite{Parajuli2022}. The application was then extended to homeland security and environmental studies \eg detection of radioactive contamination of the soil, screening for illegal transports of various materials, \etc \cite{Takahashi2012, Wahl2014, Lee2015}. Finally, Compton cameras found their application in medical imaging, with the first working device constructed in the 1980s \cite{Singh1983, Singh1983_2}. Initially, they were replacing diagnostic nuclear imaging devices such as Auger cameras and single photon emission tomography scanners (\acrshort{gl:SPECT}) \cite{Singh1983, Singh1983_2, Han2008}. For the last several years, the research has been ongoing to build a detector allowing utilization of \acrshort{gl:PG} radiation for real-time monitoring of proton therapy. Compton camera is one of the promising candidates for this task. It is suitable for observation of the far radiation sources, thus it can be placed at some distance from the patient without interfering with the therapy process. Moreover, it is suitable for the detection of the radiation matching the energy range of the \acrshort{gl:PG} radiation emitted during the therapy. Therefore, several designs of Compton cameras for real-time proton range monitoring were proposed. 

In many of the proposed setups, the scatterer is formed by a semiconductor detector, while the absorber is usually made of scintillating materials. Such detectors were proposed by the Munich group (\ce{Si} scatterer and \ce{LaBr3} absorber) and the Dresden group (two \ce{CZT} strips as the scatterer and \ce{LSO} absorber later replaced with the segmented \ce{BGO} detector) \cite{Aldawood2017, Hueso-Gonzalez2014, Golnik2016}.  The common problem of the two proposed \acrshort{gl:CC} designs was the insufficient detection efficiency in the energy range \SIrange{1.3}{4.4}{\mega\electronvolt}. This means that the registered number of the \acrshort{gl:PG} events was insufficient for the setups to operate in the clinical conditions.  

The Baltimore group constructed a multi-stage setup consisting exclusively of the commercially available \ce{CZT} detectors. Those detectors are characterized by excellent energy resolution, however, they are much slower when compared with fast scintillators. Therefore their timing resolution is relatively poor. This gives a rise to the increased background formed by accidental coincidences between the detector modules. The group performed the tests of the \acrshort{gl:CC} in clinical conditions and reported the detection of proton beam range shifts of \SIrange{2}{3}{\milli\meter}, which proves the feasibility of their approach \cite{Draeger2018}. 

Another promising multi-stage (three-stage) Compton camera is being developed in Valencia within a MACACO project. It consists of monolithic \ce{LaBr3} blocks read out with silicon photomultipliers \cite{Barrio2018, Viegas2023}. The tests with the \SI{150}{\mega\electronvolt} proton beam and tissue-like target showed sensitivity to detect distal falloff shifts of \SI{3}{\milli\meter}. However, to obtain this result a sophisticated image reconstruction algorithm including an event identification performed by the neural networks was necessary \cite{Munoz2021}.

The \acrshort{gl:CC} constructed by the Japanese group was also built solely with the scintillators. In this case, small pixels made of the \acrshort{gl:GAGG:Ce} material were used in both the scatterer and the absorber. Both modules are read out with multi-anode \acrshort{gl:PMT}s. The setup was tested with the \SI{70}{\mega\electronvolt} proton beam and a tissue-like target, however, the measurement conditions were far from clinical with a very low beam current and the measurement time of several hours. This proves, that the efficiency of this \acrshort{gl:CC} is still far from sufficient to be able to operate in clinical conditions \cite{Koide2018}. It was proposed to combine the scatterer of the Japanese \acrshort{gl:CC} and the absorber of the Munich group and therefore obtain improved performance of such a hybrid setup \cite{Liprandi2018}.  

\begin{figure}[ht]
\centering
\includegraphics[width=0.8\textwidth]{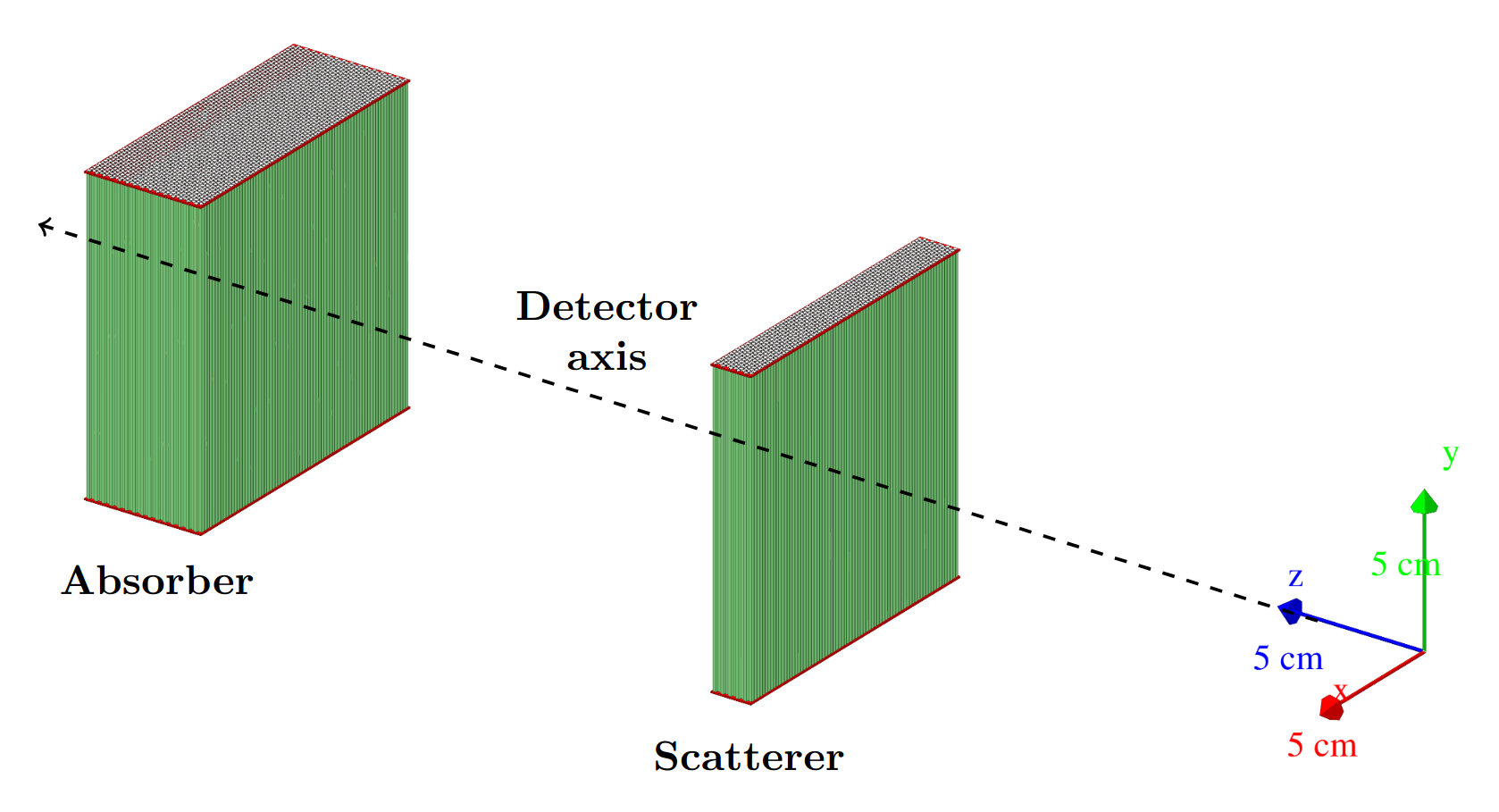}
\caption{Planned design of the \acrshort{gl:SiFi-CC} detector in the Compton camera mode, as implemented in the Monte Carlo simulations. Reprinted from \cite{Kasper2022}.}
\label{fig:cc-scheme}
\end{figure}

The Compton camera design proposed by the \acrshort{gl:SiFi-CC} collaboration is presented in \cref{fig:cc-scheme}. It consists of two modules. Both modules are made of thin, elongated fibers. The fibers are made of a heavy inorganic scintillating material. The large density and effective atomic number (\gls{gl:Zeff}) of the active part of the detector will ensure high detection efficiency for \acrshort{gl:PG} radiation. The scintillating fibers are read out by the state-of-the-art photodetectors such as silicon photomultipliers (\acrshort{gl:SiPM}s) or digital silicon photomultipliers (\acrshort{gl:dSiPM}s). The relatively fast response of the scintillator, combined with a fast read out system and electronics, will result in good timing resolution of the setup. Consequently, it will be possible to impose tight time cuts on the recorded events and better identify the Compton events. As a result, the background level of the system will be decreased. Moreover, in the final detector, it is planned to use a state-of-the-art data acquisition system (\acrshort{gl:DAQ}) based on TOFPET2 \acrshort{gl:ASIC} \cite{tofpet-2}. Integrated \acrshort{gl:FPGA}s will allow to conduct event preselection on-board, which will significantly reduce outgoing data stream.
Therefore, it is evident that the proposed design addresses the main issues encountered by the previously constructed Compton cameras.  

\subsubsection*{Coded mask}

\begin{figure}[ht]
\centering
\includegraphics[width=0.8\textwidth]{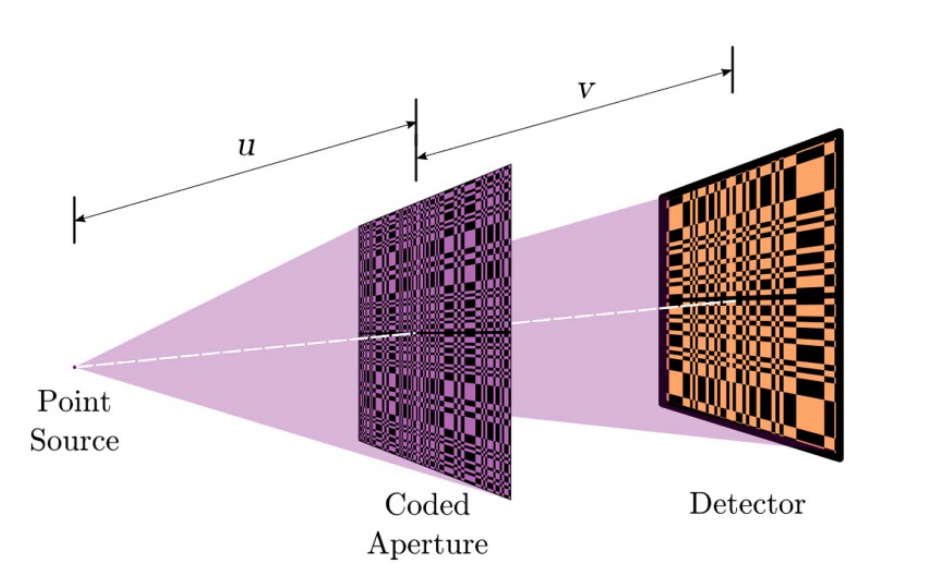}
\caption{Scheme showing principle of operation of the coded mask setup. Reprinted from \cite{Selwood2020}.}
\label{fig:cm-operation}
\end{figure}

A scheme of a detector featuring a coded mask is presented in \cref{fig:cm-operation}. Such a setup consists of two parts: a position-sensitive detector and a passive collimator of a sophisticated shape, \ie the mask. The mask is built out of elements that are transparent or opaque to the radiation of interest. The elements create a pattern, which shields the active part of the detector in a predefined way, and thus defines its response. The position resolution of the detector should match the size of the unit element of the mask and should be sensitive to the photons in the energy range of interest. The principle of operation of the coded mask is straightforward: the irradiated collimator casts a shadow on the detector surface. The pattern of the shadow is the same as this of the coded mask. Depending on the position of the radioactive source in the field of view (\acrshort{gl:FOV}) of the setup, the shadow will be shifted. During the image reconstruction, based on the distribution of hits recorded by the detector, the shadows must be deconvoluted, leading to the 2-dimensional image of the radioactive source distribution \cite{Cieslak2016, Braga2019}. Therefore, for the operation of the \acrshort{gl:CM} detector, the prior knowledge of the response of each detector element to all possible positions of the radioactive source in the \acrshort{gl:FOV} is required (system matrix). Such response information is typically obtained in the Monte Carlo simulations and is later used for image reconstruction. 

The coded mask design is an extension of a pinhole camera. A pinhole camera with an infinitely small hole would provide an excellent angular resolution of the detector, however, registered number if events would be very small and the signal to noise ratio (\acrshort{gl:SNR}) of such a setup would be very poor. To improve the \acrshort{gl:SNR} the hole should be enlarged, however, it would cause a deterioration of the angular resolution. The coded mask, with the collimator consisting of many small holes, provides a compromise between the two situations \cite{Cieslak2016}. It was first proposed in 1968 \cite{Ables1968, Dicke1968}. Similarly to the Compton camera, it was first applied in astrophysics. However, the energy range of interest for \acrshort{gl:CM} detectors is different and it mostly includes X-ray radiation and low energy $\gamma$ radiation \cite{Cieslak2016}. The great advantage of coded mask systems is the ability to achieve large \acrshort{gl:FOV}. Recently, coded mask systems also gained attention as devices for environmental control and homeland security \cite{Woolf2012}. Currently, there are no detectors featuring coded mask collimators designed for medical imaging, including proton therapy monitoring. So far only Monte Carlo designs for that purpose have been published \cite{Sun2020}. 

Within the \acrshort{gl:SiFi-CC} we propose that our detection setup could operate in the coded mask mode alternatively to the Compton camera mode. In the \acrshort{gl:CM} modality, the absorber module can be used as the active part of the detector. The mask is designed as a modified uniform redundant array (\acrshort{gl:MURA}) and made of the wolfram rods inserted in the plastic frame \cite{Gottesman1989}. The \acrshort{gl:CM} modality is developed parallel to the \acrshort{gl:CC} mode, since it does not require additional costly hardware, such as additional scintillating fibers, photodetectors, or a separate data acquisition system. The only modality-specific elements for the \acrshort{gl:CM} are the collimator, data analysis, and image reconstruction methods. 

\begin{figure}[!t]
\centering
\includegraphics[width=0.49\textwidth]{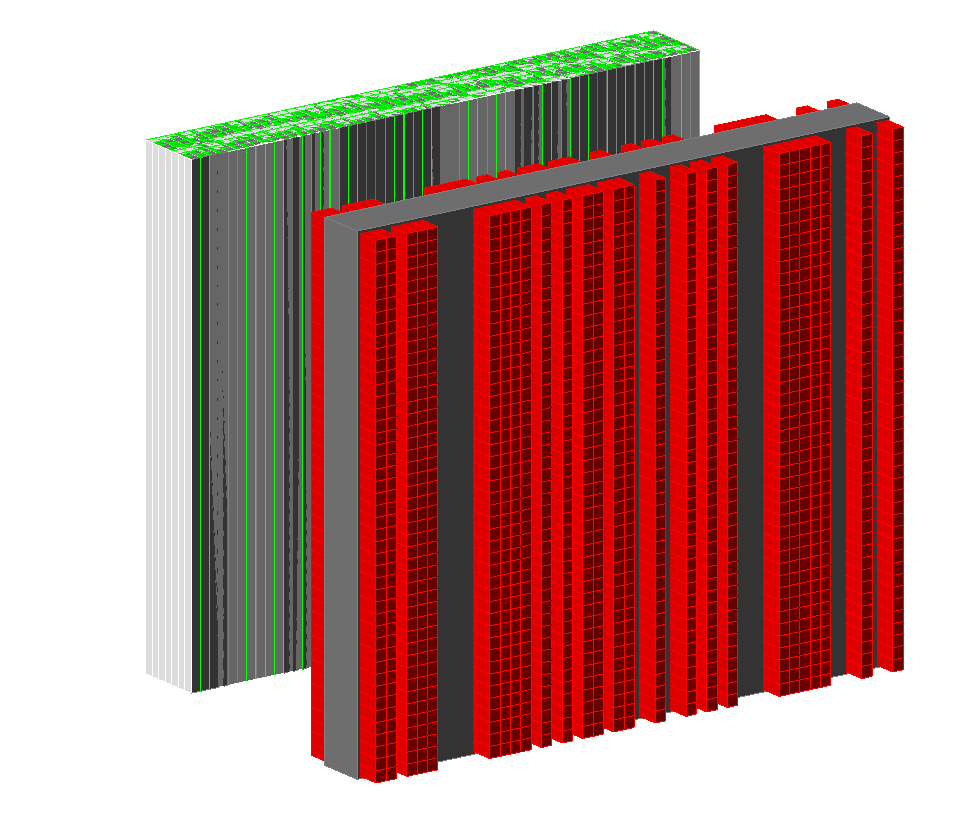}
\includegraphics[width=0.49\textwidth]{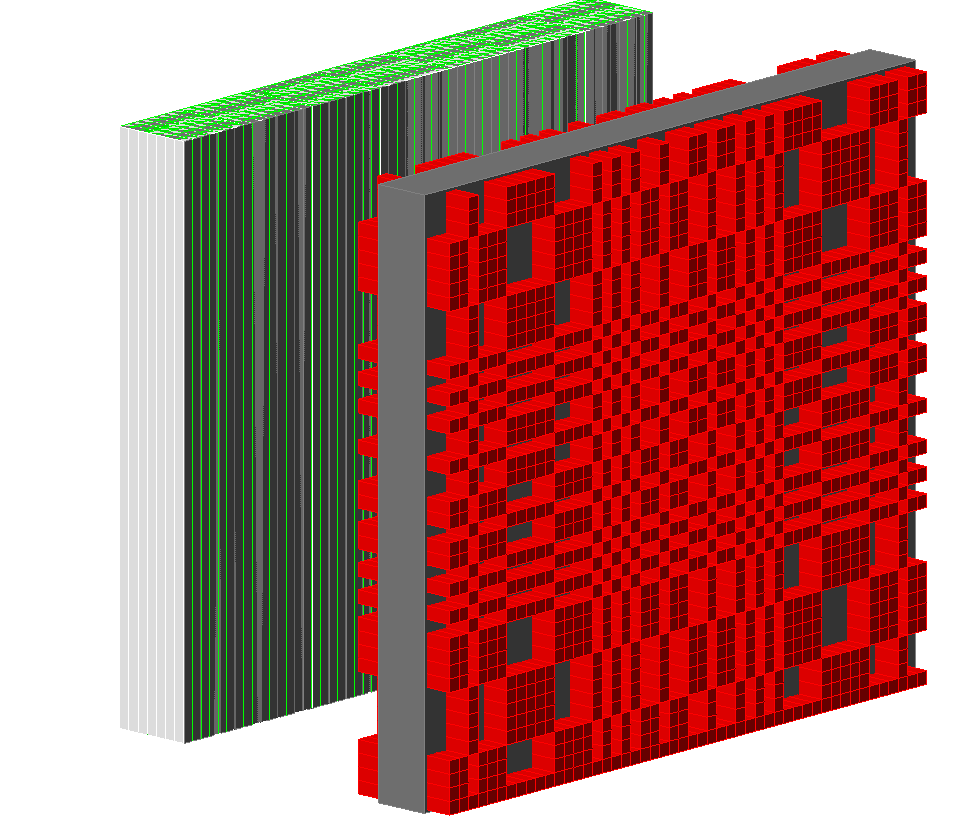}
\caption{Design of the \acrshort{gl:SiFi-CC} detector in the coded mask mode, as implemented in the Monte Carlo simulations \cite{vitalii-priv}. Left: 1D coded mask setup, right: 2D coded mask setup.}
\label{fig:cm-simulations}
\end{figure}

\vspace{1cm}
\noindent
In the presented thesis, a part of research conducted by the \acrshort{gl:SiFi-CC} collaboration is described. \Cref{chap:scintillating-materials} includes an introduction to the physics of scintillating materials and photodetectors and their role in medical imaging. It is followed by a description of the basic characteristics of scintillators and methods for their determination. In \cref{chap:single-fibers} the design optimization of the components of the proposed detection setup is presented. Finally, \cref{chap:prototype} describes construction and tests with a small-scale prototype of the first detector module. The findings are briefly summarized in \cref{chap:summary}.

\chapter{Scintillating materials}
\label{chap:scintillating-materials}

The following chapter presents the characteristics of scintillating materials and the operation of scintillating detectors. Firstly, radiation detection using scintillators is discussed, including a description of the main types of scintillators, the scintillation process, and the different types of photodetectors. The next part of the chapter presents the broad application of scintillators in medical imaging and promising prospects for application in proton therapy monitoring. Subsequent parts of the chapter focus on the properties of scintillating materials and methods for determining them. There, two models of propagation of scintillating light are presented, allowing one to determine the attenuation length of scintillating light. Other properties discussed further in the chapter are the following: light collection, decay constants, energy-, position- and timing resolutions.


\section{Radiation detection with scintillating detectors}
\label{sec:radiation-detection}

A scintillating detector always consists of two parts: scintillating material and a photodetector. The scintillating material is coupled to the photodetector either directly or with the use of a light guide. The scintillating material is the active part of the detector, which emits small flashes of light \ie scintillations, when it experiences interactions with ionizing radiation. The scintillating light reaches the photodetector, where it is converted into electric signals. The signals can then be counted or analyzed to extract information about the interaction in the detector \cite{Leo}.

The earliest example of the scintillating detector is a spinthariscope invented by William Crooks in 1903. The device allowed him to observe scintillations induced by $\alpha$ particles impinging on a \ce{ZnS} screen \cite{ORAU-museum}. 
In that case, the human eye served as the photodetector. However, scintillating detectors did not gain popularity until the 1940s, when photomultipliers were introduced. This allowed for automatic, efficient, and precise pulse counting. Today, scintillating detectors are one of the most frequently used type of detectors in nuclear, particle, and high energy physics (\acrshort{gl:HEP}) due to their radiation hardness and adjustable properties that can be tailored to various applications \cite{Leo}.

\subsection{Scintillating materials}
\label{ssec:scintillating-materials}

The scintillation process can be considered to be a type of luminescence. Luminescence occurs when certain materials exposed to different forms of energy reemit it in a form of light. In the case of scintillators, the factor that induces luminescence is ionizing radiation. 

Scintillators are very well suited for the detection of ionizing radiation and are capable of providing a variety of information. Firstly, scintillating detectors are characterized by fast response and recovery time in comparison to other detector types. The fast response allows one to obtain good timing resolution. Fast recovery time provides shortened dead time and thus enables high count-rate capability. Another favorable property of scintillators is their linear response. This means that the light output of the scintillators is proportional to the energy deposited. If the response of the coupled photodetector is also linear, the obtained electric pulses are proportional to the deposited energy as well. Therefore, scintillating detectors can serve not only as counters but also as spectrometers. Finally, for some scintillators, it is possible to distinguish which type of ionizing radiation caused the scintillation based on the shape of the registered pulse (pulse-shape discrimination)~\cite{Leo}.

In general, scintillators can be divided into several groups that differ in terms of properties, structure, and mechanism of scintillation. The main types of scintillators include organic scintillators, inorganic crystals, gaseous scintillators, and glasses.


\subsubsection*{Inorganic crystal scintillators}

Inorganic crystals are the largest and the most dynamically developing class of scintillators. This group includes materials such as halides of alkali metals, garnets, orthosilicates, and perovskites of rare-earth elements, and many more. Some of the most frequently used inorganic crystal scintillators are presented in \cref{tab:materials-literature}. Most inorganic crystal scintillators contain a small admixture of impurity called an activator or a dopant. Although the concentration of the activator is relatively small, in the order of \SI{e-3}{\per\mole}, it is essential for the scintillation process to occur.  

\begin{figure}[htbp]
\centering
\includegraphics[width=.60\textwidth]{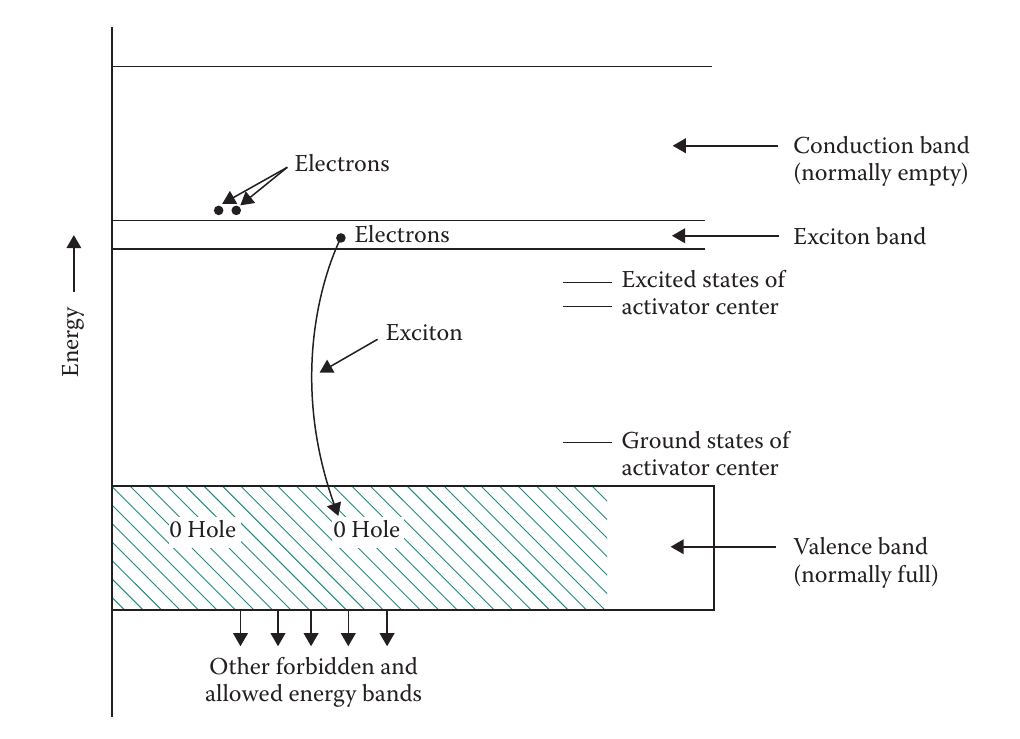}
\caption{A scheme of energy band structure in inorganic crystal scintillator. The picture comes from \cite{Tsoulfanidis}.}
\label{fig:inorganic-scintillation}
\end{figure}

The mechanism of scintillation in inorganic crystals is based on the structure of their energy bands and the transitions between them. \Cref{fig:inorganic-scintillation} presents a scheme of allowed and forbidden energy bands of the crystal. In the ground state the valence band is completely filled with electrons, and the next closest allowed band, \ie the conduction band, is empty. The incident ionizing radiation may deposit enough energy to cause the transition of an electron from the valence band to the conduction band. In that case, the electron leaves behind a hole in the valence band and both of them are free to move. However, if the energy obtained by the electron is not sufficient for transition to the conduction band, it is transferred to the exciton band instead. This very narrow band lies directly underneath the conduction band, with its upper level overlapping with the lower level of the conduction band. The electron in the exciton band remains electrostatically bound with the hole forming the exciton.  

The activator atoms create additional energy states between the valence and conduction bands. The activator atoms may exist in the ground or in one of the excited states. Excitation of the activator can be caused by photon absorption, capture of an exciton or capture of an electron and a hole. The deexcitation occurs within a time of the order of \SI{e-8}{\second} with the emission of a scintillation photon. It should be noted that the emission of the scintillating photons is primarily the result of the transitions of the activator atoms and not of the crystal lattice. However, the ionizing radiation deposits most of its energy in the lattice, which means that energy transfer between the host crystal and the activator atoms occurs.  

The most frequently used activators include cerium and thallium. In particular, the activator can be chosen specifically to obtain desired characteristics of the emitted scintillating light. Additionally, the effects of co-doping of inorganic scintillators recently became an interesting research topic. Additional doping has been shown to influence the timing properties and light output of scintillators \cite{Yamaguchi2016, Kamada2015, Kasimova2019}.  

Inorganic crystal scintillators are characterized by a response time of the order of several tens to several hundred nanoseconds. Moreover, many materials exhibit an additional elongated decay mode, as explained further in \cref{ssec:timing-properties}. Another disadvantage of certain inorganic crystals is their hygroscopicity, which means that they require dedicated housing to protect them from the humidity in the air. On the other hand, due to their high density and high effective atomic number \gls{gl:Zeff}, they have relatively high stopping power. Additionally, among all scintillators, they have some of the highest light outputs. This significantly improves their energy resolution \cite{Leo}.  

Due to their favorable properties, inorganic crystal scintillators find a wide range of applications, \eg in calorimeters for \acrshort{gl:HEP} experiments, gamma-ray spectroscopy, neutron detectors, and medical imaging. Moreover, research towards finding new types of inorganic crystal scintillators, their production and adjustment of properties, is a dynamically developing field. An excellent example is the Crystal Clear Collaboration established in 1990 at CERN. The main objective of the collaboration was \emph{developing new inorganic scintillators suitable for crystal electromagnetic calorimeters of LHC experiments}. However, since founding, they have also contributed greatly to the field of scintillator production and understanding of the scintillation process itself \cite{CCC}. 

\subsubsection*{Organic scintillators}

\begin{figure}[htbp]
\centering
\includegraphics[width=.65\textwidth]{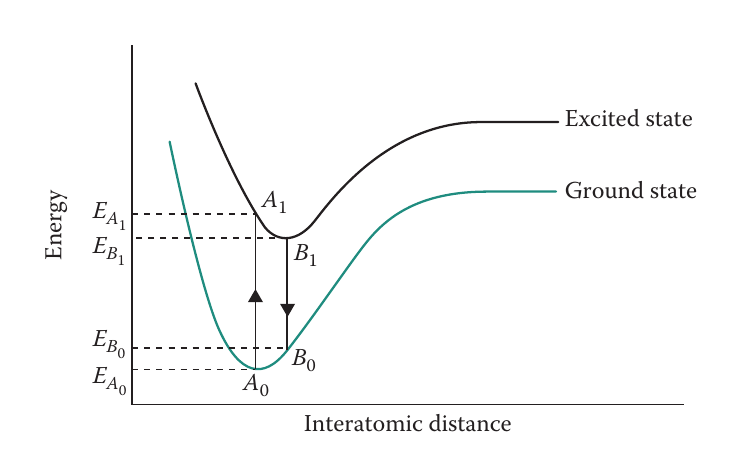}
\caption{A simplified energy diagram of a molecule. The picture comes from \cite{Tsoulfanidis}.}
\label{fig:organic-scintillation}
\end{figure}

Organic compounds exhibiting scintillation capability belong to the class of benzoid compounds. In this case, the mechanism of scintillation is based on molecular transitions. \Cref{fig:organic-scintillation} shows the energy diagram of a molecule. In the ground state $A_0$ the molecule maintains its minimum potential energy. Ionizing radiation passing through the scintillator can deposit enough energy to cause the transition of molecules to state $A_1$. Since the state $A_1$ does not correspond to the energy minimum, the molecule will release energy through vibrations and move to the state $B_1$. The molecule in the state $B_1$ can undergo deexcitation to the state $B_0$ with the emission of a photon. The energy of the emitted photon $B_1 - B_0$ is smaller than the excitation energy $A_1 - A_0$. If the two energies were equal, the absorption and emission spectra of the substance would be the same, and scintillation would not occur. 

There are three types of organic scintillators: crystals, liquids, and plastics. The most common organic crystal scintillator is anthracene. It is characterized by the highest light output among all organic scintillators (\SI{17400}{\photon\per\mega\electronvolt}). In contrast to inorganic crystals, an activator is not required for organic crystals. Moreover, any impurities are undesirable since they reduce light output. 

The organic liquid scintillators consist of a solvent and at least one solute. In the two-ingredient liquid scintillators the ionizing radiation deposits energy mostly in the solvent, but solute molecules are responsible for the emission of scintillation photons. This means that energy transfer between the solvent and the solute occurs similarly as in inorganic crystals. If another solute is added to the mixture, it acts as a wavelength shifter (\acrshort{gl:WLS}). Wavelength shifters are substances that absorb radiation of a given wavelength and reemit radiation of wavelengths larger than originally absorbed. The \acrshort{gl:WLS} is typically used to change the characteristics of the scintillating light and make it compatible with the sensitivity range of the photodetector. Organic liquids are utilized when large amounts of scintillator are required to increase detection efficiency \eg for the observation of rare processes such as in neutrino experiments. 

Plastic scintillators can be considered solid solutions of organic scintillators. Similarly to liquid scintillators, they consist of a solvent and at least one solute. Their properties are also similar to those of liquid scintillators. However, their main advantage is that there is no need for additional containers and the possibility of machining them in various shapes and sizes. In addition, they are resistant to air, humidity, and many chemicals, making them easy to handle and suitable for use under challenging conditions. A typical representative of plastic scintillators is listed in the second column of \cref{tab:materials-literature}.

The common characteristic of all organic scintillators is their rapid response time, of the order of several nanoseconds. This makes them suitable for fast timing measurements. However, their light output is significantly poorer compared to that of inorganic crystals (see \cref{tab:materials-literature}). Another disadvantage of organic scintillators is their low stopping power due to the small density and effective atomic number \Zeff. Because of this, the energy spectra obtained with the use of organic scintillators usually lack photopeaks \cite{Turner, Tsoulfanidis}.

\afterpage{%
\begin{landscape}

\begin{table}
\scriptsize

\begin{threeparttable}[h]

\caption{Properties of several selected scintillating materials, as reported by the producers or available research results. The scintillators chosen for this study are highlighted in bold font.}
\label{tab:materials-literature}

\begin{tabularx}{1.0\linewidth}{p{4.3cm}p{1.8cm}p{1.3cm}XXXXp{2.5cm}p{2.3cm}p{2.7cm}}
\toprule
Scintillator name & BC-408\tnote{2} & \ce{NaI}:\ce{Tl} & \ce{CsI}:\ce{Tl} & \ce{LaBr3}:\ce{Ce}\tnote{3} & \ce{BGO} & \ce{LuAP}:\ce{Ce} & \textbf{\ce{LYSO}:\ce{Ce}}\tnote{4} & \textbf{\ce{LuAG}:\ce{Ce}} & \textbf{\ce{GAGG}:\ce{Ce}} \\
\midrule
Chemical formula/composition & \ce{H}: \num{8.47} wt\si{\percent} \newline \ce{C}: \num{91.53} wt\si{\percent} & \ce{NaI}:\ce{Tl} & \ce{CsI}:\ce{Tl} & \ce{LaBr3}:\ce{Ce} & \ce{Bi4Ge3O12} & \ce{LuAlO3}:\ce{Ce} & \ce{(Lu,Y)2SiO5}:\ce{Ce} & \ce{Lu3Al5O12}:\ce{Ce} & \ce{Gd3Al2Ga3O12}:\ce{Ce} \\
Density [\si{\gram\per\cubic\centi\metre}] & 1.032 & 3.67 & 4.51 & 5.06--5.29 & 7.13 & 8.34 & 7.1--7.4 & 6.73 & 6.68 \\
\Zeff & - & 50 & 54 & $\sim$46 & 75 & 65 & $\sim$62 & 63 & 51 \\
Refractive index (\gls{gl:n})\tnote{1} & 1.58 & 1.85 & 1.79 & $\sim$1.9 & 2.15 & 1.94 & 1.82 & 1.84 & 1.9 @\SI{500}{\nano\meter}  \\
Maximum of emission [\si{\nano\meter}] & 425 & 410--415 & 550 & 358--380 & 480 & 365 & 420 & 535 & 520 \\
Decay constant [\si{\nano\second}] & 2.1 & 230--250 & 600--900 \par + slow \par components & 16--35 & 300 & 16--18 \par + slow \par component & $\leq$42 & 60--70 + slow \par components & 30 (\SI{25}{\percent}) \par 80 (\SI{60}{\percent}) \par 100--200 (\SI{15}{\percent}) \\
Photon yield [ $\times$ 10$^3$ \si{\photon\per\mega\electronvolt}] & 8--11.1 & 38--41 & 54--66 & 45--63 & 8-10 & 5.2--12 & 25--30 & 25 & 22--60 \\
Light attenuation length & 2100--8560 & - & - & - & - & - & $\sim$ 400 & 50--320 & 220--320 \\
Radiation length & - & 25.9 & 18.6 & 21.0 & 12.1 & 11.0 & 11.6 & 14.1 & 15.9 \\
Energy resolution at \SI{662}{\kilo\electronvolt} [\si{\percent}] & - & 5.6 & 4.4--6.6 & 2.6--3.9 & 9.0 & 9.3--15 & 8--8.7 & 6.7 & 4.9--8.3 \\
Internal activity [\si{\cps\per\gram}] & - & - & - & 0.3 & - & present & 39 & 37 & - \\
Hygroscopicity & no & yes & yes & yes & no & no & no & no & no \\
References & \cite{saint-gobain-bc-408, Gagnon2014, Watanabe2015, Almurayshid2017} & \cite{VanEijk2003, Sibczynski2017, saint-gobain-nai-tl, Balcerzyk2005, Mao2008} & \cite{VanEijk2003, Lecoq2016, Sibczynski2017, Mao2008, saint-gobain-csi-tl, Syntfeld-Kazuch2008} & \cite{VanEijk2003, Lecoq2016, Sibczynski2017, Aldawood2015, saint-gobain-labr3-ce, Mazumdar2013, Schwendimann2020, VanLoef2002} & \cite{VanEijk2003, Lecoq2016, Mao2008, Schwendimann2020, saint-gobain-bgo} &%
\cite{VanEijk2003, Lecoq2016, crytur-luap, Moszynski1997, VanEijk1997} &%
\cite{epic-crystal, meta-laser, shalom, Chewpraditkul2009, Vilardi2006, Schwendimann2020} &%
\cite{crytur, Pauwels2013, Chewpraditkul2009} &%
\cite{fomos, Dobrovolskas2019, Kasimova2019, Tamagawa2015, Seitz2016, Lowdon2019} \\
\bottomrule
\end{tabularx}

\begin{tablenotes}
\item[1] At maximum of emission, unless stated otherwise 
\item[2] Plastic scintillator, based on polyvinyltoluene
\item[3] Properties depend on the producer and doping level
\item[4] Properties depend on the producer and Lu:Y stoichiometric ratio.
\end{tablenotes}
\end{threeparttable}

\end{table}

\end{landscape}
}

\subsubsection*{Gaseous scintillators}

Some of the noble gases along with nitrogen show scintillation properties. In this case, scintillation occurs as a result of the excitation of single atoms and their return to their ground state. Gaseous scintillators are characterized by very short decay times. However, they emit light in the ultraviolet range, where the efficiency of most photodetectors is very low. Therefore, the use of \acrshort{gl:WLS} is required. In detectors featuring gaseous scintillators, \acrshort{gl:WLS} is usually introduced as a coating of the gas tank or an admixture of another gas. The low density of gases implies poor gamma detection efficiency in this type of detector. Therefore, gaseous scintillators are suitable for the detection of heavy charged particles, such as alpha particles, fission fragments, and other heavy ions. It should be noted that the light output of gaseous scintillators shows a very weak dependence on the mass and charge of the detected particles \cite{Leo, Tsoulfanidis}. 

\subsubsection*{Glass scintillators}

Glass scintillators include lithium and boron silicates activated with cerium. They are mainly used for the detection of thermal neutrons because of the high neutron cross section for boron and lithium. However, they can also be used for the detection of gamma radiation and electrons. The glass scintillators are characterized by intermediate response times, of the order of a few tens of nanoseconds. Compared to other types of scintillators, light output is relatively low, approximately below \SIrange{4300}{5200}{\photon\per\mega\electronvolt}. However, the strong point of glass scintillators is their exceptional resistance to extreme conditions. They sustain almost all organic and inorganic reagents and have high melting points. This makes them suitable for extreme conditions, in which any other scintillator would fail \cite{Leo}. 

\subsubsection*{Polycrystalline ceramic scintillators} 

Ceramic scintillators are made of powdered inorganic crystal scintillating material that is typically hot-pressed to form a solid block. They were first introduced in the 1980s as a result of an intensive search for new scintillators for medical imaging purposes. In comparison to inorganic crystal scintillators, their manufacturing process is cheaper and shorter. In addition, it is easier to produce large volumes of the scintillator with the desired shape and good uniformity. At the same time, they consist of inorganic crystals and offer favorable scintillating properties. The main drawback of most ceramic scintillators is non-transparency. This means that propagation of scintillating light at long distances is not possible and the use of thin layers is required. The technology to manufacture transparent ceramic scintillators using crystals exhibiting cubic lattice structure is well developed. The cubic crystals are suitable for the manufacturing of ceramic scintillators since they are isotropic and characterized by one refractive index. Unfortunately, most of the known inorganic crystal scintillators do not meet this requirement. Therefore, new possibilities for fabricating transparent ceramic scintillators based on non-cubic crystals are currently being investigated \cite{Wisniewski2008}.

\subsection{Photodetectors}
\label{ssec:photodetectors}

A photodetector is another integral part of a scintillating detector. Its role is to convert scintillating light into an electric signal and amplify it. Without photodetectors, the use of scintillators in experimental physics would be impractical. The oldest and still widely used type of photodetector is a photomultiplier tube. Recently, silicon-based devices are becoming increasingly popular.


\subsubsection*{Photomultiplier tube (\acrshort{gl:PMT})}

\begin{figure}[htbp]
\centering
\includegraphics[width=.80\textwidth]{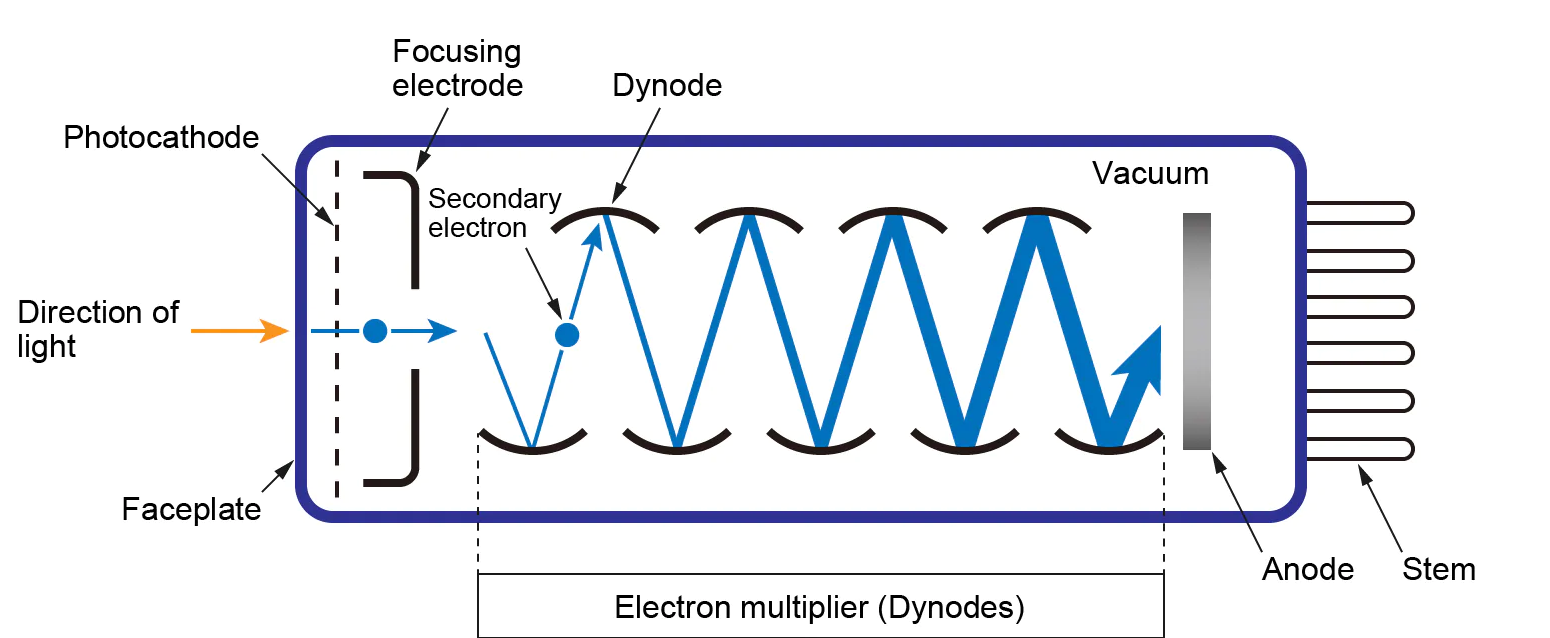}
\caption{Scheme of the scintillating detector that features the photomultiplier tube. The picture comes from \cite{matsusada}.}
\label{fig:pmt}
\end{figure}

A scheme of a typical photomultiplier is presented in \cref{fig:pmt}. It consists of a vacuum tube with a photocathode at the entrance and a series of dynodes followed by an anode serving as a charge collector. Photons emitted in the scintillator enter the photomultiplier tube and hit the photocathode, resulting in the emission of electrons. Photocathodes can be made of various materials, leading to different spectral sensitivity. For the best performance of the detector, the emission spectrum of the scintillator and the photosensitivity of the photocathode should be matched. Between the photocathode, subsequent dynodes and the anode there is an electric field with a potential of the order of \SI{-e3}{\volt}, with each dynode holding a larger potential than the previous one. Therefore, electrons emitted from the photocathode are directed towards the first dynode. Dynodes are coated with the material, which emits secondary electrons when irradiated with electrons. The secondary electrons produced at the first dynode are then directed to the second dynode, from there towards the third dynode, and so on. At each dynode, the production of secondary electrons increases, leading to the amplification of the signal. A typical photomultiplier amplifies a pulse of scintillating light by a factor of \num{e6} in a time of the order of nanoseconds. \Cref{fig:pmt} shows only a typical configuration of dynodes. In reality, several different dynode configurations are used, each characterized by a different response time and linearity range.

An important parameter of the photomultiplier response is the dark current. The predominant source of the dark current is the thermionic emission of electrons from the photocathode. The photocathode at room temperature can release approximately \SI{e5} {\per\second} electrons. Those electrons are amplified at the subsequent dynodes and disturb the detection of a signal stemming from scintillating photons. The influence of the dark current is crucial in measurements where activity of radiation source is small or energy deposits of the radiation of interest are small.
It can be reduced if the operating temperature is decreased. Another disadvantage of \acrshort{gl:PMT}s is their sensitivity to magnetic fields \cite{Leo, Tsoulfanidis}.

\subsubsection*{Silicon photomultiplier (\acrshort{gl:SiPM})}

A silicon photomultiplier is a modern solid-state detector based on single-photon avalanche diodes (\acrshort{gl:SPAD}). A \acrshort{gl:SPAD} is formed by a p-n junction with the depletion layer. When a photon is absorbed in the silicon, an electron-hole pair is created. Silicon is a suitable material for the photodetector because it efficiently absorbs photons of a wide range of wavelengths. Nevertheless, the photon detection efficiency is wavelength-dependent. A reverse bias is applied to the p-n junction, which creates an electric field across the depletion region. In the electric field, the charge carriers created by the incident photon are accelerated toward the anode and the cathode \ie current flow occurs. If the applied bias voltage is sufficiently high, the photodiode operates in Geiger mode. In this state, the charge carriers gain enough kinetic energy to cause secondary ionization in the silicone. This means that a single incident photon initiates the self-perpetuating ionization cascade, \ie the Geiger discharge. The silicon breaks down and becomes conductive. The original single electron-hole pair is amplified into a macroscopic current flow. The current is quenched by a series of resistors. They lower the reverse voltage to the value below the breakdown, thus stopping the discharge. Subsequently, the photodiode recharges to its bias voltage and is able to detect the next photon. The single \acrshort{gl:SPAD} provides binary information, since in the Geiger mode the produced signal is be the same  regardless of the number of photons simultaneously interacting in the photodiode \cite{sensl-sipm}. 

A SiPM is an array of many \acrshort{gl:SPAD}s, each coupled with its own quenching resistor. The circuit formed by a \acrshort{gl:SPAD} and its quenching resistor is called a microcell. The microcells operate independently of each other, meaning that Geiger discharge is limited to the one microcell in which it was initiated. Other microcells remain charged and ready to detect photons. The current from all microcells is summed, giving a quasi-analog output. Thus, the \acrshort{gl:SiPM} response is proportional to the number of incident photons \cite{sensl-sipm}. 

\acrshort{gl:SiPM}s are characterized by a number of parameters, the most important of which are the following \cite{sensl-sipm}:  

\begin{description}[wide=0\parindent]

\item[Fill factor] Around each microcell there is some dead space. These areas are not light-sensitive and contain quenching resistors, signal tracks, and optical and electrical isolation of microcells. The parameter describing the percentage of photon-sensitive area within the total area of the \acrshort{gl:SiPM} is called the fill factor. Higher fill factor results in improved gain and efficiency for photon detection, but on the cost of elongated recovery time and lower dynamic range.  

\item[Overvoltage] The bias point for which the Geiger discharge in the depletion layer is initiated is called the breakdown voltage $V_{\textrm{br}}$. SiPMs generally operate at the bias point \SIrange{10}{25}{\percent} higher than $V_{\textrm{br}}$. The difference between the operating voltage and the breakdown voltage is known as the overvoltage $V_{\textrm{ov}}$.

\item[Photon detection efficiency] The photon detection efficiency (\acrshort{gl:PDE}) describes the statistical probability that a single photon interacts with the microcell and triggers Geiger discharge. It is a measure of the \acrshort{gl:SiPM} sensitivity. It depends on the wavelength of the incident photons, the overvoltage, and the fill factor. 

\item[Dark count rate (\acrshort{gl:DCR})] Similarly as for \acrshort{gl:PMT}s, \acrshort{gl:SiPM}s experience a dark current. The primary source of the dark current are thermal electrons generated in the photosensitive area of the SiPM. Each thermal electron initiates a Geiger discharge and results in a dark count. The dark current consists of a series of pulses, thus it is convenient to describe it with the dark count rate. The DCR increases with the overvoltage, the active area of the SiPM, and the temperature. The signals resulting from the thermal electrons and incident photons are identical. Therefore the dark current signals have a magnitude of single photon signals. 
To reduce the DCR it is sufficient to set the measurement threshold just above the level of several-photons response.
However, it needs to be noted that dark counts contribute to the true signal. 

\item[Optical cross talk] During the discharge in the depletion layer the accelerated charge carriers emit secondary photons. These photons belong to the near-infrared range and can penetrate through silicon, causing discharges in neighboring microcells. The optical cross talk describes the probability that a single discharging microcell initiates discharge in another microcell. 

\item[Temperature dependency] Temperature affects the breakdown voltage and dark count rate of the \acrshort{gl:SiPM}. The temperature dependence of $V_{\textrm{br}}$ is typically in the order of several \si{\milli\volt\per\celsius}. If large temperature fluctuations occur, compensation of the operating voltage is required. Otherwise, it will result in changes in the effective overvoltage, and therefore in other important performance characteristics, such as \acrshort{gl:DCR}, \acrshort{gl:PDE}, gain \cite{sensl-sipm}.

\end{description}

\subsubsection*{Digital silicon photomultiplier (\acrshort{gl:dSiPM})}

Similarly to the analog SiPMs described above, digital \acrshort{gl:SiPM}s consist of an array of \acrshort{gl:SPAD}s. The main difference between the two is that \acrshort{gl:dSiPM}s are equipped with the transistor as a quenching element (active quenching) and an analog-to-digital converter (\acrshort{gl:ADC}). Replacement of the quenching resistor with an active element results in an improved recovery time of the sensor, and thus significantly shorter dead time. Additionally, power consumption is reduced \cite{DAscenzo15}. 

Each microcell that responds to the incident photon produces its own digital output. Signals produced by all microcells are captured by an on-chip counter. The final information provided by the sensor has a form of digital photon count detected in a certain time interval \cite{DAscenzo15}. 

The design of \acrshort{gl:dSiPM}s addresses the issue of the dark current. Typically, the \acrshort{gl:DCR} is not uniform over the photosensitive surface, meaning that some microcells are more prone to generate thermal electrons. The \acrshort{gl:dSiPM}s are equipped with an addressable static memory cell which can be used to disable the chosen single microcells of the sensor. This allows to block the signal transmission from particularly noisy microcells and avoid contribution of false events to the true signal. Consequently, the signal-to-noise ratio (\acrshort{gl:SNR}) of the sensor is significantly improved compared to other photosensors \cite{DAscenzo15}. 

The first \acrshort{gl:dSiPM}s were designed for applications in medical imaging. Medical imaging devices typically require information about the number of detected photons, as a measure of the energy deposit of ionizing radiation, along with the time of the signal arrival. Therefore, the \acrshort{gl:dSiPM}s were designed to provide this information specifically with great precision. Timing information is available thanks to the integrated time-to-digital converter (\acrshort{gl:TDC}). The \acrshort{gl:dSiPM}s do not use analog signal processing at any stage. As a result, they provide faster and more accurate information compared to analog \acrshort{gl:SiPM}s \cite{DAscenzo15}. However, with the \acrshort{gl:dSiPM}s, there is no access to full waveforms of incoming signals, therefore pulse-shape analysis is not possible.

\subsection{Scintillating detectors assembly}
\label{ssec:detector-assembly}

As already mentioned in this chapter, to achieve optimal performance of the scintillating detector, it is important to match the emission spectrum of the scintillator and the sensitivity of the photodetector. Another important aspect of the scintillation detector assembly is connecting the two elements in a way that ensures minimal loss of light before it reaches the photodetector. 

\begin{figure}[htbp]
\centering
\includegraphics[width=.35\textwidth]{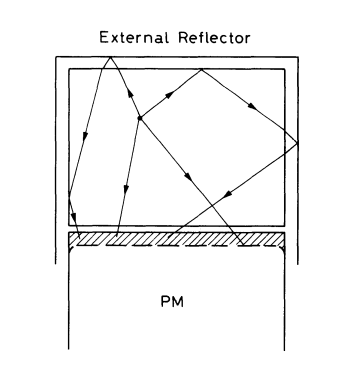}
\caption{Scheme of a scintillator surrounded by an external reflector. The picture comes from \cite{Leo}.}
\label{fig:reflections}
\end{figure}

There are two processes responsible for the loss of scintillating light: escape through the scintillator walls and absorption within the scintillator volume. Scintillating photons are emitted isotropically, thus only a small fraction reaches the photodetector directly. The remaining light propagates towards the scintillator walls, where, depending on the angle of incidence, it undergoes either total internal reflection or partial reflection and transmission. Of the two, total internal reflection is favorable since it redirects light back toward the scintillator. The most straightforward method to avoid light loss is to increase the fraction of light that undergoes total internal reflection and induce external reflection (see \cref{fig:reflections}). The first is achieved by surrounding the scintillator with a medium of the lowest possible refractive index, which results in a decrease of the Brewster angle. The most suitable and convenient candidate for such a medium is simply air. To additionally promote total internal reflection, the surfaces of scintillators are usually finely polished.

The external reflector is an additional layer of material surrounding the scintillator. It can be introduced by wrapping, coating, or casing the scintillator volume. The surface of the reflector can be specular (as shown in \cref{fig:reflections}) or diffuse. If the surface is specular, the angle of reflection is equal to the angle of incidence. In the case of diffuse surfaces, the reflection is independent of the angle of incidence. To take advantage of both, the total internal reflection, as well as external reflection, a small gap of air is often maintained between the surface of the scintillator and the reflector. 

An alternative method to increase the collection of scintillating light is to attach more than one photomultiplier to the surface of the scintillator and sum up the obtained signals. However, this solution is not suitable for all detector geometries and increases the cost and complexity of the setup. 

Absorption of scintillating light in the scintillator volume is negligible in small detectors. However, different scintillating materials show different degrees of transparency to their own scintillating light. Therefore, this effect should be taken into account when designing large detectors. In particular, a suitable material should be selected for the desired geometry. A detailed description of light propagation in scintillators can be found later in this chapter (see \cref{sec:light-propagation}).

Another crucial aspect of the scintillator detector design is assuring an efficient transport of light from the scintillator to the photodetector. In contrast to the side surfaces, total internal reflection is not desired at the surface facing the photodetector. In this case, maximum light transmission is favorable. Therefore, the gap between the scintillator and the photodetector should not be filled with air. The intermediary medium should be characterized with a refractive index possibly close to that of the scintillator and the photodetector surface \cite{Leo}. There is a variety of such materials available, including silicone greases, silicone optical interface pads, and optical glues.

In some cases, direct coupling of the scintillator to the photodetector is not possible or not desirable \eg because of lack of space, unusual geometry, or the presence of a magnetic field. Then, the scintillator and photodetector are connected via a light guide. The light guide is typically made of finely polished plexiglass. It can be machined to the desired shape and size. Alternatively, optical fibers can also be used. The principle of operation of the light guide is based on total internal reflection. The scintillating light undergoes multiple reflections within the light guide volume until it reaches the photodetector. Some light loss occurs as the scintillating light passes through the light guide, and therefore it is typically not used unless necessary. External reflectors are often added to improve the performance of the light guide, similar as for scintillators. The geometry of the light guide is also important for its performance. In particular, the output surface of the light guide should not be smaller than the input surface. Furthermore, it was shown that optimally the cross section of the light guide should be the same throughout its full length, without sudden changes in geometry\cite{Leo}.


\section{Application of scintillators in medicine}
\label{sec:scintillators-pt-monitoring}

\subsection{Medical imaging and nuclear medicine}

Scintillating materials have found wide applications not only in physics but also in many other fields \eg homeland security, industrial control, material science, and finally medical imaging and nuclear medicine. Ionizing radiation detectors incorporated into medical devices must meet rigorous criteria to achieve the required performance, \ie detection efficiency, position-, energy- and timing resolutions.  
The criteria for optimal performance are different for different imaging modalities and depend on the energy of the detected ionizing radiation \cite{Lecoq2016}.

In planar X-ray imaging, a patient is exposed to X-ray radiation with an energy of several tens of \si{keV} (\eg \SI{20}{\kilo\electronvolt} for mammography, \SI{60}{\kilo\electronvolt} for dental diagnostics). The attenuation of the radiation that penetrates the patient's body depends on the density and \Zeff of the tissues. The radiation attenuation profile, which reflects the encountered anatomical structures, is projected on the two-dimensional position-sensitive detector. Another important medical imaging technique is X-ray computed tomography(\acrshort{gl:CT}), which is an advancement of planar X-ray imaging. During \acrshort{gl:CT} examination, the patient is irradiated with the X-ray fan beam from many directions and the cross-sectional images of the body are registered. The energy of the radiation used in the CT reaches \SI{150} {\kilo\electronvolt} \cite{VanEijk2003}. Recently a dual-energy \acrshort{gl:CT} gains more attention, also in particle therapy planning. It uses two different photon energies to perform patient scan, which results in better imaging of tissues, which would be indistinguishable in traditional monoenergetic scans \cite{Tatsugami2022, Noid2020}.

Modern planar X-ray devices as well as \acrshort{gl:CT} scanners feature scintillator material arrays coupled to spectrally-matched \acrshort{gl:SiPM}s. To minimize the patient's exposure to ionizing radiation and at the same time obtain satisfactory image quality, the scintillating material should be dense to provide very good detection efficiency. The \acrshort{gl:CT} scanners record approximately a thousand projections per second, which poses stringent requirements for the timing properties of the scintillator. In particular, the afterglow \ie elongated decay mode is not desired, as it leads to artifacts in the reconstructed image. Finally, the scintillator used in medical imaging devices should have large light output improve the signal to noise ratio in a low radiation environment. 
\cite{Lecoq2016}. 

For planar X-ray imaging operating at low energies of X-ray radiation, typically thin scintillation screens made of ceramic scintillators are used. However, for modalities that require higher X-ray energies, inorganic crystal scintillators are better suited. This is due to the fact, that a thicker layer of material is required to achieve the same detection efficiency and inorganic crystals have much better transparency to scintillating light compared to ceramic materials. The only crystal scintillator currently used in CT scanners is \ce{CWO} (\ce{CdWO4}). Its main advantage is low afterglow and a small temperature coefficient. However, the material is difficult to handle due to its fragility and toxicity. The efforts to find alternative materials suitable for \acrshort{gl:CT} imaging resulted in the development of ceramic scintillators and advances in the fabrication of more transparent ceramic materials. Ceramic materials have the advantage of good performance and easy manufacturing in a variety of shapes. However, the thickness of the ceramic scintillator needs to be carefully optimized to achieve a good compromise between scintillating light transmission and detection efficiency \cite{Lecoq2016}.

In imaging methods using radionuclides, the radiopharmaceutical is introduced into the patient's body. The radiopharmaceutical agent consists of a radioactive isotope and a carrier, which has the ability to accumulate in specific tissues. Subsequently, the dedicated detection system is used to register the activity distribution. In single-photon emission computed tomography (\acrshort{gl:SPECT}) the radiopharmaceutical typically contains the \ce{^{99}Tc} isotope emitting \SI{140}{\kilo\electronvolt} radiation. A gamma camera is used to record the activity distribution. In positron emission tomography (\acrshort{gl:PET}) the radiopharmaceutical contains a $\beta^+$-emitter, typically \ce{^{18}F}. The two \SI{511}{\kilo\electronvolt} gamma quanta emitted back-to-back in the positron-electron annihilation process are registered in coincidence along the line of response. Many registered lines of response serve then as an input to reconstruct the three-dimensional distribution of activity in the patient's body \cite{VanEijk2003}.

The two crucial parameters of nuclear imaging modalities are their spatial resolution and sensitivity. Good spatial resolution is necessary for the precise determination of the emission point in the patient's tissues. Sensitivity reflects the number of useful events registered per unit of dose delivered to the patient. A higher sensitivity will result in a lower dose received by the patient for the same image quality. To improve the above two characteristics, the scintillator used in nuclear imaging devices should fulfill the following criteria: high density and \gls{gl:Zeff}, large light output, good energy resolution, and short decay constants. The majority of gamma cameras used in \acrshort{gl:SPECT} feature \ce{NaI}:{Tl} or \ce{CsI}:{Tl}. These inorganic crystals can be produced in large quantities with consistently high quality and are easy in mechanical processing. However, in recent years semiconductor detectors, such as \ce{GaAs}, \ce{CdZnTe} or \ce{CdTe} are gaining popularity in \acrshort{gl:SPECT} gamma cameras \cite{Lecoq2016}.

For \acrshort{gl:PET} scanners, the \ce{BGO} crystal has been the first choice for many years, as it had the largest \gls{gl:Zeff} out of all known scintillators. However, its main disadvantage is the long decay constant. The next generation of \acrshort{gl:PET} scanners featured phoswich detectors, resulting in a significant improvement in spatial resolution and sensitivity. A phoswich detector consists of two different scintillating materials coupled together. The difference in the decay constants allows for distinguishing between events occurring in the two volumes. The phoswich detectors are particularly useful for low-rate measurements in the presence of a background \cite{Tsoulfanidis}. Advances in inorganic crystal development resulted in a variety of new materials, many of which turned out to be suitable for \acrshort{gl:PET} imaging, \eg \ce{LSO}:{Ce}, \ce{LuAP}:{Ce}, \ce{GSO}:{Ce} or \ce{LGSO}:{Ce}. Currently, the majority of modern \acrshort{gl:PET} scanners are based on \ce{LSO}:{Ce} or its derivatives, as those materials have short decay times and large light output \cite{Lecoq2016}.

On the other hand, it needs to be noted that organic scintillators are gaining attention in the field of medical imaging, in particular \acrshort{gl:PET}. The proposed J-PET system featuring long panels of plastic scintillators may serve as an example here. Such design is cost-effective, and allows for whole-body scans with reduced dose to the patient. Excellent timing properties of the plastic scintillators allow for time-of-flight measurements, which serve as additional information included in the image reconstruction, and allow one to obtain better-quality images \cite{Moskal2020}.

\subsection{Proton therapy monitoring}
\label{subsec:scintillators-for-proton-therapy-monitoring}

As discussed in \cref{sec:pt-monitoring} and \cref{sec:sifi-cc}, scintillators are also considered for the detectors exploiting the \acrshort{gl:PG} radiation for real-time monitoring of proton therapy. Typically, the energy of the \acrshort{gl:PG} radiation emitted during patient treatment reaches up to \SI{10}{\mega\electronvolt}, with the focus on the \SIrange{2}{7}{\mega\electronvolt} range. Therefore, the energy of the detected radiation is significantly higher than that of standard medical imaging modalities. The design of a medical-purpose detector operating in this energy range with the desired performance is a challenging task, as the detection efficiency decreases with increasing energy. Additionally, for Compton cameras, such as the proposed \acrshort{gl:SiFi-CC} detector, there are additional requirements specific to the geometry and the principle of operation. The criteria which should be met by the scintillating material that constitutes the active part of the \acrshort{gl:SiFi-CC} detector to achieve the required performance can be summarized as follows:
\begin{description}[wide=0\parindent]

\item[Large density and \Zeff] - to ensure sufficient detection efficiency for the \acrshort{gl:PG} radiation emitted during proton therapy.

\item[Large light output] - light output is one of the crucial requirements for the scintillator. Large light output is desired since it will result in an improved energy resolution.

\item[Good energy resolution] - the principle of operation of a Compton camera consists in the calculation of the Compton scattering angle based on the registered energy deposits. Therefore, the energy resolution of the detector has a direct influence on the quality of the obtained images. 

\item[Short decay constant] - the decay constant of the chosen scintillating material should not exceed \SI{1}{\micro\second}. Additionally, the afterglow is not desired. It will allow suppressing the pile-up events. This feature is crucial in particular when operating in a high-radiation environment \cite{Kasper2020}. 

\item[Good time resolution] - the operation principle of a Compton camera requires building coincidences between the signals recorded at the scintillating fiber ends as well as between the two detector modules - the scatterer and the absorber. Good timing resolution, preferably below \SI{1}{\nano\second}, will allow to impose a shorter coincidence window and thus reduce the number of accidental coincidences. 

\item[Sufficient attenuation length] - the proposed \acrshort{gl:SiFi-CC} detector relies on the detection of the two correlated signals at both ends of the elongated scintillating fiber. To achieve this, the chosen scintillating material should be relatively transparent to its own scintillating light. The parameter that describes the propagation of scintillating light in the material is the attenuation length (see \cref{sec:light-propagation}). The larger the attenuation length, the farther the scintillation light reaches in the material. At the same time, a large attenuation length results in poor position resolution. Therefore, a material with the attenuation length at which both the energy and the position of the interaction can be reasonably reconstructed needs to be found. 

\item[Mechanical properties] - should allow machining and cutting into the desired shape. Unfortunately, many inorganic crystal scintillators are brittle, making them impossible to cut to form thin fibers and polish the surfaces without causing damage. 

\item[Spectral match with the \acrshort{gl:SiPM}s] - the emission spectrum of the chosen scintillating material should be compatible with the photosensitivity of the photodetector with which it will be coupled to maximize the photo-detection efficiency.

\item [Non-hygroscopicity] - hygroscopic scintillators require a dedicated casing to protect them from humidity in the environment. This makes them difficult to handle. Moreover, the casing introduces additional dead space in the detector volume. This is particularly inexpedient in a situation where the detector consists of many small elements, as in the case of the \acrshort{gl:SiFi-CC} detector. If each scintillating element required individual casing, the ratio of active to dead material would be unfavorable.

\item[No internal activity] - many of the heavy inorganic scintillators developed in recent years contain rare-earth elements, such as lutetium, yttrium, or gadolinium. Natural lutetium contains two isotopes: \ce{^{175}Lu} (abundance \SI{97.41}{\percent}) and \ce{^{176}{Lu}} (abundance \SI{2.59}{\percent}). The latter isotope is radioactive and decays via $\beta^-$ decay with the half-life of \SI{3.78e10}{\year} \cite{Rawool-Sullivan2006}. \Cref{fig:Lu-decay} (left) presents the decay scheme of \ce{^{176}Lu} and \cref{fig:Lu-decay} (right) presents its gamma energy spectrum. The intrinsic radioactivity of the scintillator contributes to the background. It is particularly problematic when operating at low event rates. 

\begin{figure}[htbp]
\centering
\includegraphics[width=.39\textwidth]{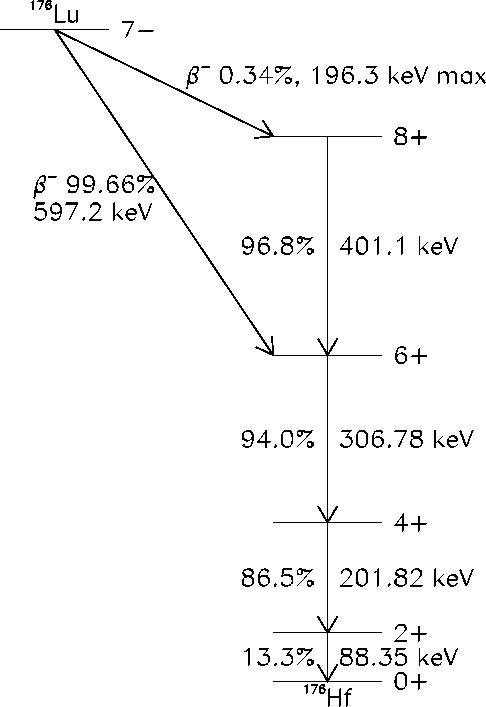}
\includegraphics[width=0.59\textwidth]{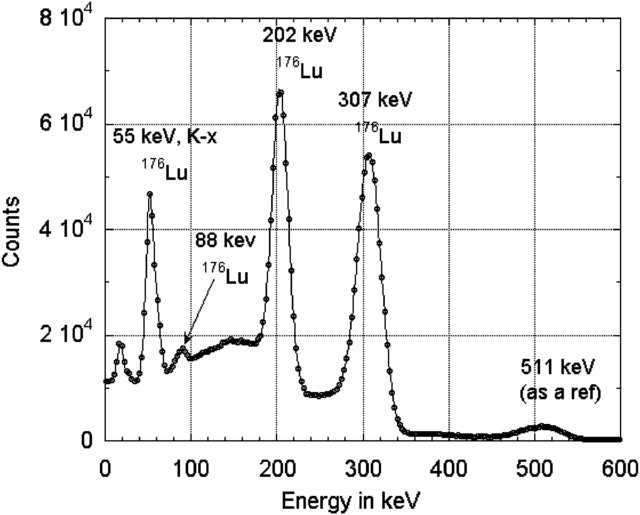}
\caption{Left: decay scheme of the \ce{^{176}Lu}. The scheme comes from \cite{Rawool-Sullivan2006}. Right: gamma energy spectrum of \ce{^{176}Lu}. The picture comes from \cite{Eriksson2005}.}
\label{fig:Lu-decay}
\end{figure}

\item[Affordable price and availability] - those aspects are non-negligible for devices with commercialization potential.
    
\end{description}

In order to achieve the required performance of the proposed SiFi-CC detector, it is important to fulfill the listed criteria. Based on the above list, three candidates for the active part of the detector were chosen: \ce{LYSO}:\ce{Ce}, \ce{LuAG}:\ce{Ce} and \ce{GAGG}:\ce{Ce}. The properties of the selected materials, as reported by the producers and in the literature, are presented in \cref{tab:materials-literature}. \Cref{chap:single-fibers} contains the description of the systematic study of the three materials, leading to the final decision on which one of them will be used for the detector construction.


\section{Light propagation in scintillators}
\label{sec:light-propagation}

Scintillating light produced during the interaction of ionizing radiation in the scintillator is emitted isotropically. Then it is attenuated as it propagates through the material. The attenuation of the scintillating light in large-bulk scintillators is exponential. However, for small and elongated scintillators, a number of factors, such as geometrical effects, surface roughness, wrapping, or coating, influence the propagation of scintillating light. An accurate description of this process is crucial for the reconstruction of the energy deposit and the position of the interaction in the scintillator. In the presented work, two models of light attenuation in the scintillator are considered, as presented further in this section. Details of the models, including mathematical formalism, were also described in \cite{Rusiecka2021}.


\subsection{Exponential light attenuation model (ELA)}
\label{ssec:ELA-model}

The most straightforward method for describing the propagation of scintillating light in the scintillator assumes simple exponential attenuation \cite{Pauwels2017, Lan2019}. According to this approach, henceforth referred to as the exponential light attenuation model (\acrshort{gl:ELA}), the signals recorded on the left $S_\textrm{l}(x)$ and the right $S_\textrm{r}(x)$ side of the scintillator can be expressed as follows: 
\begin{equation}
\label{eq:ELA}
    \begin{cases}
    S_{\textrm{l}}(x) = \xi_{\textrm{l}} S_{\textrm{0}} \exp{\left( \frac{-x}{\lambda} \right)} \\
    S_{\textrm{r}}(x) = \xi_{\textrm{r}} S_{\textrm{0}} \exp{\left( \frac{-(L-x)}{\lambda} \right)} \ .
    \end{cases}
\end{equation}
The parameter $\lambda$ represents the attenuation length, i.e. the distance in the material after which the intensity of scintillating light drops by a factor of $e$. The parameter $x$ represents the position of the interaction, $L$ is the total length of the fiber, and $S_\textrm{0}$ is the intensity of the signal at the point of interaction. $\xi_\textrm{l}$ and $\xi_\textrm{r}$ are the light transmission coefficients that are associated with the quality of the coupling at the corresponding fiber ends. It should be noted that the values of $S_\textrm{0}$ and $\xi_\textrm{i}$ are impossible to resolve based on the experimental data. Therefore, their product is determined as an effective value.

An important assumption of the \cref{eq:ELA} model is the homogeneity of the scintillating fiber, \ie the lack of defects or doping gradient. In such a situation, light propagates in a similar way in both directions along the fiber. Consequently, light attenuation is direction-independent and a single $\lambda$ parameter is sufficient to describe the propagation of scintillating light towards both ends of the scintillator at the same time.

\begin{figure}[htbp]
\centering
\includegraphics[width=.55\textwidth]{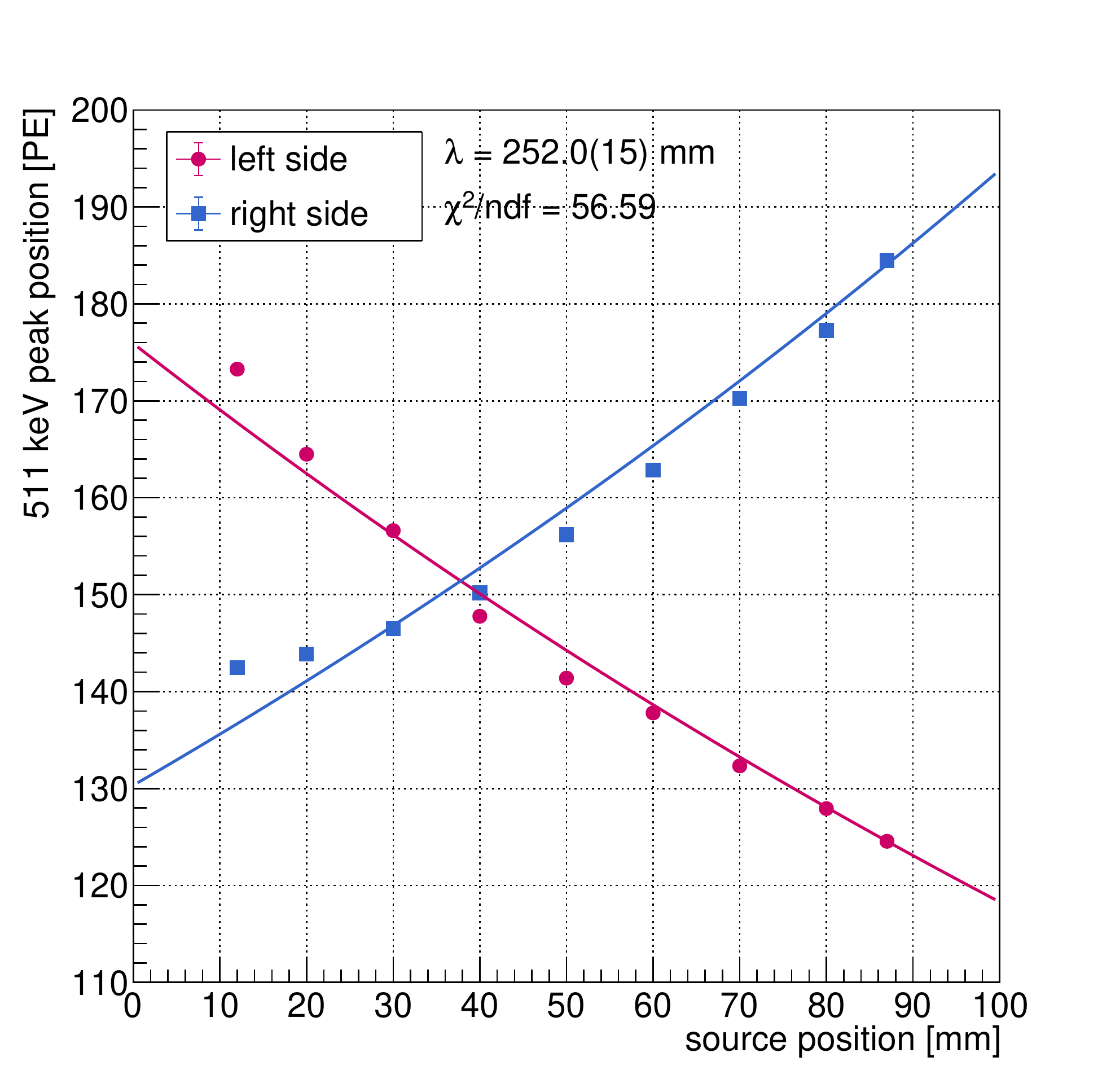}
\caption{An example of the ELA model fitted to the experimental points (error bars are contained within the point markers). The determined values of the attenuation length $\lambda$ and \chiNDF are listed. Presented data comes from the series~109 (see \cref{tab:measurements}). Picture adapted from \cite{Rusiecka2021}.}
\label{fig:ELA-fit}
\end{figure}

The attenuation length is determined based on the attenuation curve, i.e. the dependence of the intensity of the recorded signal $S_\textrm{l}$, $S_\textrm{r}$ on the position of the interaction $x$. In the presented work, the position of the \anhpeak peak in the recorded charge spectrum (denoted as \gls{gl:peakpos}) is a measure of signal intensity. Moreover, the position of the interaction in the scintillator is approximated by the position of the radioactive source placed along the scintillating fiber during the measurement. The attenuation curves are plotted independently for both fiber ends. The model equations \cref{eq:ELA} are then fitted simultaneously to the two data sets to obtain a single set of parameters (\cref{fig:ELA-fit} left).

\begin{figure}[hp]
\centering
\includegraphics[width=.99\textwidth]{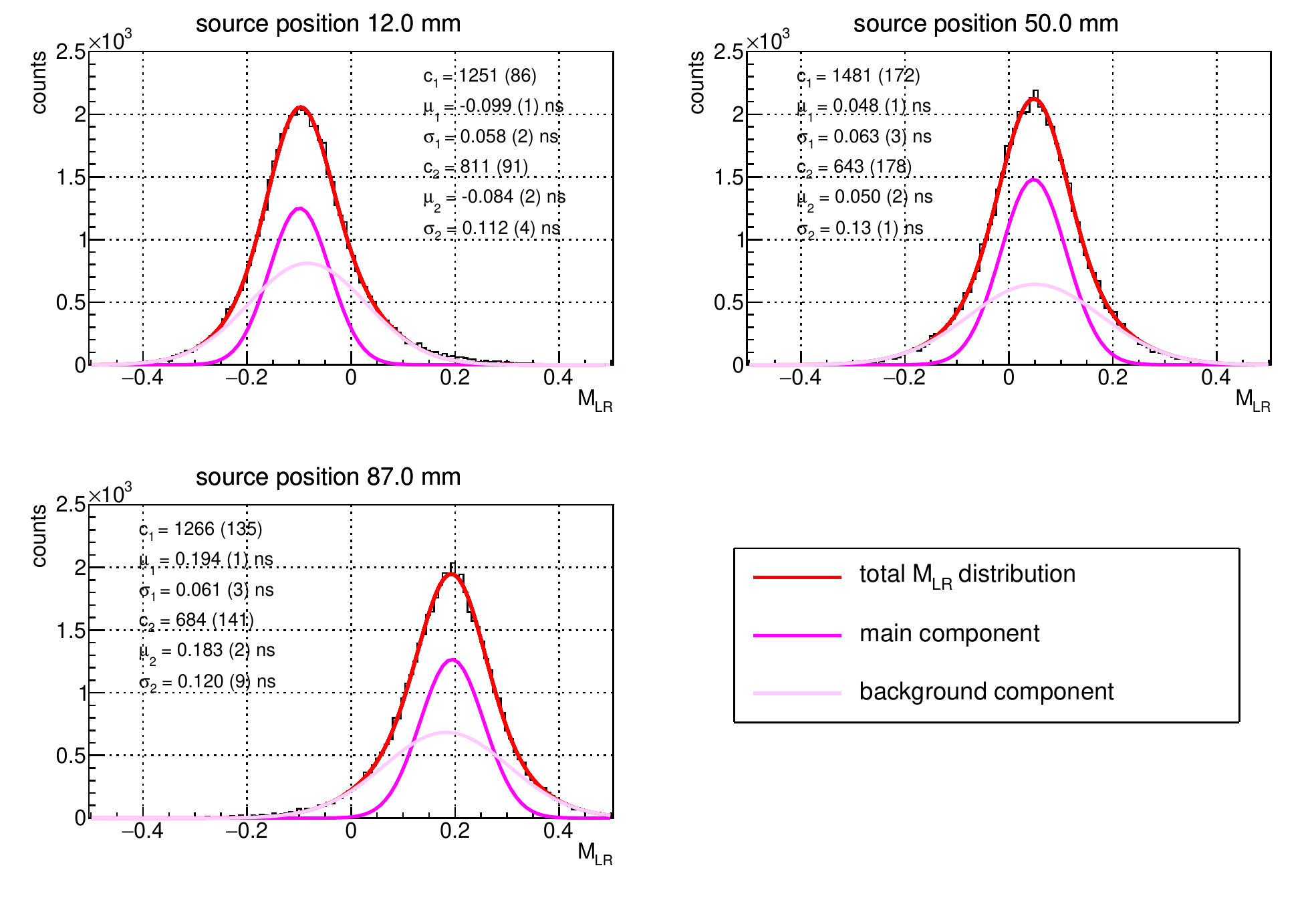}
\caption{Examples of \gls{gl:MLR} distributions for three different source positions. The sum of two Gaussian functions was fitted to each distribution. The mean value of \gls{gl:MLR} was determined as the mean of the main (narrower) component. Presented data comes from the series~109 (see \cref{tab:measurements}).}
\label{fig:MLR-examples}
\end{figure}

The description of the light propagation process can be further simplified using the quantity \gls{gl:MLR}, which combines the signals recorded at both ends of the fiber \cite{Pauwels2013}: 
\begin{equation}
\label{eq:MLR}
    M_{\textrm{LR}}(x) = \ln{\left( \sqrt{\frac{S_{\textrm{r}}(x)}{S_{\textrm{l}}(x)}} \right)} \ . 
\end{equation}
The quantity \MLR depends linearly on the position of the interaction $x$. From equations \cref{eq:ELA} and \cref{eq:MLR} the slope $a$ and the offset $b$ of the linear function can be derived: 
\begin{equation}
\label{eq:MLR-coefficients}
    \begin{cases}
    a = \frac{1}{\lambda} \\
    b = - \frac{L/2}{\lambda} + \ln{\left( \sqrt{\frac{\xi_{\textrm{r}}}{\xi_{\textrm{l}}}} \right)} \ . 
    \end{cases}
\end{equation}
In this work, the value of \gls{gl:MLR} has been calculated event by event for all recorded signal pairs. Examples of the \gls{gl:MLR} distributions for three different source positions are shown in \cref{fig:MLR-examples}. The presented distributions are fitted with the sum of two Gaussian functions corresponding to the main component and the background component. The main component originates from the annihilation peak, while the background component stems from $\gamma$ rays which underwent scattering in various elements of the experimental setup.
If the continuum is further suppressed in the measurement, \eg by the hardware coincidence tuning, a single Gaussian function is sufficient to describe the distribution. 

\begin{figure}[htbp]
\centering
\includegraphics[width=.55\textwidth]{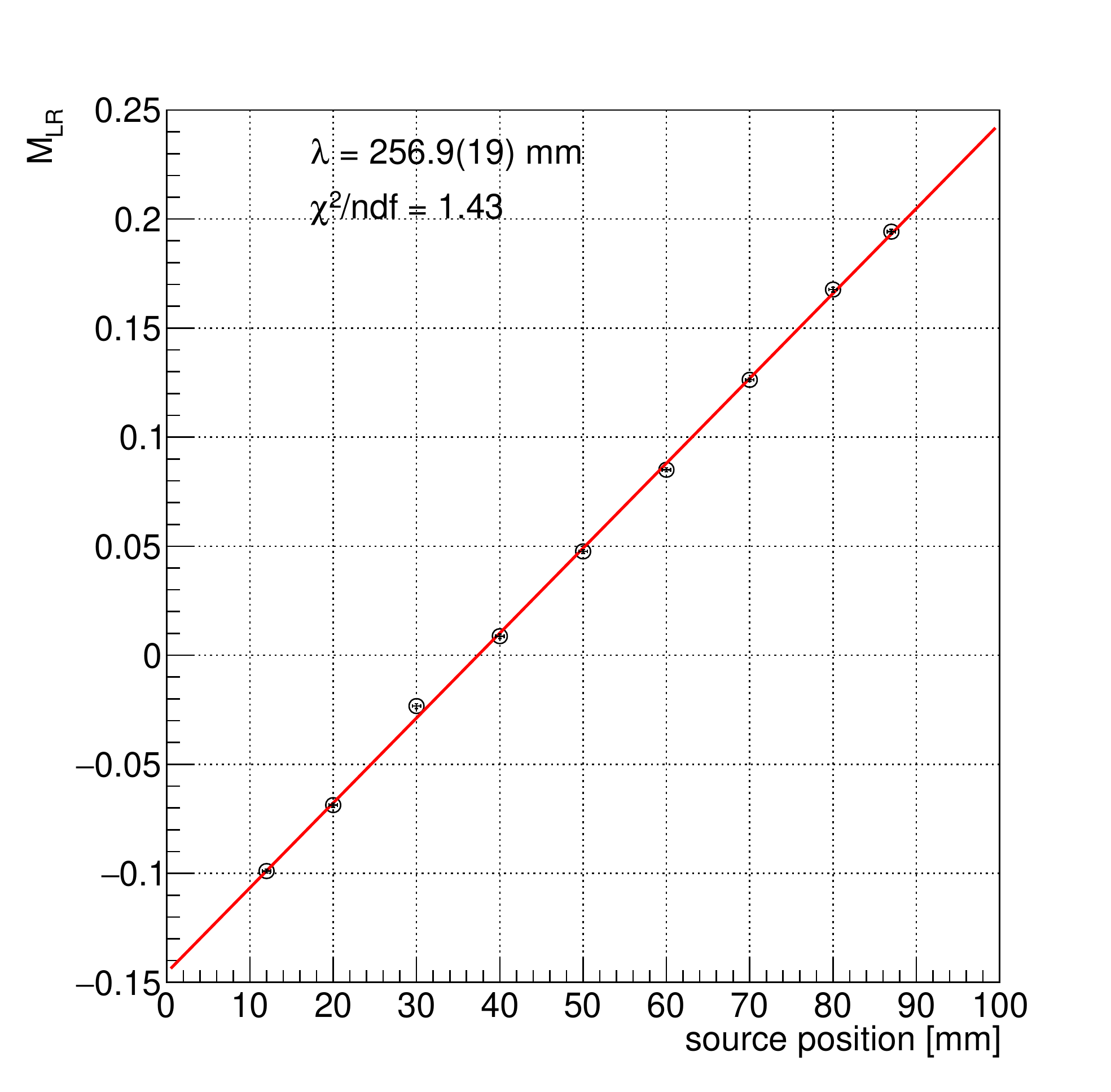}
\caption{Example of the \MLRx dependence along with the linear function fitted. The determined values of the attenuation length $\lambda$ and \chiNDF are listed. Presented data comes from the series 109 (see \cref{tab:measurements}) and corresponds to \cref{fig:ELA-fit}. Picture adapted from \cite{Rusiecka2021}.}
\label{fig:MLR-fit}
\end{figure}

The mean value of the obtained \gls{gl:MLR} distribution or a mean value of its main component is then plotted against the position of the radioactive source along the scintillating fiber, creating an \gls{gl:MLR} curve. A linear fit of the \gls{gl:MLR} curve allows to determine the attenuation length value of the investigated scintillating fiber (see \cref{fig:MLR-fit}).

\cref{fig:ELA-fit} and \cref{fig:MLR-fit} show a comparison of the \acrshort{gl:ELA} and \gls{gl:MLR} methods applied to the same set of experimental data. Based on the \chiNDF values for both approaches, it can be stated that the quality of the \gls{gl:MLR} fit is significantly better. At the same time, the attenuation length values obtained are similar and agree within $3\sigma$. This may result from the fact that in the \gls{gl:MLR} ratio some non-exponential components cancel.


\subsection[Exponential light attenuation model with light reflection (ELAR)]{Exponential light attenuation model with light reflection \\ (ELAR)}
\label{ssec:ELAR}

\begin{figure}[htbp]
\centering
\includegraphics[width=.80\textwidth]{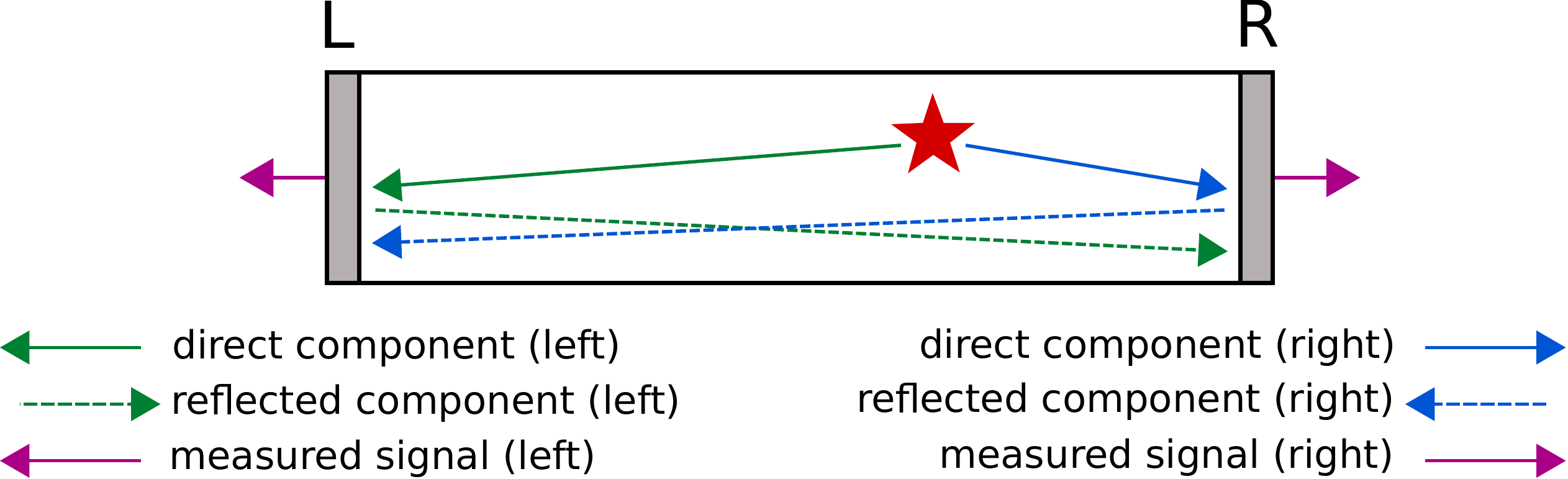}
\caption{A simplified scheme of the light propagation in the scintillating fiber according to the \acrshort{gl:ELAR} model. In reality, most of the scintillating light undergoes multiple reflections on the fiber surfaces. The star represents the point of interaction. Adapted from \cite{Rusiecka2021}.}
\label{fig:ELAR-fiber-scheme}
\end{figure}

The poor quality of the \acrshort{gl:ELA} model fit (\cref{fig:ELA-fit}) suggests that it does not accurately describe the light propagation in thin scintillating fibers. Therefore, the exponential attenuation model with light reflection (\acrshort{gl:ELAR}), based on the works \cite{Lan2019} and \cite{Taiuti1996}, was derived and tested. In this approach, it is assumed that part of the scintillating light is reflected at the end of the fiber and subsequently propagates back to be registered at the opposite side. Therefore, the signal registered at the left end (L) stems from the light emitted towards this side (direct component) and from a fraction of light emitted to the right (R), which underwent reflection (reflected component). A simplified scheme of light propagation according to the \acrshort{gl:ELAR} model is depicted in \cref{fig:ELAR-fiber-scheme}. 

Similarly as in the \acrshort{gl:ELA} model, it is assumed that scintillating light propagates similarly in all directions and that light attenuation has exponential character. Thus, the direct components $P_\textrm{l}(x)$ and $P_\textrm{r}(x)$ reaching opposite ends of the scintillating fiber are defined as follows: 
%
\begin{equation}
\label{eq:ELAR-direct-component}
    \begin{cases}
    P_{\textrm{l}}(x) = S_\textrm{0} \exp{\left( \frac{-x}{\lambda} \right)} \\
    P_{\textrm{r}}(x) = S_\textrm{0} \exp{\left( \frac{-(L-x)}{\lambda} \right)}.
    \end{cases}
\end{equation}
The reflected components $R_\textrm{l}(x)$ and $R_\textrm{r}(x)$ registered at both ends can be written as: 
\begin{equation}
\label{eq:ELAR-reflected-component}
    \begin{cases}
    R_{\textrm{l}}(x) = \eta_{\textrm{r}} P_{\textrm{r}}(x) \exp{\left( \frac{-L}{\lambda} \right)} \\
    R_{\textrm{r}}(x) = \eta_{\textrm{l}} P_{\textrm{l}}(x) \exp{\left( \frac{-L}{\lambda} \right)}.
    \end{cases}
\end{equation}
The parameters $\eta_\textrm{l}$ and $\eta_\textrm{r}$ represent the fraction of light that is reflected from the given end of the fiber. 
%
The total registered signals $S_\textrm{l}(x)$ and $S_\textrm{r}(x)$ are sums of the corresponding direct and reflected components, assuming no additional light losses, such that the fractions of the corresponding direct and reflected components sum up to unity:
\begin{equation}
\label{eq:ELAR-full}
    \begin{cases}
     S_\textrm{l}(x) = \xi_\textrm{l} S_{\textrm{0}} \left[ (1-\eta_\textrm{l}) \exp{\left( \frac{-x}{\lambda} \right)} + \eta_\textrm{r} \exp{\left( \frac{-2L+x}{\lambda} \right)} \right]\\
     S_\textrm{r}(x) = \xi_\textrm{r} S_{\textrm{0}} \left[ (1-\eta_\textrm{r}) \exp{\left( \frac{-L+x}{\lambda} \right)} + \eta_\textrm{l} \exp{\left( \frac{-L-x}{\lambda}\right)} \right].
    \end{cases}
\end{equation}
Similarly to the \acrshort{gl:ELA} model, it is not possible to resolve the values of $S_\textrm{0}$ and $\xi_\textrm{l/r}$ based on the experimental data. Therefore, the following parameterization was introduced to eliminate ambiguity and reduce the number of parameters:
\begin{equation}
\label{eq:ELAR-transformation}
    \begin{cases}
    \eta_\textrm{r}' = \frac{\eta_\textrm{r}}{1-\eta_\textrm{l}} \\
    \eta_\textrm{l}' = \frac{\eta_\textrm{l}}{1-\eta_\textrm{r}},
    \end{cases}
    \qquad\qquad
    \begin{cases}
    S_\textrm{0}' = \xi_\textrm{l} S_\textrm{0} (1-\eta_\textrm{l}) \\
    \xi = \frac{\xi_\textrm{r} (1-\eta_\textrm{r})}{\xi_\textrm{l}(1-\eta_\textrm{l})}.
    \end{cases}
\end{equation}
Then, the total signals are expressed as follows:
\begin{equation}
\label{eq:ELAR-full-simplified}
    \begin{cases}
     S_\textrm{l}(x) = S_\textrm{0}' \left[ \exp{\left( \frac{-x}{\lambda} \right)} + \eta_\textrm{r}' \exp{\left( \frac{-2L+x}{\lambda} \right)} \right] \\
     S_\textrm{r}(x) = \xi S_\textrm{0}' \left[ \exp{\left( \frac{-L+x}{\lambda} \right) } + \eta_\textrm{l}' \exp{\left( \frac{-L-x}{\lambda} \right)} \right].
    \end{cases}
\end{equation}
In this representation, the coefficient $\xi$ accounts for the possible asymmetry in light transmission between the end of the scintillator and the photodetector caused by differences in the optical coupling. 

The \acrshort{gl:ELAR} model allows to determine the attenuation length $\lambda$, as well as reconstruct the direct light components $P_\mathrm{l}$ and $P_\mathrm{r}$ based on the measured signals according to the following formulas:
\begin{equation}
\label{eq:model-primary-reconstructed}
    \begin{cases}
    P_{l}^{*} (S_{l}, S_{r}) = \frac{\exp{\left( \frac{L}{\lambda} \right)} \left( \exp{\left( \frac{L}{\lambda} \right)} \xi S_{l} - S_{r} \eta_{r}' \right)}{\xi \left( \exp{\left( \frac{2L}{\lambda} \right)} - \eta_{l}' \eta_{r}' \right)} \\ \\
    P_{r}^{*} (S_{l}, S_{r}) = - \frac{\exp{\left( \frac{L}{\lambda} \right)} \left( - \exp{\left( \frac{L}{\lambda} \right) } S_{r} + \xi S_{l} \eta_{l}' \right)}{\xi \left( \exp{\left( \frac{2L}{\lambda} \right)} - \eta_{r}' \eta_{l}' \right)}.
    \end{cases}
\end{equation}
The symbols with $*$ superscript denote values reconstructed based on the \acrshort{gl:ELAR} model. For the reconstruction of direct components, the parameters of the \acrshort{gl:ELAR} model (\cref{eq:ELAR-full-simplified}) must be known. It is possible to determine them by fitting the model to the calibration measurement. Uncertainties of the reconstructed direct components are estimated taking into account the contribution from the model fit as well as the experimental data as follows: 
\begin{equation}
\label{eq:ELAR-primary-error-I}
    \sigma_{P_{\textrm{i}}^{*}} = \sqrt{\sigma^{2}_{fP_\textrm{i}^{*}} + \left( \frac{\partial P_{\textrm{i}}^{*} (S_{\textrm{l}}, S_{\textrm{r}})}{\partial S_{\textrm{l}}} \sigma_{S_{\textrm{l}}} \right)^{2} + \left( \frac{\partial P_\textrm{i}^{*}(S_{\textrm{l}}, S_{\textrm{r}})}{\partial S_{\textrm{r}}} \sigma_{S_{\textrm{r}}} \right)^{2} },
\end{equation}
where $i$ can be substituted with $l$ or $r$. $\sigma_{fP_{i}^{*}}^{2}$ represents the variance of $P_{i}^{*}$ and can be calculated with the matrix expression for error propagation \cite{Tellinghuisen2001}: 
\begin{equation}
\label{eq:ELAR-primary-error-II}
    \sigma_{fP_{\textrm{i}}^{*}}^{2} = \textbf{g}^{\text{T}} \textbf{V} \textbf{g}.
\end{equation}
\textbf{V} is a covariance matrix and \textbf{g} is a vector containing partial derivatives of $P_{\textrm{i}}^{*}$ over its parameters. The results of the \acrshort{gl:ELAR} fit revealed correlations of some of the parameters. Therefore, the full formula including also non-diagonal matrix elements is used in the calculations. The full derivation of the \acrshort{gl:ELAR} model along with the uncertainties calculus is presented in \cref{app:ELAR}.

\begin{figure}[ht]
\centering
\includegraphics[width=.99\textwidth]{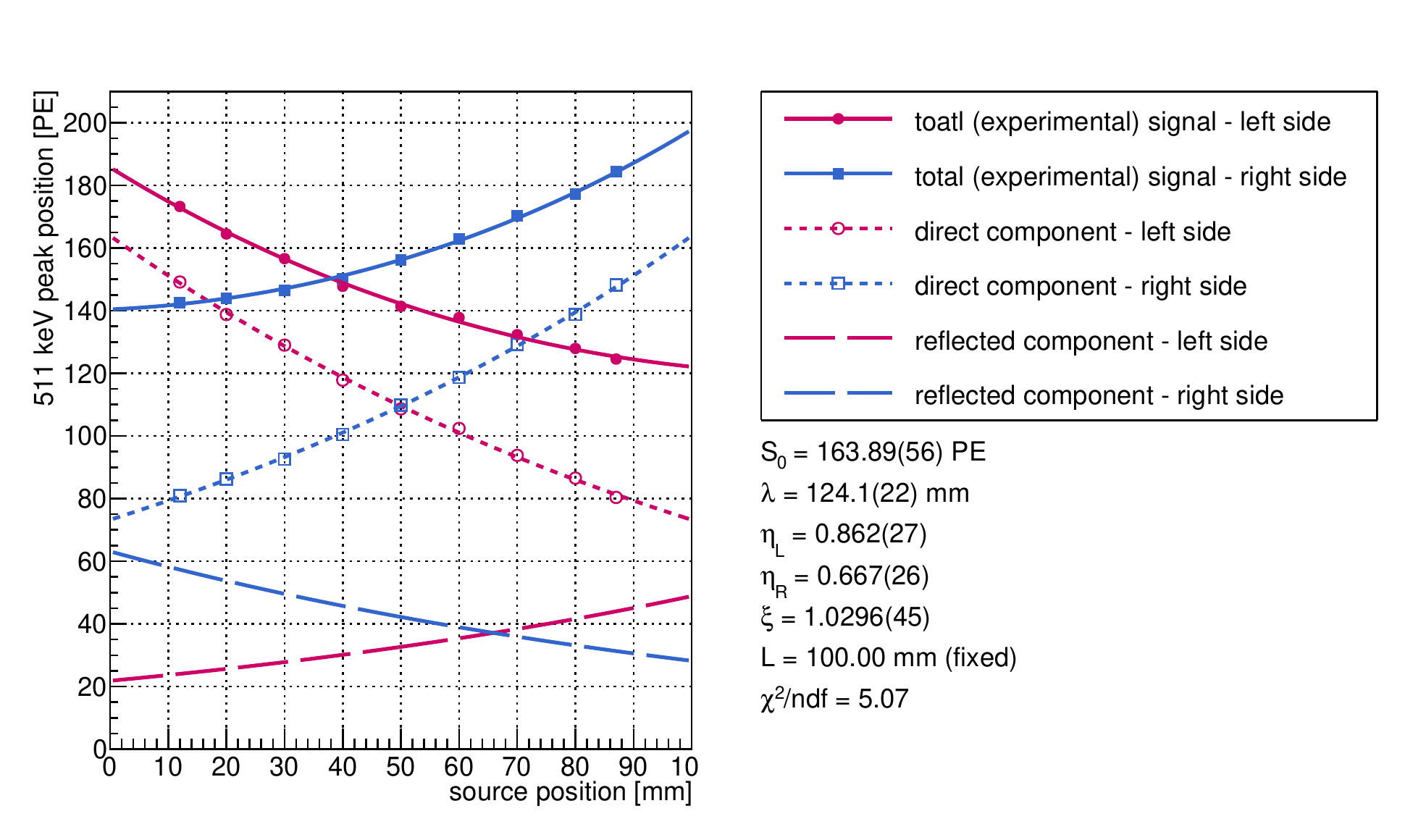}
\caption{Example of the \acrshort{gl:ELAR} model fitted to the experimental points. The attenuation curves were additionally decomposed into direct and reflected components. Based on the experimental data and fit results, the direct components for both sides of the investigated fiber were reconstructed. Presented data come from the series 109 (see \cref{tab:measurements}), just as that of \cref{fig:ELA-fit} and \cref{fig:MLR-fit}. Adapted from \cite{Rusiecka2021}.}
\label{fig:ELAR-fit}
\end{figure}

An example of the fit of the \acrshort{gl:ELAR} model to the experimental data is shown in \cref{fig:ELAR-fit}. The same set of experimental points as discussed in \cref{ssec:ELA-model} is presented here in order to compare the quality of the data description. Similarly as for the \acrshort{gl:ELA} model fit, both data sets are fitted simultaneously. Additionally, the attenuation curves were decomposed into direct and reflected components. Experimental points were also used to reconstruct the direct light component at each position of the radioactive source. The fit parameters are listed along with the value of \chiNDF. It can be observed that the proposed \acrshort{gl:ELAR} model accounting for the light reflection describes the experimental data better than the simple exponential model (\acrshort{gl:ELA}). Additionally, the reconstructed direct components reach the same intensity in the middle of the investigated fiber as expected, meaning that the model allows to correct for possible differences in coupling at the fiber ends.


\section{Characterization of scintillators}
\label{sec:charatcerization-scintillators}

Apart from the attenuation length, which describes light propagation, there are other properties that are important for characterizing the performance of the scintillator and the scintillating detector. The first of those properties is light production, which is specific to scintillating detectors. Another crucial property for position-sensitive detectors, such as the proposed SiFi-CC setup, is position resolution. Finally, energy resolution and timing properties are important for any type of ionizing radiation detector.

The following section discusses in detail the properties listed above and methods for determining them. 




\subsection{Light production}
\label{ssec:light-production}

The production of scintillating light is often characterized by a quantity called light output. It is defined as the amount of light produced at the place of the interaction by a \SI{1}{\mega\electronvolt} energy deposit. Therefore, it can be interpreted as an efficiency of conversion of the ionization energy to scintillating photons \cite{Leo}.

The light output can be determined based on the position of the chosen peak ($\mu_\textrm{E}$) in the charge spectrum expressed in photoelectrons (\acrshort{gl:PE}). The chosen peak should correspond to the known energy $E$. Moreover, the light output is a quantity that is already corrected for the attenuation of scintillating light ($\lambda$), light losses in the coupling ($C_\textrm{LL}$), photodetection efficiency ($C_\textrm{PDE}$), and crosstalk probability ($f_\textrm{CT}$) according to the following empirical formula \cite{Wronska2020}:
\begin{equation}
    LO = \frac{\mu_\textrm{E}}{E} \cdot \frac{(1 - f_\textrm{CT})}{C_\textrm{LL} C_\textrm{PDE} \exp{(x/\lambda)}}.
\end{equation}
Thus, it allows estimating the amount of produced light based on the amount of detected light. The light output is typically used to compare different scintillators in terms of their brightness. 

\begin{figure}[htbp]
\centering
\includegraphics[width=.99\textwidth]{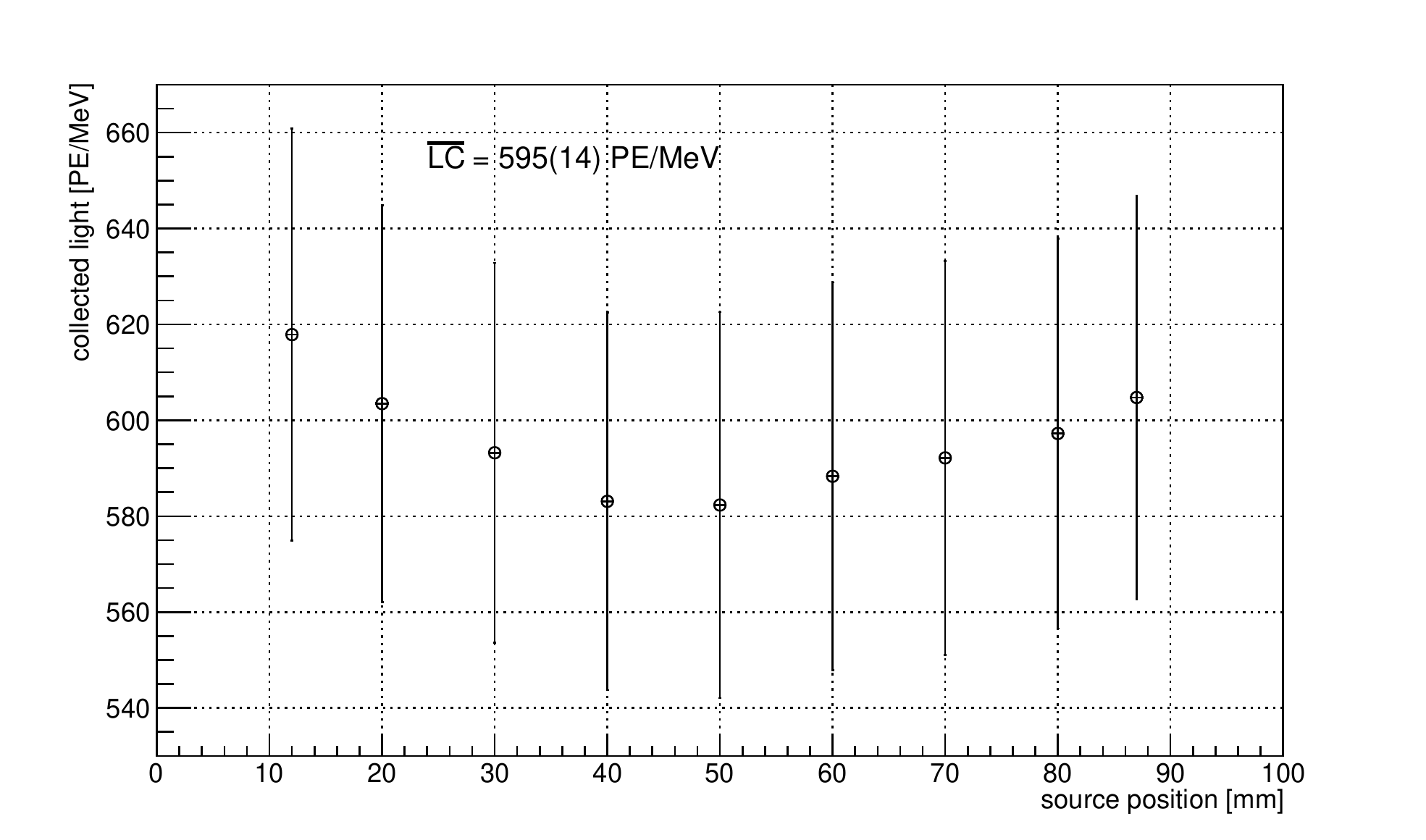}
\caption{Dependence of light collection on the position of the radioactive source along the scintillating fiber. The listed value is a weighted mean of all nine measurement points. Presented data come from series 109 (see \cref{tab:measurements}).}
\label{fig:light-collection-example}
\end{figure}

However, the performance of the \acrshort{gl:SiFi-CC} detector will depend on the amount of light detected by the system, rather than the amount of light produced by the scintillators alone. Therefore, for this study, a property called light collection was defined as follows: 
\begin{equation} 
\label{eq:light-collection} 
LC = \frac{\mu_{\anhpeak}}{\anhpeak}, 
\end{equation} 
where \peakpos is position of the annihilation peak. The values of light collection obtained for both ends of the fiber are summed to get the total characteristics of the investigated sample. Light collection quantifies the amount of scintillating light produced during the interaction of the \anhpeak $\gamma$ quanta which was registered by the experimental system, normalized per \SI{1}{\mega\electronvolt}. In contrast to light output, the light collection includes effects stemming from the quality of the applied coupling, light attenuation, and properties of the used electronics. Therefore, it is a more suitable variable for the presented study. Since the light collection is not compensated for the light losses as a result of attenuation, its value may vary with the source position along the fiber. Therefore, to characterize the investigated sample, a weighted mean of light collections obtained in all measurements in a single position scan, \ie a series, is calculated. \Cref{fig:light-collection-example} shows an example of a light collection measurement for a typical \acrshort{gl:LYSO:Ce} scintillating fiber.

\subsection{Position resolution}
\label{ssec:position-resolution}

The precise determination of the position of the interaction point in the position-sensitive detectors is crucial for their performance. The hit position is often reconstructed based on the time difference between two correlated signals registered at both ends of an elongated detector. However, in the case of the investigated scintillating fibers, the obtained timing resolution is insufficient (see \cref{ssec:timing-properties}). Therefore, an alternative approach exploiting the ratio of charges carried by the signals registered at both ends of the fiber was used. This method of position reconstruction is related to the description of light propagation in scintillators and the attenuation of scintillating light (\cref{sec:light-propagation}). In the presented work, two models describing the propagation of scintillating light are presented, yielding the corresponding two methods for position reconstruction. 

\begin{figure}[htbp]
\centering
\includegraphics[width=\linewidth]{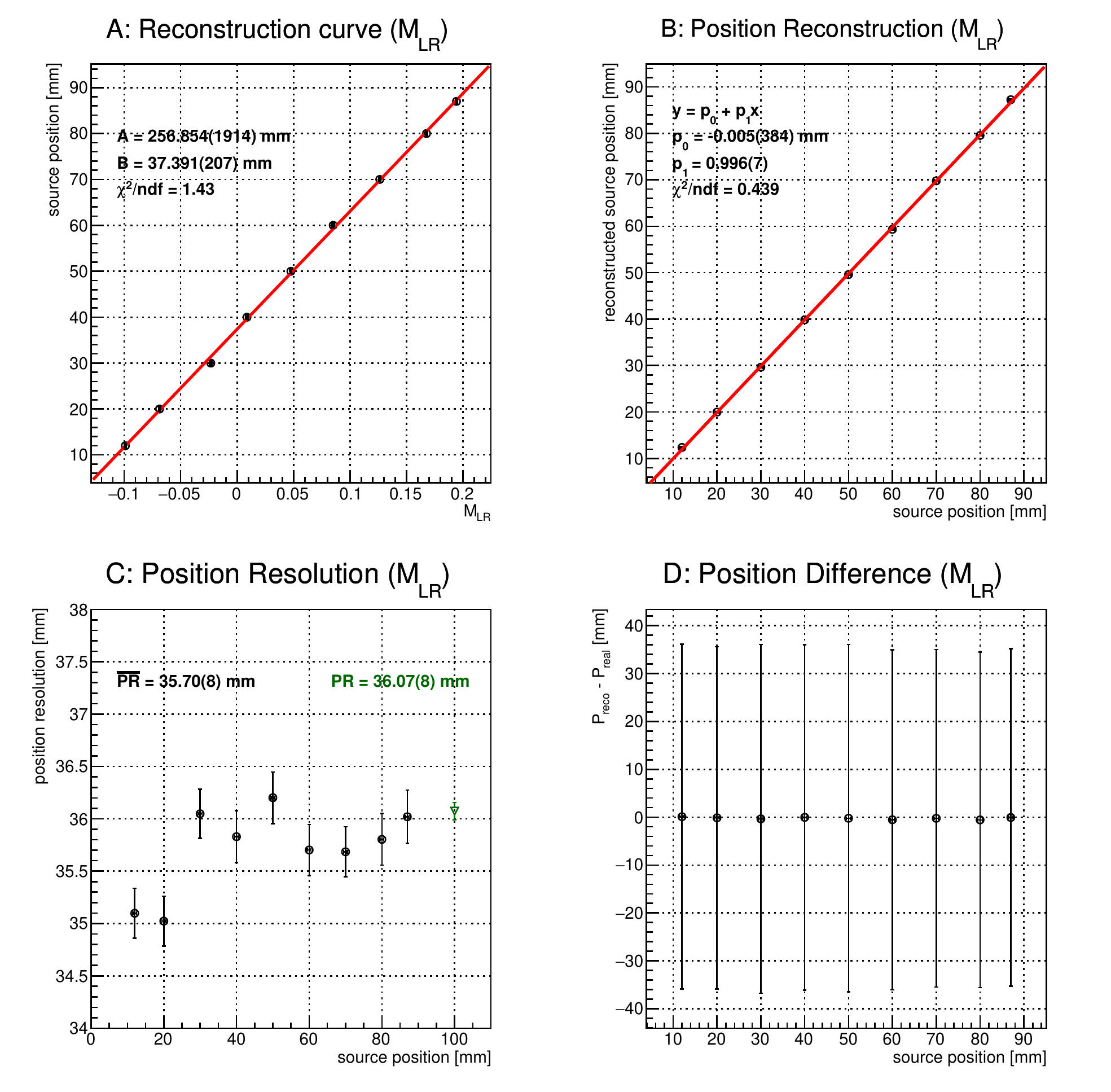}
\caption{Summary of position reconstruction with the \gls{gl:MLR} method. A - calibration curve; B - reconstructed interaction position versus the true position of the radioactive source; C - obtained position resolution for different positions of the radioactive source (black points) and integrated position resolution (green triangle). The average position resolution ($\overline{PR}$) and the integrated position resolution ($PR$) are listed;  D - position reconstruction residuals. The vertical error bars represent the position resolution obtained for the corresponding measurement. Presented data come from the series 109 (see \cref{tab:measurements}).}
\label{fig:position-reco-mlr}
\end{figure}

\begin{figure}[htbp]
\centering
\includegraphics[width=.44\textwidth]{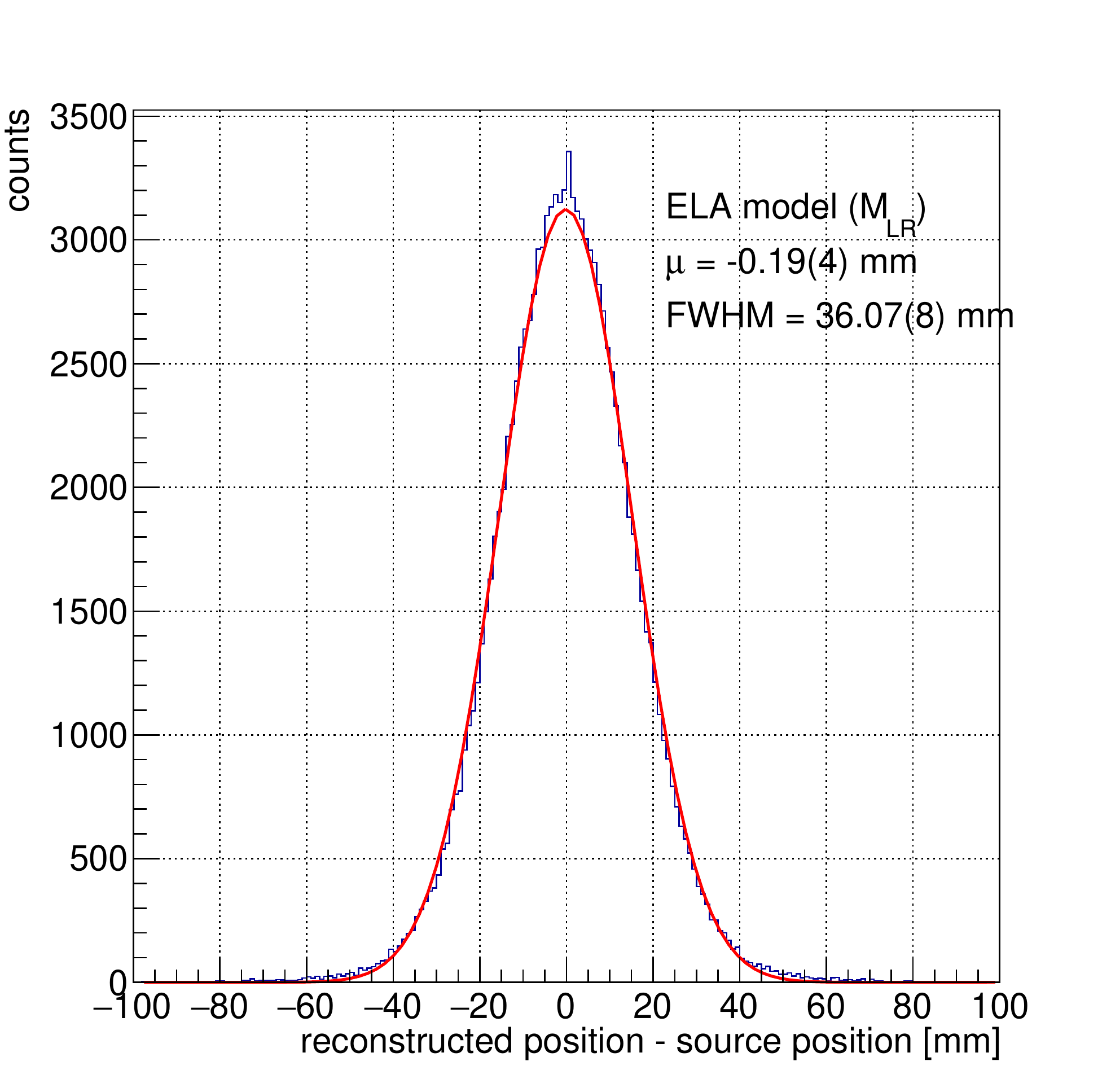}
\includegraphics[width=.44\textwidth]{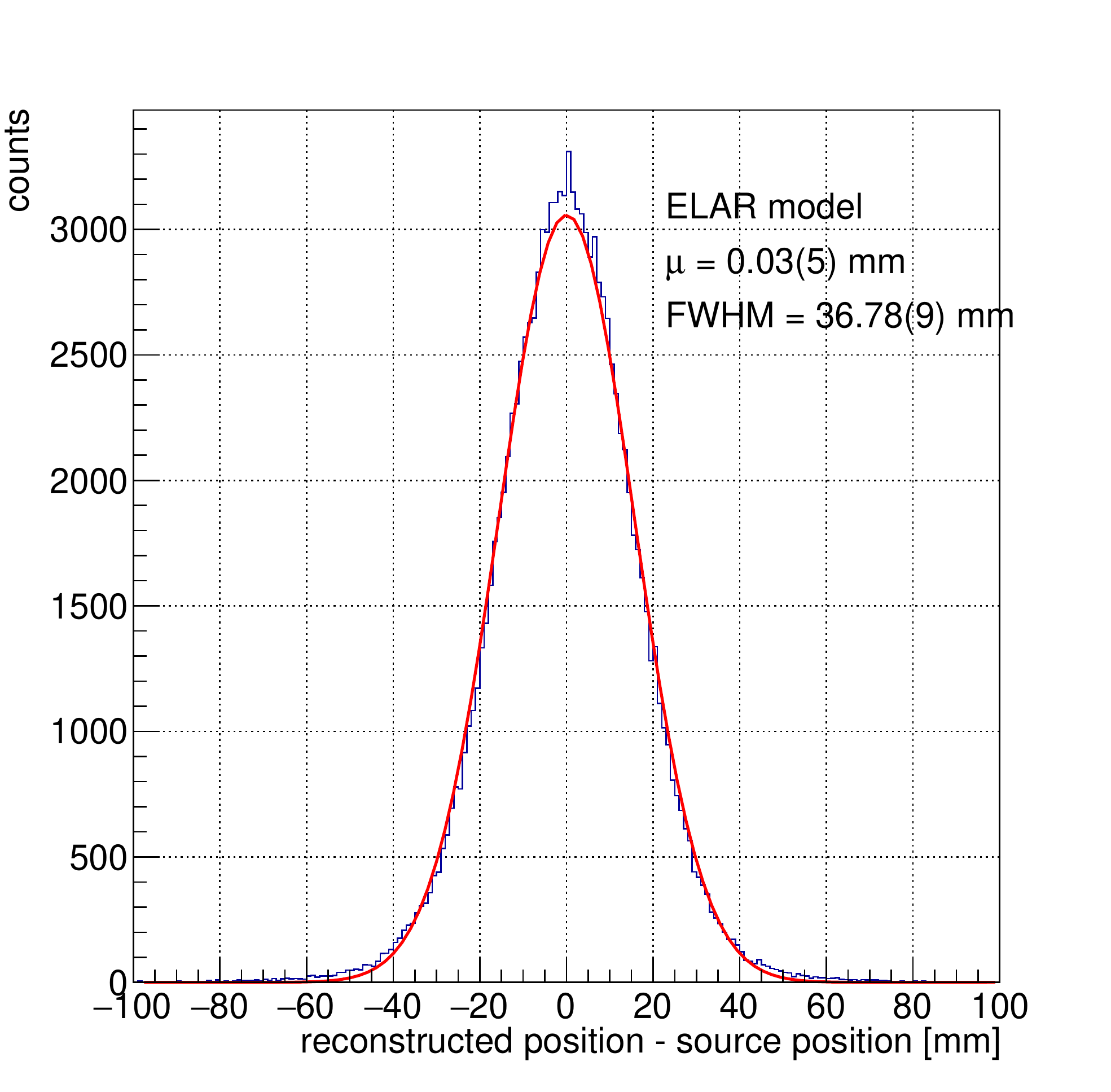}
\caption{Distributions of the residuals of the reconstructed position \gls{gl:Xreco} $-$ \gls{gl:Xreal} for all measurements in the series. Left: position reconstruction was performed with the \gls{gl:MLR} method. Right: position reconstruction was performed using data recalculated according to the \acrshort{gl:ELAR} model. Presented data comes from the series 109 (see \cref{tab:measurements}).}
\label{fig:position-reco-dist-examples}
\end{figure}

The first method of position reconstruction uses previously defined \gls{gl:MLR} quantity (\cref{eq:MLR}). In the first step, the dependence of the known source position $X_\textrm{real}$ on mean \gls{gl:MLR} is plotted and parameterized with a linear function to serve as a calibration curve (\cref{fig:position-reco-mlr} A). Subsequently, the positions of the interactions are reconstructed event-by-event. Only events forming the annihilation peak are taken into account for position reconstruction. The obtained position distribution is fitted with the Gaussian function. This allows to extract the source position in the measurement $X_\textrm{reco}$ (mean of the distribution) and the position resolution (\acrshort{gl:FWHM} of the distribution). An example of reconstructed position distributions for all measurements in a chosen experimental series is presented in \cref{app:energy-position-reconstruction}, \cref{fig:position_reconstruction_all_mlr}. The dependence of the reconstructed interaction position on the true position of the radioactive source is shown in \cref{fig:position-reco-mlr} B. \Cref{fig:position-reco-mlr} D shows that the differences between the reconstructed and expected interaction positions are negligible compared to the obtained position resolution. The distribution of residuals \gls{gl:Xreco} $-$ \gls{gl:Xreal} for all measurements in the series is additionally investigated, providing the integrated position resolution for the examined fiber (see \cref{fig:position-reco-dist-examples}). Position resolutions for all measurements in the experimental series, as well as integrated position resolution, are presented in \cref{fig:position-reco-mlr} C.

\begin{figure}[htbp]
\centering
\includegraphics[width=\linewidth]{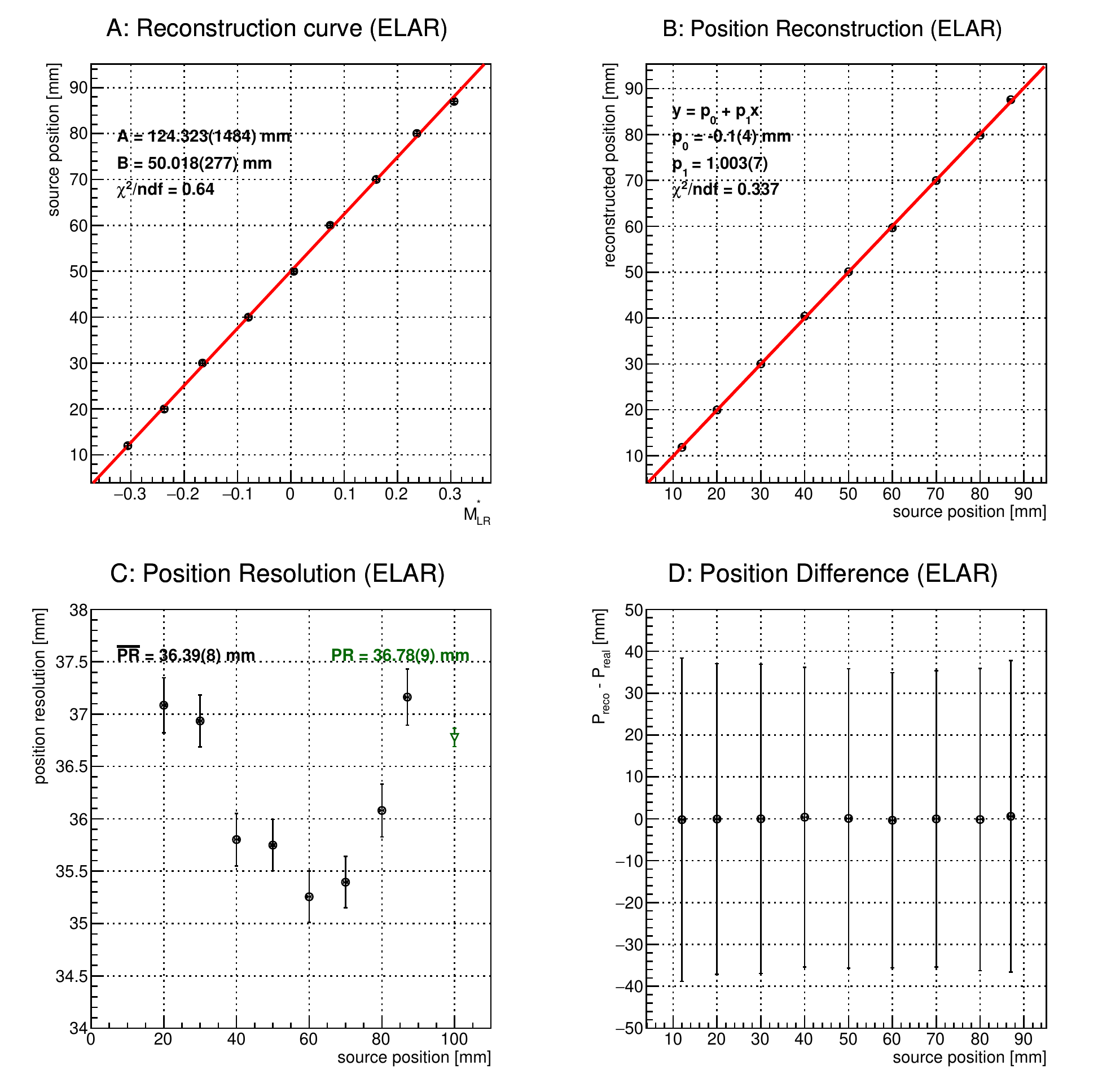}
\caption{Summary of position reconstruction using data recalculated according to the \acrshort{gl:ELAR} model. A - calibration curve; B - reconstructed interaction position versus the true position of the radioactive source; C - obtained position resolution for different positions of the radioactive source (black points) and integrated position resolution (green triangle). The average position resolution ($\overline{PR}$) and the integrated position resolution ($PR$) are listed;  D - position reconstruction residuals. The vertical error bars represent the position resolution obtained for the corresponding measurement. Presented data come from the series 109 (see \cref{tab:measurements}).}
\label{fig:position-reco-elar}
\end{figure}

Taking advantage of the exponential attenuation model that accounts for light reflection (\acrshort{gl:ELAR}), the second method of position reconstruction was proposed. In this method, the experimental data are recalculated using \cref{eq:model-primary-reconstructed} and the quantity \gls{gl:MLRstar} is defined as the ratio of direct light reaching the ends of the fiber:
\begin{equation}
\label{eq:MLR-corrected}
    M_\textrm{LR}^{*} = \ln{\left( \sqrt{\frac{P_\textrm{r}^{*}(S_\textrm{l}, S_\textrm{r})}{P_\textrm{l}^{*}(S_\textrm{l}, S_\textrm{r})}} \right)}.
\end{equation}
Similarly as in the previous method, known source position $X_\textrm{real}$ is plotted as a function of the mean value of the \MLRstar distribution. A linear function fitted to this dependency serves as a calibration curve for position reconstruction. Otherwise, the procedure is the same as in the \MLR-based approach. \Cref{fig:position-reco-elar} shows subsequent steps and results of position reconstruction: calibration curve (A), reconstructed source positions (B) and their residuals (D), as well as obtained position resolution (C). An example of reconstructed position distributions for all measurements in a chosen experimental series is presented in \cref{app:energy-position-reconstruction}, \cref{fig:position_reconstruction_all_elar}. For this approach, the integrated position resolution is also determined (\cref{fig:position-reco-dist-examples}, right).

Comparison of \cref{fig:position-reco-mlr} and \cref{fig:position-reco-elar} as well as cumulative $X_\textrm{real} - X_\textrm{real}$ distributions (\cref{fig:position-reco-dist-examples}) show that the performance of both methods is very similar. In both cases obtained differences between the expected and reconstructed source positions are well below position resolutions. Additionally, both methods yield similar position resolutions. 


\subsection{Energy resolution}
\label{ssec:energy-resolution}

An example of charge spectra registered with the scintillating fiber for two different positions of the radioactive source is shown in \cref{fig:raw-spectra-fitting}. As explained in \cref{sec:sf-data-preprocessing} the obtained spectra are position dependent, which is connected with the attenuation of the scintillating light inside the fiber. Therefore, energy calibration of the charge spectra is not straightforward. Similarly as for the position reconstruction, two methods of energy reconstruction are proposed in this work, as a consequence of two models of scintillating light propagation.

\begin{figure}[htbp]
\centering
\includegraphics[width=.49\textwidth]{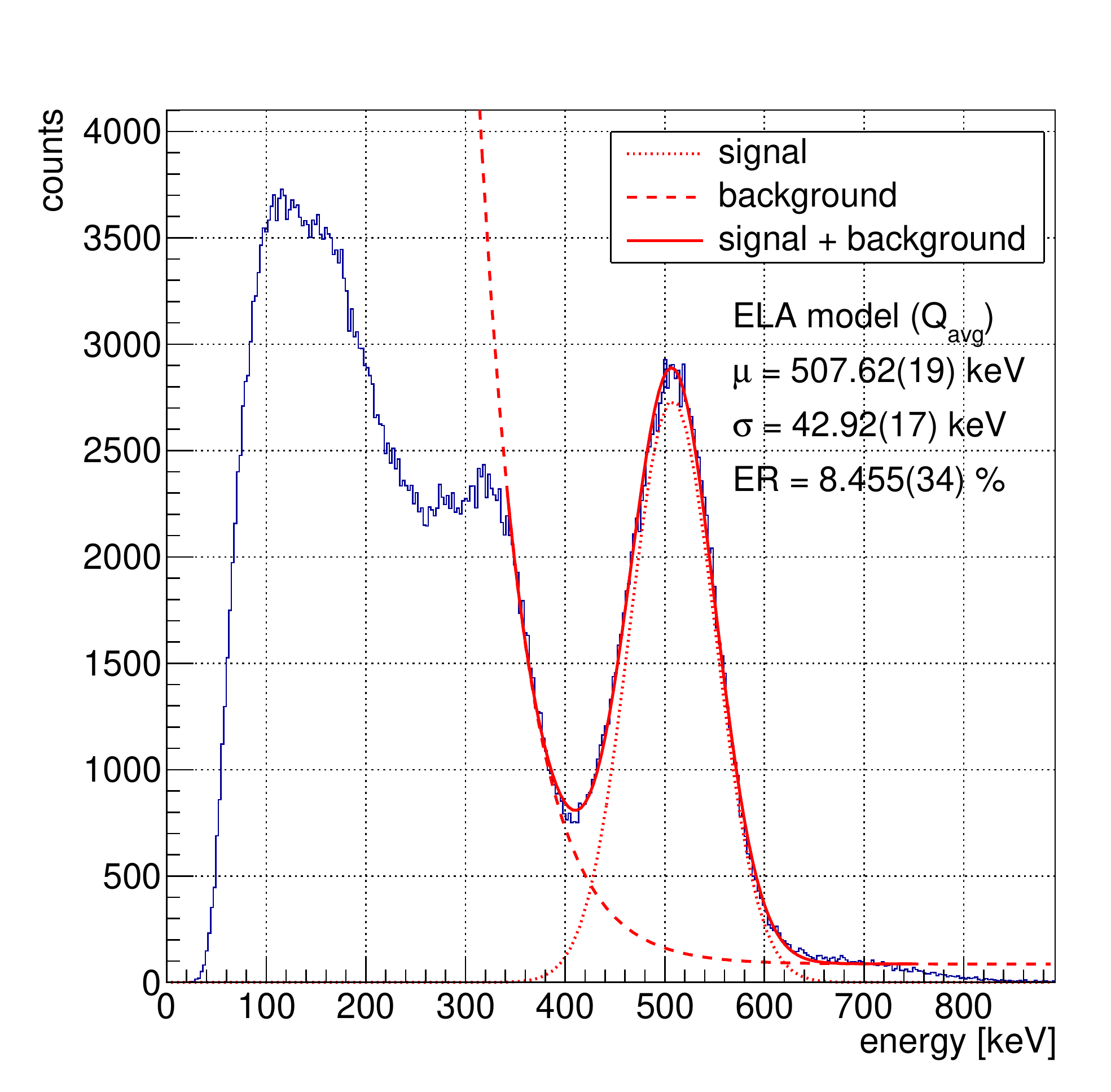}
\includegraphics[width=.49\textwidth]{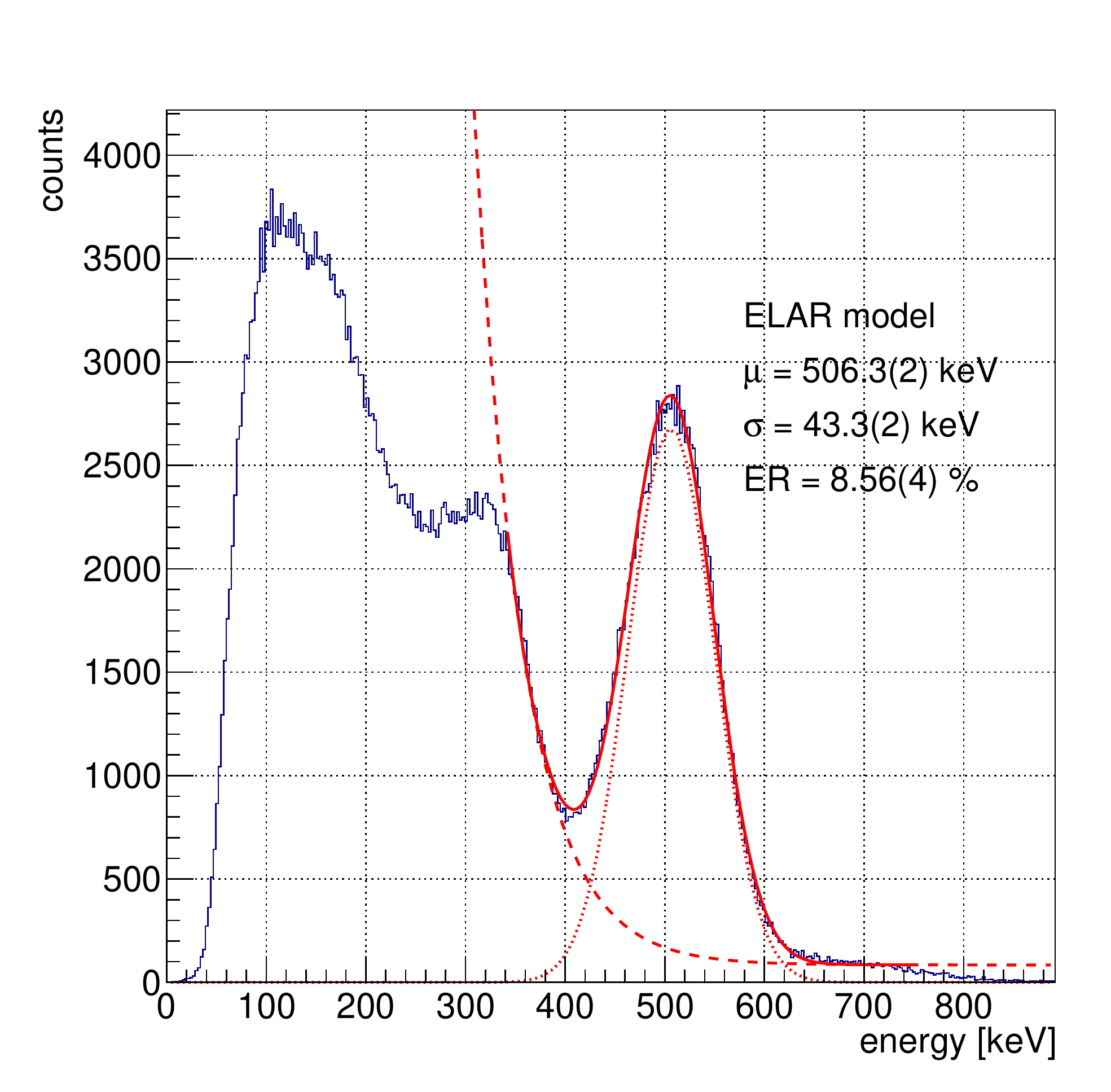}
\caption{Examples of summed energy spectra, containing all events recorded in every measurement along the fiber. Energy was reconstructed with two methods: the \gls{gl:Qavg}/\acrshort{gl:ELA} model (left) and the \acrshort{gl:ELAR} model (right). Both spectra were fitted as described in \cref{sec:sf-data-preprocessing} to determine the parameters of \anhpeak peak and calculate the energy resolution. Presented data come from the series 109 (see \cref{tab:measurements}). Figure adapted from \cite{Rusiecka2021}.}
\label{fig:energy-reconstruction-spectra}
\end{figure}

\begin{figure}[htbp]
\centering
\includegraphics[width=.99\textwidth]{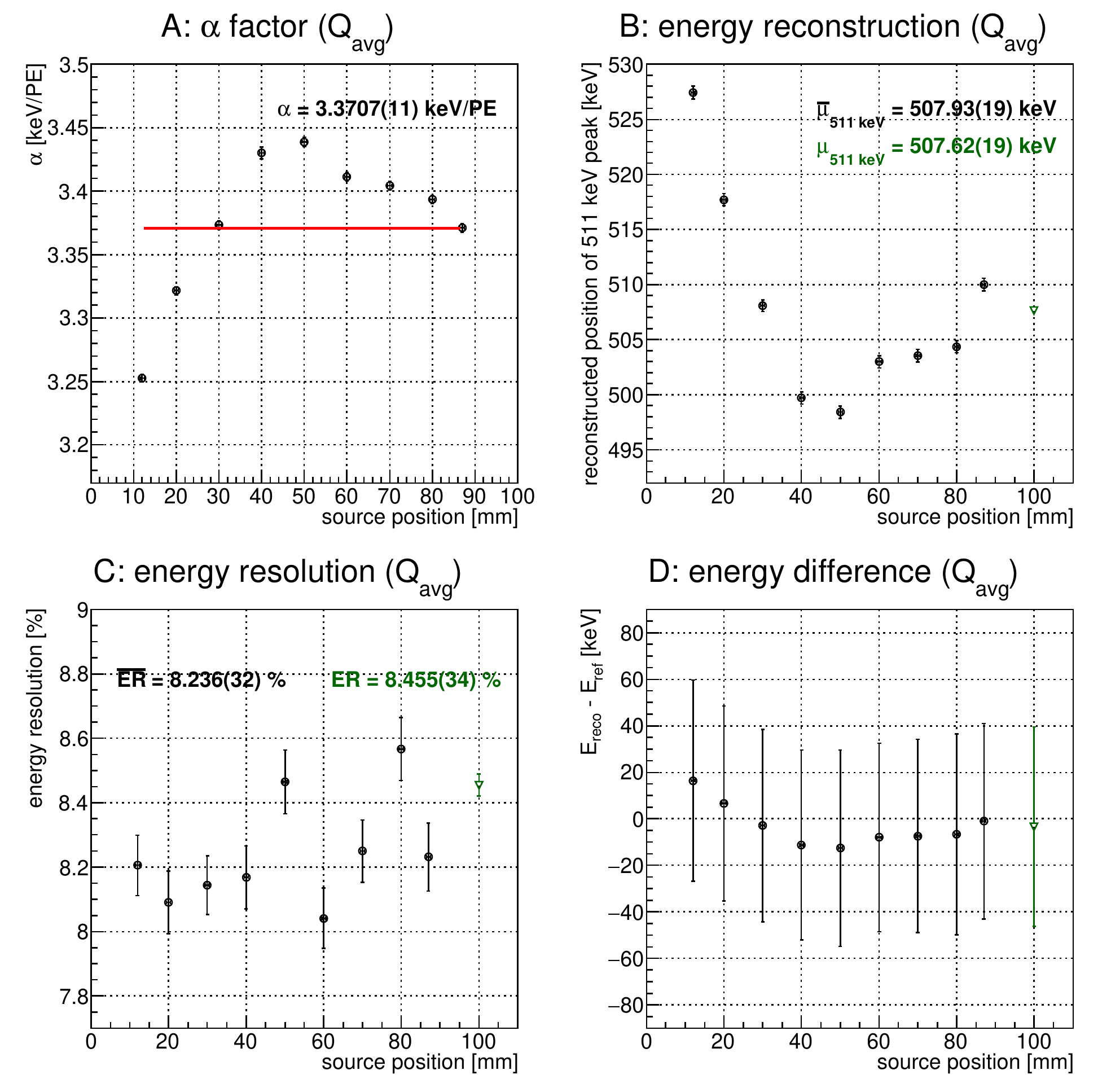}
\caption{Summary of energy reconstruction with the \gls{gl:Qavg} method. A - determination of the $\alpha$ coefficient; red line denotes the value of the $\alpha$ coefficient for the investigated fiber; B - positions of the reconstructed \anhpeak peak for all source positions (black points) and for the cumulative spectrum (green triangle); $\bar{\mu}_\textrm{511~keV}$ denotes the average reconstructed peak position calculated as weighted mean, \peakpos denotes the integrated peak position; C - energy resolution for all measurements in the series ($\overline{ER}$) and for the cumulative spectrum ($ER$); D - difference between the reconstructed and expected positions of the \anhpeak peak for all measurements in the series (black points) and for the cumulative spectrum (green triangle); vertical error bars correspond to the $\sigma$ of the \anhpeak peak. Presented data come from the series 109 (see \cref{tab:measurements}).}
\label{fig:energy-reconstruction-summary-qavg}
\end{figure}

The first approach to energy reconstruction uses a geometric mean of signals registered at both ends of the fiber:
\begin{equation}
\label{eq:qavg}
    Q_{\textrm{avg}} = \sqrt{S_\textrm{l}(x) S_\textrm{r}(x)}\ .
\end{equation}
The \gls{gl:Qavg} quantity in the model is position-independent, which allows limiting the effect of scintillating light attenuation on energy reconstruction \cite{Liu2004}. Then, the energy can be reconstructed based on measured charges using the following formula:
\begin{equation}
\label{eq:energy-reco-qavg}
    E(x) = \alpha \cdot \sqrt{S_\mathrm{l}(x) S_\mathrm{r}(x)}\ ,
\end{equation}
where $\alpha$ is a calibration factor calculated for the reference energy $E_\textrm{ref}$:
\begin{equation}
\label{eq:alpha-qavg}
    \alpha = \frac{E_\textrm{ref}}{\sqrt{\mu_{\textrm{l}} (E_\textrm{ref}) \cdot \mu_{\textrm{r}} (E_\textrm{ref})}}\ .
\end{equation}
The $\mu_{\textrm{i}\ E_\textrm{ref}}$ parameter is the position of the peak corresponding to the reference energy in the charge spectrum at the $i$-th side of the fiber. In this work $E_{\textrm{ref}}=\SI{511}{\kilo\electronvolt}$. The $\alpha$ coefficient is calculated independently for all measurements in the calibration series, \ie for different positions of radioactive source along the fiber. The final $\alpha$ coefficient, which becomes a calibration factor for the investigated fiber, is calculated as a weighted mean of all values obtained in the series (\cref{fig:energy-reconstruction-summary-qavg} A). Using the formula \cref{eq:energy-reco-qavg}, the energy reconstruction is performed event-by-event for all measurements in the series. The reconstructed energy spectra for a chosen experimental series are shown in \cref{app:energy-position-reconstruction}, \cref{fig:energy_reconstruction_all_spectra_qavg}. Each spectrum is fitted as described in \cref{{sec:sf-data-preprocessing}} in order to determine the position of the reconstructed annihilation peak and its width reflecting the energy resolution. Additionally, an overall spectrum containing all reconstructed energies in a series is filled to determine the energy resolution integrated over the entire fiber length (\cref{fig:energy-reconstruction-spectra} left). The energy resolution is determined as:
\begin{equation}
    ER = \frac{\sigma_{E}}{E} \cdot 100\% \ .
\end{equation}
The energy resolution values for all measurements in a chosen experimental series, as well as in the cumulative spectrum, are shown in \cref{fig:energy-reconstruction-summary-qavg}~C. \Cref{fig:energy-reconstruction-summary-qavg}~B and \Cref{fig:energy-reconstruction-summary-qavg}~C show that the difference between the reconstructed and expected energies of the \anhpeak peak vary strongly for different positions of the radioactive source and can reach approximately \SI{20} {\kilo\electronvolt}, which is comparable with the obtained energy resolution. Strong variability of the reconstructed energy is a direct consequence of similar variability observed for the $\alpha$ coefficient (\cref{fig:energy-reconstruction-summary-qavg}~A). 

\begin{figure}[htbp]
\centering
\includegraphics[width=.99\textwidth]{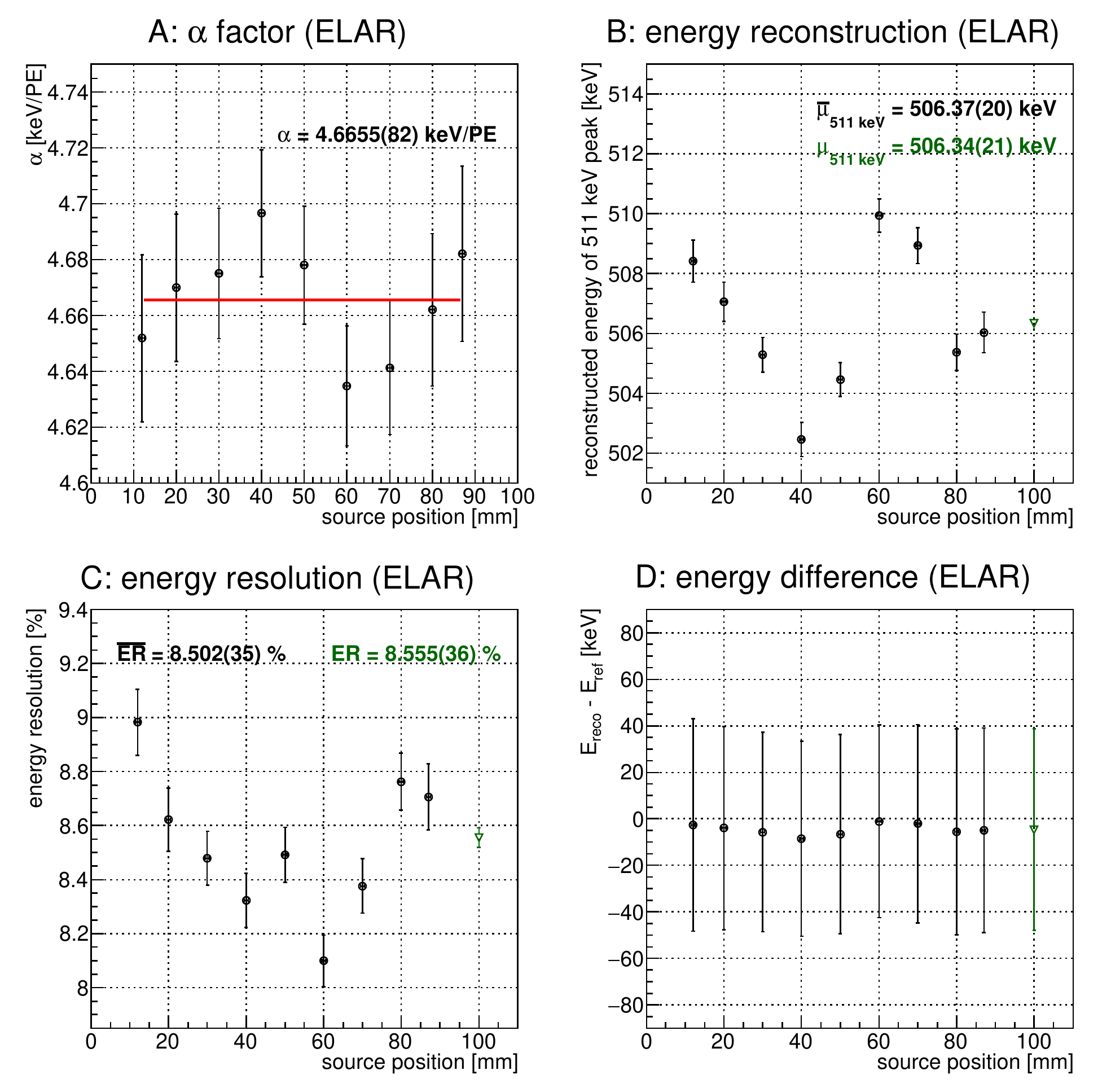}
\caption{Summary of energy reconstruction using data recalculated according to the \acrshort{gl:ELAR} model. A - determination of the $\alpha$ coefficient; red line denotes the value of the $\alpha$ coefficient for the investigated fiber; B - positions of the reconstructed \anhpeak peak for all source positions (black points) and for the cumulative spectrum (green triangle); $\bar{\mu}_\textrm{511~keV}$ denotes the average reconstructed peak position calculated as a weighted mean, \peakpos denotes the integrated peak position; C - energy resolution for all measurements in the series ($\overline{ER}$) and for the cumulative spectrum ($ER$); D - the difference between the reconstructed and expected positions of the \anhpeak peak for all measurements in the series (black points) and for the cumulative spectrum (green triangle); vertical error bars correspond to the standard deviation of the \anhpeak peak. Presented data come from the series 109 (see \cref{tab:measurements})}
\label{fig:energy-reconstruction-summary-elar}
\end{figure}

The issue of position-dependent energy reconstruction is addressed with the second proposed method. In this approach, the experimental data is recalculated according to the \acrshort{gl:ELAR} model equations (\cref{eq:model-primary-reconstructed}). The reconstructed primary components of the registered signals are used to calculate \gls{gl:Qavgstar} quantity, analogously to \cref{eq:qavg}:
\begin{equation}
\label{eq:qavg-corrected}
    Q_{\textrm{avg}}^* = \sqrt{P_\textrm{l}^{*}(S_\textrm{l}, S_\textrm{r}) P_\textrm{r}^{*}(S_\textrm{l}, S_\textrm{r})} \ .
\end{equation}
Then, the energy reconstruction is carried out following the formula:
\begin{equation}
\label{eq:energy-reco-elar}
    E^{*} = \alpha^{*} \sqrt{P_\textrm{l}^{*}(S_\textrm{l}, S_\textrm{r}) P_\textrm{r}^{*}(S_\textrm{l}, S_\textrm{r})} \,
\end{equation}
where $\alpha^{*}$ is defined analogously to \cref{eq:alpha-qavg}. The remaining steps of the energy reconstruction procedure are the same as described in the preceding paragraph. Reconstructed energy spectra for a chosen experimental series are presented in \cref{app:energy-position-reconstruction}, \cref{fig:energy_reconstruction_all_spectra_elar}. The overall energy spectrum, containing all reconstructed events was also filled (\cref{fig:energy-reconstruction-spectra}, right). \Cref{fig:energy-reconstruction-summary-elar} shows a summary of the energy reconstruction procedure for a chosen series. It can be observed that the variability of the $\alpha^{*}$ coefficients with the position of the radioactive source is significantly smaller than for the \Qavg method (\cref{fig:energy-reconstruction-summary-elar}~A). This is reflected in a more stable energy reconstruction (\cref{fig:energy-reconstruction-summary-elar}~B). As a result, the differences between the reconstructed and expected position of the \anhpeak peak are well below the obtained energy resolution. Finally, the energy resolutions are comparable with those obtained with the \gls{gl:Qavg} method (\cref{fig:energy-reconstruction-summary-elar}~C). Although the \acrshort{gl:ELAR} method does not yield any significant improvement in the energy resolution, it eliminates the position-dependence and thus leads to smaller residuals, which is a clear advantage over the \gls{gl:Qavg} method.


\subsection{Timing properties}
\label{ssec:timing-properties}


\subsubsection*{Decay constants}

Scintillation light pulses are typically characterized by a fast increase in intensity (pulse rise time) followed by an exponential decrease. The most straightforward method to parameterize the evolution of the falling slope of the signal in time ($t$) uses a simple single exponential function:
\begin{equation}
\label{eq:single-decay}
 f(t) = A \cdot \exp{\left( \frac{-(t - t_\textrm{0})}{\tau} \right)} + BL,
\end{equation}
where $A$ is an amplitude, $t_\textrm{0}$ is time offset, $\tau$ is the decay constant, and \acrshort{gl:BL} denotes the baseline level. According to \cref{eq:single-decay}, the decay constant can be defined as the time after which the intensity of the light pulse drops $e$ times. For many scintillators, this approach is sufficient to describe the scintillation process. In particular, this is the case for one of the materials investigated in the presented work - \acrshort{gl:LYSO:Ce}. 

As explained in \cref{sec:radiation-detection}, the emission of scintillating photons is primarily the result of the transitions of the activator atoms. Therefore, the decay time of the scintillator is determined by the lifetime of the excited states of the activator atoms. This means that all scintillators containing the same type of activator will be characterized with similar decay constants, \eg cerium-doped scintillators have decay times of the order of approximately \SI{40}{\nano\second} \cite{Tavernier}. 

Some scintillators exhibit a more complex decay pattern. In those cases, a more accurate description can be given by the decay function consisting of two exponential terms:
\begin{equation}
\label{eq:double-decay}
 f(t) = A_\textrm{f} \cdot \exp{\left(\frac{-(t - t_\textrm{0})}{\tau_\textrm{f}}\right)} + A_\textrm{s} \cdot \exp{\left(\frac{-(t - t_\textrm{0})}{\tau_\textrm{s}}\right)} + BL,
\end{equation}
where $A_{\textrm{f}}$, $A_{\textrm{s}}$ are amplitudes of the fast and slow components respectively, and $\tau_{\textrm{f}}$, $\tau_{\textrm{s}}$ are fast and slow decay constants. The intensities of the two components $I_\textrm{f}$ and $I_\textrm{s}$ can be calculated as follows\cite{Iwanowska2013}: 
\begin{equation}
\label{eq:components-intensity}
\begin{split}
 & I_\textrm{f} = \frac{\tau_\textrm{f} \cdot A_\textrm{f}}{\tau_\textrm{f} \cdot A_\textrm{f} + \tau_\textrm{s} \cdot A_\textrm{s}} \cdot 100\% \\
 & I_\textrm{s} = 100 \% - I_\textrm{f}
\end{split}
\end{equation}
The remaining two scintillating materials investigated in this work (\ce{GAGG}:\ce{Ce} and \ce{LuAG}:\ce{Ce}) exhibit such double decay pattern and thus are characterized by the two decay constants. 

The existence of two or more decay modes can be explained by the presence of multiple types of activator centers. This can mean atoms of different elements, but also atoms of the same element located in different lattice environments, \eg in neighborhoods of other lattice defects. Another explanation for this phenomenon is the trapping of electrons and holes in the lattice defects which delays the transition to the activator centers \cite{Tavernier}. 

\begin{figure}[ht]
\centering
\includegraphics[width=.99\textwidth]{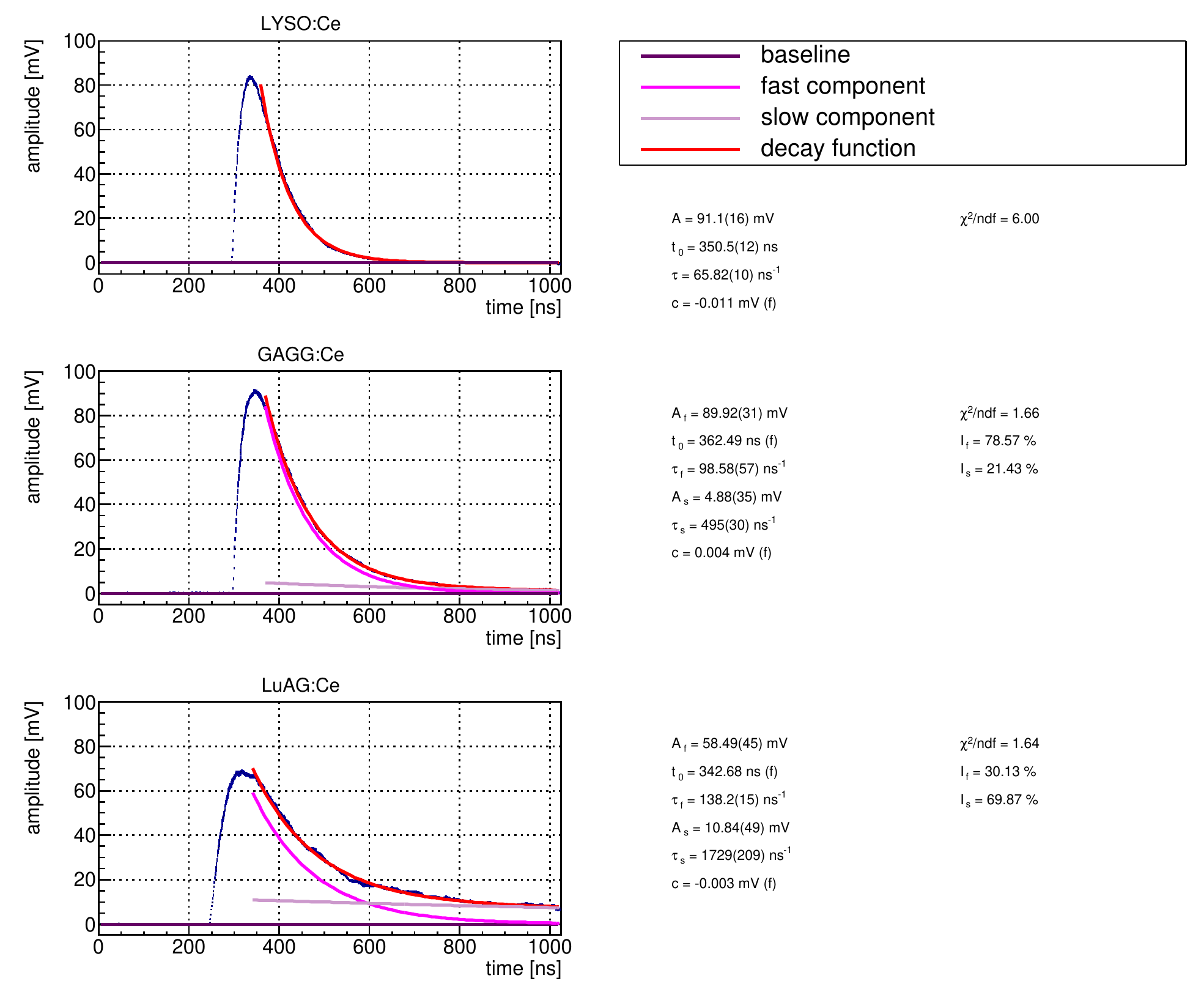}
\caption{Examples of the averaged waveforms of three different scintillators included in the presented study. Single or double decay functions are fitted to the falling slopes of the corresponding signals. The parameters of the fitted functions are listed. Letter $f$ means that the parameter was fixed during the fitting. Presented data come from the following experimental series: 43 (\ce{LYSO}:\ce{Ce}), 45 (\ce{GAGG}:\ce{Ce}), and 64 (\ce{LuAG}:\ce{Ce}), see \cref{tab:measurements}.}
\label{fig:timing-constants-example}
\end{figure}

Examples of signals recorded for the three investigated scintillators, along with fitted decay functions, are presented in \cref{fig:timing-constants-example}. It should be noted that, in order to improve the fit quality and reduce the influence of statistical fluctuations, the signals have been averaged, taking into account their amplitude and time when they begin. Obtained parameters of the decay process are also listed in \cref{fig:timing-constants-example}. The literature values of the decay constants for the three investigated materials are listed in \cref{tab:materials-literature}. Comparison with the experimental values shows that the measured decay constants are significantly larger than expected. This is because the decay constants determined experimentally are convolutions of the scintillating material constants and the time constants of the \acrshort{gl:SiPM} readout circuits. It is important to determine such effective decay constants, since the timing capacity of the future detector will depend not only on the used scintillating material but also on electronic components. 

\subsubsection*{Timing resolution}

\begin{figure}[tbp]
\centering
\includegraphics[width=.99\textwidth]{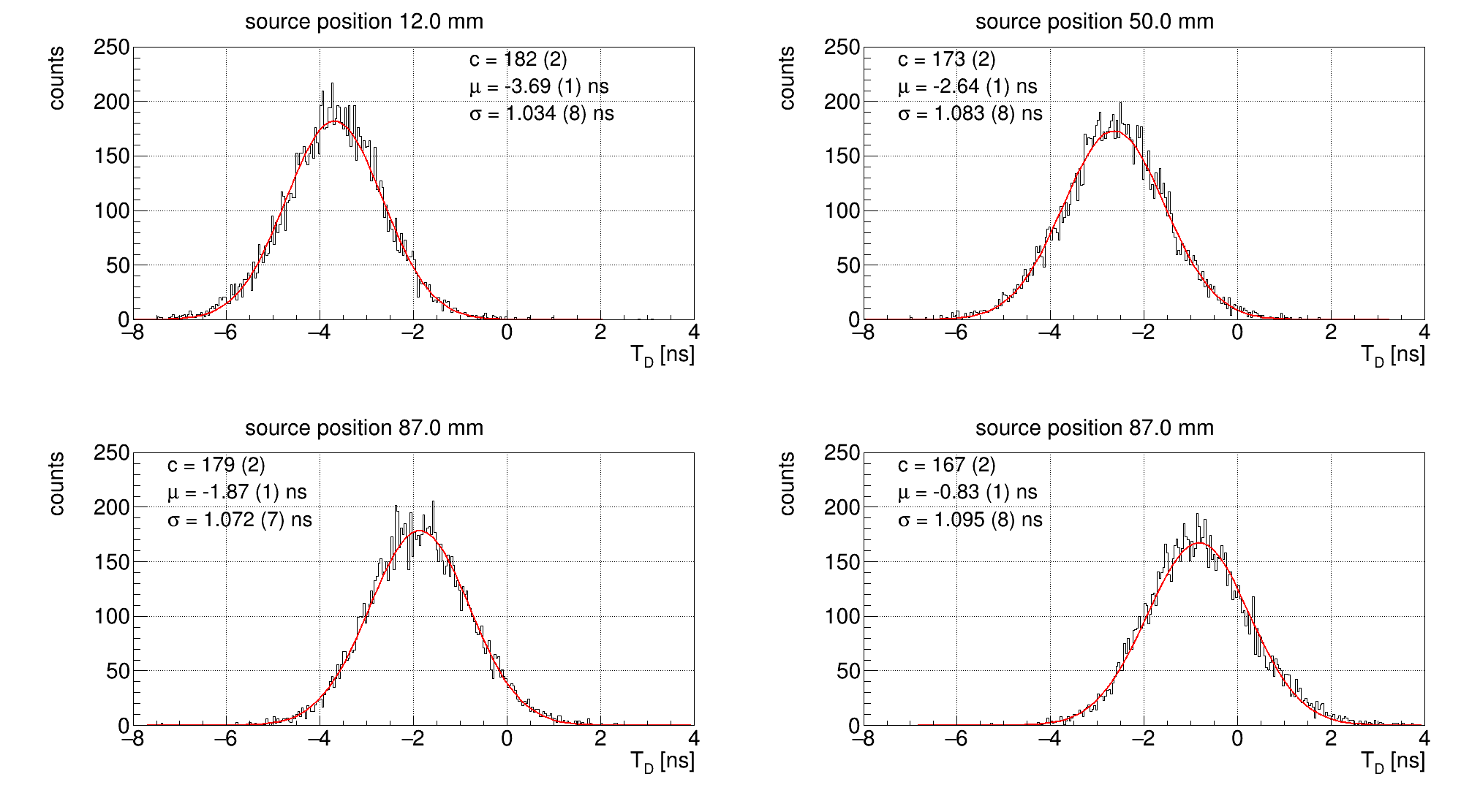}
\caption{Timing distributions obtained for the chosen positions of the radioactive source along the \acrshort{gl:LYSO:Ce} fiber. Distributions were fitted with the Gaussian function to determine the timing resolution. Presented data come from series 109 (see \cref{tab:measurements}).}
\label{fig:tdiff-examples}
\end{figure}

\begin{figure}[!ht]
\centering
\includegraphics[width=.99\textwidth]{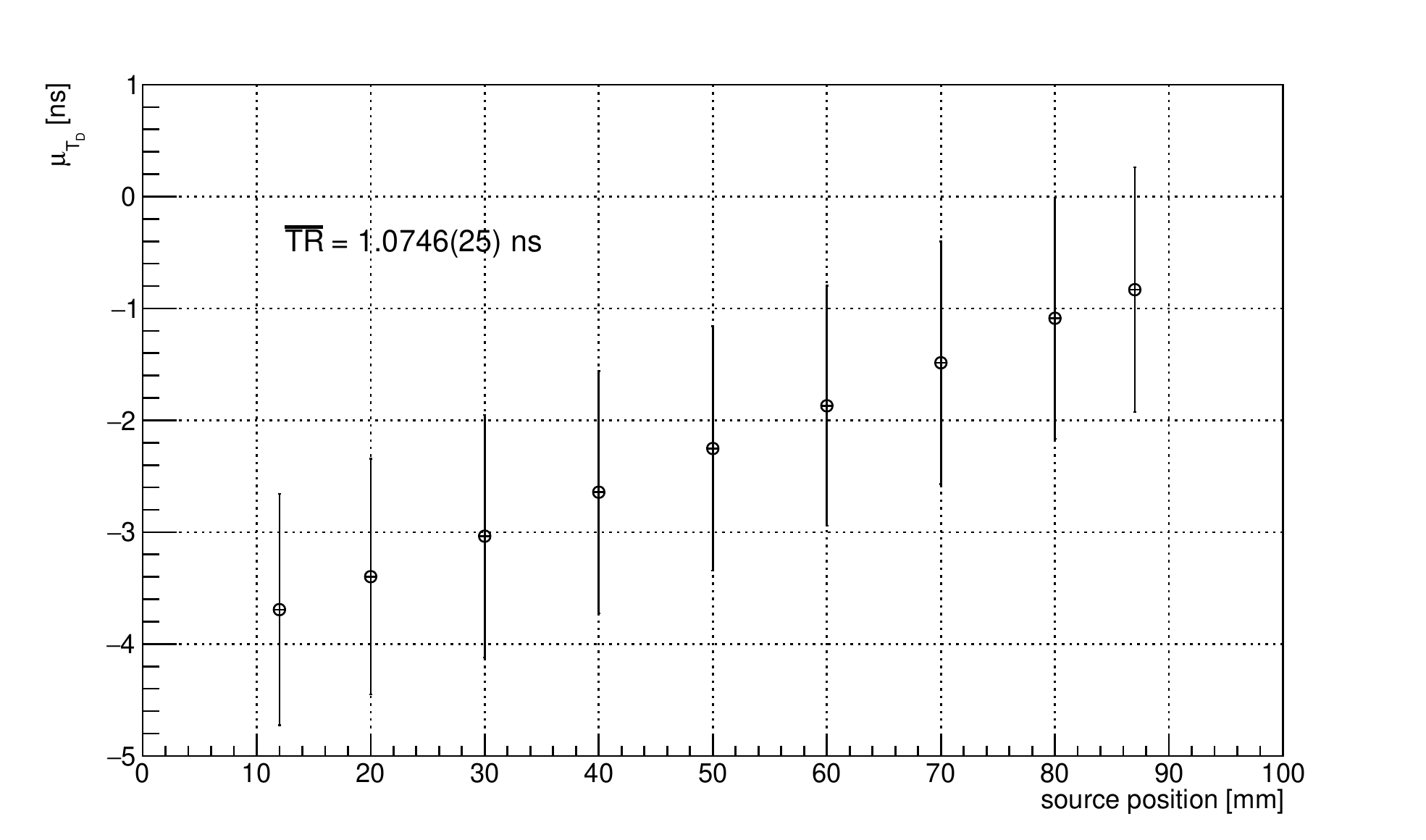}
\caption{Dependence of $\mu_{\textrm{T}_\textrm{D}}$ on the source position for series 109 (see \cref{tab:measurements}). Vertical error bars represent the timing resolution of a given measurement. Adapted from \cite{Rusiecka2021}.}
\label{fig:timing-resolution-example}
\end{figure}

Timing resolution is related to the time difference $T_\textrm{D}$ between two correlated signals being registered \eg at the opposite sides of the scintillating fiber:
\begin{equation}
    T_\textrm{D} = T_\textrm{0,L} - T_\textrm{0,R}.
\end{equation}
In the presented study, the quantity $T_\textrm{D}$ is calculated event by event, creating time difference distributions. The method of determination of the signal arrival time is described in \cref{sec:sf-data-preprocessing}. Only events that form the annihilation peak  are included in the analysis to minimize the influence of the walk effect. The obtained $T_D$ distributions are subsequently fitted with the Gaussian function. The Gaussian mean $\mu_{\textrm{T}_\textrm{D}}$ is associated with the average time difference of the correlated signals at the two ends of the fiber. The timing resolution is determined as the standard deviation $\sigma_{\textrm{T}_{\textrm{D}}}$ of that distribution. Examples of timing distributions along with the obtained fit parameters are presented in \cref{fig:tdiff-examples}. 
The mean timing resolution for the investigated fiber is calculated as a weighted mean of all measurements in the experimental series. \Cref{fig:timing-resolution-example} presents an example of the $\mu_{T_D}$ dependency on the source position along the \acrshort{gl:LYSO:Ce} fiber. The vertical error bars represent the obtained timing resolution. The time difference between the registration of the two correlated signals at the ends of the elongated detector is usually exploited to reconstruct the interaction position. In this case, however, the achieved timing resolution is clearly insufficient to reconstruct the interaction position based on the time difference only. From \cref{fig:timing-constants-example} it can be seen that timing resolution is independent on the source position along the investigated fiber. It should be noted that the dependency depicted in \cref{fig:timing-resolution-example} does not reach \num{0} value for the middle point as expected. This is due to a constant delay of approximately \SI{2}{\nano\second} unintentionally introduced in the setup for one of the channels. 
%

\chapter{Single-fiber studies}
\label{chap:single-fibers}

The following chapter presents studies of single scintillating fibers. Firstly, the experimental setup, measurement procedure, and data processing are described. The subsequent part of the chapter presents obtained results, including attenuation length, light collection, energy-, position- and timing resolution. The investigated samples are categorized in terms of the scintillating material, fiber vendor, surface modification (coating or wrapping), as well as type and size of the optical coupling. This allowed to conduct a thorough comparative study focusing on different aspects of the construction of the scintillating detector. 

The results presented in this chapter had a significant impact on the design optimization of the \acrshort{gl:SiFi-CC} detector. Most of them were described in the article published in the Journal of Instrumentation \cite{Rusiecka2021}.


\section{Materials and methods}
\label{sec:sf-materials-methods}


\subsection{Samples of scintillators}

As stated in \cref{subsec:scintillators-for-proton-therapy-monitoring}, the following three inorganic scintillators were chosen as potential candidates for the active part of the \acrshort{gl:SiFi-CC} detector: \acrshort{gl:LYSO:Ce}, \acrshort{gl:LuAG:Ce} and \acrshort{gl:GAGG:Ce} (\acrshort{gl:GAGG:Ce,Mg}). The properties of these materials, as reported in the literature and by the producers, are summarized in \cref{tab:materials-literature}. Additional samples of \acrshort{gl:GAGG:Ce} material co-doped with \ce{Mg2+} ions were purchased since the addition of magnesium ions was reported to shorten both decay constants of that scintillator \cite{Kamada2015, Lucchini2016}.

\begin{figure}[ht]
\centering
\includegraphics[width=.60\textwidth]{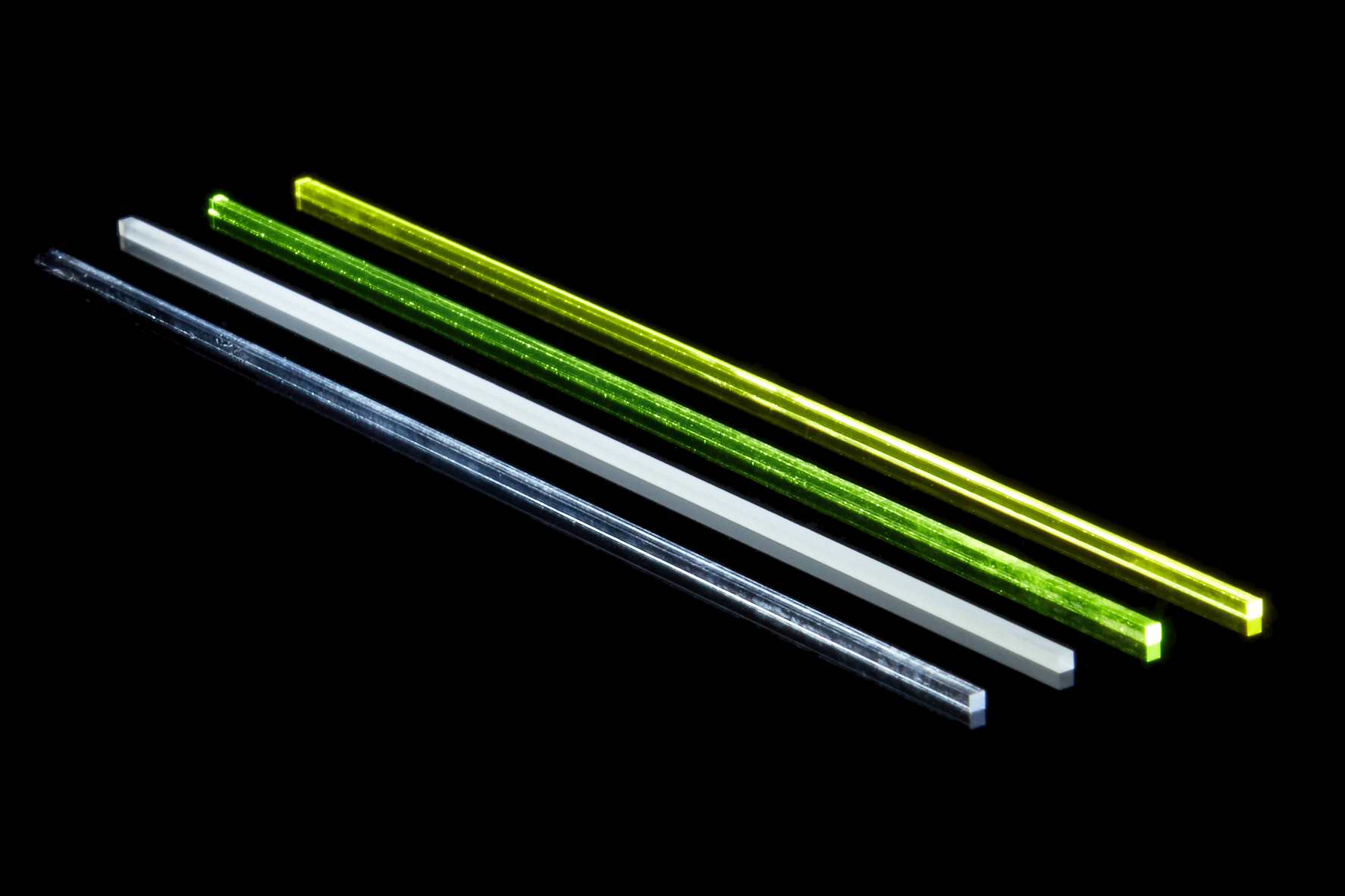}
\caption{Samples of the investigated scintillating fibers. From the left: \acrshort{gl:LYSO:Ce}, \acrshort{gl:LYSO:Ce} (unpolished), \acrshort{gl:LuAG:Ce}, \acrshort{gl:GAGG:Ce}. Photo: dr Damian Gil. Adapted from \cite{Rusiecka2021}.}
\label{fig:fibers-photo}
\end{figure}

All investigated samples had an elongated cuboid shape with a square \numproduct{1 x 1}~\si{\milli\meter\squared} cross section and a length of \SI{100} {\milli\meter}. \Cref{fig:fibers-photo} shows examples of the three material samples. Samples were purchased from different vendors to verify whether possible differences in the production process have a significant impact on scintillator performance. The \acrshort{gl:GAGG:Ce,Mg} samples were a subject to chemical polishing process, while all remaining samples were mechanically polished. The producer suggested chemical polishing as other customers reported improved performance of the scintillators. \Cref{tab:fibers} lists all investigated scintillator samples along with the producer and polishing method. 

\begin{figure}[ht]
\centering
\includegraphics[height=4.5cm]{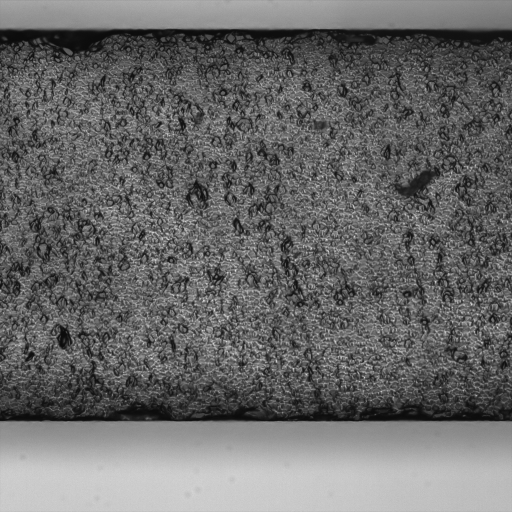} \quad
\includegraphics[height=4.5cm]{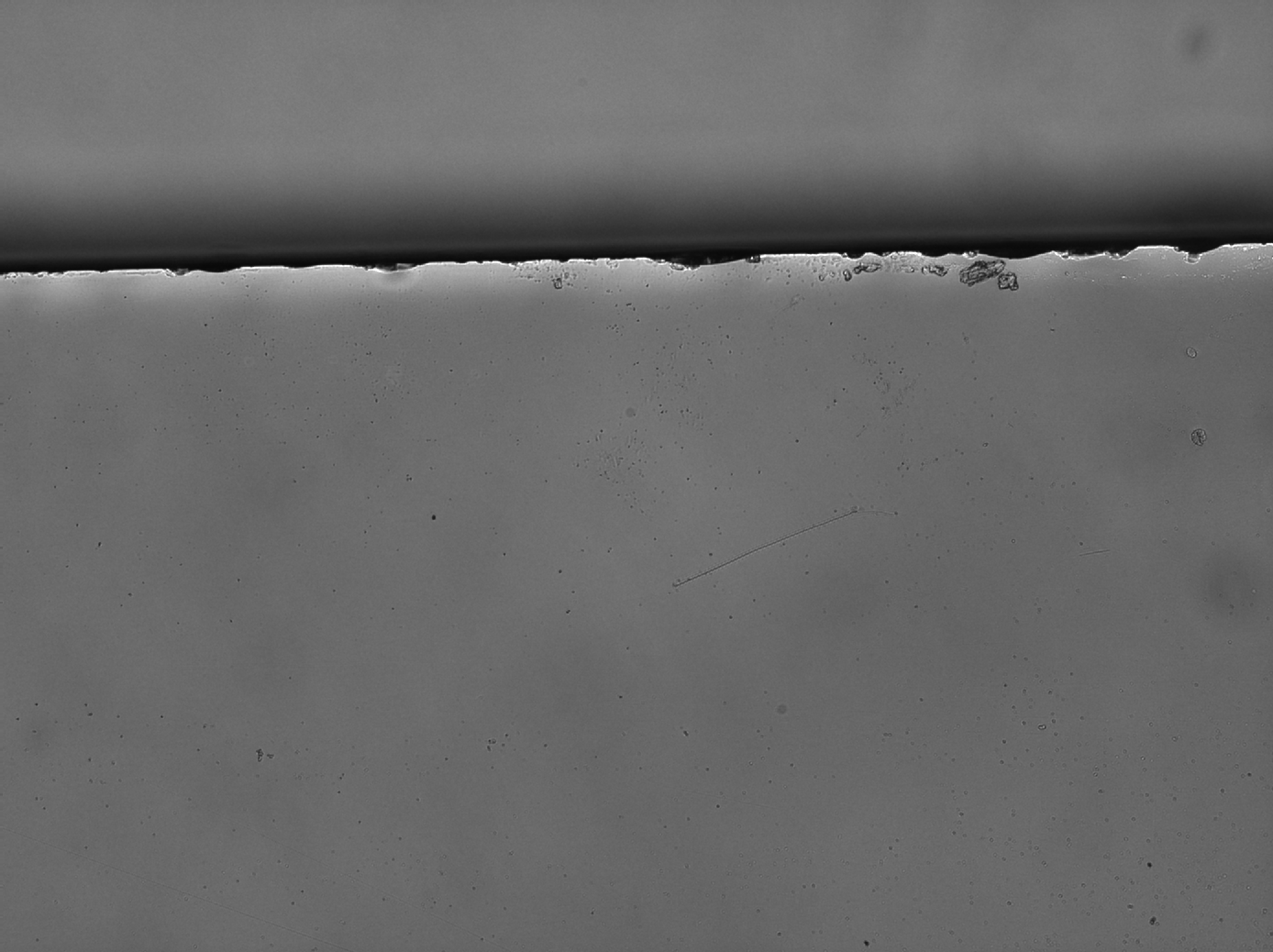} \\ \vspace{0.2cm}
\includegraphics[height=4.5cm]{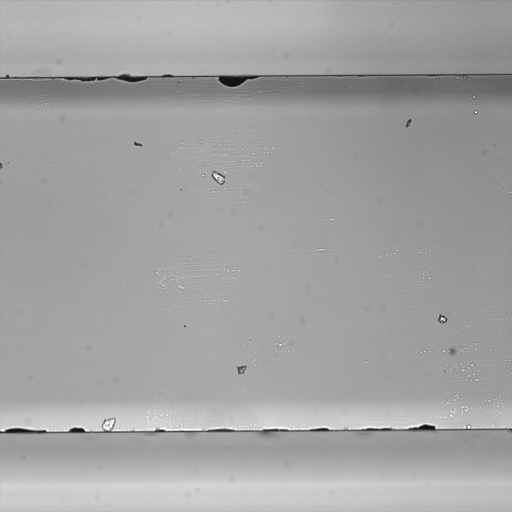} \quad
\includegraphics[height=4.5cm]{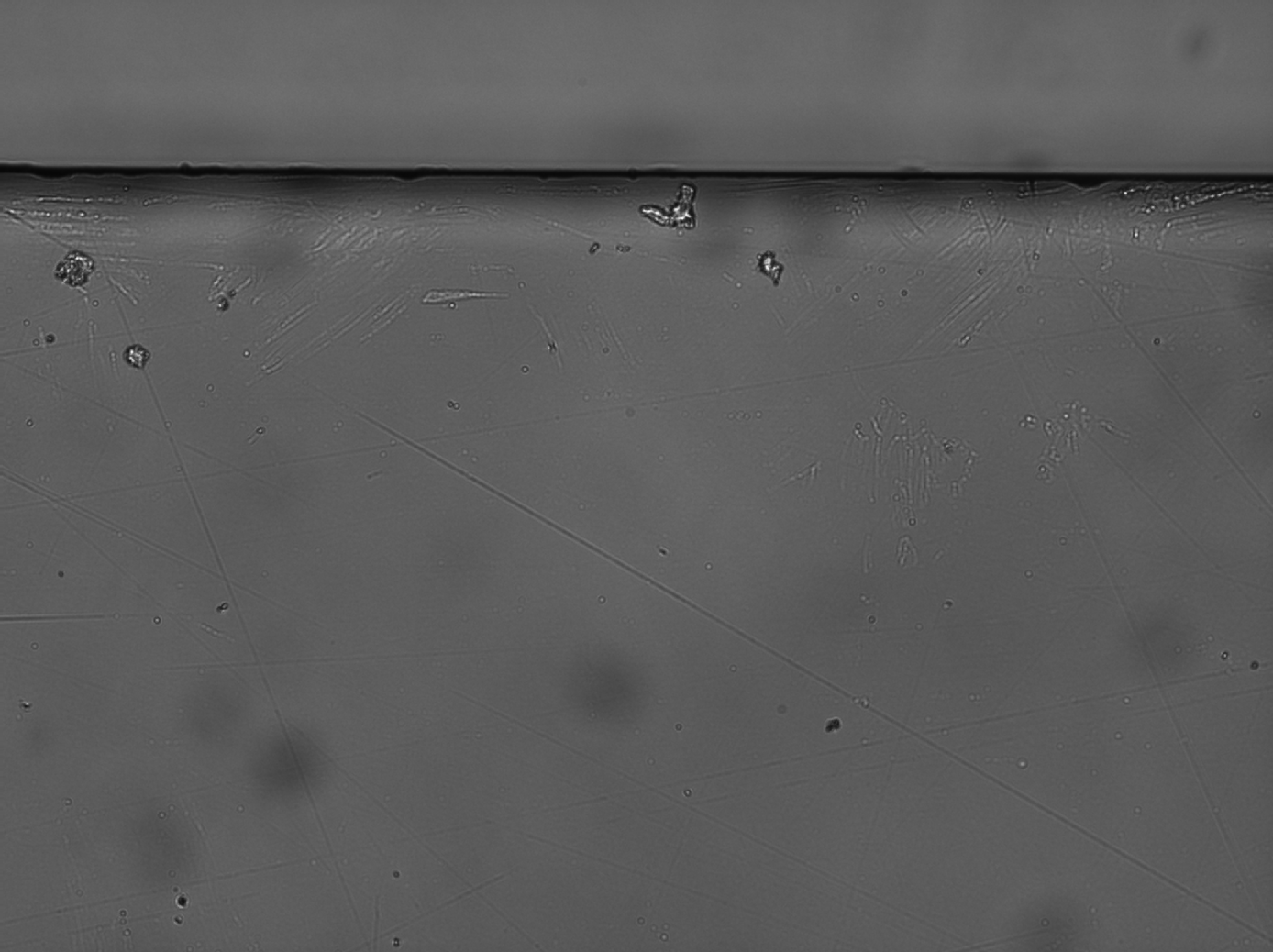}
\caption{Microscopic images of the surface of the investigated scintillating fibers. Top left:  \acrshort{gl:GAGG:Ce,Mg} chemically polished (C\&A Corporation), top right: \acrshort{gl:LYSO:Ce} mechanically polished (Meta Laser), bottom left: \acrshort{gl:LuAG:Ce} mechanically polished, bottom right: \acrshort{gl:LYSO:Ce} mechanically polished (Epic Crystal). Pictures were taken with the Axio Observer Z1 Zeiss microscope operating in bright-field mode. Photos: dr Zbigniew Baster, mgr Tomasz Ko\l{}odziej.}
\label{fig:fibers-microscope}
\end{figure}

The selected samples were additionally photographed with the use of an optical microscope to evaluate the surface quality (see \cref{fig:fibers-microscope}). Observations of the fibers that were mechanically polished revealed microscopic cracks and scratches on the surfaces of all fibers, most probably created during the polishing process and further handling. The surface of the chemically polished scintillator differs significantly from those polished mechanically, with a granular and opaque appearance.

\begin{table}[!ht]
\centering
\caption{List of investigated scintillator samples. All samples had elongated fiber-like shape with the dimensions of \numproduct{1 x 1 x 100}~\si{\cubic\milli\meter}.}
\label{tab:fibers}
\begin{tabular}{p{3.7cm} p{4cm} p{4cm} p{3cm}}
\toprule
Material & Producer & Number of samples & Polishing \\ \midrule
\acrshort{gl:LuAG:Ce} & Crytur & 2 & mechanical \\ \midrule
\acrshort{gl:GAGG:Ce} & Fomos Materials & 2 & mechanical \\ \midrule
\acrshort{gl:GAGG:Ce,Mg} & C\&A Corporation & 2 & chemical \\ \midrule
\acrshort{gl:LYSO:Ce} & Epic Crystal, \newline Meta Laser, \newline Tianle, \newline Shalom EO & 2 \newline 10 \newline 2 \newline 5 & mechanical \\ \bottomrule
\end{tabular}
\end{table}

Selected \acrshort{gl:LYSO:Ce} samples were coated or wrapped in various materials to investigate the influence of surface modifications. The following types of wrappings and coatings prepared on site were tested: 
\begin{itemize}[topsep=0pt,itemsep=0pt,partopsep=0pt, parsep=0pt]
    \item bright aluminum foil (bright side facing toward the fiber),
    \item mat aluminum foil (mat side facing toward the fiber),
    \item metalized Mylar foil,
    \item Teflon,
    \item light guide coating,
    \item  black heat shrink,
    \item AlZn spray paint.
\end{itemize}
In addition, the following samples of \acrshort{gl:LYSO:Ce} fibers wrapped by the producer were purchased: 
\begin{itemize}[topsep=0pt,itemsep=0pt,partopsep=0pt, parsep=0pt]
    \item wrapped in ESR reflector and aluminum foil, 
    \item painted with \baso-based paint and wrapped in aluminum foil.
\end{itemize}


\subsection{Experimental setup and data acquisition}
\label{ssec:sf-setup-daq}

To study the properties of scintillating fibers, a dedicated test bench was constructed. The experimental setup as well as the measurement protocol were changing throughout the experiment, as it was adapted and perfected. Two main setup configurations were used for majority of the performed measurements, as described below.

\subsubsection*{Configuration I}

\begin{figure}[htb]
\centering
\includegraphics[width=.80\textwidth]{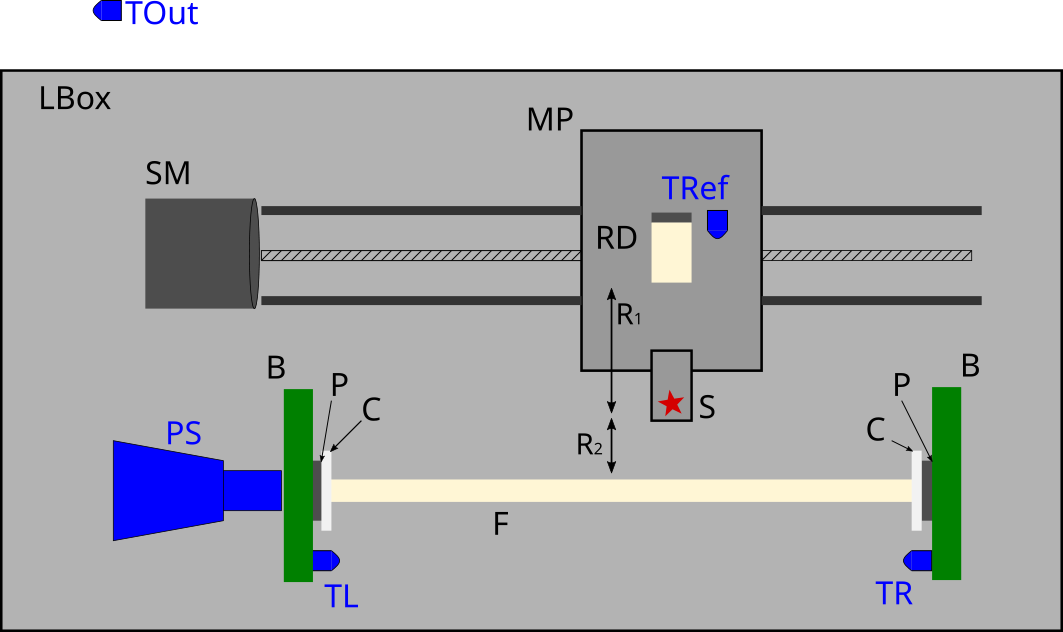}
\caption{Scheme of the experimental setup. Symbols in the picture denote: F - fiber, C - optical coupling, P - \acrshort{gl:SiPM}, B - \acrshort{gl:PCB} housing the \acrshort{gl:SiPM}, S - radioactive source, RD - reference detector, TRef - temperature sensor for the reference detector, MP - moving platform for the electronic collimation system, SM - stepping motor, LBox - light tight box, $R_1$ - distance between the reference detector and the radioactive source, $R_2$ - distance between the investigated scintillating fiber and the radioactive source. The blue elements were introduced in configuration II of the setup: TR - temperature sensor (right), TL - temperature sensor (left), TOut - temperature sensor placed outside of the light tight box, PS - positioning micrometer screw. Scheme not to scale. Adapted from \cite{Rusiecka2021}.}
\label{fig:setup}
\end{figure}

The scheme of the test bench used at this stage of the experiment is shown in \cref{fig:setup}. The entire setup was placed in a light-tight box. The investigated fiber was mounted in a holder and coupled at both ends to \acrshort{gl:SiPM}s. At this stage Hamamatsu \acrshort{gl:SiPM}s (S13360-3050VE, \numproduct{3 x 3}~\si{\milli\meter\squared}, microcell size \SI{50}{\micro\meter}) with dedicated evaluation boards (C12332-01) were used. The detailed specification of \acrshort{gl:SiPM}s can be found in \cref{tab:sipms} in \cref{app:photodetectors}. The choice of \acrshort{gl:SiPM}s was motivated by their relatively high \acrshort{gl:PDE} for wavelengths between \SI{420}{\nano\meter} and \SI{530}{\nano\meter}, matching well the emission spectra of the investigated scintillation materials. The evaluation boards housing the \acrshort{gl:SiPM}s were equipped with temperature sensors. The information from these sensors was used to correct the operating voltage of the \acrshort{gl:SiPM}s. To ensure good optical contact between the investigated fiber and the photodetector, a silicone grease was used (Saint Gobain BC-630, \gls{gl:n} $=1.47$).


\subsubsection*{Configuration II}

At this stage of the experiment, Hamamatsu \acrshort{gl:SiPM}s were replaced due to malfunction with SensL \acrshort{gl:SiPM}s (C series, \numproduct{3 x 3}~\si{\milli\meter\squared}, microcell size \SI{20}{\micro\meter}) operated on custom boards \cite{Pooth2015a} (see \cref{tab:sipms} in \cref{app:photodetectors}). Additionally, other improvements were introduced to better control the experimental conditions, as shown in \cref{fig:setup}. One of them was a micrometer screw, mounted to allow for precise tightening of the fiber between the layers of coupling and \acrshort{gl:SiPM}s, ensuring stable and repetitive connections.  

Similarly to configuration I, the boards housing the \acrshort{gl:SiPM}s were equipped with temperature sensors to carry out voltage correction. However, it was not possible to read out the temperature values during the measurements. Therefore, independent temperature sensors based on a 1-wire interface were added to the setup, which was another important modification in configuration II. Two sensors were installed near the ends of the investigated fiber, one next to the reference detector, and finally one sensor was placed outside the light-tight box. The temperature information was automatically saved once per minute. This allowed for additional control of the experimental conditions. 


In most of the measurements performed with this setup configuration, the Eljen EJ-560 silicone rubber optical interface (\gls{gl:n} $=1.43$) was used as a coupling medium between the fiber and the \acrshort{gl:SiPM}s. However, additional measurements with Saint Gobain silicone grease and without a coupling medium were performed in order to conduct comparative studies of different coupling types. Additionally, custom-made silicone rubber pads of different sizes were tested in this configuration (see \cref{ssec:sf-diff-couplings}).

\subsubsection*{Electronic collimation}

Both configurations of the experimental setup featured an electronic collimator placed on a moving and remotely controlled platform. The design of the collimator was based on the work by Anfr\'{e} \etal \cite{Anfre2007}. In this study, the collimator consisted of a SensL \acrshort{gl:SiPM} (C series, \numproduct{3 x 3}~\si{\milli\meter\squared}, microcell size \SI{20}{\micro\meter}) with slow and fast output channels and a \acrshort{gl:LYSO:Ce} crystal (\numproduct{2 x 3 x 20}~\si{\cubic\milli\meter}) attached to it. The \Na source was placed between the investigated fiber and the reference detector in the electronic collimator as depicted in \cref{fig:setup}. The principle of operation of the electronic collimator is based on two \anhpeak photons emitted back-to-back as a result of positron annihilation in the \Na source. Such a collimation system allows for irradiation of the investigated fiber at a desired position, with the position and area of irradiation precisely determined by the geometry of the setup. In the presented experimental setup the distances $R_1$ and $R_2$ were respectively \SI{10}{\milli\meter} and \SI{5}{\milli\meter}, which resulted in a \SI{1}{\milli\meter} irradiation window on the fiber. Additionally, the use of the electronic collimator causes a significant reduction of the background stemming from scattered photons.

In the preliminary stage of the study, the test bench was equipped with a lead collimator with a \SI{2}{\milli\meter} slit. The results obtained in the measurements with the lead collimator were described in \cite{Rusiecka2019}. The introduction of the electronic collimation system significantly improved the quality of recorded gamma spectra, and thus the lead collimator was no longer used after preliminary measurements.

\subsubsection*{Data acquisition system (\acrshort{gl:DAQ})}

The data were acquired using a CAEN Desktop Digitizer DT5742 which samples incoming signals and saves full waveforms. The sampling frequency of \SI{1}{\giga\hertz} was used and \num{1024} samples were saved in a single event. The acquisition was externally triggered by a triple coincidence of signals incoming from both ends of the investigated fiber and the reference detector in the electronic collimator. Detailed diagrams of analog signal processing chain in single-fiber measurements are presented in \cref{app:signal-processing} (\cref{fig:electronics-hamamatsu-1}, \cref{fig:electronics-hamamatsu-2}, \cref{fig:electronics-sensl-1}, \cref{fig:electronics-sensl-2}). 

The main disadvantage of the Desktop Digitizer module was a relatively small dynamic range of \SI{1}{\volt}, which was further reduced by the baseline offset to approximately \SI{700}{\milli\volt}. Due to this limitation, it was often necessary to reduce the gain of the detection system to avoid signal saturation. Gain reduction was achieved by tuning the operating voltage of the \acrshort{gl:SiPM}s (configurations I and II) or introducing hardware changes \eg attenuators (configuration I). Registered signals were stored on a disk and subsequently analyzed using custom written software as described in detail in \cref{sec:sf-data-preprocessing}.


\subsection{Measurement procedure and list of runs}
\label{ssec:sf-measurements-lor}

The experimental sequence was organized into series. Each series consisted of several measurements recorded at different positions of the \Na source along the fiber. Position \SI{0}{\milli\meter} corresponded to the left end of the fiber and position \SI{100}{\milli\meter} to the right end. In each measurement full waveforms of incoming signals were recorded. During the data preprocessing, key characteristics of waveforms were extracted (see \cref{sec:sf-data-preprocessing}), allowing to plot charge, amplitude and time spectra. As a result, a single measurement consisted of three sets of spectra: one for each end of the investigated fiber and one for the reference detector in the electronic collimator. 

\begin{figure}[ht]
\centering
\includegraphics[width=.49\textwidth]{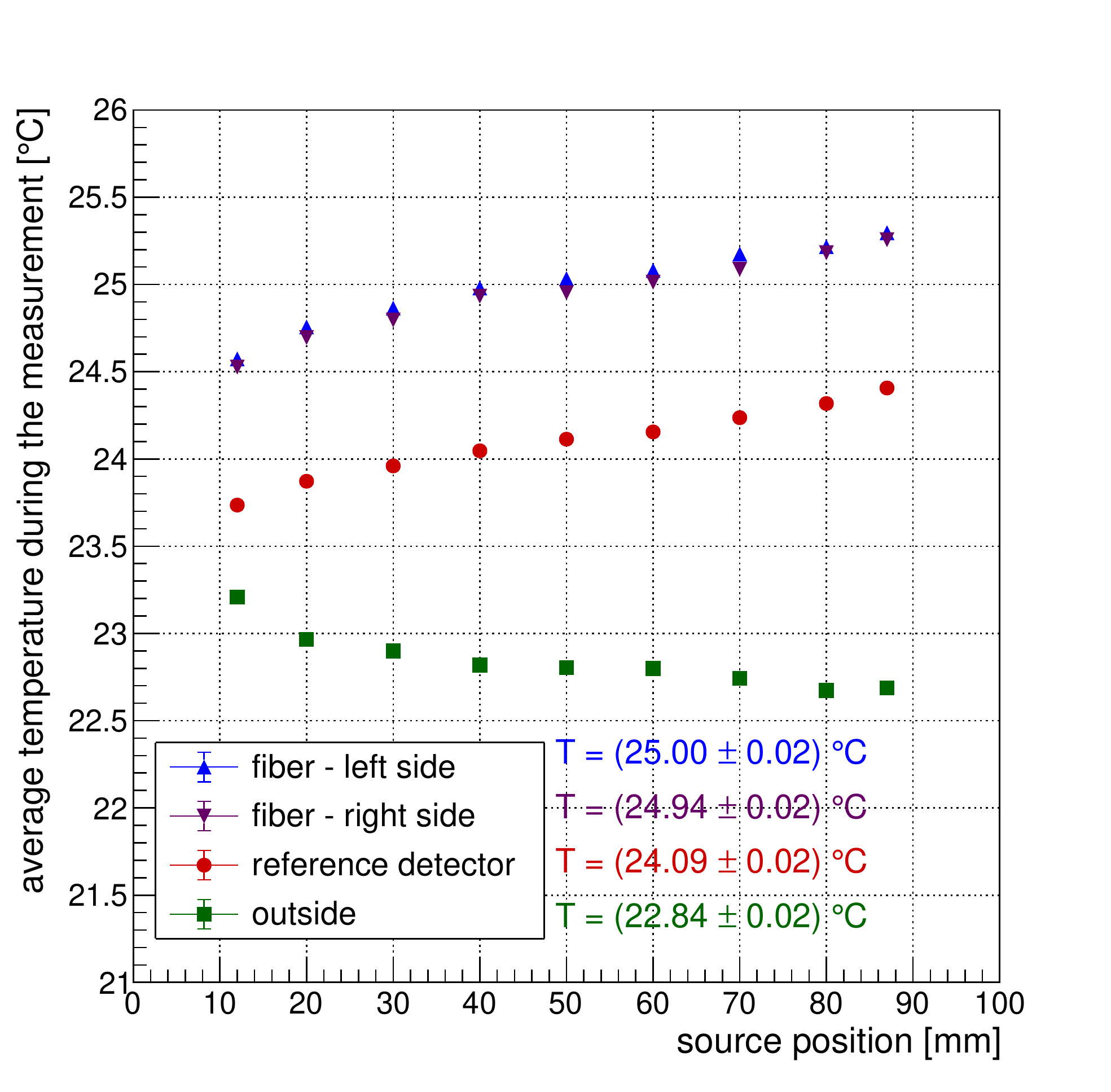}
\includegraphics[width=.49\textwidth]{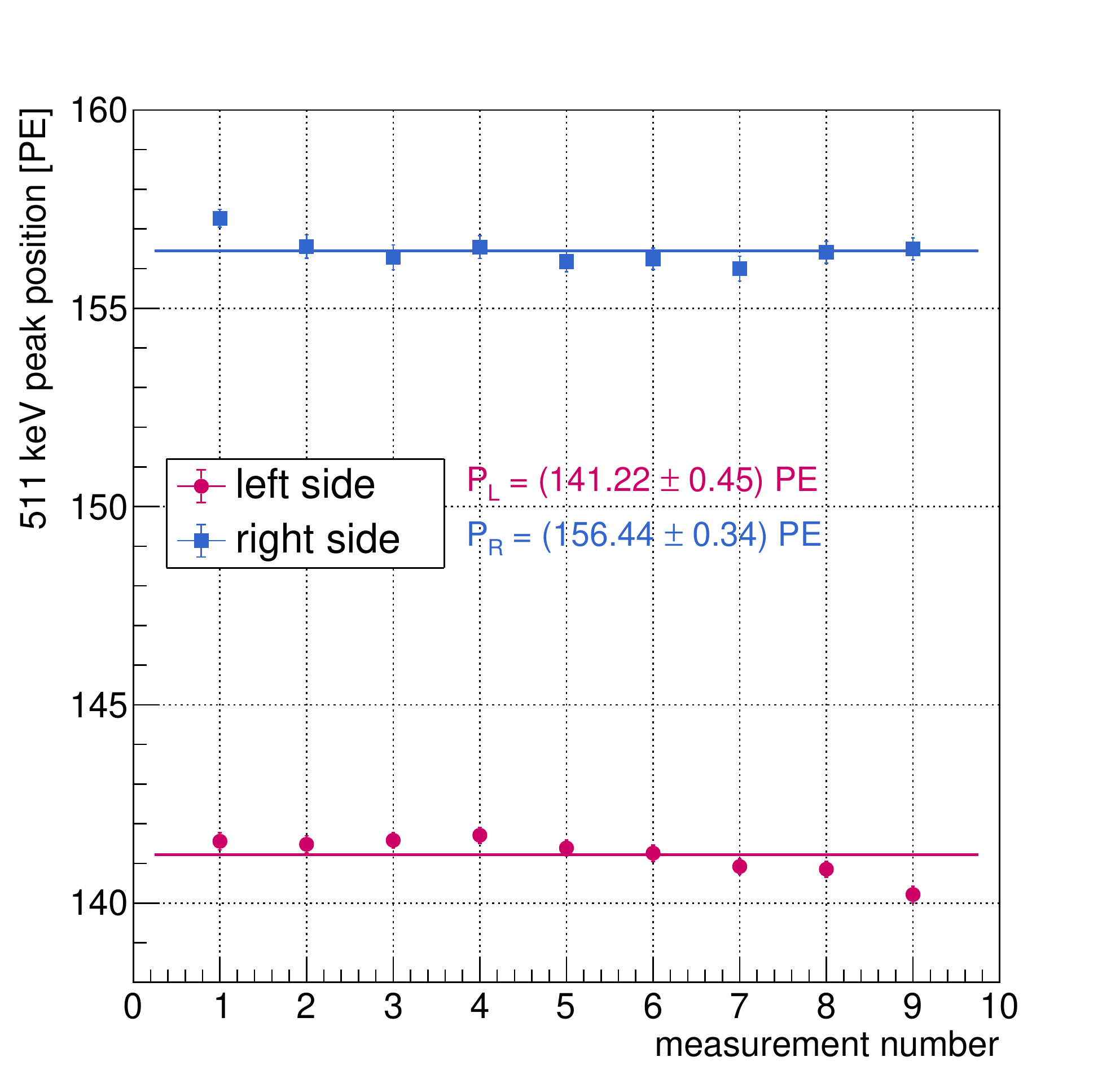}
\caption{Left: example of temperature measurement for all temperature sensors placed in the experimental setup. Each point corresponds to an average temperature during the measurement, listed temperature values denote average temperatures throughout the entire series. Right: monitoring of \anhpeak peak position stability. Listed values denote average position of the \anhpeak peak and the uncertainty calculated as a standard deviation. Both graphs refer to series 109 (see \cref{tab:measurements}).}
\label{fig:measurement-monitoring}
\end{figure}

Some inconsistencies were observed in the data recorded with configuration I of the test bench. Therefore, in the configuration II extra precautions were taken to control measurement stability and experimental conditions. Every second measurement was recorded in the middle of the fiber to verify the stability of the annihilation peak position. Such a sequence resulted in twin series: an experimental and a monitoring one. \Cref{fig:measurement-monitoring} (right) shows an example of stability monitoring throughout the experimental series. Additionally, each fiber was examined twice, in two opposite orientations, to observe possible effects stemming from differences in coupling at the fiber ends and confirm the reproducibility of the obtained results. The information from the temperature sensors installed in the setup was also plotted. \Cref{fig:measurement-monitoring} (left) presents the average temperatures during the measurements registered by all four sensors. Comparing the panels of~\cref{fig:measurement-monitoring} one sees that the hardware temperature correction worked very well.

In total \num{93} measurement series recorded under various conditions were analyzed. Of these, \num{15} were recorded in configuration I and the remaining \num{78} in configuration II. \Cref{tab:measurements} presents the list of analyzed series along with the experimental conditions under which they were recorded. The recorded data allowed to compare different scintillating materials, different vendors, types and sizes of \acrshort{gl:SiPM}-fiber coupling, wrapping or coating of scintillating fibers, used \acrshort{gl:SiPM}s and front-end electronics.

\begin{landscape}
\centering
\begin{longtable}{|p{2cm}|p{2.8cm}|p{2.2cm}|p{2.3cm}|p{3.7cm}|p{2.6cm}|p{7cm}|} 
\caption{Summary of conducted single-fiber measurements. The measurement series IDs are used hereafter in the figures. The numbers in bold denote the total number of series taken under the same conditions.} 
\label{tab:measurements} \\
\hline
Material & Producer & SiPMs & Configura-tion & Coupling & Coating/ \newline wrapping & Measurement series \\ \hline
\acrshort{gl:LYSO:Ce} & Epic Crystal & Hamamatsu & I & Silicone gel & - & 30, 31, 32 \textbf{(3)} \\
\acrshort{gl:LYSO:Ce} & Meta Laser & Hamamatsu & I & Silicone gel & - & 34 -- 43 \textbf{(10)} \\
\acrshort{gl:GAGG:Ce} & Fomos & Hamamatsu & I & Silicone gel & - & 45, 46 \textbf{(2)} \\
\acrshort{gl:LuAG:Ce} & Crytur & Hamamatsu & I & Silicone gel & - & 63, 64, 66, 67 \textbf{(4)} \\ \hline
\acrshort{gl:LYSO:Ce} & Epic Crystal & SensL & II & Silicone pads Eljen & - & 117, 119, 121, 123 \textbf{(4)} \\
\acrshort{gl:LYSO:Ce} & Meta Laser & SensL & II & Silicone pads Eljen & - & 109, 111, 125, 127, 129, 131, 133, 135, 137, 139, 141, 143 \textbf{(12)} \\
\acrshort{gl:LYSO:Ce} & Shalom & SensL & II & Silicone pads Eljen & - & 113, 115 \textbf{(2)} \\
\acrshort{gl:LYSO:Ce} & Tianle & SensL & II & Silicone pads Eljen & - & 98, 100, 102, 104, 106, 201, 203, 229 \textbf{(8)} \\ \hline
\acrshort{gl:LuAG:Ce} & Crytur & SensL & II & Silicone pads Eljen & - & 189, 191, 193, 195 \textbf{(4)} \\ \hline
\acrshort{gl:LYSO:Ce} & Tianle & SensL & II & Silicone gel & - & 225, 227 \textbf{(2)} \\
\acrshort{gl:LYSO:Ce} & Tianle & SensL & II & Air gaps & - & 205, 207 \textbf{(2)} \\ \hline
\acrshort{gl:LYSO:Ce} & Meta Laser/ \newline Shalom & SensL & II & Silicone pads Eljen & Teflon & 145, 147, 149, 151, 153, 155 \textbf{(6)} \\
\acrshort{gl:LYSO:Ce} & Meta Laser & SensL & II & Silicone pads Eljen & Mylar & 165, 167, 169, 171 \textbf{(4)} \\
\acrshort{gl:LYSO:Ce} & Epic Crystal/ \newline Meta Laser & SensL & II & Silicone pads Eljen & Al (mat) & 157, 159, 161, 163 \textbf{(4)} \\
\acrshort{gl:LYSO:Ce} & Epic Crystal/ \newline Meta Laser & SensL & II & Silicone pads Eljen& Al (bright) & 173, 175, 177, 179 \textbf{(4)} \\
\acrshort{gl:LYSO:Ce} & Tianle & SensL & II & Silicone pads Eljen & Light guide \newline coating & 209, 211 \textbf{(2)} \\
\acrshort{gl:LYSO:Ce} & Meta Laser/ \newline Shalom & SensL & II & Silicone pads Eljen & Heat shrink & 219, 221 \textbf{(2)} \\
\acrshort{gl:LYSO:Ce} & Shalom & SensL & II & Silicone pads Eljen & ESR + Al & 231, 233 \textbf{(2)} \\
\acrshort{gl:LYSO:Ce} & Shalom & SensL & II & Silicone pads Eljen & White paint + \newline Al & 235, 237, 239, 241 \textbf{(4)} \\ \hline
\acrshort{gl:LYSO:Ce} & Shalom & SensL & II & Silicone pads custom (\si{0.5} $\times$ \si{5} $\times$ \SI{5}{\cubic\milli\meter}) & Al (bright) & 243, 245, 247, 249 \textbf{(4)} \\
\acrshort{gl:LYSO:Ce} & Shalom & SensL & II & Silicone pads custom (\si{0.5} $\times$ \si{7} $\times$ \SI{7}{\cubic\milli\meter}) & Al (bright) & 251, 253 \textbf{(2)} \\
\acrshort{gl:LYSO:Ce} & Shalom & SensL & II & Silicone pads custom (\si{0.5} $\times$ \si{10} $\times$ \SI{10}{\cubic\milli\meter}) & Al (bright) & 255, 257 \textbf{(2)} \\
\acrshort{gl:LYSO:Ce} & Shalom & SensL & II & Silicone pads custom (\si{0.5} $\times$ \si{3} $\times$ \SI{3}{\cubic\milli\meter}) & Al (bright) & 259, 261 \textbf{(2)} \\
\hline
\end{longtable}
\end{landscape}


\subsection{Data preprocessing}
\label{sec:sf-data-preprocessing}

Before the determination of key characteristics of scintillating fibers, the recorded waveforms required additional preprocessing, as described below.

\subsubsection*{Extraction of signals features}

\begin{figure}[ht]
\centering
\includegraphics[width=.99\textwidth]{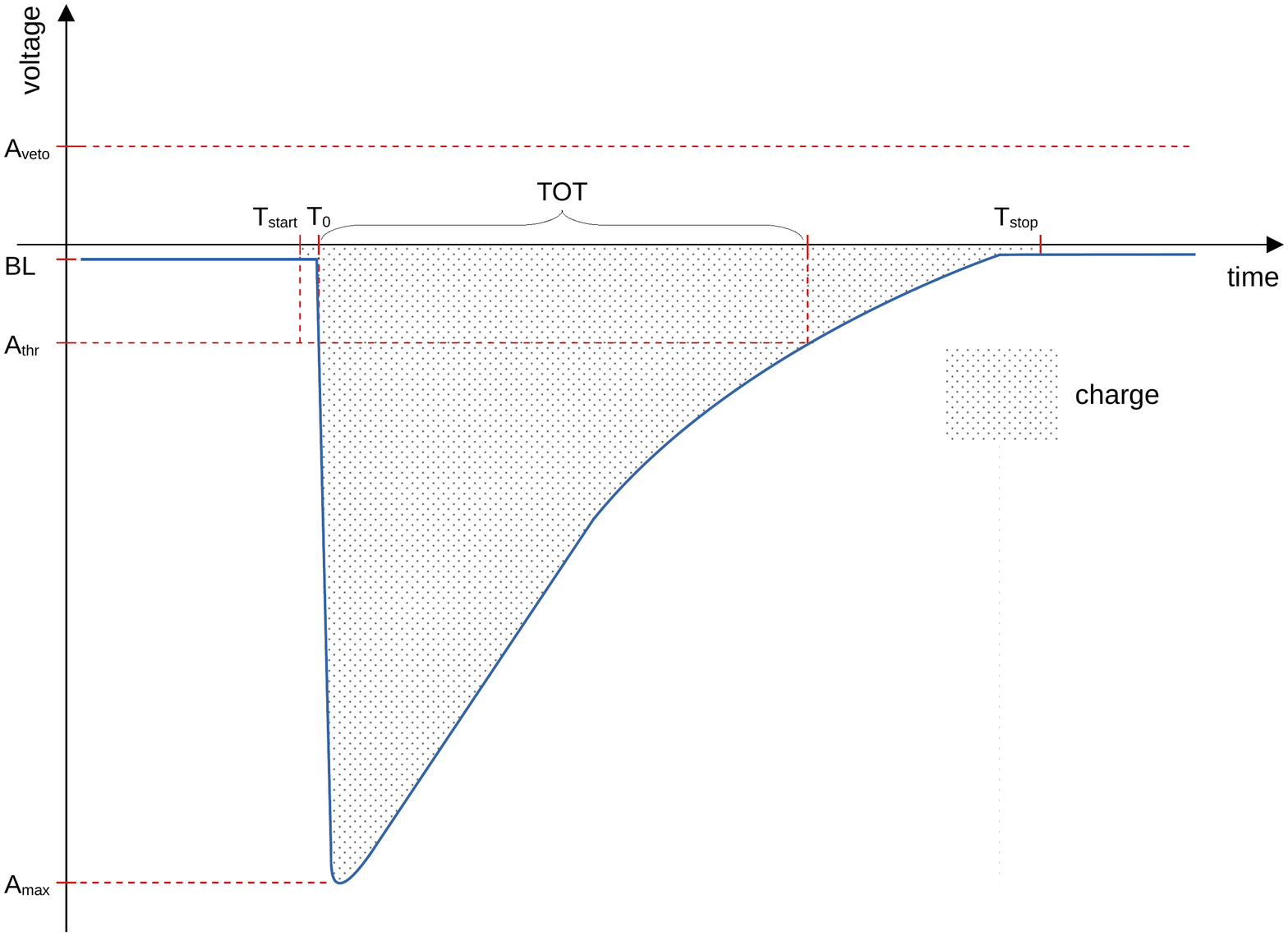}
\caption{An analog signal scheme with determined signal features marked.}
\label{fig:signal-scheme}
\end{figure}

Firstly, the saved signals were analyzed using custom written C++ software based on the ROOT framework \cite{root}. Its main task was to extract the basic features of the registered signals as depicted in \cref{fig:signal-scheme} and listed below:
\begin{itemize}
\item \textbf{Signal arrival time (\gls{gl:tzero})} - time at which the signal reaches the threshold value ($A_{\textrm{thr}}$) for the first time. For different types of front-end electronics, the characteristics of registered signals differed \eg in terms of polarity or amplitude. Therefore, the $A_{\textrm{thr}}$ values were adjusted according to the configuration type.
%
\item \textbf{Amplitude ($A_{\textrm{max}}$)} - determined as the largest (for positive signal polarity) or the smallest (for negative signal polarity, as in the picture \cref{fig:signal-scheme}) sample in the acquisition window.
\item \textbf{Time over threshold \acrshort{gl:TOT}} - duration of the signal determined as the difference between the last time point at which the signal reaches $A_{\textrm{thr}}$ and \gls{gl:tzero}.
\item \textbf{Baseline level (\acrshort{gl:BL})} - calculated as an average of the first \num{50} samples in the acquisition window for each signal separately. This enabled precise baseline subtraction. Additionally, the standard deviation of the baseline was calculated.
\item \textbf{Charge} - signal integral calculated in the range from $T_{\textrm{start}}$ to $T_{\textrm{stop}}$. The start time ($T_{\textrm{start}}$) was calculated as \tzero $-$ \SI{20}{\nano\second}, ensuring that the entire signal was captured in the integral. The stop time ($T_{\textrm{stop}}$) was calculated as \tzero $+$ \emph{integration time}. The integration time was chosen for each scintillating material: \SI{400}{\nano\second} for \acrshort{gl:LYSO:Ce}, \SI{600}{\nano\second} for \acrshort{gl:GAGG:Ce} and \SI{800}{\nano\second} for \acrshort{gl:LuAG:Ce}.

\end{itemize}

An additional veto threshold $A_{\textrm{veto}}$ was defined as depicted in \cref{fig:signal-scheme}. If any sample in the acquisition window crossed the veto threshold value, the event was flagged as potential noise. The veto threshold was always chosen to be of opposite polarity to the expected signal polarity. Along with the standard deviation of the \acrshort{gl:BL}, it enabled filtering out signals that contained random noise or oscillations that occurred during measurements. The source of the noise signals was identified to be the power supply wires of the stepping motor and the ribbon cables of the 1-wire temperature sensors. 

The software used for the extraction of signal characteristics is available in a public repository \cite{sifi-framework}.

\subsubsection*{Charge calibration}

\begin{figure}[ht]
\centering
\includegraphics[width=.49\textwidth]{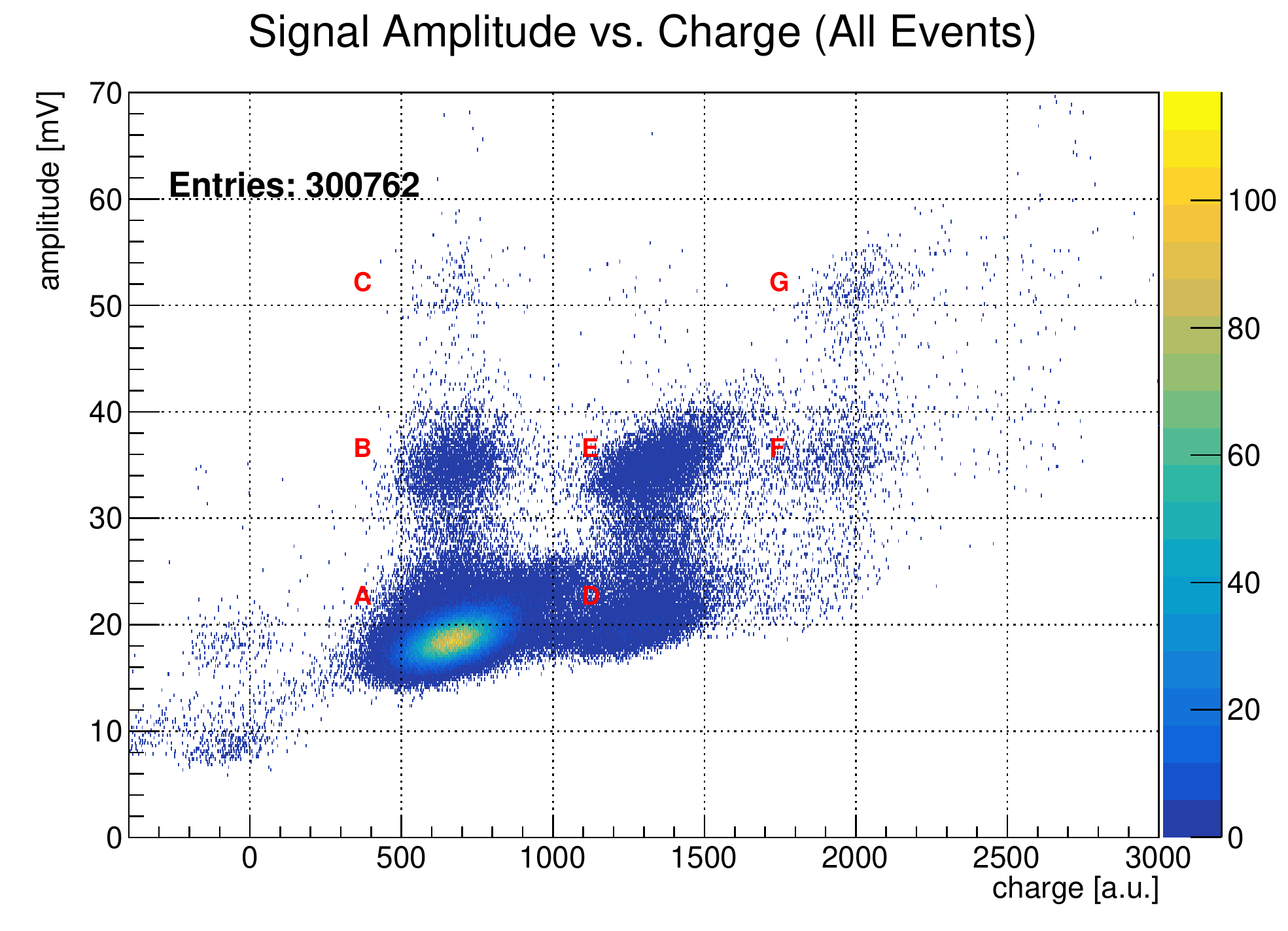}
\includegraphics[width=.49\textwidth]{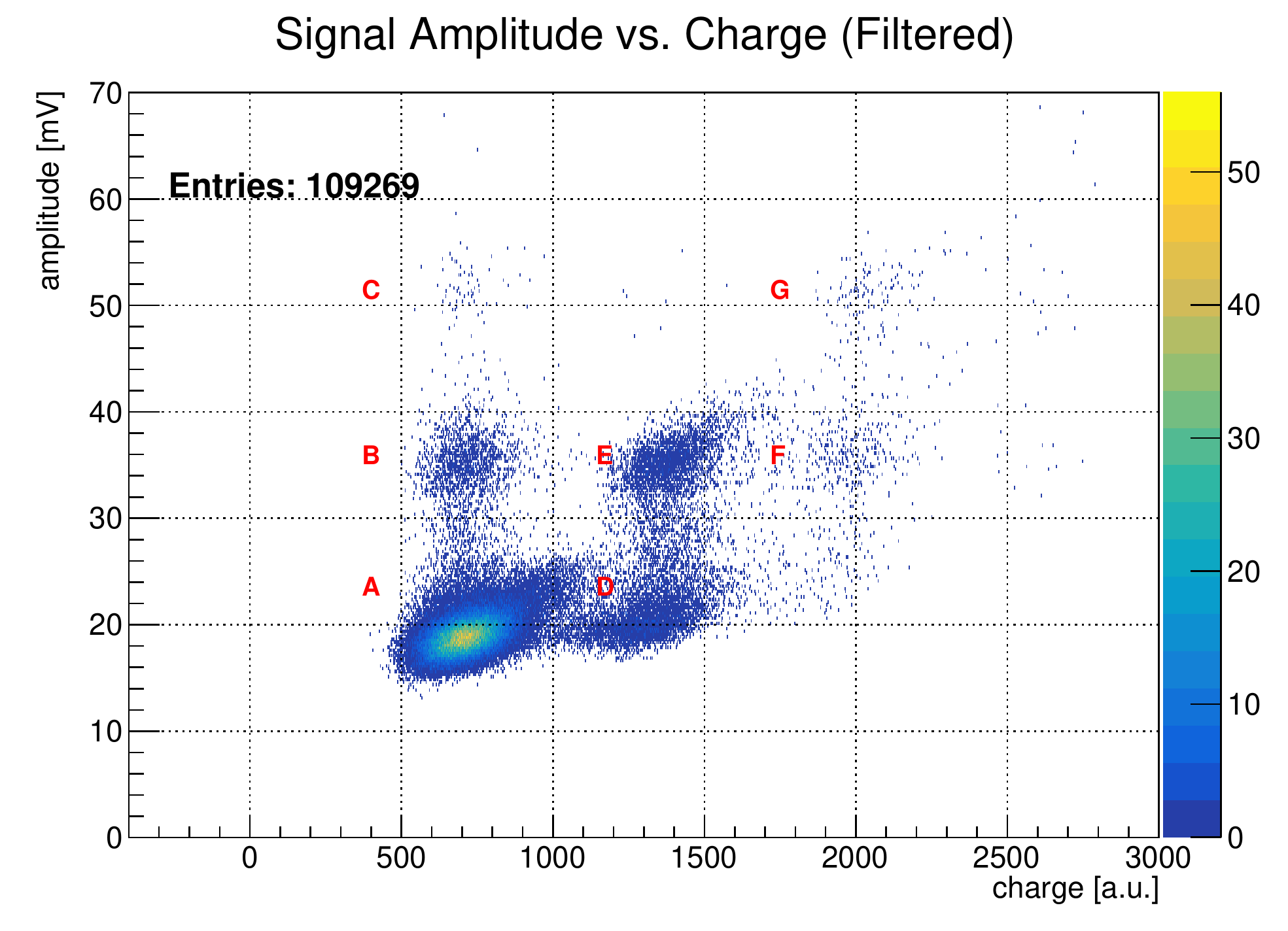}
\caption{Amplitude--charge correlation histograms for one of the channels reading out the investigated fiber. Left: histogram containing all recorded events. Right: histogram after application of filtering cuts. As a result of filtering the events which were flagged as potential noise were rejected. Red letters denote the distinguished signal groups described in the text. Presented data comes from calibration series 200 (see \cref{tab:calibrations-summary}).}
\label{fig:calib-maps}
\end{figure}

In the first stage of data preprocessing, the charge of recorded signals was recalculated to photoelectrons (\acrshort{gl:PE}).
Only data registered by the \acrshort{gl:SiPM}s connected with the investigated fiber were calibrated. The reference channel data were not calibrated since this channel was used solely for triggering and did not contribute to fiber characterization directly. For the purpose of \acrshort{gl:PE} calibration, the measurements of the \acrshort{gl:SiPM}s dark current for all used gain settings were performed. The features of recorded signals were determined using the same software as used for fiber measurements. The integration time for charge extraction was set at \SI{150}{\nano\second}. Subsequently, correlation histograms of signals amplitude and charge were plotted for each channel. \Cref{fig:calib-maps} presents such correlation histograms for a chosen channel in one of the calibration measurements. In the first histogram (left), all recorded events are plotted, and in the second (right) the filtering cuts were applied. Filtering cuts rejected events which were flagged as possible noise based on a large standard deviation of the baseline and reaching veto threshold $A_{\textrm{veto}}$. In \cref{fig:calib-maps} it can be seen that data quality improved significantly, proving that the defined criteria for the identification of noise signals were efficient. Therefore, filtered correlation histograms were further used for the determination of calibration coefficients. 

\begin{figure}[htp]
\centering
\includegraphics[width=.95\textwidth]{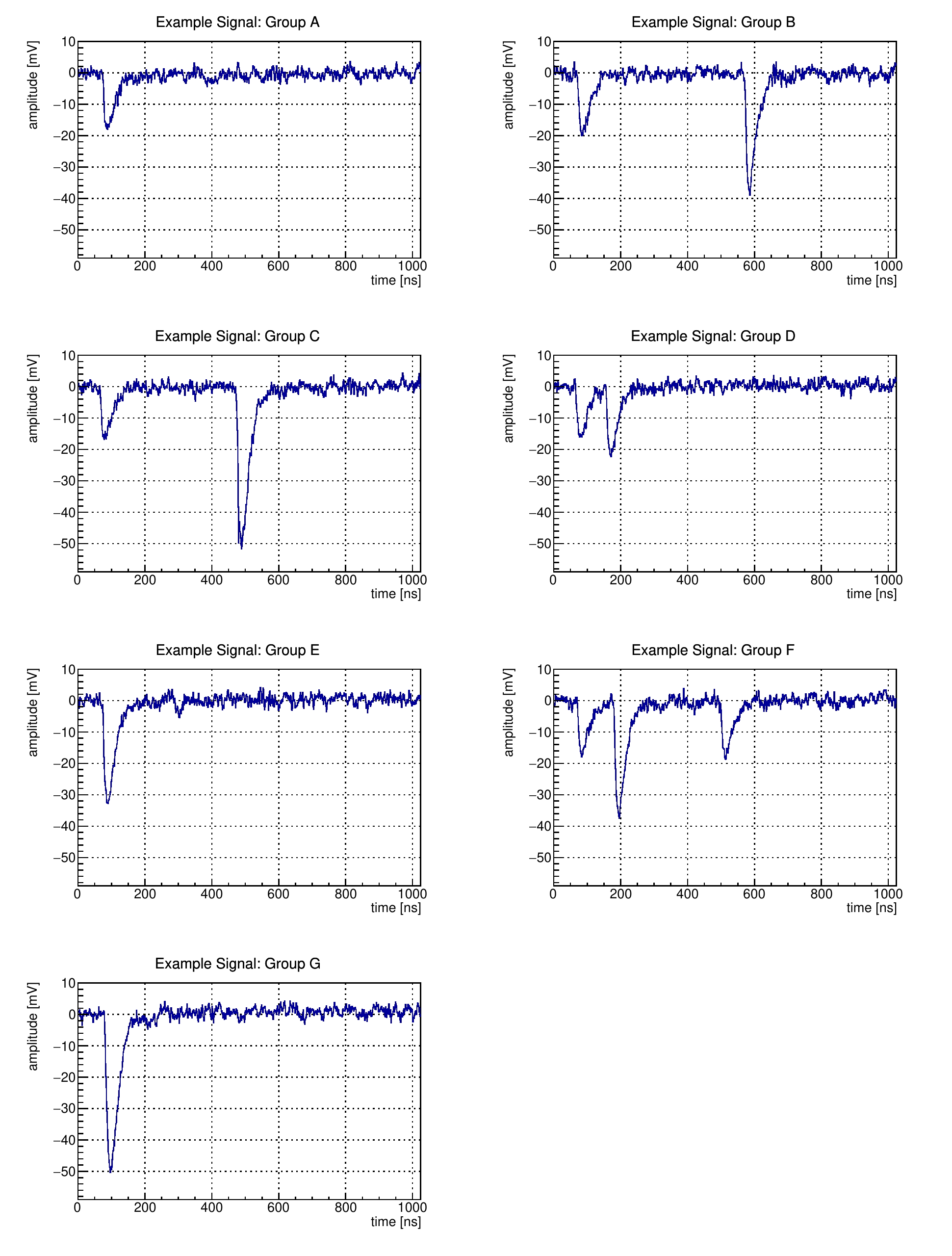}
\caption{Examples of signals registered during calibration measurement (series 200, see \cref{tab:calibrations-summary}).}
\label{fig:signals-examples}
\end{figure}

In the correlation histograms in \cref{fig:calib-maps} several distinct groups of signals can be identified:
\begin{enumerate}[label=(\Alph*)]

\item signals of amplitude approximately \SI{15}{\milli\volt} -- \SI{25}{\milli\volt} and charge approximately \SI{500}{\au} -- \SI{1000}{\au}, which correspond to single photoelectron (\SI{1}{\pe} signals). An example of such an event is shown in \cref{fig:signals-examples}. 

\item signals of amplitude approximately \SI{30}{\milli\volt} -- \SI{40}{\milli\volt} and charge approximately \SI{500}{\au} -- \SI{1000}{\au}, which correspond to amplitude of \SI{2}{\pe} signal and charge of \SI{1}{\pe} signal. This event topology corresponds to the situation in which more than one signal was captured in the acquisition window. In such a situation, the earlier signal was integrated, but the amplitude of the largest recorded signal was captured. This is a consequence of the operation of the algorithm for extraction of the signal features, \ie integration starts when the threshold $A_{\textrm{thr}}$ is achieved for the first time in the acquisition window, and the amplitude is determined as the highest recorded sample in the whole acquisition window. An example of such a situation is shown in \cref{fig:signals-examples}. The extraction algorithm works correctly in measurements with scintillating fibers, since scintillating signals are much longer than dark current signals and only one signal fits in the acquisition window. In case of dark current measurements, this difficulty was solved by choosing specific signal groups for analysis based on charge - amplitude correlation histograms as presented. 

\item signals of amplitude \SI{45}{\milli\volt} -- \SI{60}{\milli\volt} and charge \SI{500}{\au} -- \SI{1000}{\au} which correspond to amplitude of \SI{3}{\pe} signals and charge of \SI{1}{\pe} signal. This signal topology is analogous to the situation described in point B. An example event is shown in \cref{fig:signals-examples}. 

\item signals of amplitude \SI{16}{\milli\volt} -- \SI{28}{\milli\volt} and charge \SI{1050}{\au} -- \SI{1600}{\au}, \ie amplitude corresponding to \SI{1}{\pe} signals and charge of \SI{2}{\pe} signals. This event topology was observed when a pile-up occured, meaning that two single photoelectron signals were integrated as one, leading to doubled charge. An example of such an event is presented in \cref{fig:signals-examples}. 

\item signals of amplitude \SI{30}{\milli\volt} -- \SI{45}{\milli\volt} and charge \SI{1050}{\au} -- \SI{1600}{\au}, which corresponds to \SI{2}{\pe} signals (see \cref{fig:signals-examples}). 

\item signals of amplitude \SI{30}{\milli\volt} -- \SI{45}{\milli\volt} and charge \SI{1750}{\au} -- \SI{2300}{\au}, which corresponds to amplitude of \SI{2}{\pe} signals and charge of \SI{3}{\pe} signals. This event topology is characteristic of a pile-up of \SI{1}{\pe} and \SI{2}{\pe} signals. In this case summed integral of both signals and the amplitude of the higher one are obtained. An example of such an event is presented in \cref{fig:signals-examples}. 

\item signals of amplitude \SI{45}{\milli\volt} -- \SI{60}{\milli\volt} and charge \SI{1750}{\au} -- \SI{2300}{\au}, which corresponds to \SI{3}{\pe} signals (see \cref{fig:signals-examples}). 

\end{enumerate}

\begin{figure}[!ht]
\centering
\includegraphics[width=.49\textwidth]{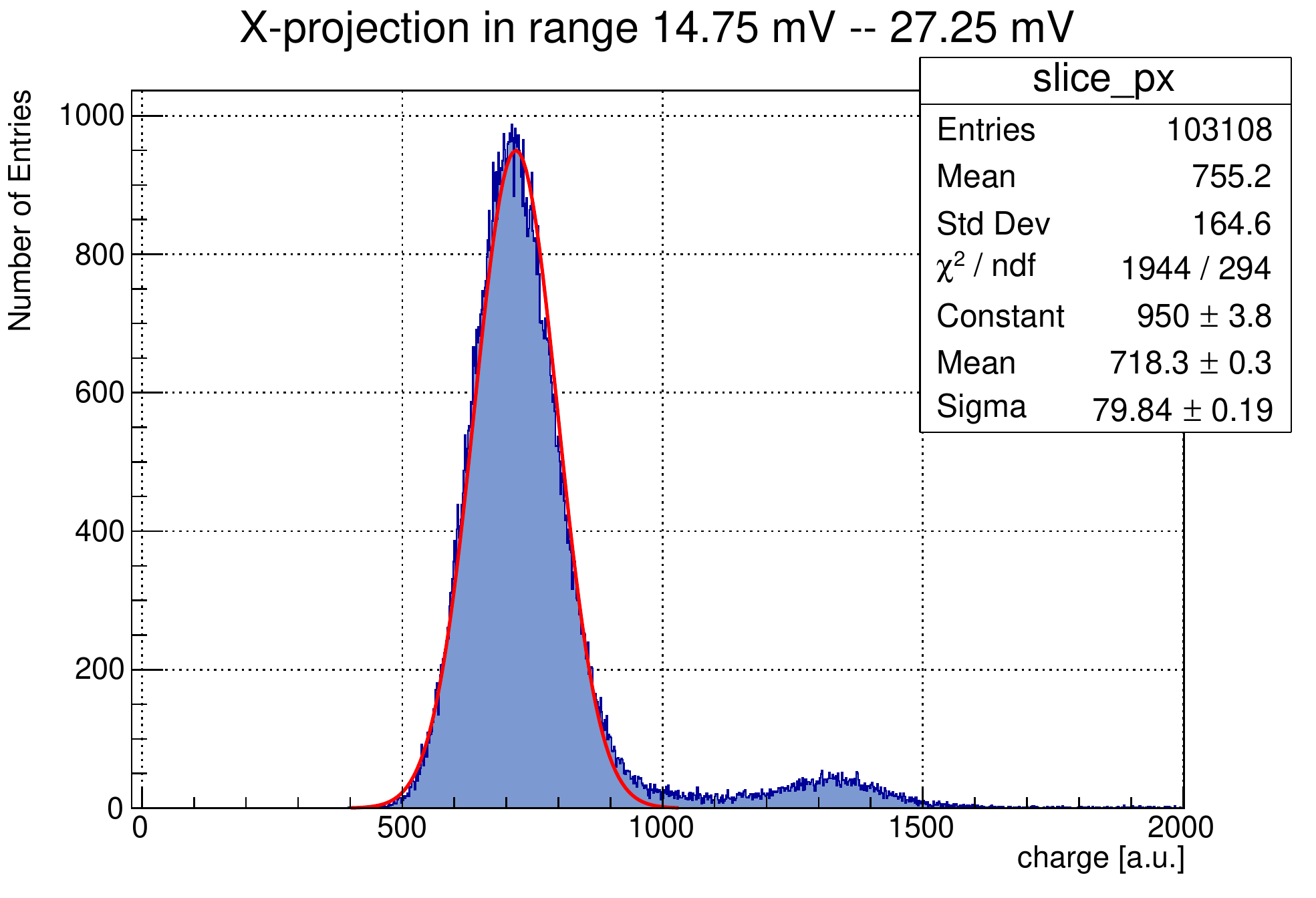}
\includegraphics[width=.49\textwidth]{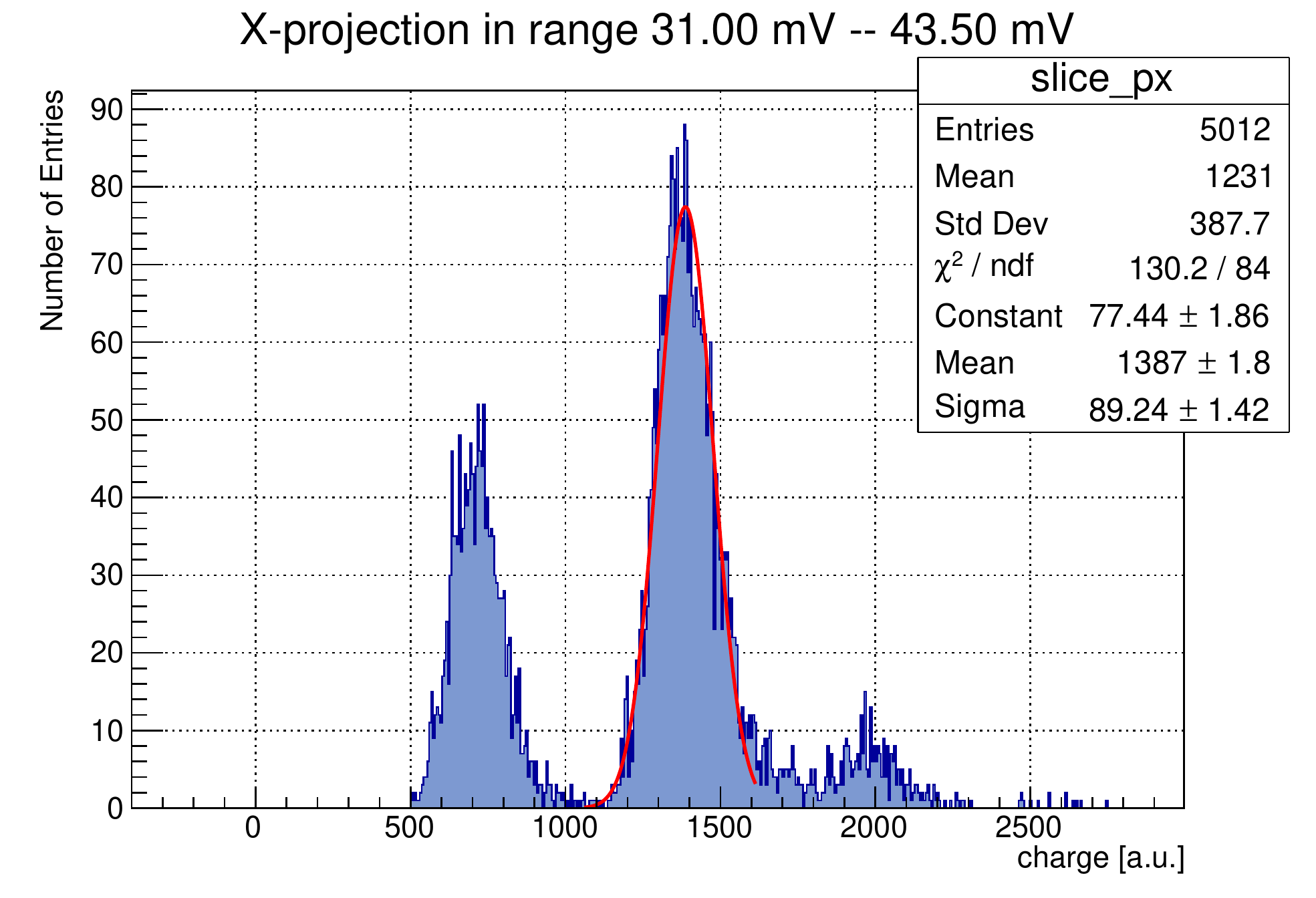}
\caption{Projections of the filtered correlation histogram on the X-axis. Left: projection taken in the range \SI{14.75}{\milli\volt} -- \SI{27.25}{\milli\volt}, which includes signals group A. The peak corresponding to the group A was fitted with the Gaussian function and obtained fit parameters are listed. Right: projection taken in the range \SI{31.00}{\milli\volt} -- \SI{43.50}{\milli\volt}, which includes signals group E. The peak corresponding to the group E was fitted with the Gaussian function and obtained fit parameters are listed.}
\label{fig:x-projections}
\end{figure}

For the determination of calibration coefficients, signals groups A and E were taken into account. Group G was rejected due to too small statistics. For both analyzed groups, a projection on the X-axis of the histogram was taken. The resulting distribution was subsequently fitted with the Gaussian function to determine a mean charge value corresponding to the given \acrshort{gl:PE} values (see \cref{fig:x-projections}). For calibration measurements in which more than two valid points were obtained, the points were plotted and fitted with the linear function. The calibration coefficient was then determined as the slope of the linear function. For series in which only two valid points were identified, the calibration factor was calculated as $\frac{1}{2}$ of the charge value corresponding to \SI{2}{PE}, which is the case for the presented calibration measurement. Having the calibration factor $f$, the charge was recalculated as follows:
\begin{equation}
    Q_\mathrm{PE} = \frac{Q_\mathrm{int}}{f},
\end{equation}
where $Q_\mathrm{PE}$ is charge expressed in \si{pe} and $Q_\mathrm{au}$ is signal integral obtained during data preprocessing.

Using obtained calibration coefficients all charge spectra were calibrated to \acrshort{gl:PE}. For further analysis only calibrated charge spectra were used.\Cref{tab:calibrations-summary} summarizes the list of performed calibration measurements and obtained calibration coefficients. The remaining recorded quantities (amplitude, \gls{gl:tzero}, \acrshort{gl:TOT}) did not require calibration, since they were stored directly in physical units. 

\begin{table}[!ht]
\centering
\caption{Details of the \acrshort{gl:PE} calibrations for all single-fiber measurements.}
\label{tab:calibrations-summary}
\begin{threeparttable}[!ht]
\begin{tabular}{|p{1.4cm}|p{2.0cm}|p{3.9cm}|p{3cm}|p{2.8cm}|}
\hline
Series number & Valid for \newline series & Operating voltage [\si{\volt}] \newline (overvoltage [\si{\volt}]) & Threshold & Calibration factors [\si{\volt\nano\second\revPE}] \\ \hline
\multicolumn{5}{|c|}{\acrshort{gl:SiPM}s: Hamamatsu, signal polarity: positive} \\ \hline
28 & 30 -- 32 & \SI{53.1}{\volt} (\SI{1.5}{\volt}) & \acrshort{gl:BL} + \SI{7}{\adc} & $f_L =$ \num{64.62} \newline $f_R =$ \num{60.56} \\ \hline
33\tnote{1} & 34 -- 43 & \SI{53.4}{\volt} (\SI{1.8}{\volt}) & \acrshort{gl:BL} + \SI{7}{\adc} & $f_{L} =$ \num{25.265} \newline $f_{R} =$ \num{24.493} \\ \hline
44\tnote{2} & 45 -- 46  & \SI{53.4}{\volt} (\SI{1.8}{\volt}) & L: \acrshort{gl:BL} + \SI{5}{\adc} \newline R: \acrshort{gl:BL} + \SI{6}{\adc} & $f_{L} =$ \num{36.82} \newline $f_{R} =$ \num{38.75} \\ \hline
65 & 63 -- 64, \newline 66 -- 67  & \SI{54.6}{\volt} (\SI{3}{\volt}) & \acrshort{gl:BL} + \SI{7}{\adc} &  $f_{L} =$ \num{61.02} \newline $f_{R} =$ \num{61.59} \\ \hline
\multicolumn{5}{|c|}{\acrshort{gl:SiPM}s: SensL, signal polarity: negative} \\ \hline
197 &  98 -- 107, 109 -- 180, 201 -- 222, 229 -- 261 & \SI{26.0}{\volt} (\SI{1.55}{\volt}) & L: \acrshort{gl:BL} - \SI{12}{\adc}, \acrshort{gl:BL} - \SI{25}{\adc} \newline R: \acrshort{gl:BL} - \SI{12}{\adc}, \acrshort{gl:BL} - \SI{23}{\adc} & $f_L =$ \num{246.57} \newline $f_{R} =$ \num{237.59} \\ \hline
224 & 225 -- 228 & \SI{25.8}{\volt} (\SI{1.35}{\volt}) & \acrshort{gl:BL} - \SI{16}{\adc}, \newline \acrshort{gl:BL} - \SI{20}{\adc} & $f_{L} =$ \num{209.71} \newline $f_{R} =$ \num{210.21} \\ \hline
199 & 181 -- 188 & \SI{26.8}{\volt} (\SI{2.35}{\volt}) & \acrshort{gl:BL} - \SI{14}{\adc} & $f_{L} =$ \num{371.75} \newline $f_{R} =$ \num{349.40} \\ \hline
200 & 189 -- 196 & \SI{28.7}{\volt} (\SI{4.25}{\volt}) & \acrshort{gl:BL} - \SI{20}{\adc} & $f_{L} =$ \num{693.5} \newline $f_{R} =$ \num{655.0} \\ \hline
\end{tabular}
\begin{tablenotes}
\item[1] For series 34 -- 43 and corresponding calibration series 33 additional attenuating wires were introduced in the setup. The wires were attenuating incoming signals by a factor of $\sim$ \si{4}.
\item[2] Starting with series 44 the attenuating wires were removed from the setup, however, the evaluation boards were modified in order to decrease signal amplification.
\end{tablenotes}
\end{threeparttable}
\end{table}

\subsubsection*{Annihilation peak parameterization}

\begin{figure}[!ht]
\centering
\includegraphics[width=.49\textwidth]{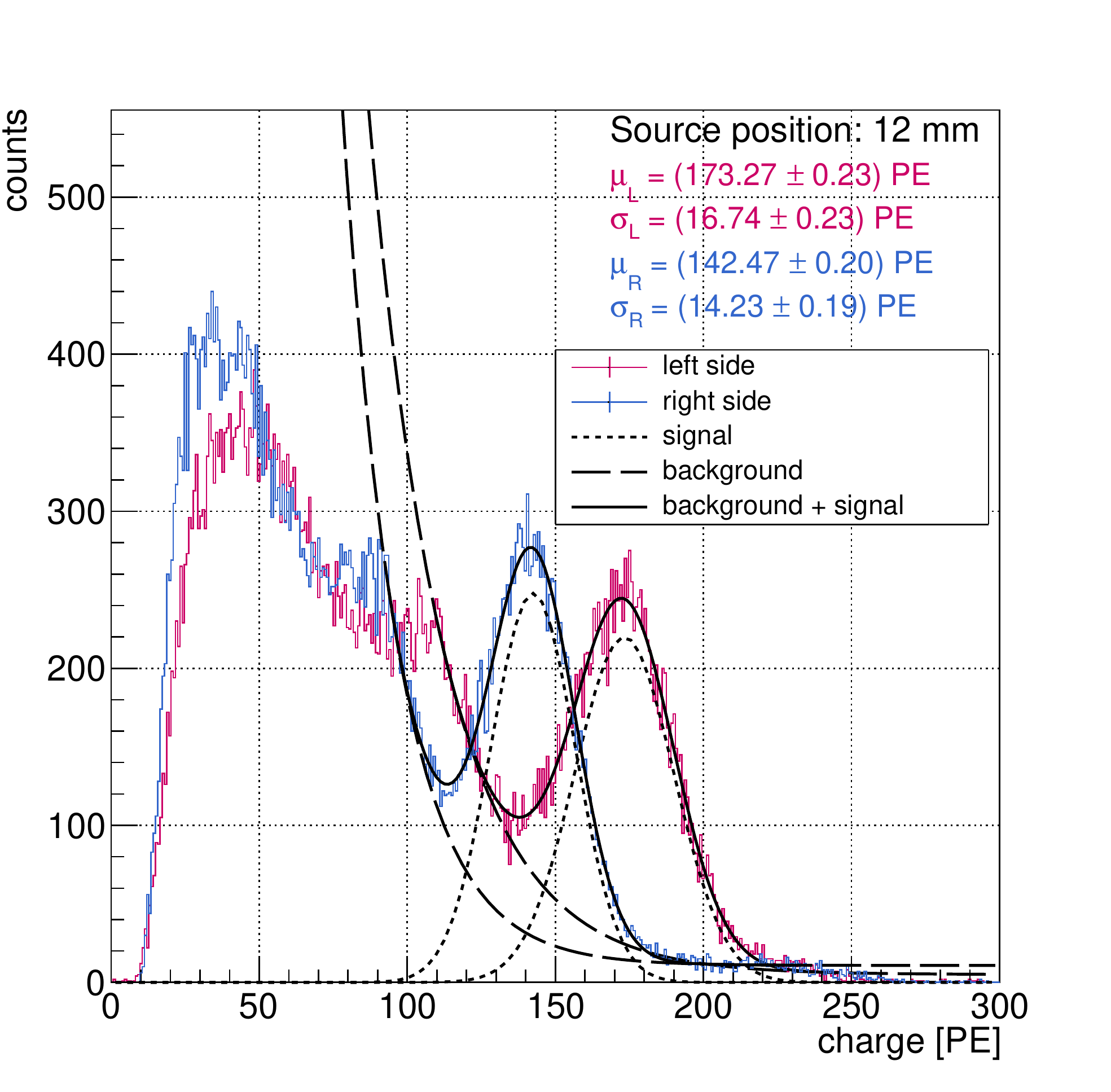}
\includegraphics[width=.49\textwidth]{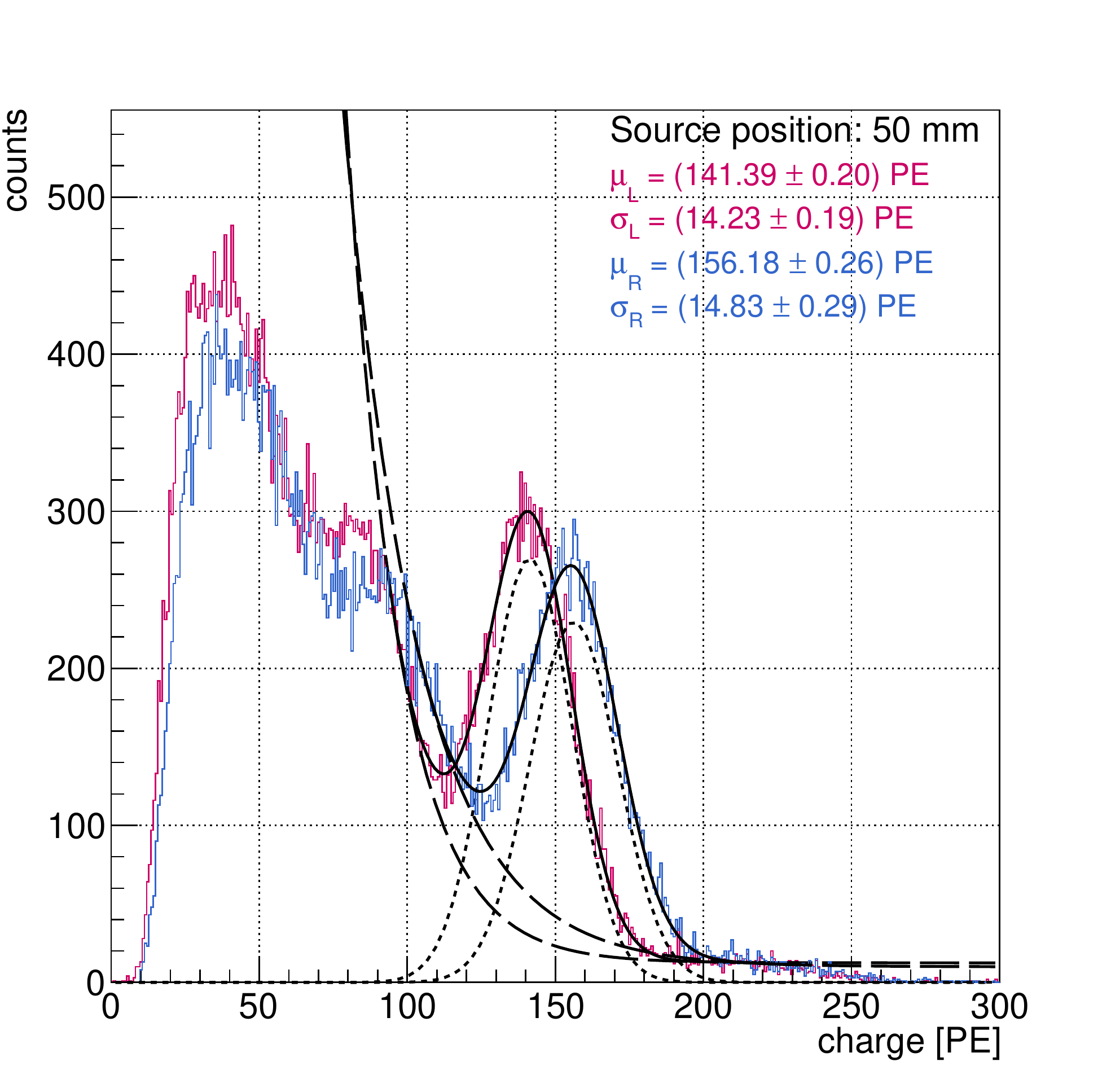}
\caption{Examples of charge spectra recorded at both ends of the investigated \acrshort{gl:LYSO:Ce} fiber (series 109) and for two different source positions: \SI{12}{\milli\meter} (left) and \SI{50}{\milli\meter} (right). Function ~\cref{eq:fitted-function} was fitted to each spectrum, resulting in the parameterization of the \SI{511}{\kilo\electronvolt} peaks. The parameters obtained from the fits are listed in each panel. Adapted from \cite{Rusiecka2021}.} 
\label{fig:spectrum-example}
\label{fig:raw-spectra-fitting}
\end{figure}

Most of the further data analysis was based on the \anhpeak peak, therefore, it was necessary to fit the obtained \acrshort{gl:PE} spectra and parameterize the peak shape and the surrounding background. For that purpose, the following function was used: 
\begin{equation}
\label{eq:fitted-function}
    f(x) = c \cdot \exp{\left(- \frac{\left( x - \mu\right)^2}{2\sigma^2} \right)} + p_1 \cdot \exp{\left( (x - p_2) \cdot p_3 \right)} + p_4,
\end{equation}
where the parameters of the Gaussian function describe the \anhpeak peak: $c$ - intensity, $\mu$ - position (charge, energy), $\sigma$ - standard deviation. The parameters $p_1$ - $p_4$ describe background. Examples of charge spectra recorded at both ends of the fiber and at two different positions of the radioactive source can be seen in \cref{fig:spectrum-example}, where also the signal-background decomposition according to \cref{eq:fitted-function} is shown. It can be observed that, depending on the source position along the fiber, the spectra change their shape and the annihilation peak shifts as expected: towards larger charges for smaller source-\acrshort{gl:SiPM} distances. This is a consequence of the attenuation of scintillation light in the elongated fiber. Thus, for the source positioned in the middle of the fiber, the charge spectra from both ends are expected to overlap. The observed differences, such as in \cref{fig:spectrum-example}, can be associated with differences in the coupling or internal structural defects of the fiber.

\newpage


\section{Characterization results}
\label{sec:sf-results}

The obtained charge, amplitude, and time spectra constituted the input for further analysis and allowed to characterize the performance of each investigated fiber in combination with coupling, wrapping, or coating. During data analysis, properties such as attenuation length, energy resolution, timing resolution, position resolution, and light collection were determined. All properties and methods for their determination were described in \cref{sec:light-propagation} and \cref{sec:charatcerization-scintillators}.

\subsubsection*{Quality of data description by light attenuation models}

\begin{sidewaysfigure}[hp]
\centering
\includegraphics[height=.45\textwidth]{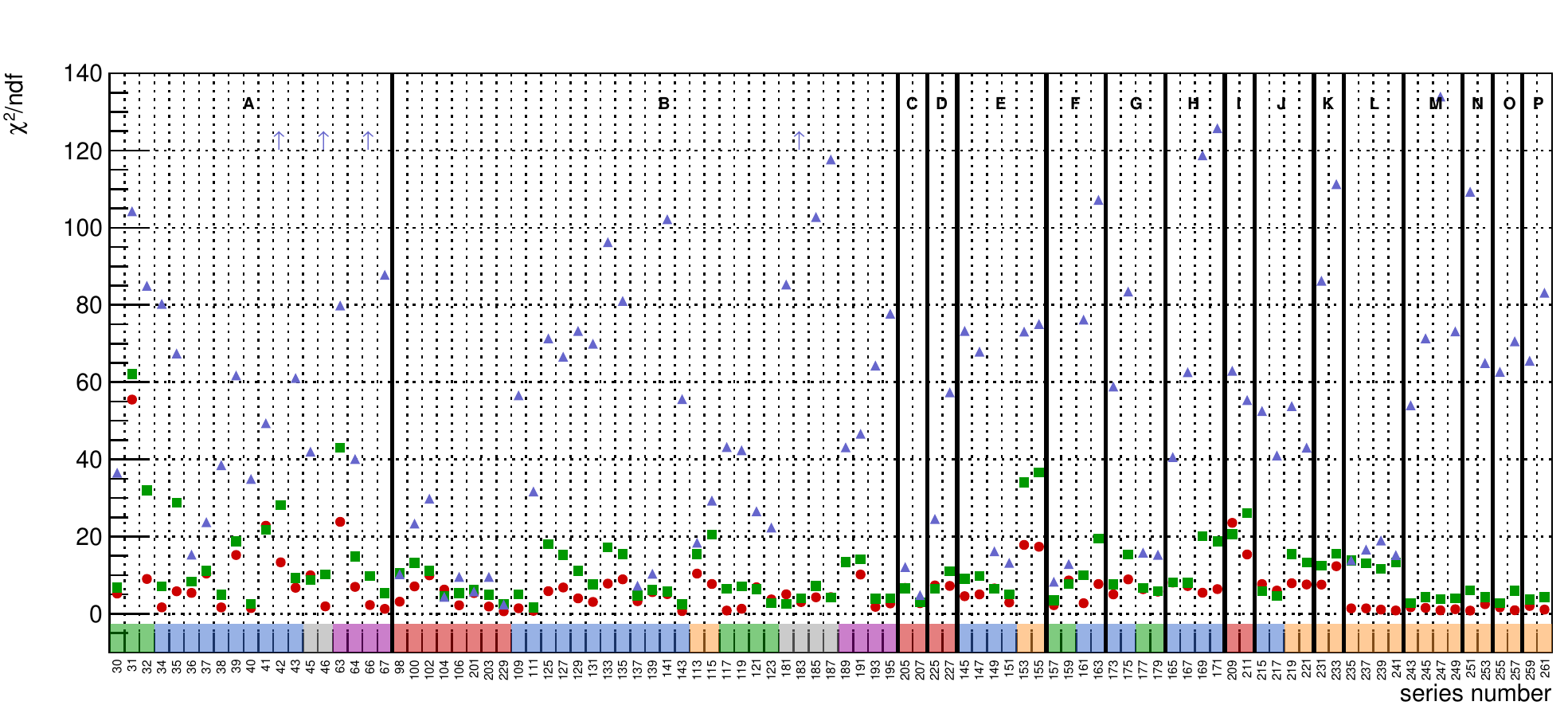}
\includegraphics[height=.12\textwidth]{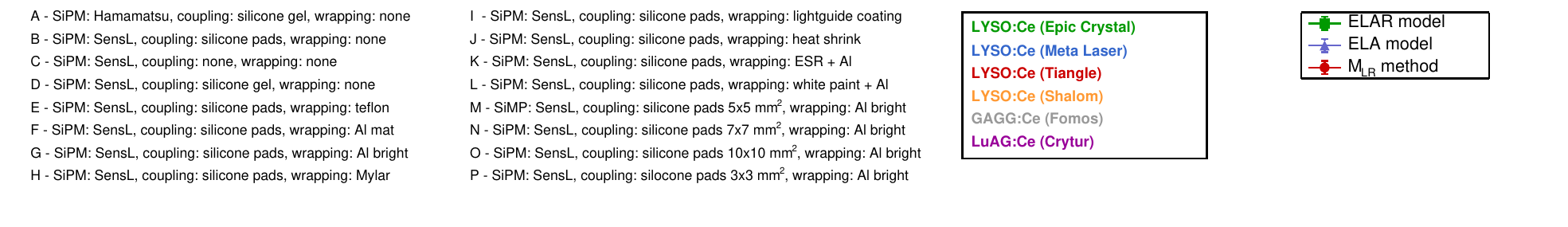}
\caption{Reduced $\chi^2$ values of \acrshort{gl:ELA}, \acrshort{gl:ELAR} and \gls{gl:MLR} models fitting for all analyzed measurements. The value of \chiNDF of the \acrshort{gl:ELA} fit for series 183 was too large to be shown in the graph (180.54) and was denoted with an arrow.}
\label{fig:single-fib-chi2}
\end{sidewaysfigure}

All three methods of light attenuation determination: \acrshort{gl:ELA}, \gls{gl:MLR} and \acrshort{gl:ELAR} were applied to the experimental data. To evaluate the performance of the models, the values of \chiNDF for each method were plotted for all experimental series (see \cref{fig:single-fib-chi2}). It can be seen that the \gls{gl:MLR} method yields the best performance, with the lowest \chiNDF values for most of the experimental series (0.65--55.49). The \acrshort{gl:ELAR} method shows similar performance, with comparable or slightly higher values of \chiNDF (1.64 -- 62.08). In most cases, the \acrshort{gl:ELA} model shows the worst performance, with the \chiNDF values ranging from 2.39 to 351.48. This comparison hints at the \gls{gl:MLR} and \acrshort{gl:ELAR} methods as the more reliable for determining the attenuation length for the investigated scintillating fibers.

\vspace{0.5cm}

\begin{sidewaysfigure}[hp]
\centering
\includegraphics[height=.45\textwidth]{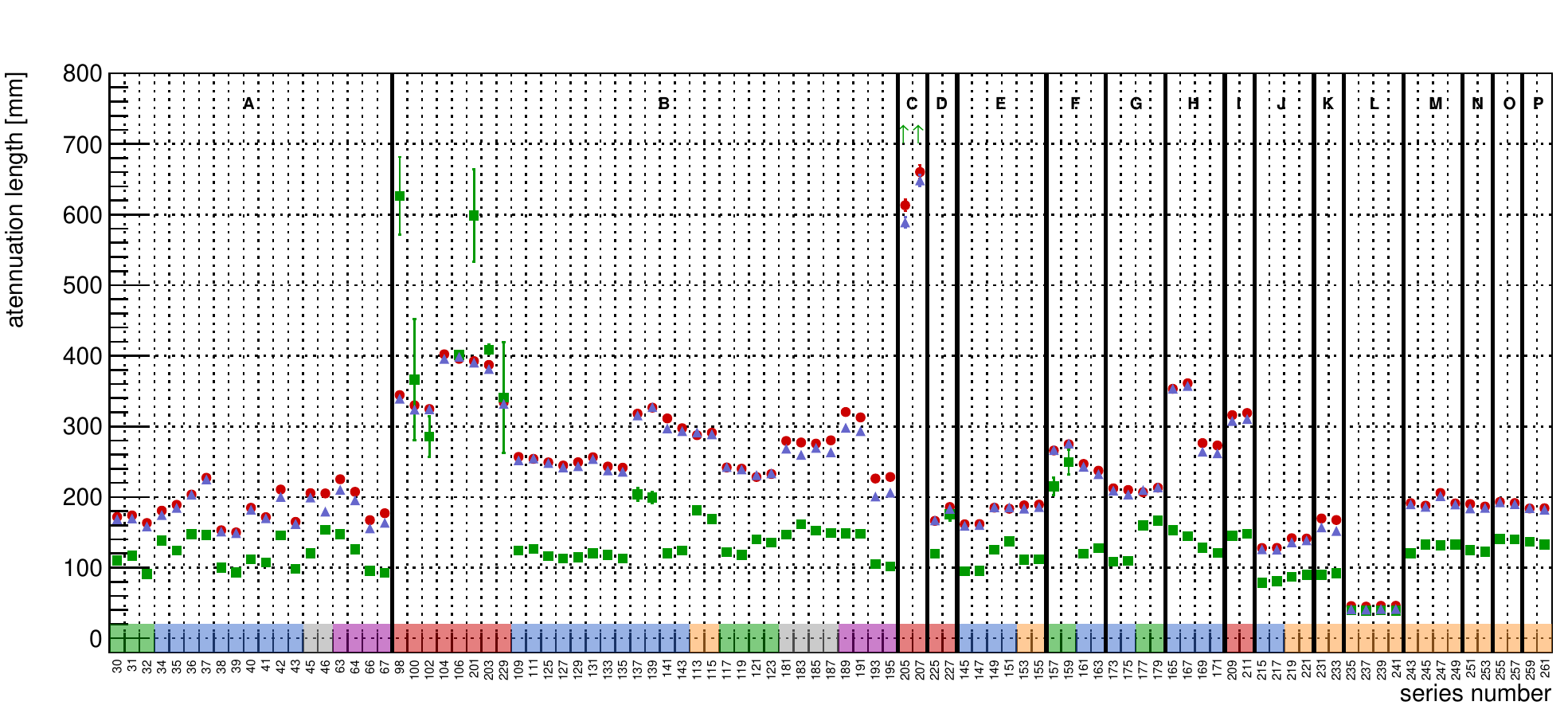}
\includegraphics[height=.12\textwidth]{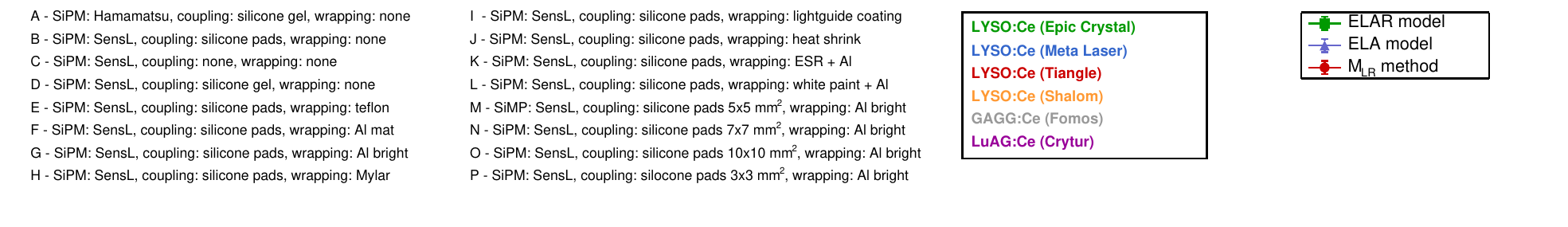}
\caption{Values of attenuation length determined with the \acrshort{gl:ELA}, \acrshort{gl:ELAR} and \gls{gl:MLR} methods for all analyzed experimental series. The attenuation lengths determined with the \acrshort{gl:ELAR} model for series 205 and 207 were too large to fit in the graph and were indicated with arrows. The values of the attenuation length for these series are close to the imposed limit ($10^4$) with uncertainties close to \SI{100}{\percent}.}
\label{fig:single-fib-attenuation}
\end{sidewaysfigure}

\subsubsection*{Attenuation length}

The values of attenuation length obtained with the three methods (\acrshort{gl:ELA}, \gls{gl:MLR} and \acrshort{gl:ELAR}) were plotted for all analyzed experimental series (see \cref{fig:single-fib-attenuation}). The results obtained with the \acrshort{gl:ELA} and \gls{gl:MLR} methods are in agreement. At the same time, the values obtained with the \acrshort{gl:ELAR} method are significantly smaller for most of the series. This is a consequence of the different assumed light propagation pattern in this model. Thus, the values should only be used together with the formalism of the model within which they were obtained.

Furthermore, the attenuation length values obtained with the \acrshort{gl:ELA} and \gls{gl:MLR} methods in measurements with the Hamamatsu \acrshort{gl:SiPM}s are lower and more scattered than those obtained in measurements with the SensL \acrshort{gl:SiPM}s. This can be caused by different couplings used: silicone gel in the Hamamatsu measurements and silicone pads in the SensL measurements. Such an observation suggests that silicone gel coupling is unstable in consecutive measurements, which leads to non-reproducible results. This effect is even stronger in the analysis of light collection (see \cref{fig:single-fib-lcol}). 

It can be seen from \cref{fig:single-fib-attenuation}, that the results of the \acrshort{gl:ELAR} model for Tianle fibers in groups B and C raise doubts. The obtained values are unusually large for this scintillating material, differ strongly from results obtained for other samples, and have very large uncertainties. Additionally, results obtained for series 205 and 207 suggest that the fitting limit ($10^4$) was reached. The fact that uncertain results were obtained only for the Tianle fibers under specific experimental conditions may suggest that the failure of the \acrshort{gl:ELAR} model is caused by some feature of this fiber-coupling combination. It should be noted that Tianle fibers, especially when no coupling is applied, are characterized by the largest attenuation lengths. Additionally, it can be seen in \cref{fig:single-fib-chi2}, that for these series, the performance of the \acrshort{gl:ELA} method is satisfactory. Therefore, it appears that the proposed \acrshort{gl:ELAR} model does not perform well since it introduces additional degrees of freedom, which are not required for a satisfactory data description. Therefore, in further comparisons the results obtained with the use of the \acrshort{gl:ELAR} model for Tianle fibers in groups B and C are ignored. 

Differences in attenuation length that are visible in particular for different couplings and some of the wrappings and coatings are further discussed in \cref{sec:sf-comparative-studies}.

\subsubsection*{Position resolution}

\begin{sidewaysfigure}[!hp]
\centering
\includegraphics[height=.45\textwidth]{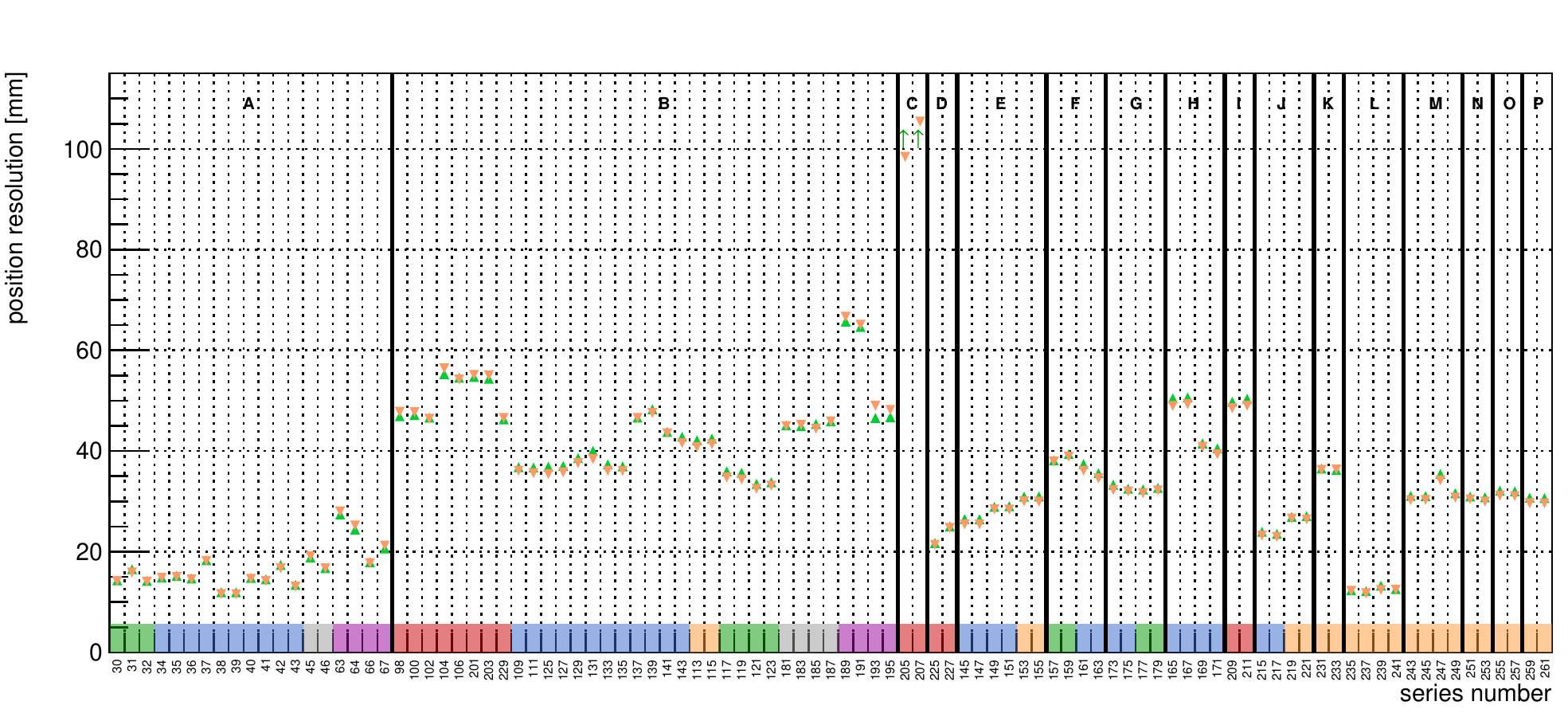}
\includegraphics[height=.12\textwidth]{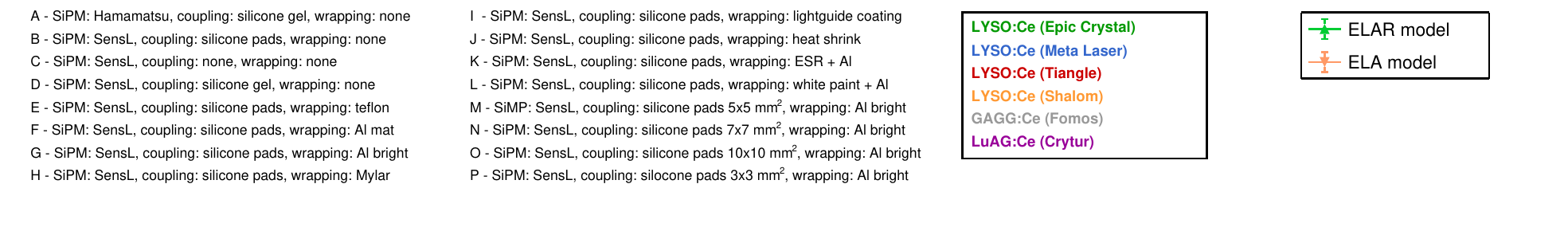}
\caption{Position resolution determined using \acrshort{gl:ELAR} model, and the \gls{gl:MLR} method of \acrshort{gl:ELA} model for all analyzed measurement series. The position resolutions determined with the \acrshort{gl:ELAR} model for series 205 and 207 were too large to fit in the graph and were denoted with arrows. Values of position resolution for those series were in the order of \SI{2e3}{\milli\meter} with uncertainties of \num{80}--\SI{90}{\percent}. This is connected with the large attenuation length obtained for those series (see \cref{fig:single-fib-attenuation}).}
\label{fig:single-fib-positionres}
\end{sidewaysfigure}

The position reconstruction and determination of position resolution were carried out using two previously described methods: using \gls{gl:MLR} quantity, and using \gls{gl:MLRstar} obtained with the \acrshort{gl:ELAR} parameterization. The values of the position resolution determined with both methods, were plotted for all experimental series (see \cref{fig:single-fib-positionres}). The plotted values are integrated position resolutions, \ie they were obtained from the distributions of residuals \gls{gl:Xreco} $-$ \gls{gl:Xreal} for all measurements in the series. The differences that can be observed for various materials, producers, couplings, and wrappings are further discussed in \cref{sec:sf-comparative-studies}. It is necessary to note that position resolution is correlated with the attenuation length, \ie  the fiber characterized with the large attenuation length will also have large (\ie poor) position resolution. Therefore, the pattern of \cref{fig:single-fib-attenuation} is reproduced in that figure. Moreover, it can be seen that the two methods yield a very similar position resolution for each analyzed series.

\subsubsection*{Energy resolution}

\begin{sidewaysfigure}[!hp]
\centering
\includegraphics[height=.45\textwidth]{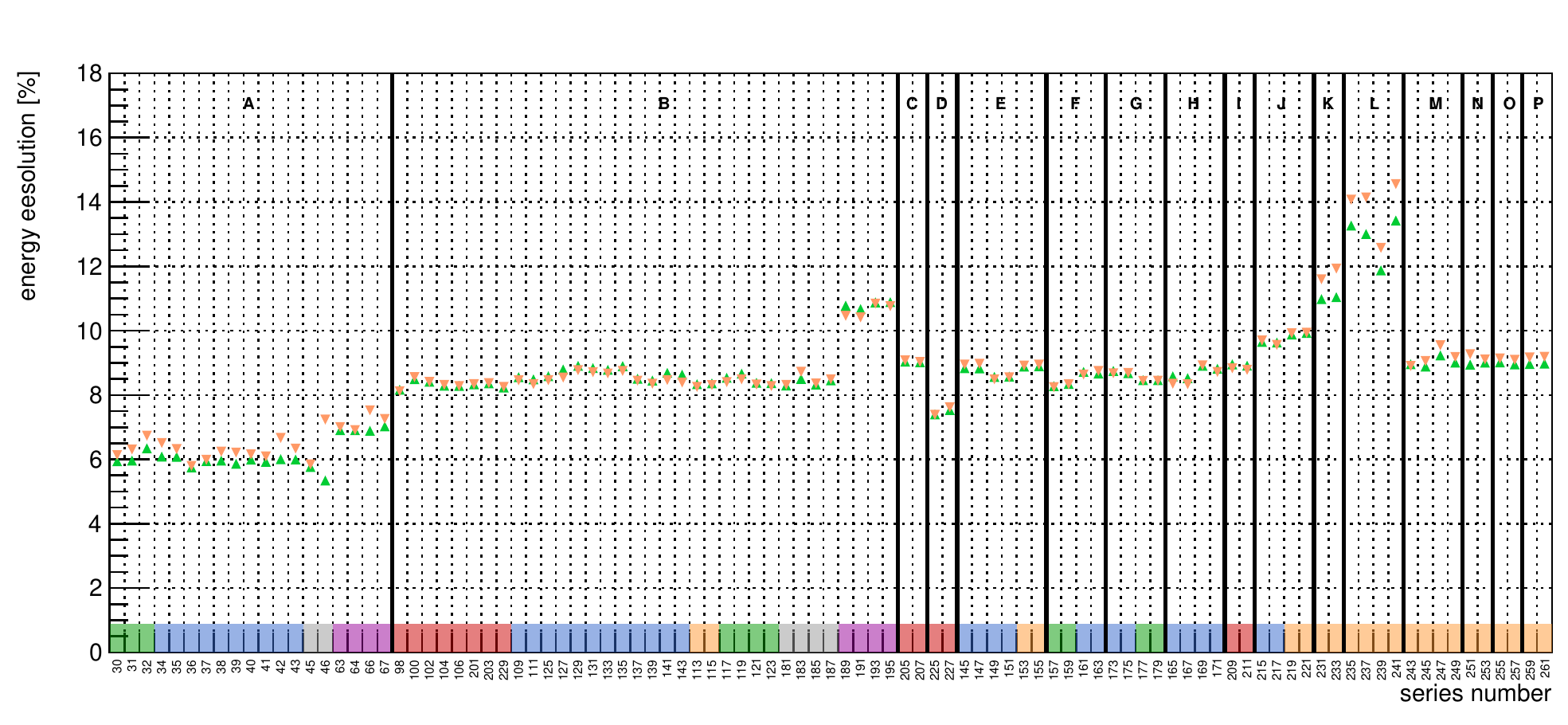}
\includegraphics[height=.12\textwidth]{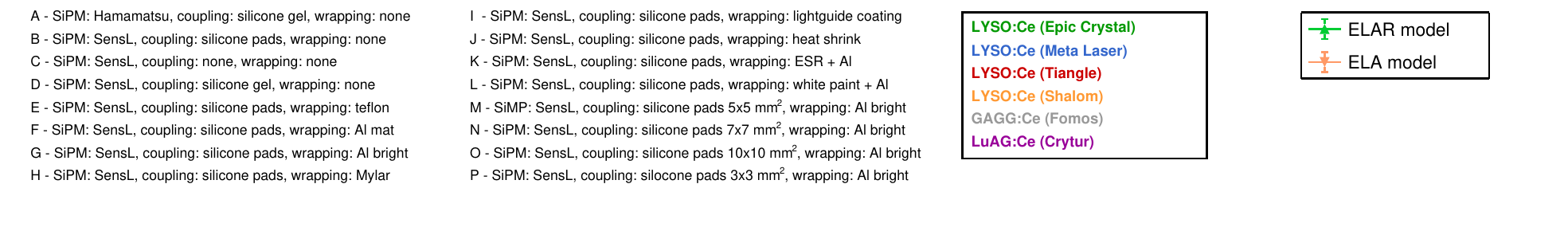}
\caption{Energy resolution determined using both \acrshort{gl:ELA} and \acrshort{gl:ELAR} method for all analyzed measurement series.}
\label{fig:single-fib-energyres}
\end{sidewaysfigure}

Similarly as for the position, the energy reconstruction and energy resolution determination were done with the two methods: using the \gls{gl:Qavg} quantity, and using the \gls{gl:Qavgstar} quantity, obtained with the \acrshort{gl:ELAR} parameterization. The results obtained with the two methods were plotted for all recorded series in \cref{fig:single-fib-energyres} allowing for comparison of energy resolution among investigated samples. Each energy resolution value was determined basing on the summed reconstructed energy spectra. Differences that can be observed for different materials, couplings, and wrappings are discussed further in \cref{sec:sf-comparative-studies}. As visible in \cref{fig:single-fib-energyres}, energy resolution is better in the measurements with the Hamamatsu \acrshort{gl:SiPM}s compared to the measurements with the SensL \acrshort{gl:SiPM}s. This is a consequence of larger amount of light collected in those measurements, which is demonstrated in detail in \cref{fig:single-fib-lcol}. The two methods of energy reconstruction yield very similar results in terms of energy resolution. A significant difference is visible only for fibers wrapped in ESR+Al and painted with the \baso-based paint+Al. 

\subsubsection*{Timing resolution}

\begin{sidewaysfigure}[!hp]
\centering
\includegraphics[height=.45\textwidth]{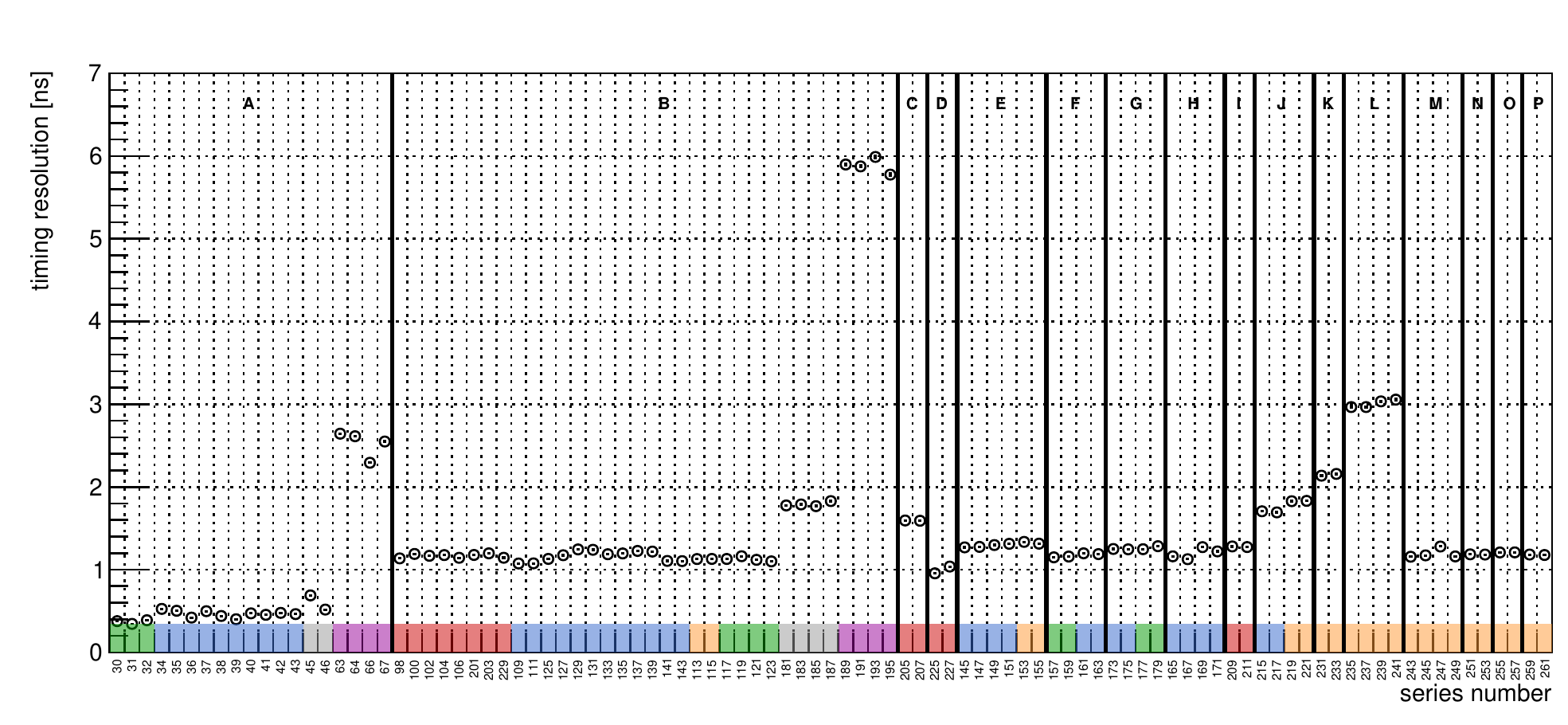}
\includegraphics[height=.12\textwidth]{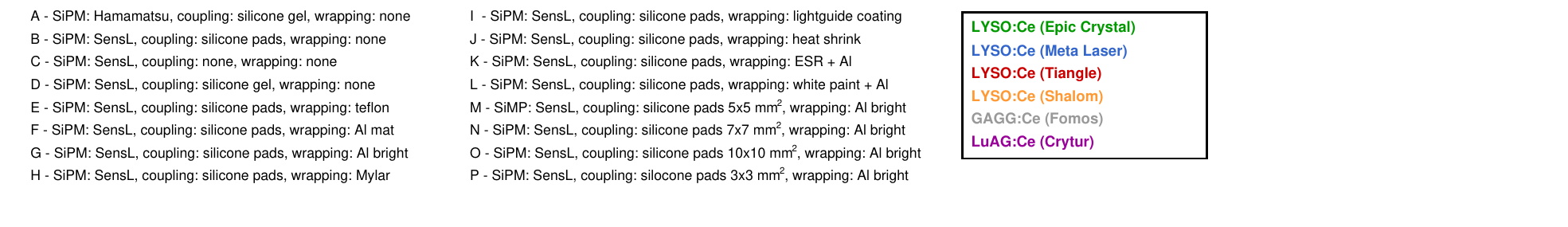}
\caption{Obtained timing resolution values for all analyzed measurement series.}
\label{fig:single-fib-timeres}
\end{sidewaysfigure}

The obtained values of timing resolution for all experimental series are plotted in \cref{fig:single-fib-timeres}. It can be observed that the type of scintillating material, the coupling, as well as the type of coating and wrapping influence the timing characteristics. These differences are further discussed in \cref{sec:sf-comparative-studies}. Additionally, the timing resolution is significantly better in measurements with the Hamamatsu \acrshort{gl:SiPM}s, which is a consequence of their superior timing characteristics.

\subsubsection*{Light collection}

\begin{sidewaysfigure}[!hp]
\centering
\includegraphics[height=.45\textwidth]{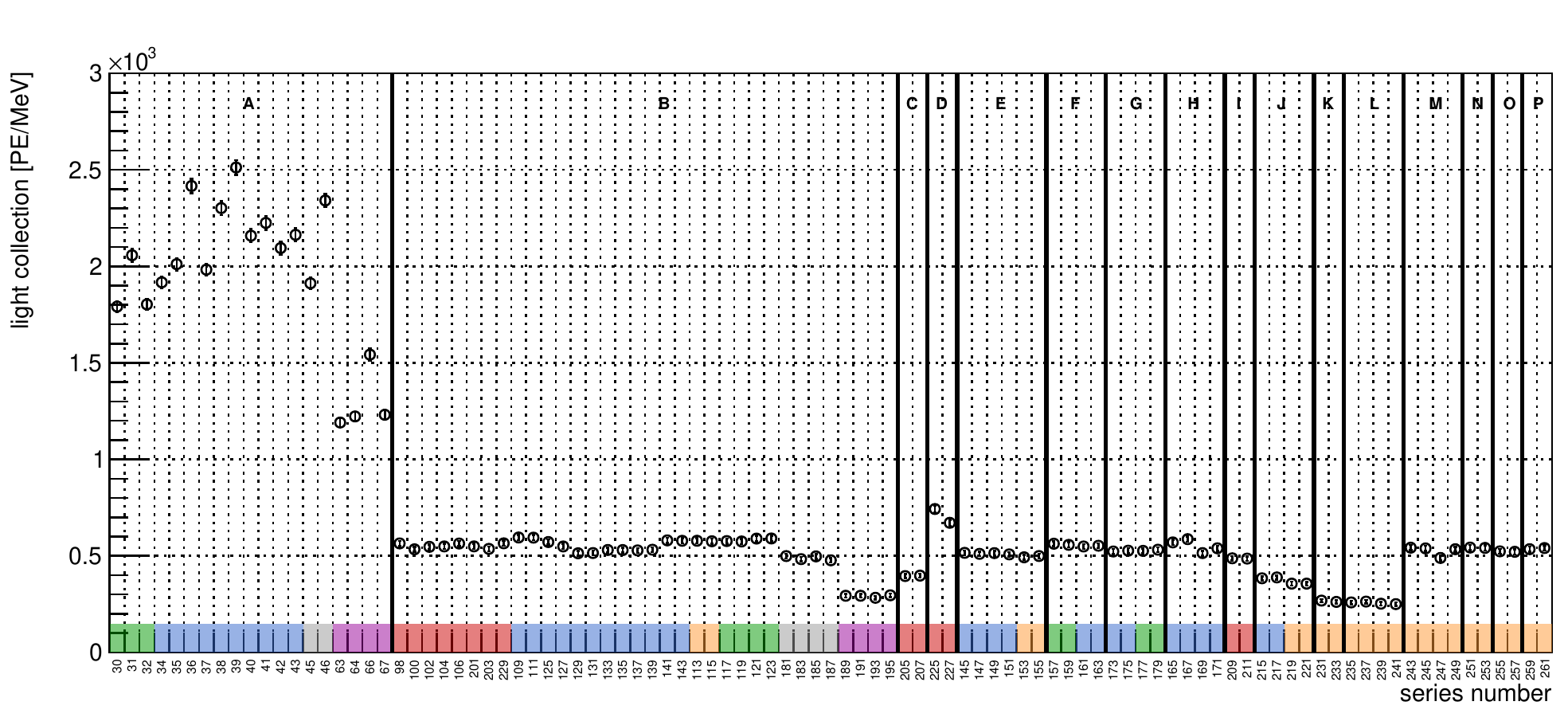}
\includegraphics[height=.12\textwidth]{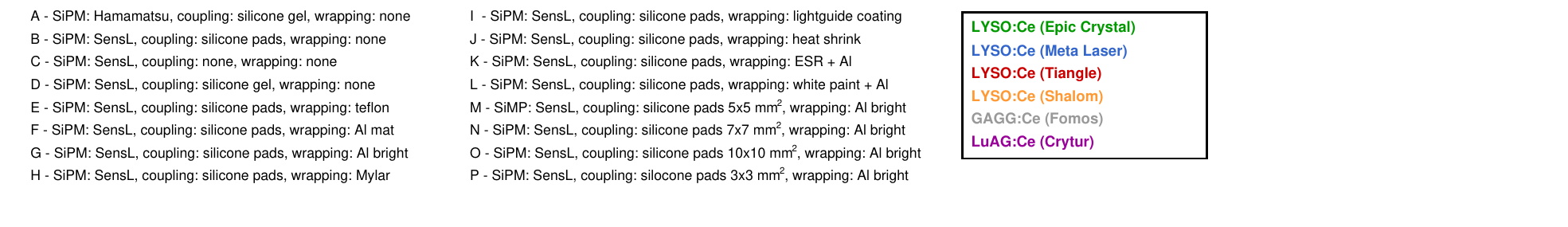}
\caption{Values of light collection for all analyzed measurement series.}
\label{fig:single-fib-lcol}
\end{sidewaysfigure}

The amount of light recorded from each of the investigated samples was parameterized using the light collection quantity. The obtained values are presented in \cref{fig:single-fib-lcol}. It can be observed that the light collection determined for measurements with the Hamamatsu \acrshort{gl:SiPM}s is significantly larger than that for measurements with the SensL \acrshort{gl:SiPM}s. This is caused by the larger photodetection efficiency (\acrshort{gl:PDE}) of Hamatatsu \acrshort{gl:SiPM}s and the use of silicone gel as the coupling. However, although the light collection is larger in the Hamamatsu measurements, the values exhibit a large spread, while those in the SensL measurements are much more stable. As mentioned previously, this can be attributed to different types of coupling used. The silicone pads always have the same thickness and are relatively durable, thus providing a stable and reproducible coupling. In case of the silicone gel, the layer of the applied gel can vary strongly between the measurements when the fiber is removed from the experimental setup and later installed again. Moreover, the layer can deteriorate during the measurement, \eg as a result of heat or a leak. The observed inconsistencies in light collection were a direct reason for the introduction of improvements in the experimental setup, such as installation of positioning millimeter screw and independent temperature sensors (see \cref{ssec:sf-setup-daq}).

Due to a large spread in the Hamamatsu measurements resulting from the poor quality of coupling, these data were not used in further comparisons for properties that strongly depend on the number of registered photoelectrons, since they could lead to misleading conclusions about the scintillating materials.

In measurements with the SensL \acrshort{gl:SiPM}s, some differences in the light collection obtained for different materials, types of couplings, and some wrappings and coatings can be observed. They are discussed in detail in \cref{sec:sf-comparative-studies}.


\section{Comparative studies}
\label{sec:sf-comparative-studies}

All recorded series were grouped taking into account setup elements and measurement conditions, \ie type and producer of the scintillating fiber, the coupling medium, the wrapping or coating of the fiber, the \acrshort{gl:SiPM}s used. This allowed to compare performance of the investigated fibers in various configurations. The following properties were compared:  

\begin{itemize} 
\item The attenuation length determined with the \gls{gl:MLR} and \acrshort{gl:ELAR} model. Attenuation length determined using the \acrshort{gl:ELA} method was not included in the comparison due to the poor quality of the fits. Since it was observed that the \acrshort{gl:ELAR} model does not perform well for Tianle \acrshort{gl:LYSO:Ce} fibers, those results were discarded in comparative studies.
\item Position resolution - since there was no significant difference between the two described methods of determination of position resolution, the method based on the \gls{gl:MLR} variable within the \acrshort{gl:ELA} model was used in the comparative studies. 
\item Energy resolution - similarly to above, the results obtained with the \acrshort{gl:ELA} model and the \gls{gl:Qavg} variable were used for comparison.
\item Timing resolution. 
\item Light collection.  
\end{itemize} 

Due to the poor quality of coupling in the experimental series with the Hamamatsu \acrshort{gl:SiPM}s and the resulting instability of measurements, these data were not included in the comparison of most of the properties. The exception is the timing resolution, which is not affected strongly by the coupling instability. Thus, in the following section, the presented results are for the SensL \acrshort{gl:SiPM}s, unless stated otherwise.

If several measurement series were recorded under the same set of conditions, the final result listed represents a weighted mean of single results. The weight in the weighted mean was the inverse square of the uncertainty. There are no uncertainties listed since the calculated statistical uncertainties were much smaller than the achieved measurement precision. 


\subsection{Different scintillating materials}
\label{ssec:sf-diff-materials}

\cref{tab:diff-materials} presents a comparison of properties for scintillating fibers made of three different materials: \acrshort{gl:LYSO:Ce}, \acrshort{gl:GAGG:Ce} and \acrshort{gl:LuAG:Ce}. Samples of \acrshort{gl:GAGG:Ce,Mg} scintillating fibers were not examined. Chemical polishing of these samples caused strong attenuation of the scintillating light due to absorption at the crystal surface, and it was not possible to register correlated signals at both ends of the sample. The lack of clear \anhpeak peak in the registered spectra made it impossible to extract the fiber characteristics. Chemical polishing of the scintillator surface was previously reported to improve light output and position resolution \cite{Shao2002}. However, in the cited study, much shorter crystals were examined. Our observations suggest that this type of surface treatment is not suitable for elongated, fiber-like shape of scintillators.

From \cref{tab:diff-materials} it can be seen that all three materials are very similar in terms of attenuation length (the difference is more pronounced for the \acrshort{gl:GAGG:Ce} and \acrshort{gl:ELAR} method). The remaining properties prove that \acrshort{gl:LYSO:Ce} is superior to other materials with the best values of position-, energy-, and timing resolution as well as the largest light collection. \acrshort{gl:GAGG:Ce}, even though the obtained results are very close to those of \acrshort{gl:LYSO:Ce}, performs slightly worse in comparison, especially in terms of timing resolution. The performance of \acrshort{gl:LuAG:Ce} is the worst of all three materials, with significantly worse resolutions and light collection. Additionally, using Hamamatsu \acrshort{gl:SiPM}s with better timing properties yielded a significant improvement of timing resolution.

\begin{table}[ht]
\centering
\caption{Comparison of the properties of different scintillating materials. Values marked with a bullet ($\bullet$) were obtained in measurements in configuration I (Hamamatsu \acrshort{gl:SiPM}s).}
\label{tab:diff-materials} 
\smallskip
\begin{tabularx}{1.0\textwidth}{p{2.5cm}XXXXXX}
\hline
Material & Att. length (\gls{gl:MLR}) [\si{\milli\meter}] & Att. length (\acrshort{gl:ELAR}) [\si{\milli\meter}] & Position res. [\si{\milli\meter}] & Energy res. at \newline \SI{511}{\kilo\electronvolt} [\si{\percent}] & Timing res. [\si{\nano\second}] & Light collection [\si{\pe\per\mega\electronvolt}] \\
\hline
\acrshort{gl:LYSO:Ce} & 271 & 121 & 40 & 8.43 & 1.15\par 0.43$^{\bullet}$ & 556 \\
\acrshort{gl:GAGG:Ce} \newline \footnotesize{(Fomos)} & 278 & 151 & 45 & 8.47 & 1.79\par 0.58$^{\bullet}$ & 489 \\
\acrshort{gl:GAGG:Ce,Mg} \footnotesize{(C\&A Corp.)} & - & - & - & - & - & - \\
\acrshort{gl:LuAG:Ce} & 265 & 116 & 56 & 10.57 & 5.88\par 2.50$^{\bullet}$ & 291 \\
\hline
\end{tabularx}
\end{table}


\subsection{Different vendors}
\label{ssec:sf-diff-vendors}

\cref{tab:diff-producers} lists properties of \acrshort{gl:LYSO:Ce} fibers produced by different manufacturers. The biggest differences are visible in the attenuation length, and thus also in the position resolution. The fibers delivered by Epic Crystal company are characterized by the smallest attenuation length and position resolution, whereas the fibers produced by Tianle have the largest attenuation length and position resolution. The remaining properties are relatively similar for all manufacturers. The appearance of large differences in the attenuation length and position resolution values, while other characteristics remain comparable, can be explained by different manufacturing methods and crystal treatment. In particular, cutting and polishing techniques and precision can affect light propagation and absorption in the fibers.

\begin{table}[ht]
\centering
\caption{Comparison of the properties of \acrshort{gl:LYSO:Ce} fibers purchased from different producers. Values marked with a bullet ($\bullet$) were obtained in measurements with Hamamatsu \acrshort{gl:SiPM}s.}
\label{tab:diff-producers} 
\smallskip
\begin{tabularx}{1.0\textwidth}{p{2.5cm}XXXXXX}
\hline
Producer & Att. length (\gls{gl:MLR}) [\si{\milli\meter}] & Att. length (\acrshort{gl:ELAR}) [\si{\milli\meter}] & Position res. [\si{\milli\meter}] & Energy res. at \newline \SI{511}{\kilo\electronvolt} [\si{\percent}] & Timing res. [\si{\nano\second}] & Light collection [\si{\pe\per\mega\electronvolt}] \\
\hline
Epic \newline Crystal & 235 & 124 & 34 & 8.37 & 1.13\par 0.37$^{\bullet}$ & 583 \\
Meta \newline Laser & 262 & 119 & 38 & 8.52 & 1.16\par 0.46$^{\bullet}$ & 549 \\
Shalom & 289 & 174 & 41 & 8.28 & 1.13 & 577 \\
Tianle & 356 & --  & 50 & 8.35 & 1.17 & 551 \\
\hline
\end{tabularx}
\end{table}


\subsection{Different fiber coating and wrapping}
\label{ssec:sf-diff-wrapping}

Wrapping or coating a scintillator is usually necessary for correct operation of the scintillating detector. It prevents light crosstalk between adjacent scintillating elements or the surrounding environment. In scintillating detectors built from many crystals, such as the \acrshort{gl:SiFi-CC} detector, a suitable wrapping or coating ensures that propagation of the scintillating light is limited only to the element in which it was produced. This allows a more precise determination of the interaction point in the scintillator. However, it should be taken into account that the introduction of wrapping or coating modifies the fiber surface and consequently influences the propagation of scintillating light. This effect will be enhanced for the strongly elongated, fiber-like scintillator shape because multiple reflections on the crystal surfaces will occur before the scintillating light reaches the photodetector. For that reason, the influence of different types of coatings and wrappings was investigated in this optimization study. Measurements were performed with samples of bare fibers that were subsequently wrapped or coated with the chosen materials and reexamined. There were two exceptions from this procedure: (1) fibers wrapped in ESR+Al and (2) painted with white \baso-based paint with an additional layer of Al wrapping. Those samples were prepared by the producer, therefore it was not possible to examine them before they were wrapped and painted. The results obtained in those experimental series were compared with measurements carried out with different unwrapped and not painted samples produced by the same manufacturer (Shalom EO). 

\begin{table}[ht]
\centering
\caption{Comparison of properties of bare \acrshort{gl:LYSO:Ce} fibers and fibers wrapped or coated with different materials.}
\label{tab:diff-wrapping}
\smallskip
\begin{tabularx}{1.0\textwidth}{p{2.5cm}XXXXXX}
\hline
Wrapping/ coating & Att. length (\gls{gl:MLR}) [\si{\milli\meter}] & Att. length (\acrshort{gl:ELAR}) [\si{\milli\meter}] & Position res. [\si{\milli\meter}] & Energy res. at \newline \SI{511}{\kilo\electronvolt} [\si{\percent}] & Timing res. [\si{\nano\second}] & Light collection [\si{\pe\per\mega\electronvolt}] \\
\hline
Teflon              & 176 & 103 & 28 & 8.79 & 1.30 & 507 \\
none                & 303 & 130 & 43 & 8.37 & 1.15 & 561 \\ \hline
Mylar               & 303 & 130 & 44 & 8.55 & 1.19 & 551 \\
none                & 249 & 119 & 36 & 8.53 & 1.12 & 559 \\ \hline
Al (bright)         & 211 & 113 & 32 & 8.56 & 1.26 & 527 \\
Al (mat)            & 254 & 125 & 37 & 8.47 & 1.17 & 555 \\
none                & 240 & 122 & 35 & 8.51 & 1.17 & 548 \\ \hline
Light guide \newline coating & 317 & --  & 49 & 8.81 & 1.28 & 487 \\
none                & 332 & --  & 47 & 8.42 & 1.16 & 548 \\ \hline
Heat shrink         & 134 & 83  & 25 & 9.76 & 1.76 & 371 \\
none                & 303 & 183 & 44 & 8.34 & 1.17 & 552 \\ \hline
ESR + Al            & 169 & 91  & 36 & 11.73 & 2.15 & 265 \\
none                & 289 & 174 & 41 & 8.28 & 1.13 & 577 \\ \hline
White paint \newline + Al    & 46  & 39  & 12 & 13.85 & 3.01 & 256 \\
none                & 289 & 174 & 41 & 8.28 & 1.13 & 577 \\ \hline
AlZn spray          & --  & --  & -- & --    & --   & --  \\
\hline
\end{tabularx}
\end{table}

\cref{tab:diff-wrapping} presents a comparison of the properties of bare scintillating fibers and fibers coated or wrapped with different materials. There are no results listed for one of the proposed coatings - the AlZn spray. Similarly as for chemically polished fibers, the sample covered with AlZn spray showed very strong signal attenuation, resulting in the absence of a clear annihilation peak in the registered spectra. Therefore, this measurement series was eliminated from the comparative studies. The influence of the remaining materials on the properties of the fibers is the following: 
\begin{itemize} 
\item Teflon: causes shortening of the attenuation length and hence improves the position resolution. The light collection is reduced, leading to a deterioration of the energy resolution. Timing resolution is also worse when compared to unwrapped fibers. 
\item Mylar: increases the attenuation length, causing worsening of the position resolution; the remaining properties also deteriorate, although the effect is smaller. 
\item Aluminum foil: depending on the wrapping method quite different results are observed. If the bright side of the foil faces the investigated fiber, the attenuation length is decreased, and the position resolution is improved; at the same time, smaller light collection is observed, which results in poorer energy resolution; the timing resolution also deteriorates. If the mat side of the foil faces the investigated fiber, the results are opposite, namely, the attenuation length is larger and the position resolution worsens, and the light collection and energy resolution are slightly improved; the timing resolution remains unchanged. It should be noted that all described changes in the properties are relatively small in comparison to those caused by Teflon or Mylar.  
\item Light guide coating: the influence on attenuation length and position resolution is rather small; light collection, energy- and timing resolution deteriorated slightly after application of the coating. 
\item Heat shrink: affects the performance of the investigated scintillator samples in a way very similar to Teflon; the observed effect is even stronger than in the case of Teflon wrapping;  
\item ESR + Al: affects the investigated fibers similarly to Teflon and heat shrink; the effect of wrapping is particularly strong for light collection, energy- and timing resolution. 
\item White paint + Al: affects the investigated fiber similarly to Teflon, heat shrink, and ESR+Al; the effect is definitely the strongest of all listed surface modifications; significant shortening of the attenuation length results in excellent position resolution. However, a large reduction of light collection causes a worsening of energy resolution. The timing resolution is also noticeably poorer.   
\end{itemize} 


\subsection{Optical coupling studies}
\label{ssec:sf-diff-couplings}

To investigate the influence of the coupling, measurements were conducted with the same \acrshort{gl:LYSO:Ce} fiber attached to the \acrshort{gl:SiPM}s with the use of different coupling media: Saint Gobain silicone gel, Eljen EJ-560 silicone rubber interface pads and air gap (no coupling). The results of the comparison presented in \cref{tab:diff-couplings} prove that the coupling significantly influences the performance of the detection system. The fiber coupled to the \acrshort{gl:SiPM}s with the silicone gel showed the best properties: reduced attenuation length resulting in good position resolution and large light collection causing improvement of the energy resolution. Coupling with the use of silicone pads leads to an increase of the attenuation length and the reduction of the light collection, which results in worse position and energy resolution. The timing resolution for both types of coupling is similar. The lack of coupling medium between the scintillating fiber and \acrshort{gl:SiPM}s causes a significant deterioration of the system properties: large attenuation length results in position resolution comparable with the length of the fiber itself. The light collection, energy, and timing resolutions are also significantly worse in comparison to the previous coupling types. 

\begin{table}[ht]
\centering
\caption{Comparison of the properties of the LYSO:Ce fiber attached to the \acrshort{gl:SiPM}s with different types of coupling. Comparison was performed for a selected Tianle fiber.}
\label{tab:diff-couplings} 
\smallskip
\begin{tabularx}{1.0\textwidth}{p{2.5cm}XXXXXX}
\hline
Coupling & Att. length (\gls{gl:MLR}) [\si{\milli\meter}] & Att. length (\acrshort{gl:ELAR}) [\si{\milli\meter}] & Position res. [\si{\milli\meter}] & Energy res. at \newline \SI{511}{\kilo\electronvolt} [\si{\percent}] & Timing res. [\si{\nano\second}] & Light collection [\si{\pe\per\mega\electronvolt}] \\
\hline
Silicone gel & 175 & -- & 23  & 7.48 & 0.99 & 704 \\
Silicone pad & 376 & -- & 53  & 8.31 & 1.17 & 552 \\
Air gap      & 633 & -- & 102 & 9.05 & 1.59 & 397 \\
\hline
\end{tabularx}
\end{table}

\begin{figure}[ht]
\centering
\includegraphics[width=0.75\textwidth]{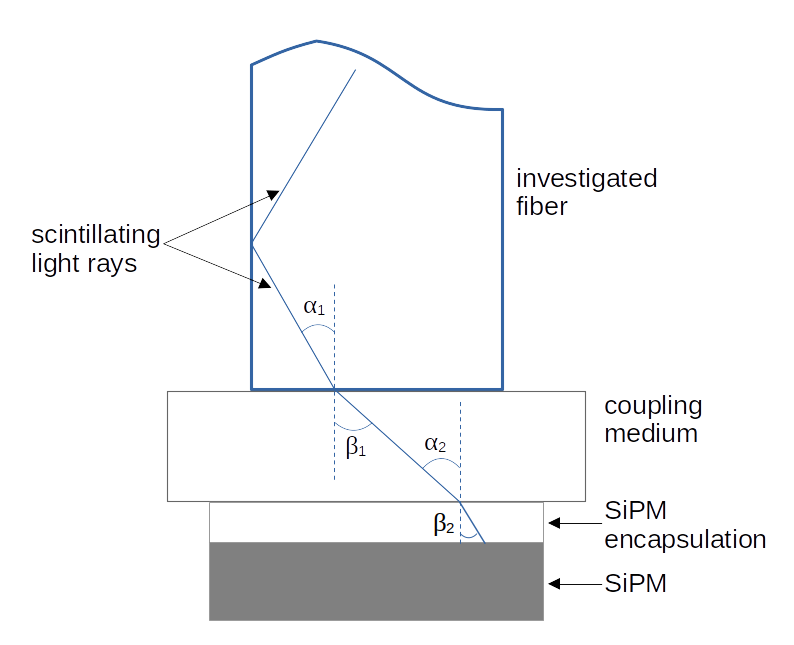}
\caption{Schematic representation of the scintillating light passing through the investigated fiber, the layer of coupling, and finally the encapsulation of the \acrshort{gl:SiPM}. $\alpha_1$ and $\alpha_2$ denote angles of incidence and $\beta_1$ and $\beta_2$ denote angles of refraction.}
\label{fig:coupling-scheme}
\end{figure}

The observed large differences in attenuation lengths can be explained by changes in collection angles caused by different types of coupling, which was previously described in \cite{Pauwels2013}. \Cref{fig:coupling-scheme} shows a schematic representation of scintillating light propagating through subsequent elements of the setup: scintillating fiber, coupling medium, and encapsulation of the \acrshort{gl:SiPM} protecting the photosensitive elements of the sensor. In this system, scintillating light undergoes refraction at two surfaces - boundary of fiber and coupling and boundary of coupling and \acrshort{gl:SiPM} encapsulation. Knowing the refractive indices of each medium, values of refraction angles in the system for all three types of tested coupling were calculated, assuming initial incidence angle $\alpha_1 = 30\deg$ (see \cref{tab:coupling-example}). It can be seen that due to large difference in refractive indices between the \acrshort{gl:LYSO:Ce} scintillator and air the refraction angle on the first boundary is the largest for this combination. Refraction angles $\beta_1$ for silicone pads and silicone gel are very similar.  

Except from the refraction angle, another factor defining the light collection angle is the thickness of each layer. The thin layer of silicone gel ($d_{\textrm{gel}} \approx \SI{0.4}{\milli\meter}$) increases the collection angle allowing to detect light leaving the fiber at a wide range of angles. This decreases the attenuation length and increases the light collection at the same time. Even a slightly wider air gap ($d_{\textrm{air}} \approx \SI{0.6}{\milli\meter}$) decreases the collection angle, which favors light traveling straight or within a small cone. Light which travels in a straight line or within a narrow cone in the fiber experiences a shorter optical path in comparison to light which undergoes multiple internal reflections and leaves the fiber at a large angle. It should be additionally noted, that significant light losses occur during reflections, when fraction of light is transmitted through the scintillator surface. This means, that light traveling at small angles experiences weaker attenuation. As a consequence the attenuation length increases and light collection decreases. The silicone pads of a fixed \SI{1}{\milli\meter} thickness due to their optical properties give intermediate results.

\begin{table}[!ht]
    \caption{The refractive indices for \acrshort{gl:LYSO:Ce}, \acrshort{gl:SiPM} encapsulation, and different types of coupling used in this study. Based on the listed refractive indices, the values of the example refraction angles were calculated, assuming $\alpha_1 = 30\deg$ (see \cref{fig:coupling-scheme}).}
    \label{tab:coupling-example}
    \begin{minipage}{.3\linewidth}
      \centering
        \begin{tabular}{p{0.7\linewidth}}
            Refractive indices \\ \hline
            $n_{\textrm{LYSO:Ce}}$ = 1.82 \\
            $n_{\textrm{pad}}$ = 1.43 \\
            $n_{\textrm{gel}}$ = 1.47 \\
            $n_{\textrm{air}}$ = 1.00  \\
            $n_{\textrm{SiPM}}$ = 1.59 \\
        \end{tabular}
    \end{minipage}%
    \begin{minipage}{.7\linewidth}
      \centering
        \begin{tabular}{p{0.27\linewidth} | p{0.13\linewidth} | p{0.13\linewidth} | p{0.13\linewidth}} 
         \multicolumn{4}{c}{Angles indicated in \cref{fig:coupling-scheme}, assuming $\alpha_1 = 30\deg$}  \\ \hline
         Material & $\beta_1$ [\si{\degree}] & $\alpha_2$ [\si{\degree}] & $\beta_2$ [\si{\degree}] \\ \hline
         Silicone pad & $39$ & $39$ & $35$ \\
        Silicone gel & $38$ & $38$ & $35$ \\
        Air & $65$ & $65$ & $35$ \\
\end{tabular}
\vspace{0.7cm}
    \end{minipage} 
\end{table}

The processes occurring on the second media boundary are less significant because of smaller difference in refractive indices and small thickness of the encapsulant layer, typically about \num{0.1} -- \SI{0.2}{\milli\meter}.

The above comparison shows that silicone gel as a coupling medium can improve the performance of the scintillating detector. However, as demonstrated in \cref{fig:single-fib-lcol}, it can be highly unstable and hard to reproduce in consecutive measurements after the reassembly of the detector. Therefore, this type of coupling is not suitable for certain types of detector designs. In that case, silicone pads appear to be a reasonable compromise between optimal performance and stability and durability of the system. 

The silicon pads can be cut into the desired shape and size depending on the detector design. In the case of the \acrshort{gl:SiFi-CC} detector, where many scintillating elements are read out by arrays of \acrshort{gl:SiPM}s, there are two possibilities for optical coupling: (1) individual, independent coupling of each element, \eg by using dedicated rasters and (2) large-area coupling without distinguishing of single elements. In the first type the scintillating light would be registered only by the \acrshort{gl:SiPM}s assigned to a specific fiber, while the second design would allow for light sharing between neighboring SiPMs. To estimate, whether the light sharing would influence the performance of a single element significantly, a test with different sizes of silicone pads was conducted. Four different pad areas were tested, ranging from \num{3} $\times$ \SI{3}{\milli\meter\squared} to \num{10} $\times$ \SI{10}{\milli\meter\squared}, as listed in \cref{tab:diff-couplings-sizes}. The pads used in this study were produced on site, using Elastosil RT 604 room temperature curing silicone rubber \cite{elastosil}. The thickness of the pads was \SI{0.5}{\milli\meter}. The obtained results show that there is no significant influence of the coupling area on the scintillator performance. 

\begin{table}[ht]
\centering
\caption{Comparison of properties of the \acrshort{gl:LYSO:Ce} fiber attached to the \acrshort{gl:SiPM}s with the use of silicone pads of different areas. Comparison was made for a selected Shalom EO fiber.}
\label{tab:diff-couplings-sizes} 
\smallskip
\begin{tabularx}{1.0\textwidth}{p{2.5cm}XXXXXX}
\hline
Coupling size & Att. length (\gls{gl:MLR}) [\si{\milli\meter}] & Att. length (\acrshort{gl:ELAR}) [\si{\milli\meter}] & Position res. [\si{\milli\meter}] & Energy res. at \newline \SI{511}{\kilo\electronvolt} [\si{\percent}] & Timing res. [\si{\nano\second}] & Light collection [\si{\pe\per\mega\electronvolt}] \\
\hline
\num{3} $\times$ \SI{3}{\milli\meter\squared} & 184 & 134 & 30 & 9.18 & 1.18 & 538 \\
\num{5} $\times$ \SI{5}{\milli\meter\squared} & 193 & 128 & 31 & 9.16 & 1.19 & 525 \\
\num{7} $\times$ \SI{7}{\milli\meter\squared} & 188 & 124 & 30 & 9.19 & 1.18 & 542 \\
\num{10} $\times$ \SI{10}{\milli\meter\squared} & 192 & 140 & 31 & 9.12 & 1.21 & 522 \\
\hline
\end{tabularx}
\end{table}

\chapter{Prototype studies}
\label{chap:prototype}

The single fiber study described in detail in \cref{chap:single-fibers} proved that LYSO:Ce is the best candidate for the active part of the \acrshort{gl:SiFi-CC} detector. Among all of the tested materials used for fiber wrapping or coating the Al foil (bright side facing the scintillator surface) was determined to ensure satisfactory properties. Additionally, silicone pads were chosen as the optimal coupling interface between fiber and \acrshort{gl:SiPM}s, as they offer a reasonable compromise between scintillator performance and system stability. Based on these conclusions, a small-scale prototype (\acrshort{gl:SSP}) of the \acrshort{gl:SiFi-CC} detector was constructed and tested with two different read-out and \acrshort{gl:DAQ} systems. In the first experiment, the \acrshort{gl:LYSO:Ce} crystals in the prototype were coupled to the \acrshort{gl:SiPM}s matching the fiber size and thus ensuring "one-to-one" read-out (\cref{subsec:krk-tests}). In the second experiment, the prototype was coupled with the large-area digital photosensor, allowing for light sharing based read out (\cref{sec:pmi-tests}). The following chapter presents a description of the prototype construction, two experimental setups tested, and the results of the detector characterization in both configurations. 

\section{Module construction}
\label{sec:module-construction}

\begin{figure}[ht]
    \centering
    \includegraphics[height=9cm]{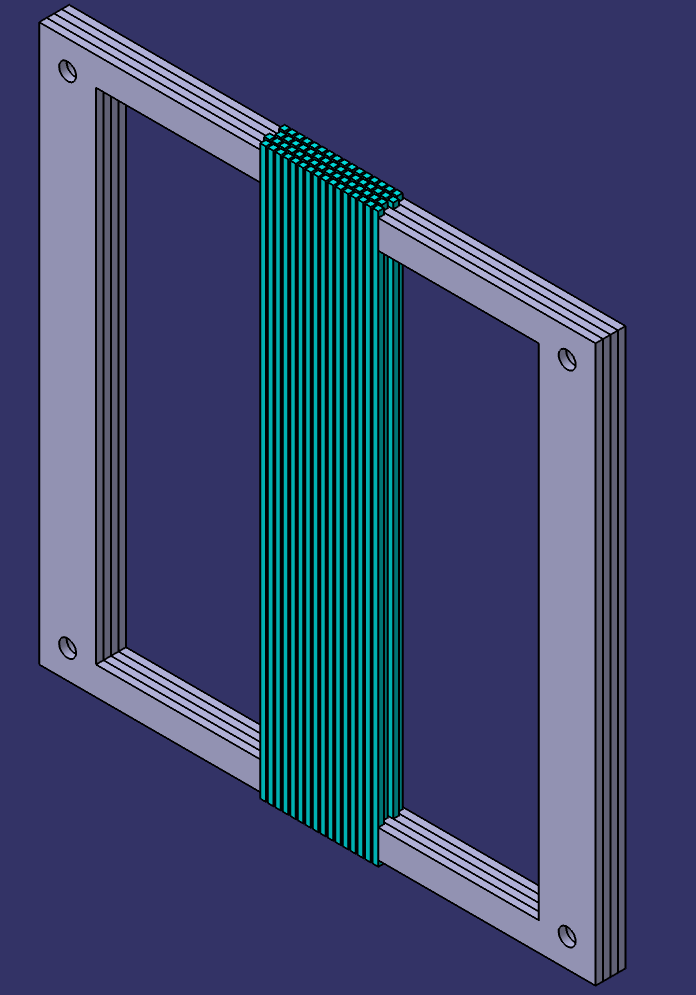}
    \includegraphics[height=9cm]{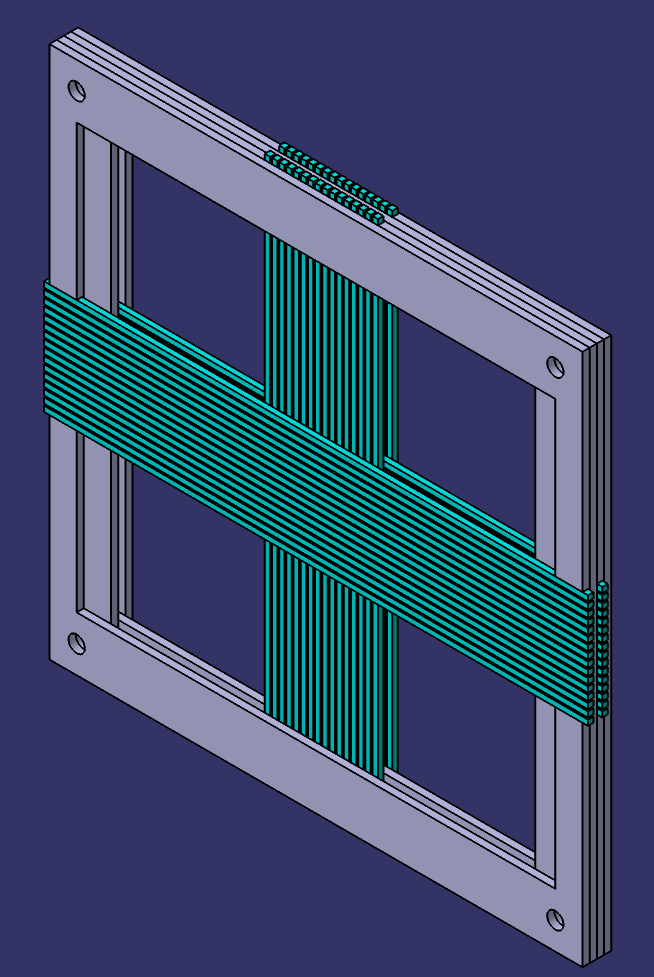}
    \caption{Different geometries of the detector prototype. Left: aligned 4x16 geometry. Right: crossed 4x16 geometry.}
    \label{fig:prototype-scheme}
\end{figure}

\begin{figure}[!ht]
    \centering
    \includegraphics[width=0.7\textwidth]{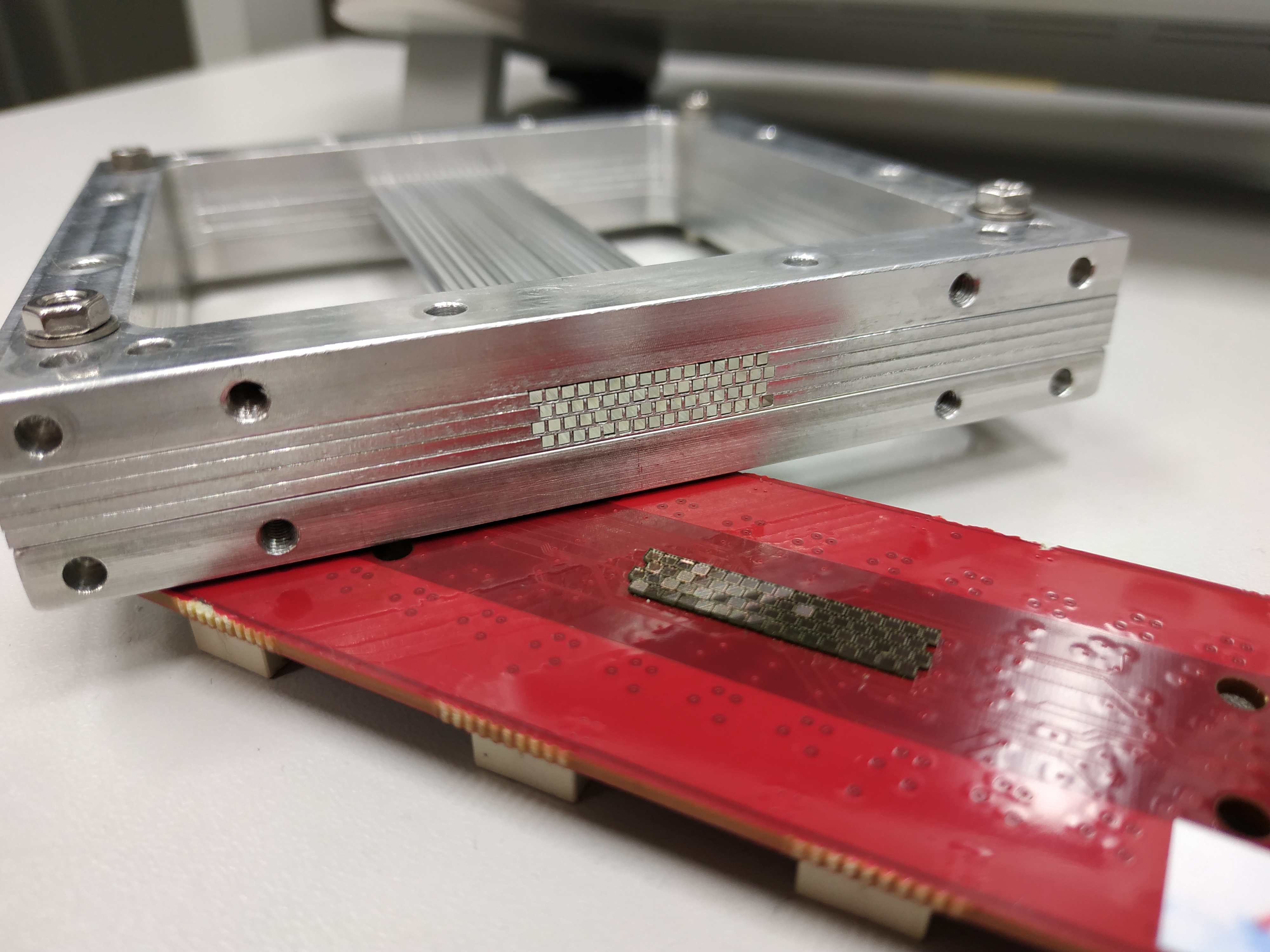}
    \caption{Small-scale prototype assembled into four parallel layers. The half-pitch shift between neighboring fiber layers is visible. The \acrshort{gl:PCB} board with the \acrshort{gl:SiPM}s, which is a part of one of the readout systems, is also visible in the photo.}
    \label{fig:prototype}
\end{figure}

The small-scale prototype was made of 64 \acrshort{gl:LYSO:Ce} scintillating fibers wrapped in aluminum foil (bright side facing the fiber surface). The fiber wrapping was done at the Jagiellonian University (\acrshort{gl:JU}) technical workshop. All crystals were mechanically polished on all six surfaces by the manufacturer Shalom EO. The dimensions of fibers were \num{1}$~\times$~\num{1}~$\times$~\SI{100}{\cubic\milli\meter}. The fibers in \acrshort{gl:SSP} were organized in independent layers. The fiber pitch was \SI{1.36}{\milli\meter}, which was a result of the fiber wrapping, as well as the dimensions of the \acrshort{gl:SiPM}s used in the read-out system (see \cref{subsec:krk-tests}). To maintain such geometry without the use of glue or other substances, the fibers were inserted into custom made aluminum frames. The frames were assembled to form a complete prototype. The frames housing the fibers can be assembled in different configurations, resulting in different detector geometries:

\begin{enumerate}
\item \textbf{aligned 4x16} - the prototype consisted of four layers with 16 fibers in each layer. All fibers were parallel, as shown in \cref{fig:prototype-scheme} (left). The neighboring layers were shifted with respect to each other by half of the fiber pitch. The shift was proposed to compensate for the dead space between the fibers and thus improve the detection efficiency of the setup. In the simulation study, this geometry was proven to be optimal, therefore it was used for most of the conducted tests \cite{Kasper2022}. It is also a subject of this thesis.

\item \textbf{crossed 4x16} - similarly as above, the detector consisted of four layers of 16 fibers each. The fibers in the neighboring layers were perpendicular to each other (see \cref{fig:prototype-scheme}, right). There was no additional shift between the fiber layers. Rotation of every second layer was proposed to improve the position resolution of the detector. The position resolution in two dimensions is determined by the fibers size, so it is in the order of \SI{1}{\milli\meter}. However, the hit position along the fiber has to be reconstructed based on the registered charge ratio (see \cref{ssec:position-resolution}). Position resolution along the fiber resulting from the reconstruction procedure is in the order of \SI{30}{\milli\meter}, which is one of the main limitations of the proposed detector. In the situation when the incident prompt gamma induces interactions in neighboring layers, the hit position could be determined more accurately taking advantage of the crossed geometry. However, the extensive simulation study showed that improved hit position determination is outweighed by the significant decrease in detection efficiency, making image reconstruction impossible \cite{Kasper2022}. Therefore, only preliminary tests of this geometry were conducted, and they are not presented in this work. 

\item \textbf{aligned 2x32} - this additional geometry was constructed for preliminary tests in the \acrshort{gl:CM} modality conducted at RWTH Aachen University (see \cref{sec:pmi-tests}). In this assembly, the scintillating fibers were organized in two parallel layers with 32 crystals in each layer. There was no shift between the layers. This geometry required the preparation of new dedicated aluminum frames. 
\end{enumerate}
The picture of the prototype, with one of the to examined readout boards, is shown in \cref{fig:prototype}.

The presented detector design featuring single fibers organized into independent layers offered flexibility to test different geometries. The same components were used for every configuration, allowing significant cost reduction, which is important in the early stage of detector development and optimization. In particular, the same scintillating fibers were reused in many detector assemblies, and they are among the most costly components used.  

\begin{figure}[ht]
\centering
\vspace{0.4cm}
\includegraphics[width=.49\textwidth]{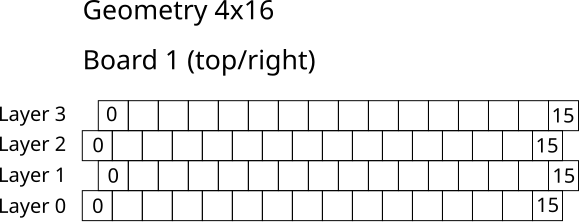} 
\includegraphics[width=.49\textwidth]{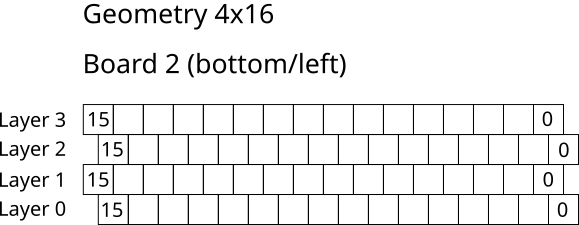} 
\caption{Scheme of fiber numbering in the prototype assembled in 4 parallel layers (top and bottom view). To refer to a chosen fiber a layer (L) number and fiber (F) number are given, \eg L0F5, meaning fiber number five in layer zero.}
\label{fig:prototype-numbering}
\end{figure}

As mentioned above, the aligned 4x16 geometry is the focus of this work. To be able to refer to a chosen fiber in the detector, a fiber numbering scheme was agreed upon. \Cref{fig:prototype-numbering} presents the fiber numbering scheme in the discussed aligned 4x16 geometry from two perspectives - top and bottom. For the result representation in the following part of this thesis, presented numbering convention, assuming top view on the detector, will be used. 

\section{Module characterization - "one-to-one" readout}
\label{subsec:krk-tests}

The \acrshort{gl:SSP} was tested with two different photosensors and data acquisition systems. This section describes the first examined experimental setup. The measurements were conducted in the Hadron Physics Department of Jagiellonian University, and hereinafter will be referred to as \acrshort{gl:JU} measurements. 

\subsection{Experimental setup}
\label{susec:krk-tests-setup}

\begin{figure}[!hp]
\centering
\includegraphics[width=0.92\textwidth]{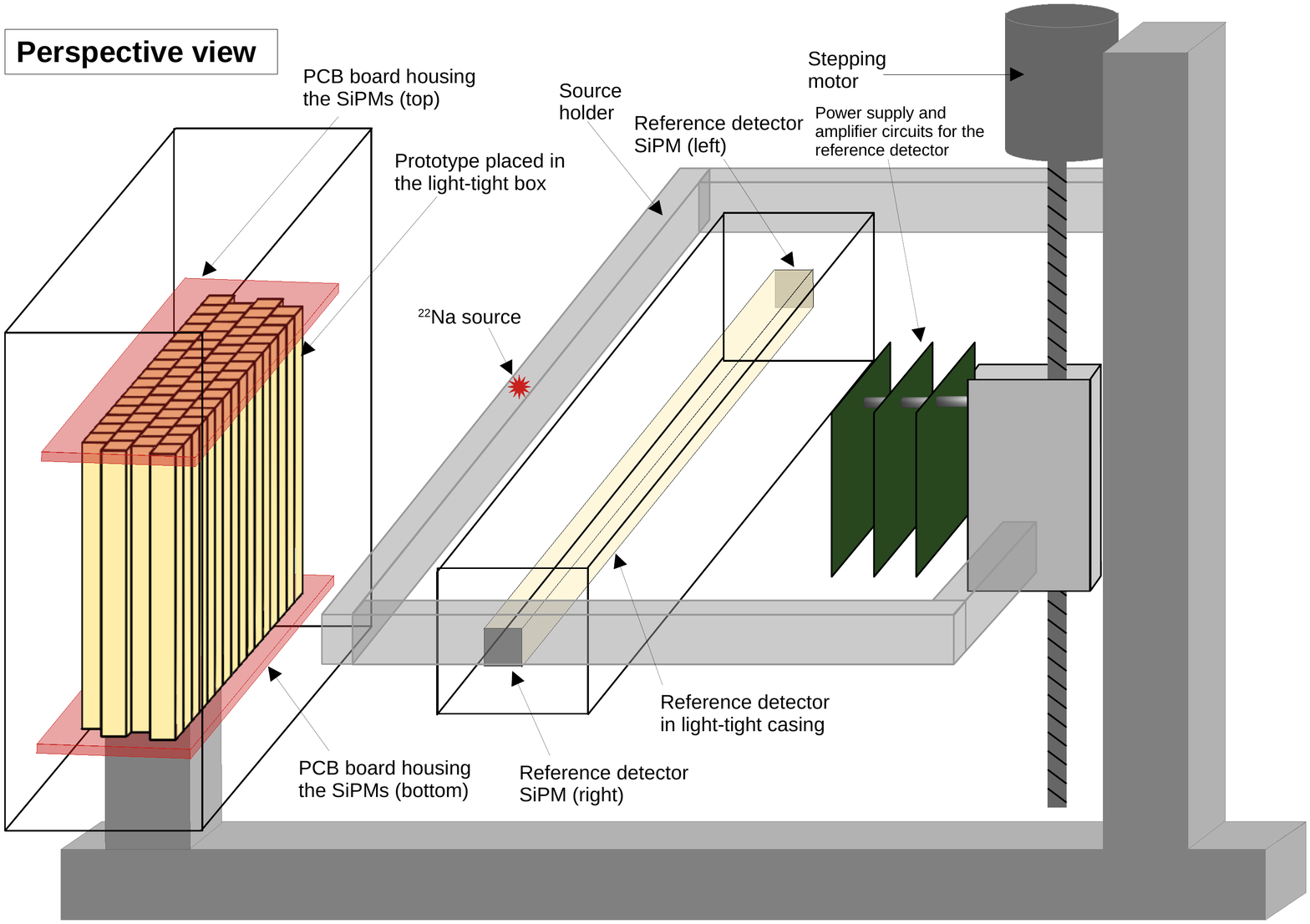}
\includegraphics[width=0.92\textwidth]{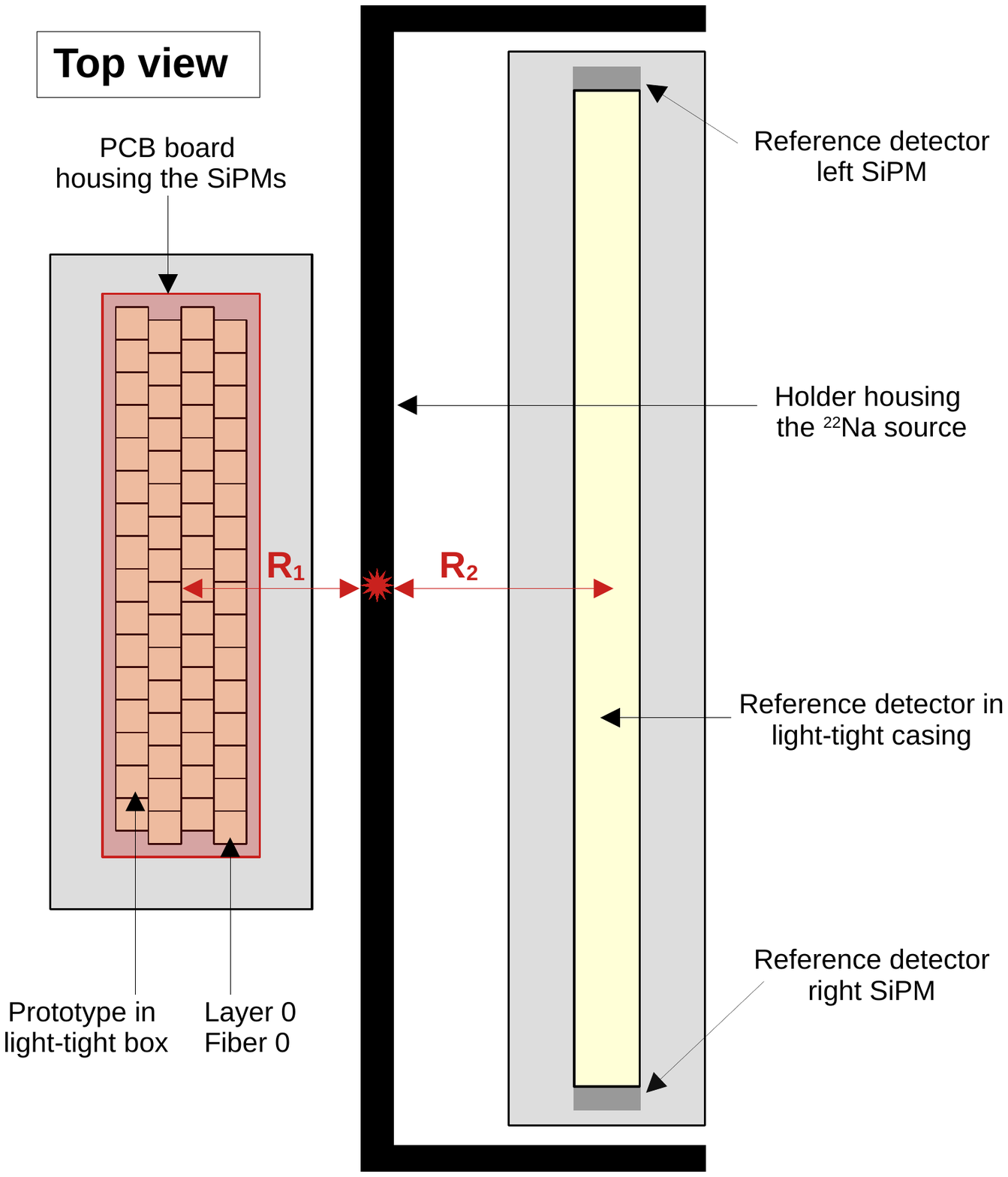}
\caption{Scheme of the experimental setup used in \acrshort{gl:JU} measurements (not to scale).}
\label{fig:ju-scheme}
\end{figure}

In order to conduct measurements with the \acrshort{gl:SSP} a dedicated test bench was constructed. A scheme of the experimental setup is presented in \cref{fig:ju-scheme}. The prototype module, consisting of assembled scintillating fibers coupled with the photodetectors, was placed in a light-tight box. The box was placed on the aluminum bar in front of the remotely controlled electronic collimator system. Two temperature sensors based on the 1-wire interface were additionally included in order to control experimental conditions. The first sensor was placed inside the light-tight box, and the other was placed outside. Each distinctive part of the presented setup is described in detail further in this section. 

\subsubsection*{Electronic collimation}

In this setup the electronic collimator was significantly different from the one used in single-fiber measurements (see \cref{fig:setup}). It consisted of a \acrshort{gl:LYSO:Ce} fiber wrapped in ESR and Al foil by the producer Shalom EO. The fiber was mechanically polished on all six surfaces and it had the following dimensions: \num{3} $\times$ \num{3} $\times$ \SI{100}{\cubic\milli\meter}. At both ends of this fiber Hamamatsu \acrshort{gl:SiPM}s S13360-3050CS were attached. The active area of used \acrshort{gl:SiPM}s matched the size of the \acrshort{gl:LYSO:Ce} fiber in the collimator. The \acrshort{gl:SiPM}s were operated using amplifier boards and a power supply board. The scintillating fiber coupled with the photomultipliers was encapsulated in a light-tight box. The entire system of electronic collimator was placed on the rail controlled by a stepping motor. The motor allowed for remotely controlled movement of the collimator along the investigated prototype detector. Between the collimator box and the prototype box a \Na radioactive source was placed on a fixed arm. The arm was attached to the electronic collimation system such that it moved along with it. Signals registered at both ends of the reference fiber along with the logic sums of signals from the top and bottom sides of the detector, were used to build a four-fold coincidence. The coincidence signal was subsequently fed to the \acrshort{gl:DAQ} in order to trigger data recording. A detailed scheme of signal processing in the setup along with the list of exploited modules are presented further in this section. This design allowed for a position scan of the prototype, \ie measurements at desired, well defined positions along the detector, similarly as in the case of single-fiber measurements. The distances in the setup resulted in irradiation of approximately \SI{3}{\milli\meter} fragment along the \acrshort{gl:SSP} ($R_1 \approx$~\SI{30.5}{\milli\meter} and $R_2 \approx$~\SI{61.4}{\milli\meter}, see the bottom part of \cref{fig:ju-scheme}). The application of an elongated crystal in the electronic collimator, rather than a small scintillator as in the single-fiber test bench, resulted in collection of annihilation gammas from various directions in the horizontal plane. This allowed us to characterize many fibers in the prototype at the same time. 

The construction of the experimental setup was a joint effort of RWTH Aachen University (mechanical support structure, stepping motor and electronics operating the electronic collimation) and Jagiellonian University (active part of the electronic collimator, holder for the radioactive source and integration with the prototype).

\subsubsection*{Readout electronics}

In order to register the scintillating light produced in the prototype detector, custom front-end electronics (\acrshort{gl:FEE}) featuring photodetectors was designed. The intention of this design was to ensure one-to-one coupling, meaning that each scintillating fiber would be read out by a pair of \acrshort{gl:SiPM}s, one at each fiber end. This would lead to a straightforward identification of the crystal in which an interaction occurred. The read-out system consisted of printed circuit boards (\acrshort{gl:PCB}s) with soldered single \acrshort{gl:SiPM}s. For this design, KETEK PM1125-WB-B0 and PM1125-WB-C0 \acrshort{gl:SiPM}s were chosen, as their dimensions fit the scintillating fibers in the prototype and thus allowed one-to-one coupling. Therefore, compared to the Hamamatsu and SensL \acrshort{gl:SiPM}s used previously in single-fiber tests, the active area of KETEK \acrshort{gl:SiPM}s was smaller. The \acrshort{gl:PDE} of KETEK \acrshort{gl:SiPM}s was similar to that of Hamamatsu \acrshort{gl:SiPM}s. A complete comparison of the properties of all used \acrshort{gl:SiPM}s models can be found in \cref{app:photodetectors}.

\begin{figure}[!ht]
\centering
\includegraphics[width=.49\textwidth]{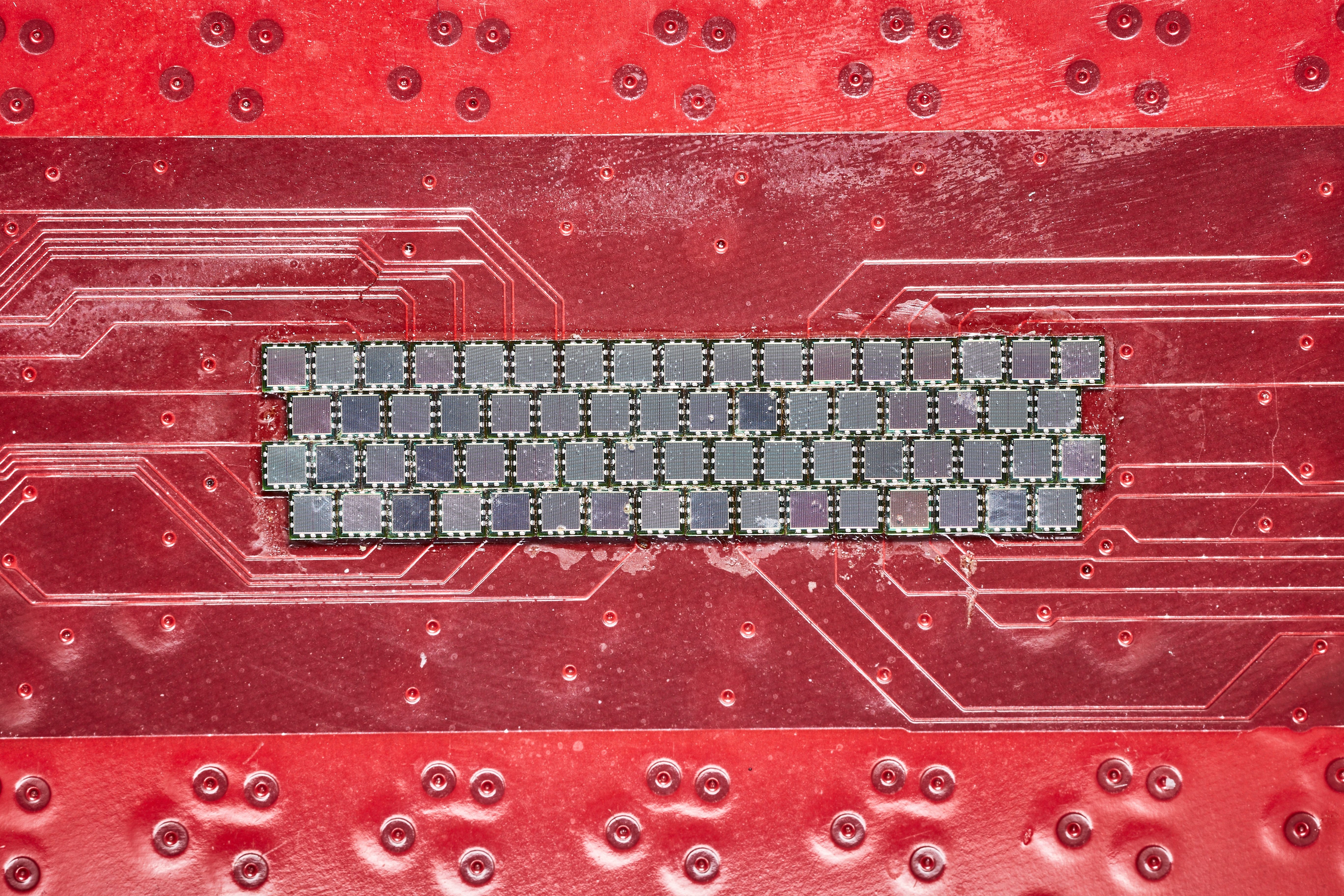} 
\includegraphics[width=.49\textwidth]{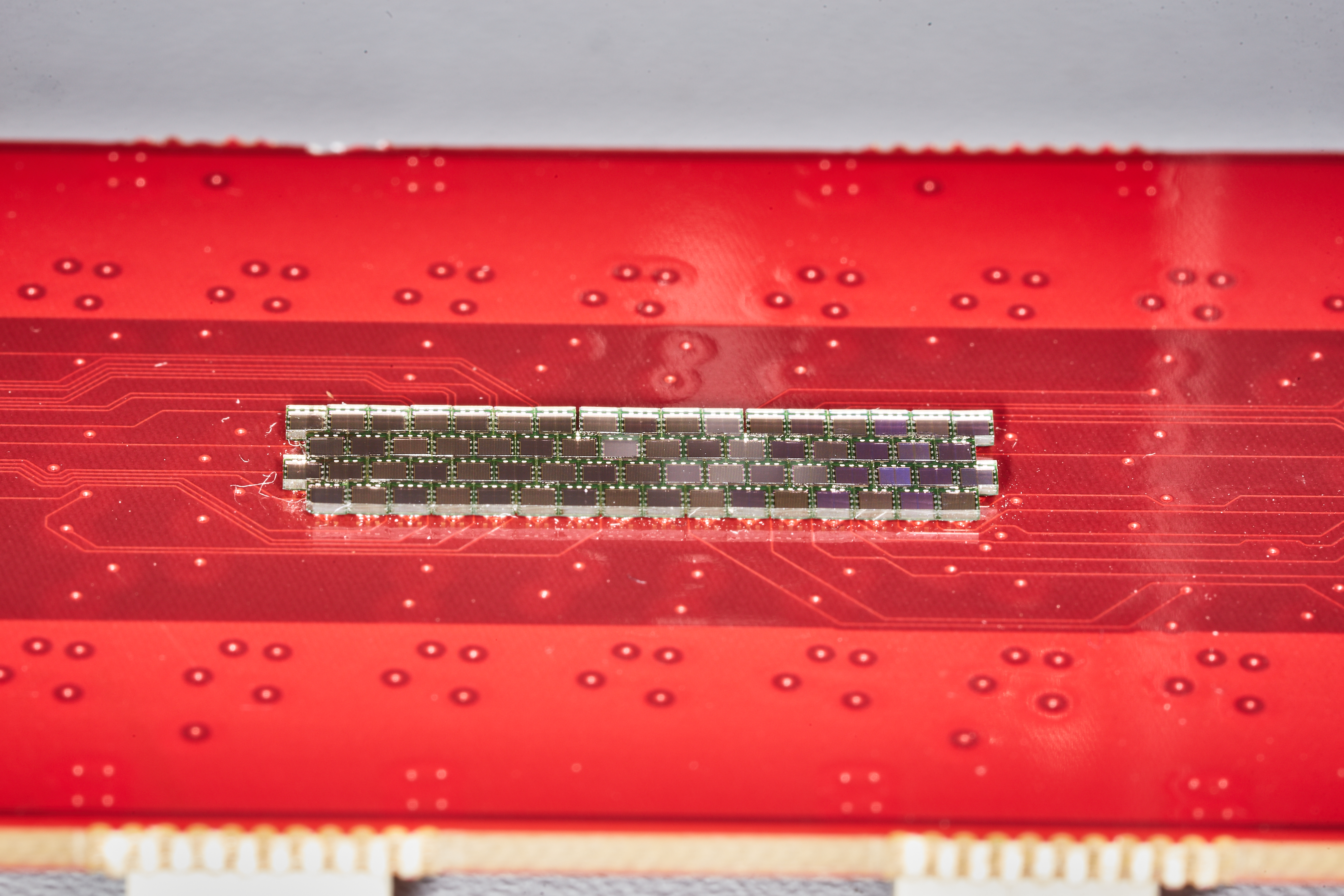} 
\caption{The \acrshort{gl:PCB} with Ketek SiPMs soldered. Left: top view, right: perspective view. The size of each SiPM is \num{1.32} $\times$ \SI{1.32}{\milli\meter\squared}, with \num{1} $\times$ \SI{1}{\milli\meter\squared} photosensitive area. The placement of the SiPMs on the board corresponds to the assembly of the fibers in the prototype, allowing one-to-one coupling of the elements.}
\label{fig:red-boards}
\end{figure}

The \acrshort{gl:SiPM}s were connected to the flat flexible cable \acrshort{gl:FFC} connectors on the back of the \acrshort{gl:PCB}, which ensured both power supply and signal readout. Each \acrshort{gl:FFC} connector served four \acrshort{gl:SiPM}s. Two of such read-out boards were prepared for the aligned 4x16 geometry, one for the top side of the detector and another for the bottom side. \Cref{fig:red-boards} shows pictures of one of the readout boards with \acrshort{gl:SiPM}s. 

To ensure good optical contact between the \acrshort{gl:LYSO:Ce} scintillating crystals in the prototype and the \acrshort{gl:SiPM}s optical coupling in the form of optical pads was used. Large-area optical interface pads were produced on site in the \acrshort{gl:JU} clean room using Elastosil RT 604 room temperature curing silicone rubber. This type of coupling was also tested in single-fiber measurements, as mentioned in \cref{ssec:sf-diff-couplings}. Used pads had a thickness of \SI{0.4}{\milli\meter} and an area covering the entire top and bottom surface of the prototype.

\subsubsection*{Signals processing and \acrshort{gl:DAQ} system}

Data acquisition was carried out with the CAEN Desktop Digitizer DT5742, which was previously used for single-fiber measurements (see \cref{ssec:sf-setup-daq}). The operation of this \acrshort{gl:DAQ} system was already tested and well understood, which was an advantage during tests of a completely new detector setup under laboratory conditions. Additionally, the software tools developed for single-fiber tests could be reused with only minor modifications. The main disadvantage of the Desktop Digitizer was a limited number of channels - only 16 channels available.

To operate the setup with the Desktop Digitizer, an additional interface board was required. 
It allowed for splitting signals from each \acrshort{gl:FFC} connector into individual channels, which were subsequently plugged in to the Desktop Digitizer. The power supply for the \acrshort{gl:SiPM}s was also provided via the interface board. 

\begin{sidewaysfigure}[hp]
\centering
\includegraphics[height=.65\textwidth]{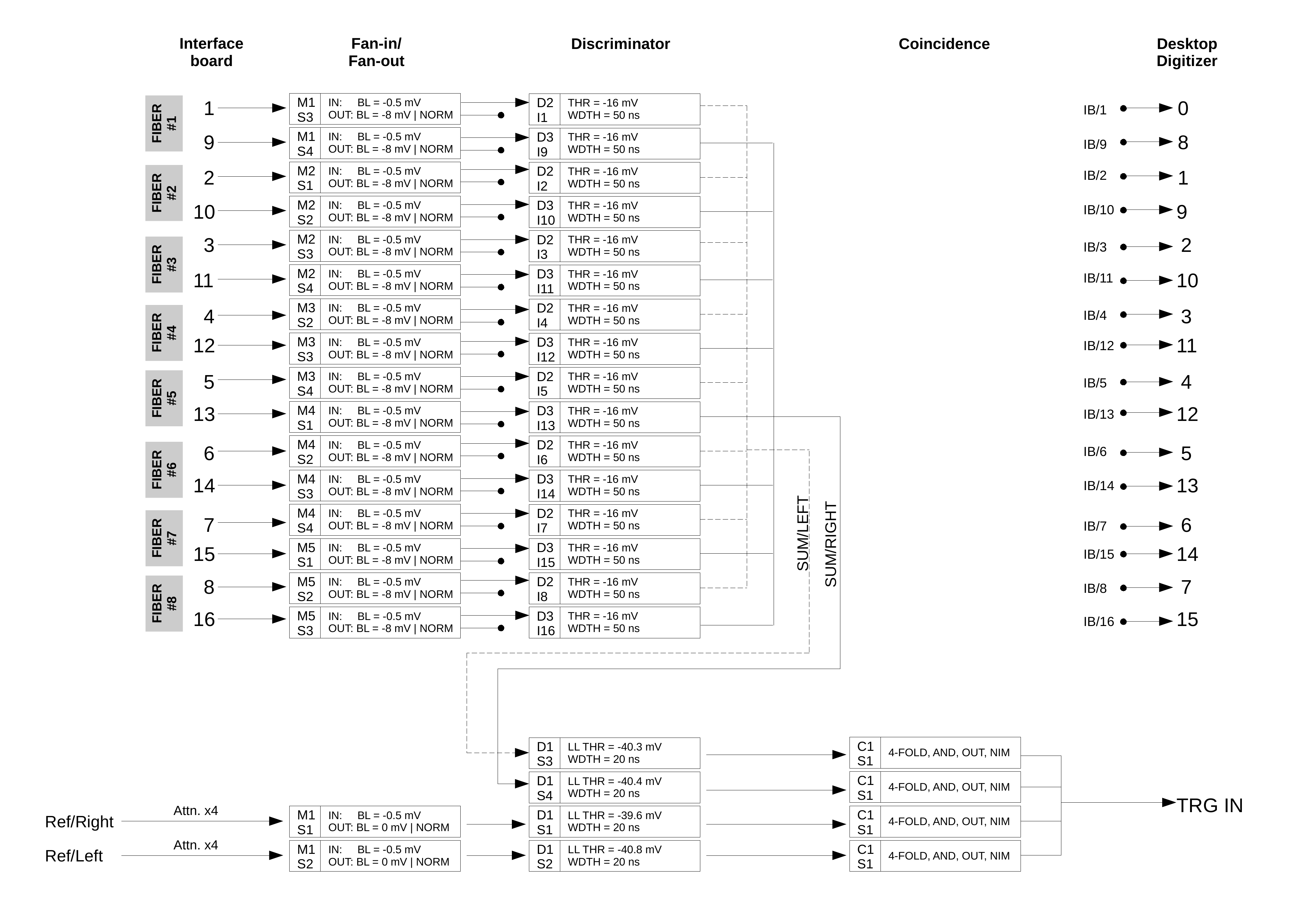}
\caption{Scheme of the signal processing chain in measurements with the prototype (\acrshort{gl:JU}).}
\label{fig:prototype-electronics}
\end{sidewaysfigure}

As mentioned above, the \acrshort{gl:DAQ} was triggered by the four-fold coincidence signal. To trigger the data acquisition, a hit in thee reference detector and at least one hit in the \acrshort{gl:SSP} were required. 
The detailed scheme of the analog signal processing chain in the \acrshort{gl:JU} setup along with the list of used electronic modules is presented in \cref{fig:prototype-electronics} and \cref{tab:prototype-electronics}.

\begin{table}[ht]
\centering
\caption{List of electronic modules used for signal processing in \acrshort{gl:JU} prototype measurements. The symbols correspond to the notation in \cref{fig:prototype-electronics}.}
\begin{tabular}{|c|c|c|c|}
\hline
Module & Model & Crate & Symbol \\ \hline
Linear fan-in/fan-out & Le Croy 428F & NIM & M1 -- M5 \\
Discriminator & Phillips Scientific 730 & NIM & D1 \\
Discriminator & Le Croy 3412E & Camac & D2, D3 \\
Coincidence unit & CAEN V976 & VME & C1 \\
\hline
\end{tabular}
\label{tab:prototype-electronics}
\end{table}

\subsubsection*{Measurement procedure}

Due to the limited number of channels available in the Desktop Digitizer, the prototype characterization was performed in batches, with eight fibers examined in each batch. As mentioned previously, \acrshort{gl:SiPM}s on the \acrshort{gl:PCB}s were operated by a single \acrshort{gl:FFC} connector in groups of four. The connection design caused that data from some \acrshort{gl:SiPM}s reading out opposite sides of the same fiber were  recorded in different batches. As a result, six of the 64 fibers in the prototype were not characterized. However, data from the remaining 58 fibers proved to be sufficient to characterize the performance of the prototype. 

Each measurement batch consisted of a position scan of eight fibers in the prototype. During the position scan, measurements were performed at nine positions along the prototype detector, at \SI{10}{\milli\meter} intervals. In each measurement, correlated signals were recorded at both ends of the investigated fibers. 

\subsubsection*{Data preprocessing}

In the first stage of data preprocessing, similarly to the single-fiber measurements, recorded waveforms were analyzed to extract their key characteristics (see \cref{sec:sf-data-preprocessing}). Experience gained during an extensive single-fiber study allowed simplifying data analysis, therefore only time \gls{gl:tzero} and signal integral (charge) were determined. 

During the measurements with the prototype in the \acrshort{gl:JU} setup, no clear calibration measurements were recorded. There were no amplifiers installed on the \acrshort{gl:FEE} boards. As a result, small single-photoelectron signals were not distinguishable from noise due to a small gain. This prevented the determination of the calibration factors and carrying out the \acrshort{gl:PE} calibration. Therefore, further data analysis was performed using uncalibrated charge spectra. This is particularly important when evaluating light collection results. The \acrshort{gl:PE} calibration does not influence the remaining properties of scintillating fibers discussed in this study. 

The obtained uncalibrated charge spectra were subsequently fitted with the \cref{eq:fitted-function}, similarly to the charge spectra obtained in the single-fiber measurements (see \cref{sec:sf-data-preprocessing}). The resulting parameterization of the annihilation peak was used in further data analysis. 

\subsection{Characterization results}
\label{subsec:krk-tests-results}

The characterization of the \acrshort{gl:SSP} included the following aspects: 
\begin{itemize}[topsep=0pt,itemsep=0pt,partopsep=0pt, parsep=0pt]
\item application of the \acrshort{gl:ELA} (\gls{gl:MLR} method) and \acrshort{gl:ELAR} models of scintillating light propagation and determination of attenuation length;
\item energy reconstruction using methods based on the \gls{gl:Qavg} and \gls{gl:Qavgstar} parameters. Based on the reconstructed energy spectra the energy resolution was determined;
\item reconstruction of hit position in the scintillating fiber. In one of the approaches, the \gls{gl:MLR} parameterization was used. Another used approach utilized the \gls{gl:MLRstar} parameter. Reconstructed position spectra additionally served for determination of position resolution;
\item determination of light collection; 
\item analysis of timing properties. 
\end{itemize}
The listed characteristics were determined for all fibers in the prototype for which good quality data was recorded. In case of the discussed \acrshort{gl:JU} measurement 56 fibers in total fulfilled this condition, with six fibers rejected due to limitations of the \acrshort{gl:DAQ} system and another two discarded due to poor data quality caused by the small \acrshort{gl:SNR} due to the lack of amplifiers in the system.

Data analysis was performed using custom written framework SiFiDetectorAnalysis, which is available in the public repository \cite{sifidetana}. The framework was written based on the software prepared previously for single-fiber tests and used methods described in \cref{sec:light-propagation} and \cref{sec:charatcerization-scintillators}. Compared to single-fiber analysis software, SiFiDetectorAnalysis was better optimized for prototype data analysis. Namely, it allowed for easy parallelization of computations and usage of multiple cores, it included only methods that were previously tested and proved to perform well, and it was integrated with the dedicated database. The SiFiDetectorAnalysis framework was designed to be universal, so it can also be used in the future for measurements with the final \acrshort{gl:SiFi-CC} detector or expanded with other analysis methods. 

Further in this section, obtained results are presented and discussed in detail. In particular, for each fiber property, a 2D histogram representing the prototype geometry was plotted. They represent spatial distributions of properties within the prototype. The empty bins in the 2D histograms indicate fibers that were not characterized or were discarded from the analysis. The half-pitch shift between layers of the prototype was neglected in this representation for simplicity. Additionally, obtained values of various characteristics were plotted on 1D histograms to extract their mean values and standard deviations. The weighted means of all determined propertied were also calculated for better comparison with the results of single-fiber characterization. In single-fiber experiments, due to small number of series recorded in the same conditions, it was not feasible to build distributions of determined values, therefore only weighted means were calculated.    

\subsubsection*{Accuracy of the light propagation models}

Light propagation models are described in detail in \cref{sec:light-propagation}. For characterization of the \acrshort{gl:SSP} only \gls{gl:MLR} implementation of the \acrshort{gl:ELA} model and the \acrshort{gl:ELAR} model were used. The choice was motivated by the fact that these two models performed the best in the single-fiber tests (see \cref{sec:sf-results}). For analysis of the prototype data, the implementation of the \gls{gl:MLR} method was simplified. In order to avoid fit of the double Gaussian function (\cref{fig:MLR-examples}), which often was unsuccessful and required tuning, only events contained in the annihilation peak were taken into account. This allowed to simplify fit to a single Gaussian function.

\begin{figure}[!hp]
\centering
\includegraphics[width=0.49\textwidth]{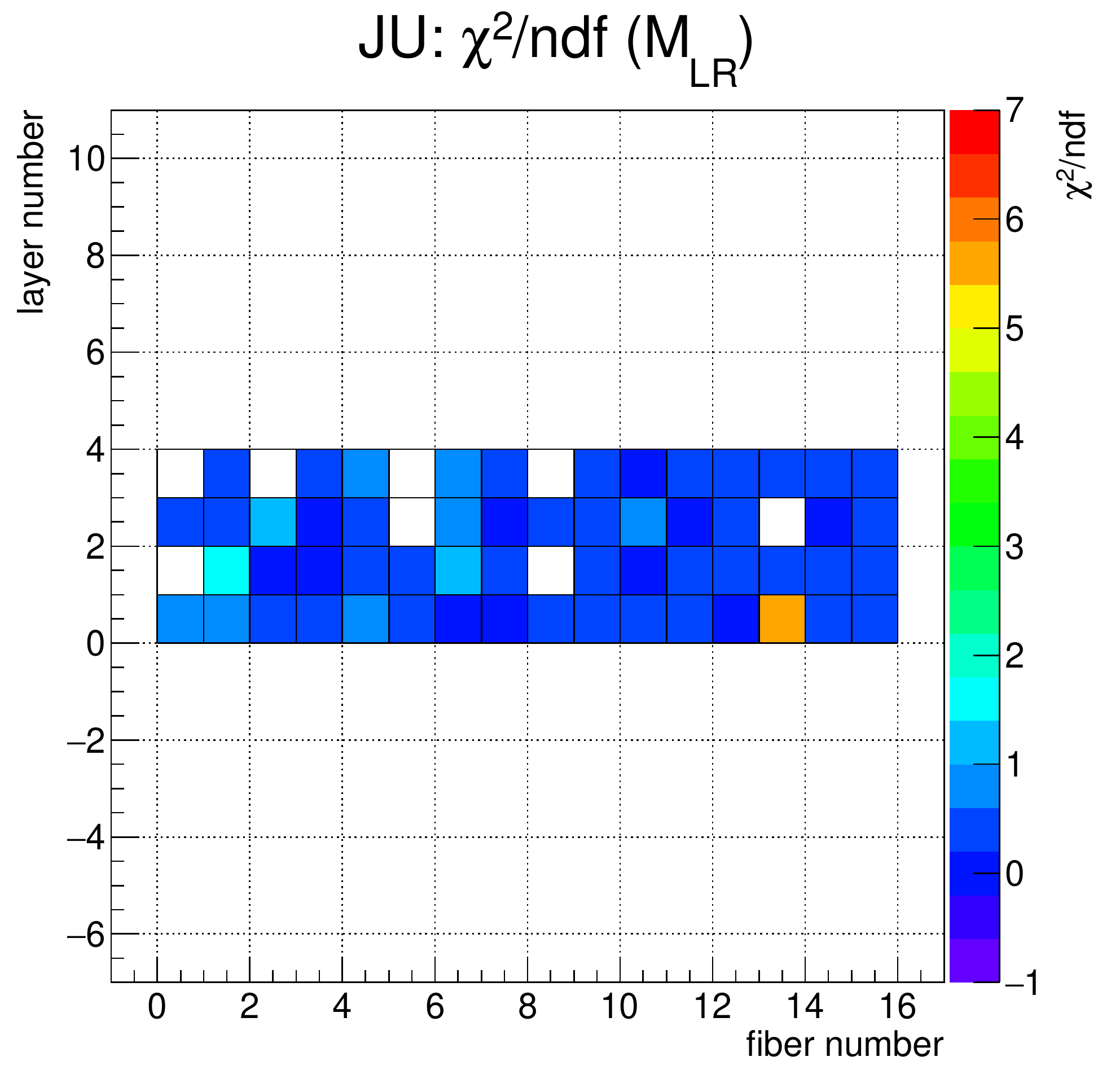}
\includegraphics[width=0.49\textwidth]{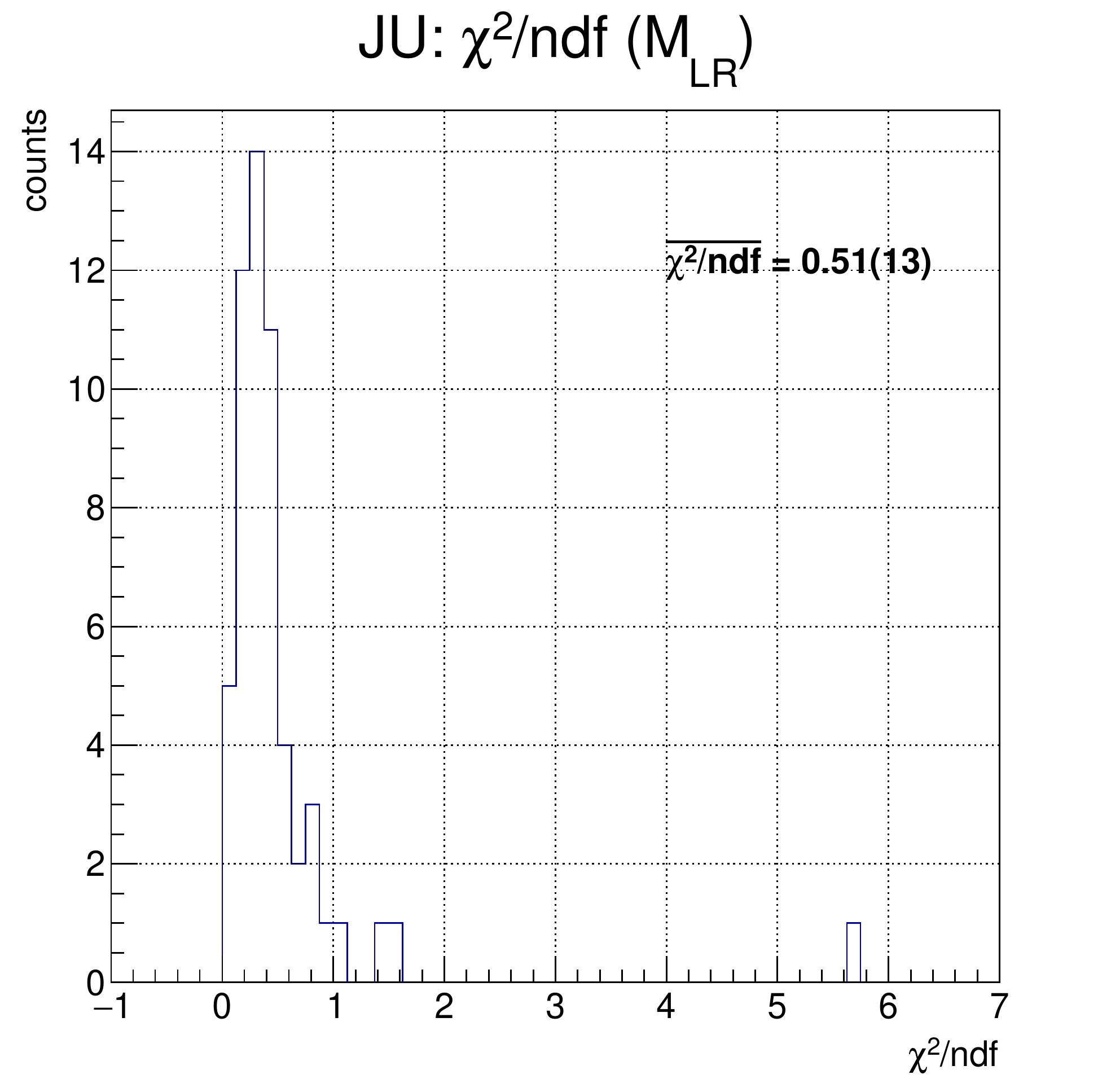}
\includegraphics[width=0.49\textwidth]{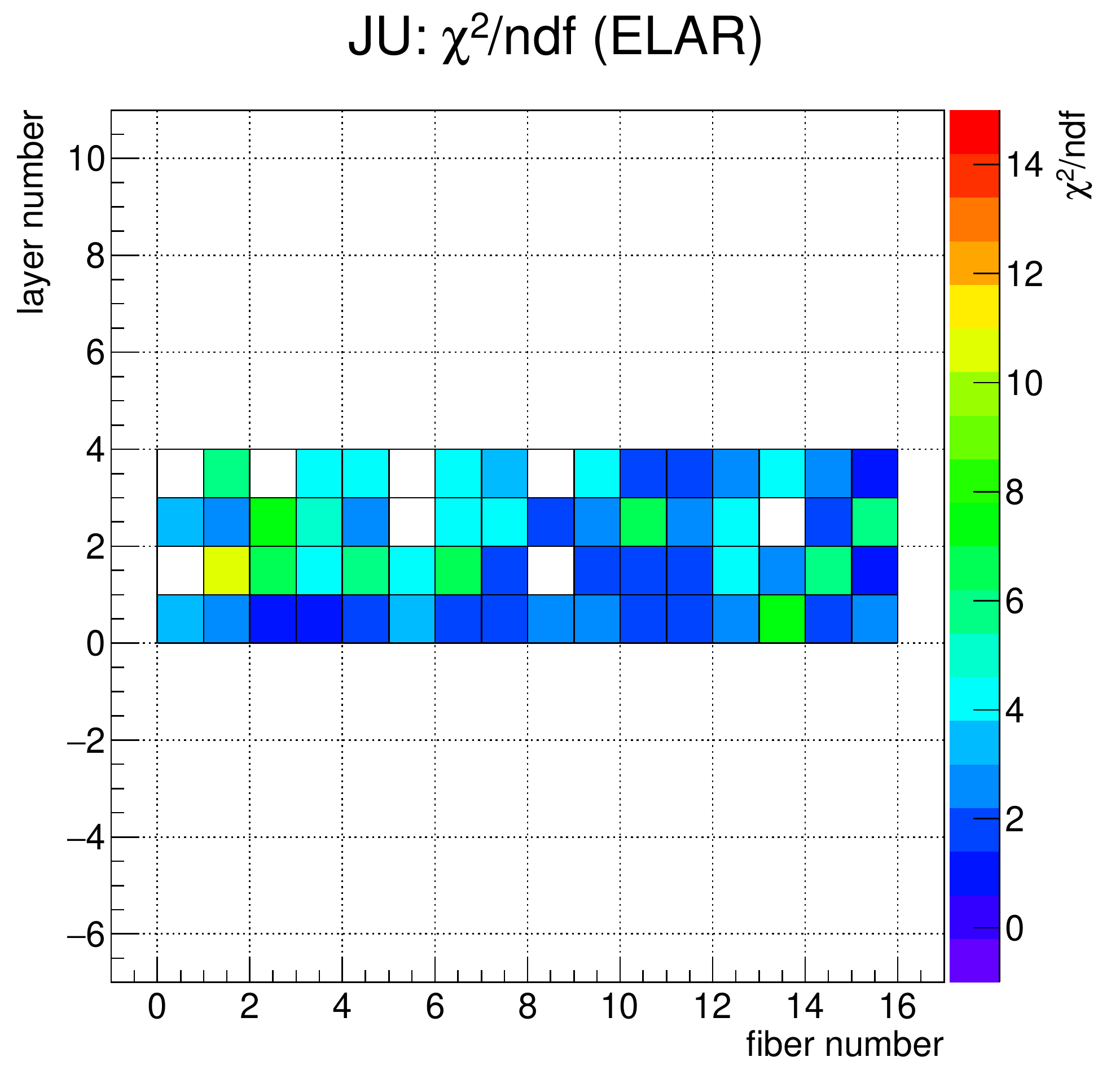}
\includegraphics[width=0.49\textwidth]{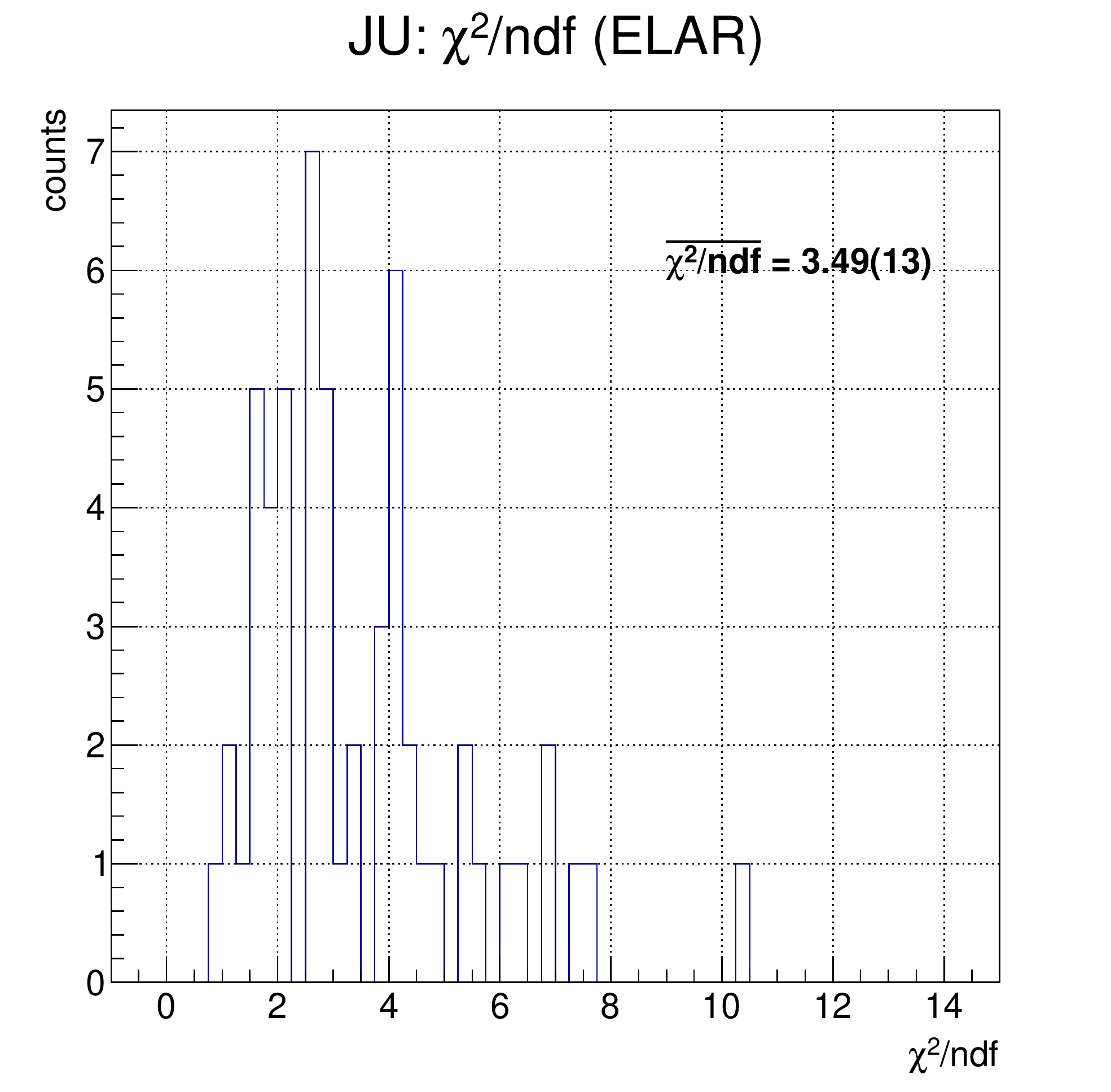}
\caption{Values of \chiNDF obtained in \gls{gl:MLR} (top row) and \acrshort{gl:ELAR} fits (bottom row). Left histograms present spatial distribution of the obtained values within the prototype, while the right plots show their statistical distributions. The weighted mean is listed in both histograms.}
\label{fig:ju-chindf}
\end{figure}

\Cref{fig:ju-chindf} presents \chiNDF values of the light attenuation function fits for both methods and all investigated fibers in the prototype. It can be observed that the quality of the \gls{gl:MLR} function fit is very homogeneous, with nearly all \chiNDF values between \num{0.05} and \num{1.60}. The only fiber for which the fit performed significantly worse is fiber L0F13 (\cref{fig:ju-chindf}, upper row). 

The obtained \chiNDF values demonstrate that the performance of the \acrshort{gl:ELAR} model is worse compared to the \gls{gl:MLR} fit. The values of \chiNDF for \acrshort{gl:ELAR} model range from \num{0.91} to \num{10.29}. Additionally, the histogram reflecting the prototype geometry shows that the spatial distribution of \chiNDF values is not as homogeneous as in the case of the \gls{gl:MLR} function fit (\cref{fig:ju-chindf}, lower left).

\begin{figure}[htbp]
\centering
\includegraphics[width=0.49\textwidth]{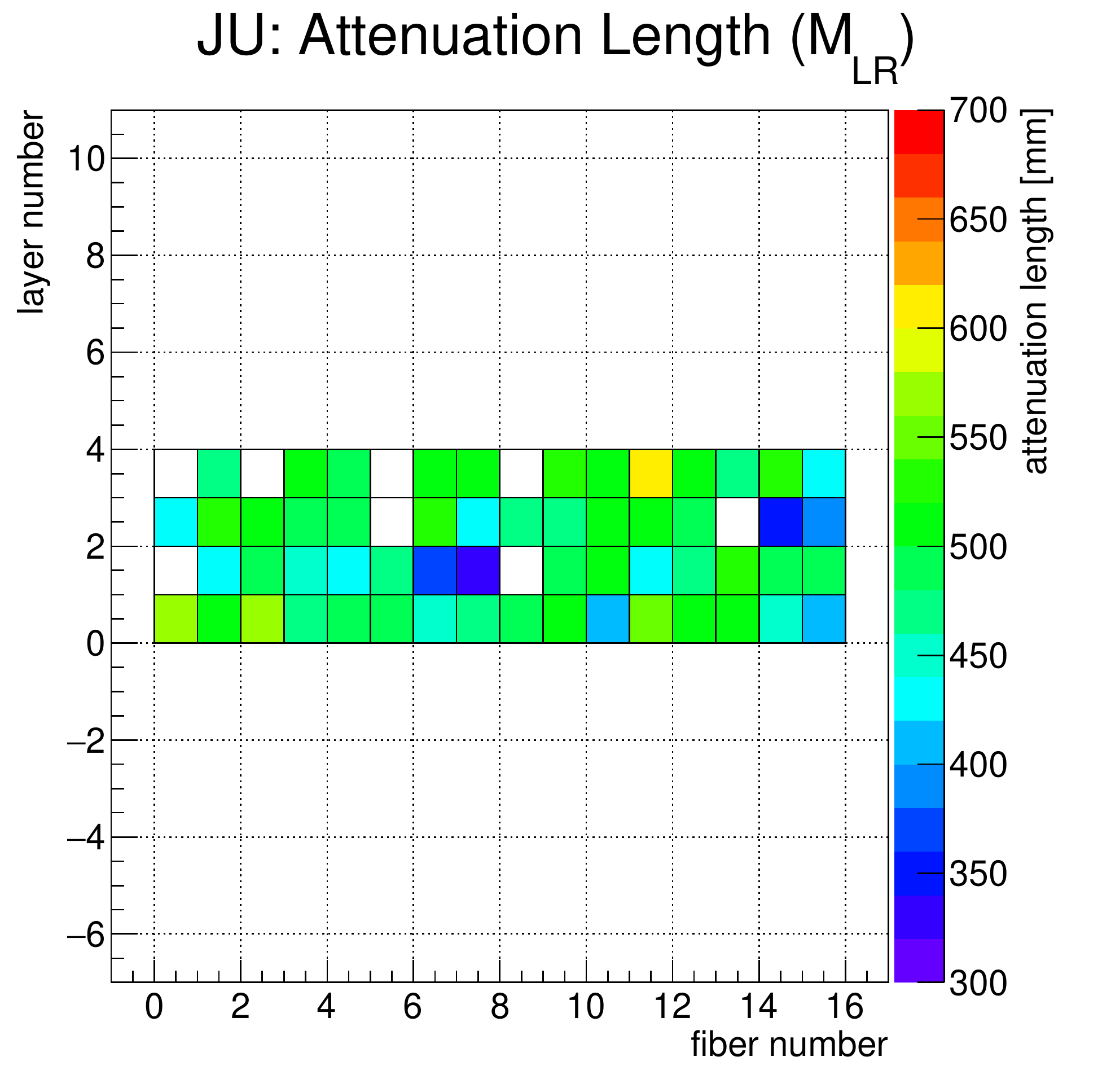}
\includegraphics[width=0.49\textwidth]{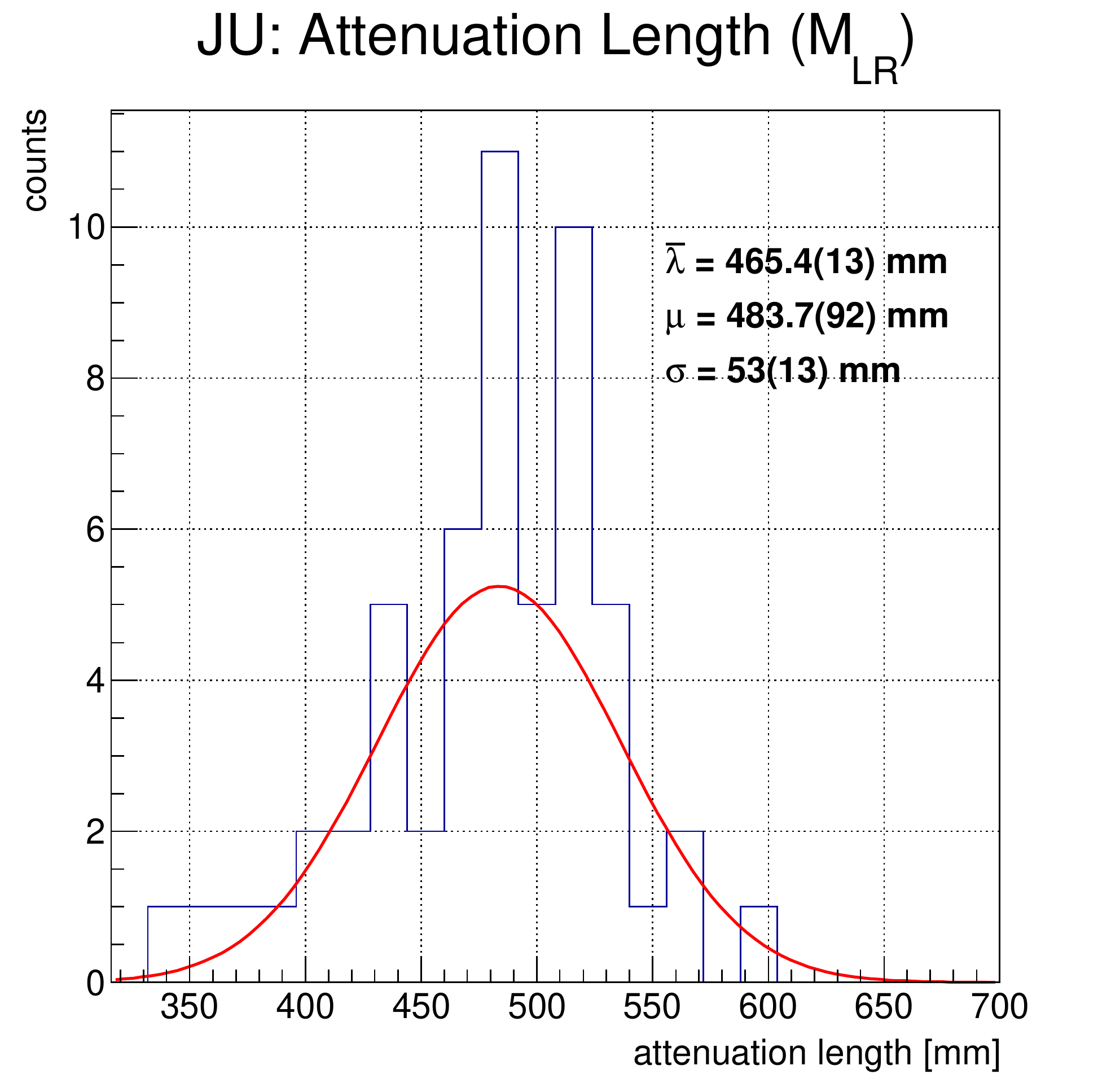}
\includegraphics[width=0.49\textwidth]{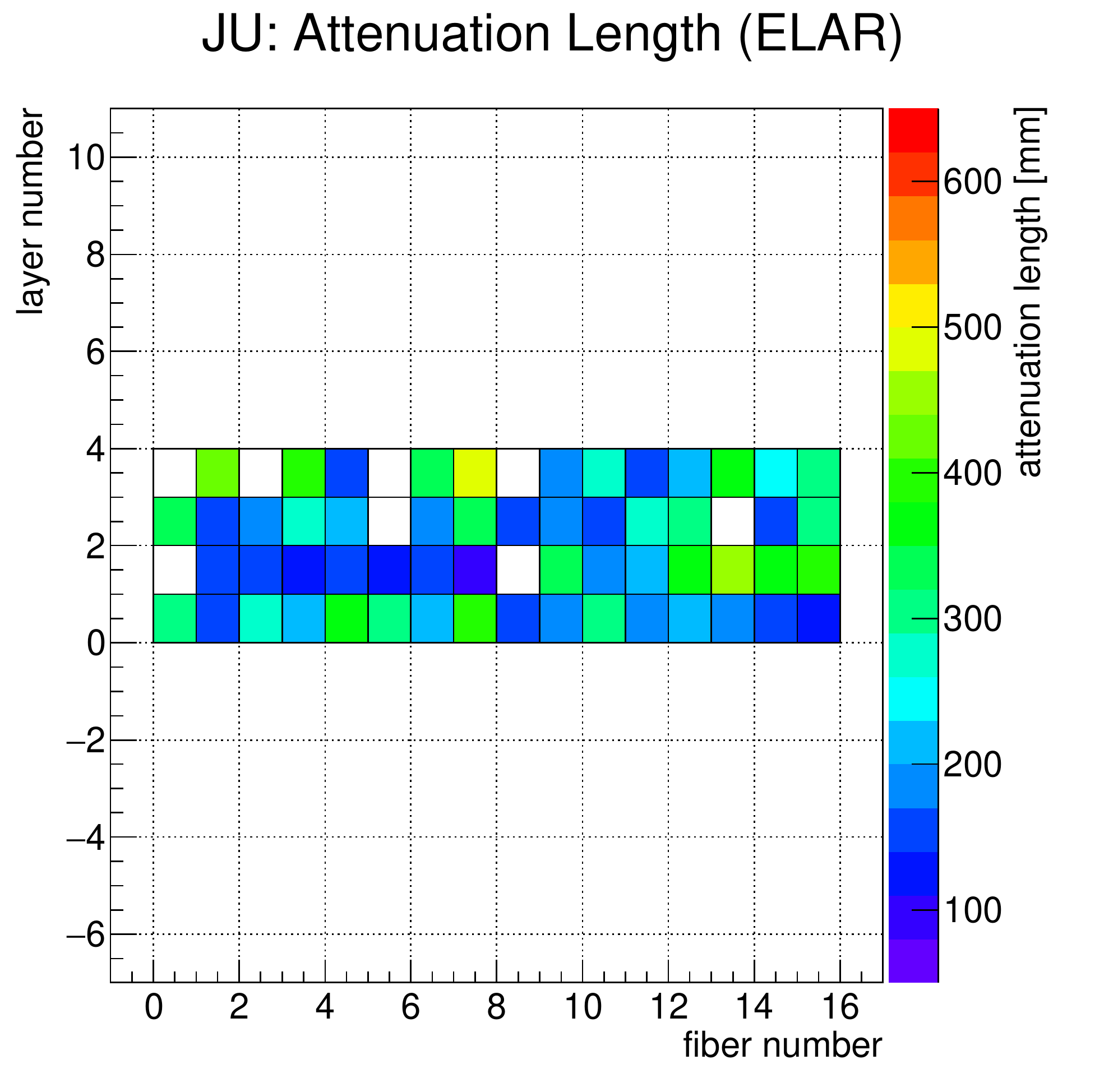}
\includegraphics[width=0.49\textwidth]{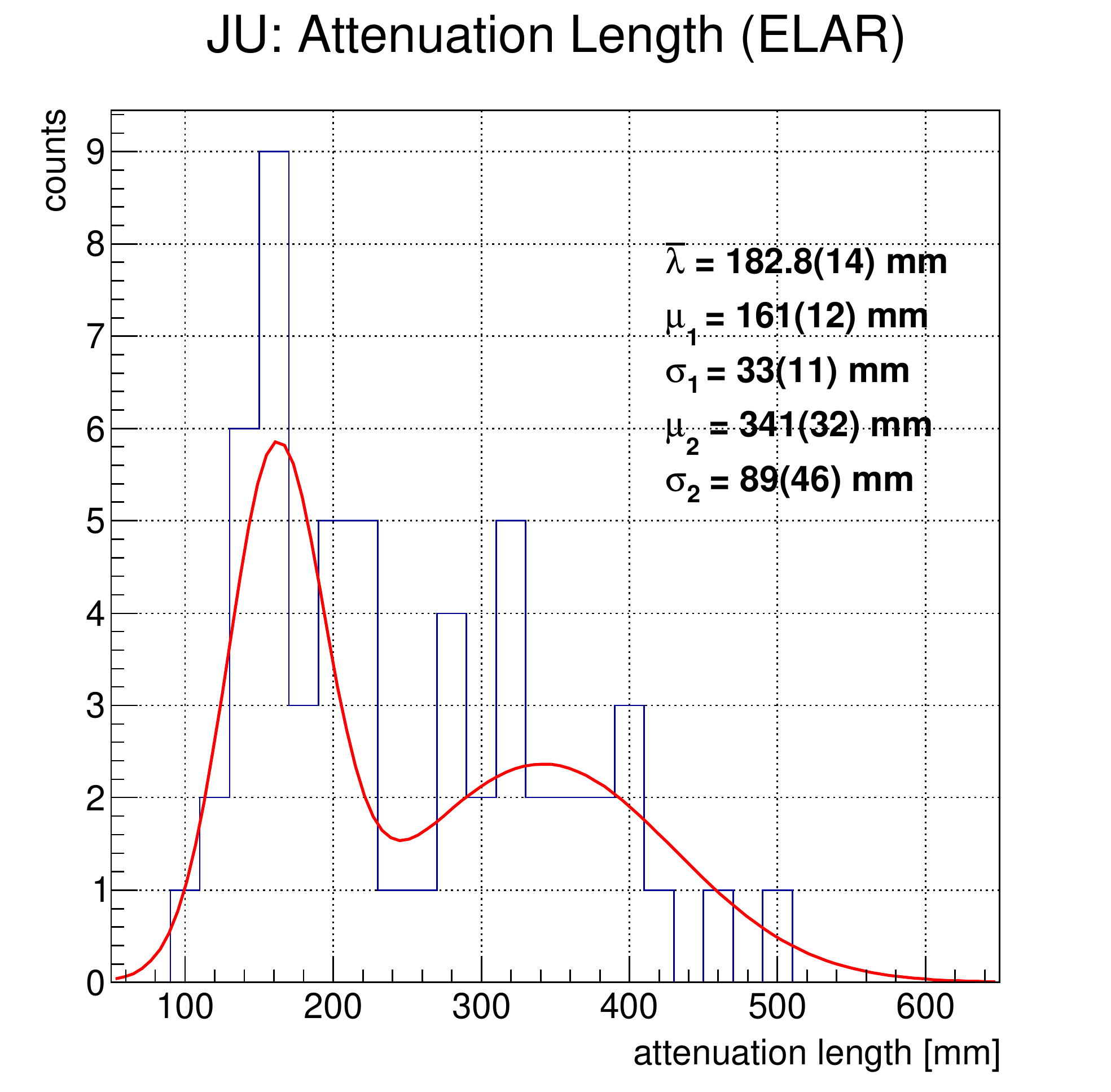}
\caption{Attenuation lengths obtained with \gls{gl:MLR} (top row) and \acrshort{gl:ELAR} methods (bottom row). Left histograms present spatial distribution of the obtained values within the prototype, while the right plots show their statistical distributions. The histograms were additionally fitted with a Gaussian function (\gls{gl:MLR}) or a double Gaussian function (\acrshort{gl:ELAR}). Parameters of the fitted functions and weighted means are listed in both histograms.}
\label{fig:ju-attenuation}
\end{figure}

\subsubsection*{Attenuation length}

Results of the light propagation analysis using the \gls{gl:MLR} method are presented in the upper part of \cref{fig:ju-attenuation}. The obtained values of attenuation length range from \SI{334}{\milli\meter} to \SI{601}{\milli\meter} and can be well described with the Gaussian distribution. The average attenuation length, calculated as a weighted mean, is \SI{465}{\milli\meter}, which is significantly more than previously determined for an analogous fiber configuration in single-fiber tests (\SI{211}{\milli\meter}, see \cref{tab:diff-wrapping}). This effect can be attributed to different \acrshort{gl:SiPM}s sizes used in the two setups. 
Non-uniform spatial distribution of the attenuation length within the prototype can be caused by differences in the fiber surface treatment during the production process, quality of coupling and wrapping. 

The results of the light propagation analysis using \acrshort{gl:ELAR} are presented in the lower part of \cref{fig:ju-attenuation}. The obtained values of the attenuation length range from \SI{103}{\milli\meter} to \SI{495}{\milli\meter}. However, in contrast to the \gls{gl:MLR} method, the distribution of obtained attenuation lengths shows two clear peaks, first at \SI{161}{\milli\meter} and second at \SI{341}{\milli\meter}. There is no physically justified reason for this shape of the distribution, which suggests that part of the obtained results are not accurate. Measurements performed for the analogous fiber configuration in single-fiber studies yielded \acrshort{gl:ELAR}-based attenuation length of \SI{113}{\milli\meter} (see \cref{tab:diff-wrapping}). This hints towards the lower attenuation length values as the correct ones. 

\begin{figure}[t]
\centering
\includegraphics[width=0.99\textwidth]{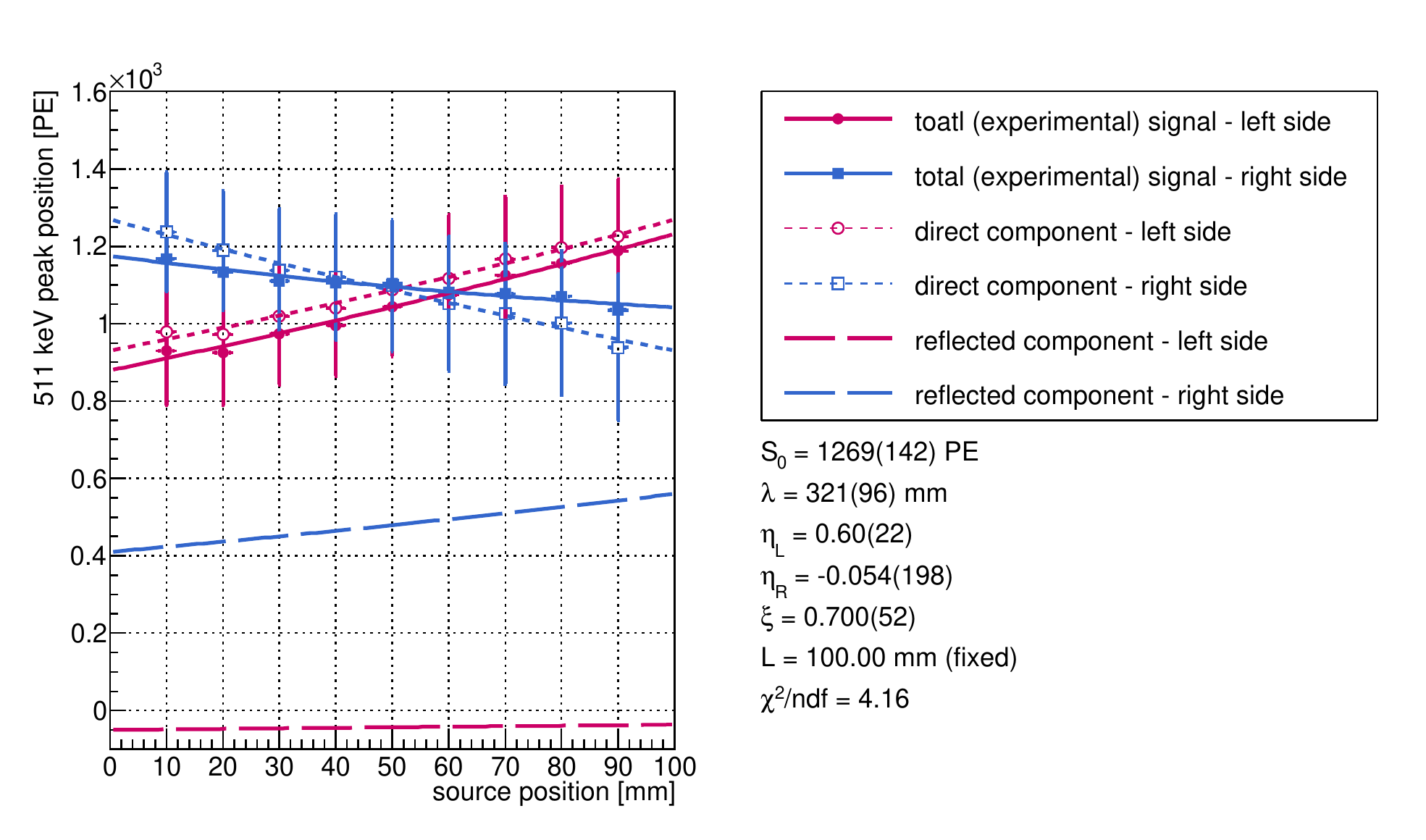}
\caption{An example of a dubious fit of the \acrshort{gl:ELAR} model. When the light attenuation in the scintillator is weak the data variability is too small to correctly determine all the parameters of \acrshort{gl:ELAR} model fit. In that situation the \acrshort{gl:ELA} model is more suited. Presented data was recorded for the fiber L3F6.}
\label{fig:ju-bad-elar}
\end{figure}

Closer inspection of the \acrshort{gl:ELAR} functions revealed inconsistencies in the fits that yielded high attenuation length values. An example of such a fit is presented in \cref{fig:ju-bad-elar}. Firstly, the error bars of the reconstructed data points representing the direct components are large compared to successful fits (see \cref{fig:ELAR-fit}). Secondly, one of the $\eta$ parameters is negative, although it is consistent with \num{0} within its uncertainty. As a result, the corresponding reconstructed reflected component of the signal is also negative. Consequently, one of the reconstructed primary components exceeds the total recorded signal for the positions between \SI{0}{\milli\meter} and \SI{50}{\milli\meter}. All the listed observations are not consistent with the physical interpretation of the \acrshort{gl:ELAR} model. In all the fits that yielded a high attenuation length, at least one of the described features can be observed. Another common factor for all uncertain fits was the shape of the attenuation curves, resembling a linear rather than an exponential dependence. This shape indicates weak attenuation of the scintillating light. Therefore, it can be concluded that the \acrshort{gl:ELAR} model is not suitable for situations in which weak light attenuation is expected. In such situations, the physical interpretation of the model parameters is no longer valid and they should be treated as purely abstract.

\subsubsection*{Energy reconstruction}

\begin{figure}[hp]
\centering
\includegraphics[width=0.49\textwidth]{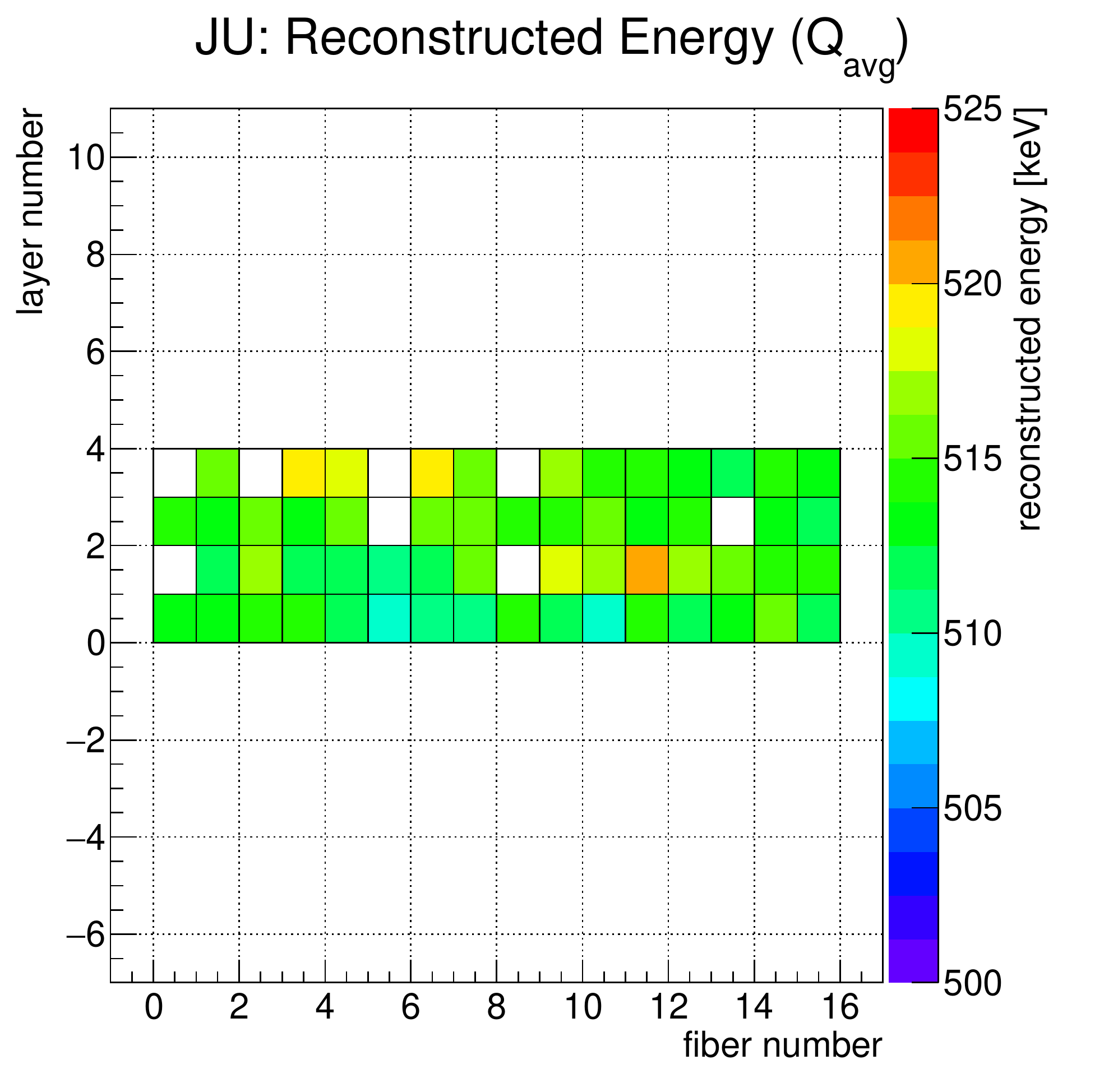}
\includegraphics[width=0.49\textwidth]{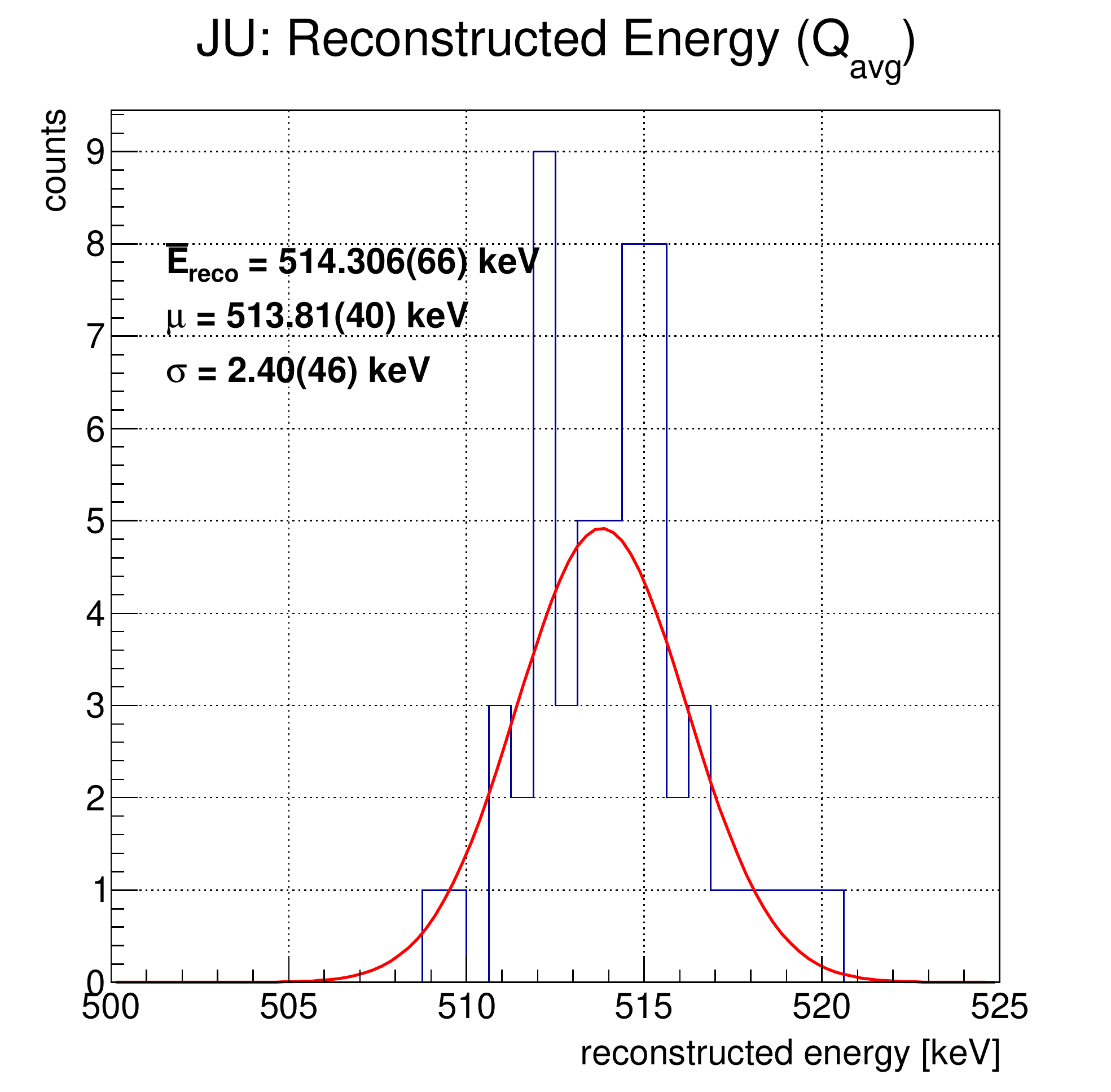}
\includegraphics[width=0.49\textwidth]{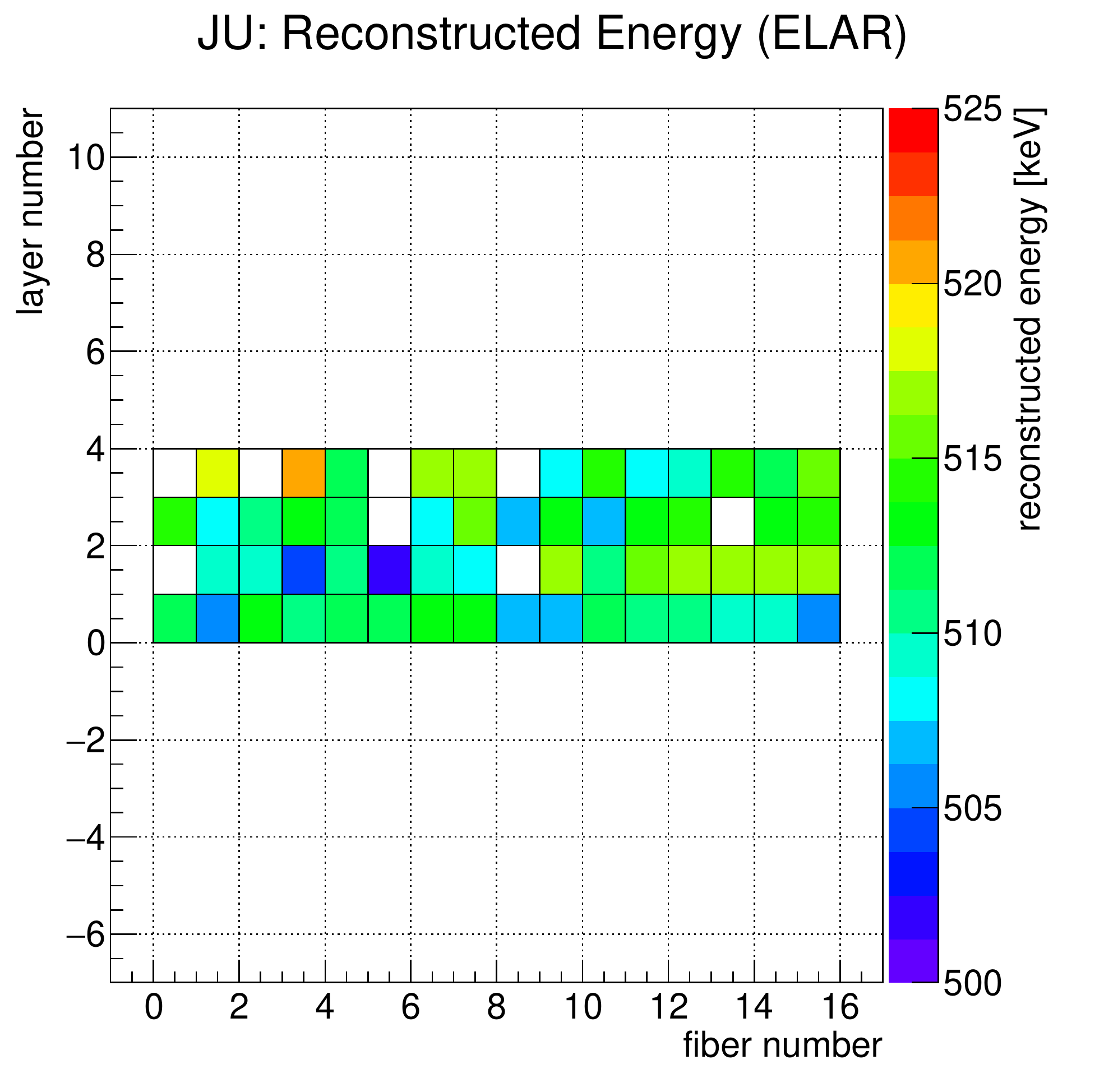}
\includegraphics[width=0.49\textwidth]{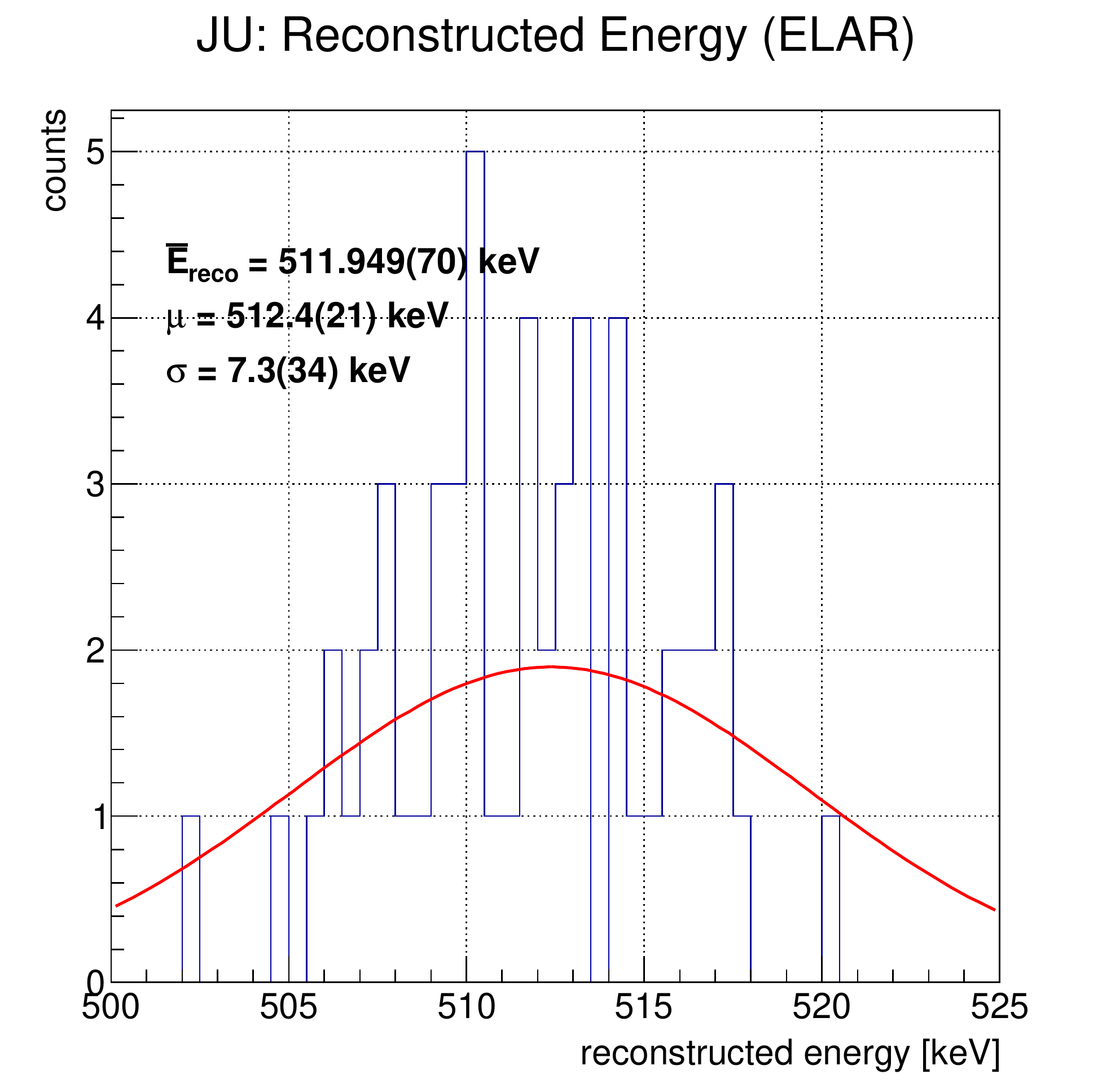}
\caption{Energies of the annihilation peak reconstructed with the two methods: \gls{gl:Qavg} (top row) and \acrshort{gl:ELAR} (bottom row). Left histograms present the values of the reconstructed peak centers for different fibers in the prototype, while the right plots show their statistical distributions and their parametrizations as Gaussian functions. Fit parameters and weighted means are also listed.}
\label{fig:ju-energyreco}
\end{figure}

The energy spectra reconstructed using both described previously methods (\cref{ssec:energy-resolution}) were assessed in terms of mean energy of the annihilation peak (\gls{gl:peakpos}) and energy resolution. \Cref{fig:ju-energyreco} presents results of the determination of \gls{gl:peakpos}. The average energy of the annihilation peak reconstructed using ELAR-based method (\SI{512}{\kilo\electronvolt}) is closer to the expected value than the average energy obtained with \gls{gl:Qavg} method (\SI{514}{\kilo\electronvolt}). However, the spread of the obtained values is noticeably broader for the \acrshort{gl:ELAR} method, with the standard deviation approximately three times larger than for the \gls{gl:Qavg} method. This effect can be attributed to the inconsistencies in the fitting of the \acrshort{gl:ELAR} model described in the previous paragraphs. 

\begin{figure}[hp]
\centering
\includegraphics[width=0.49\textwidth]{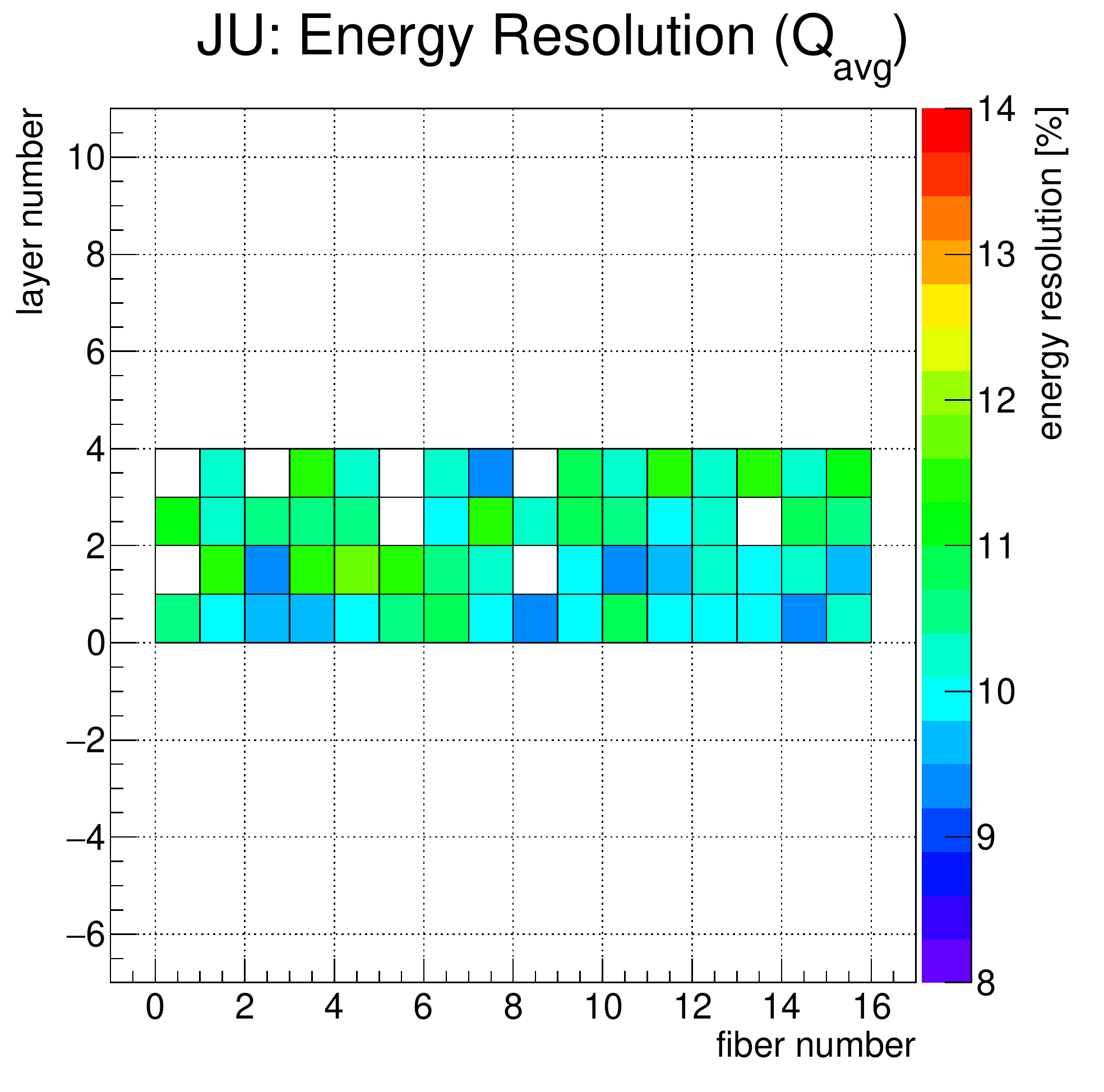}
\includegraphics[width=0.49\textwidth]{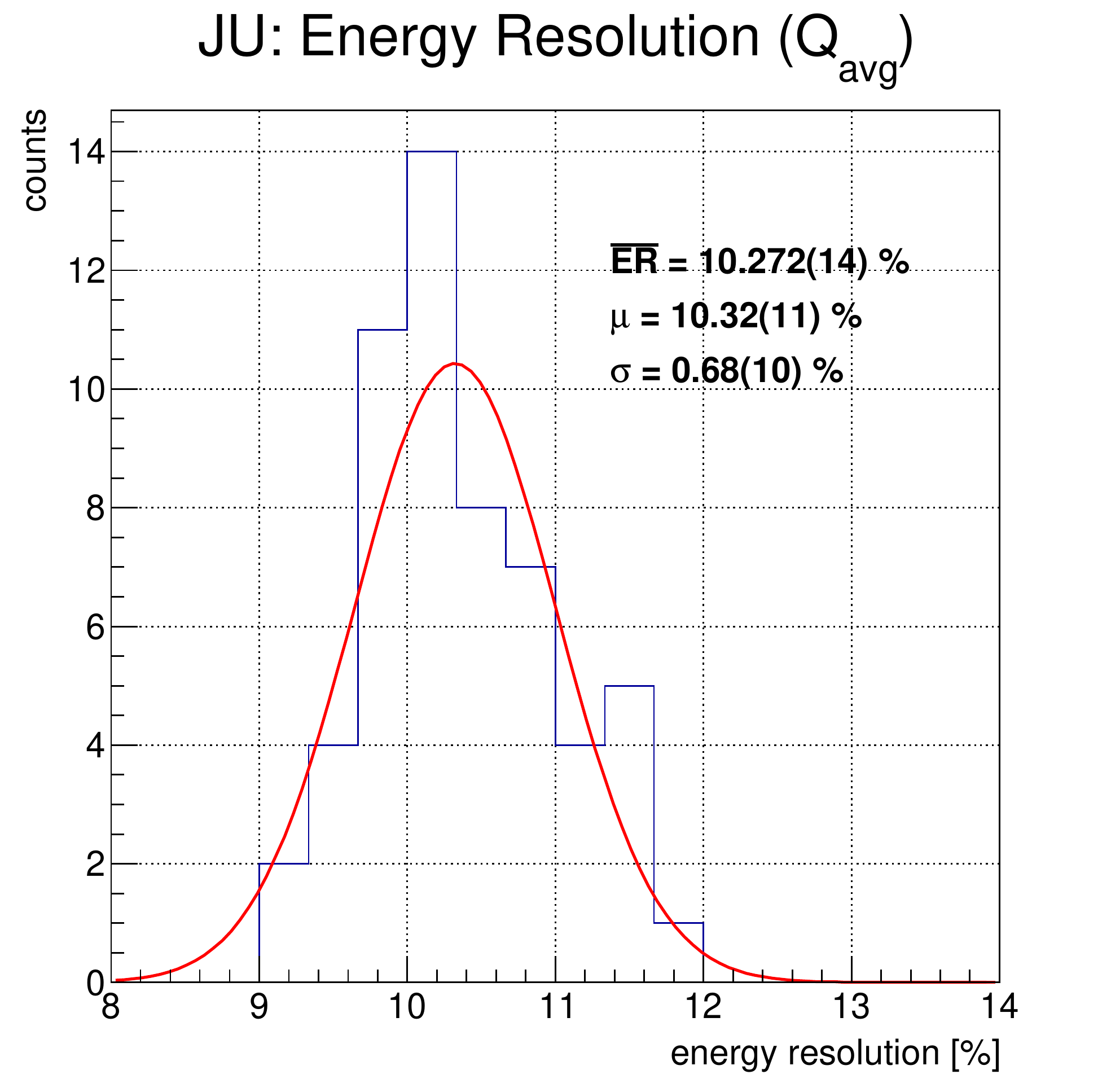}
\includegraphics[width=0.49\textwidth]{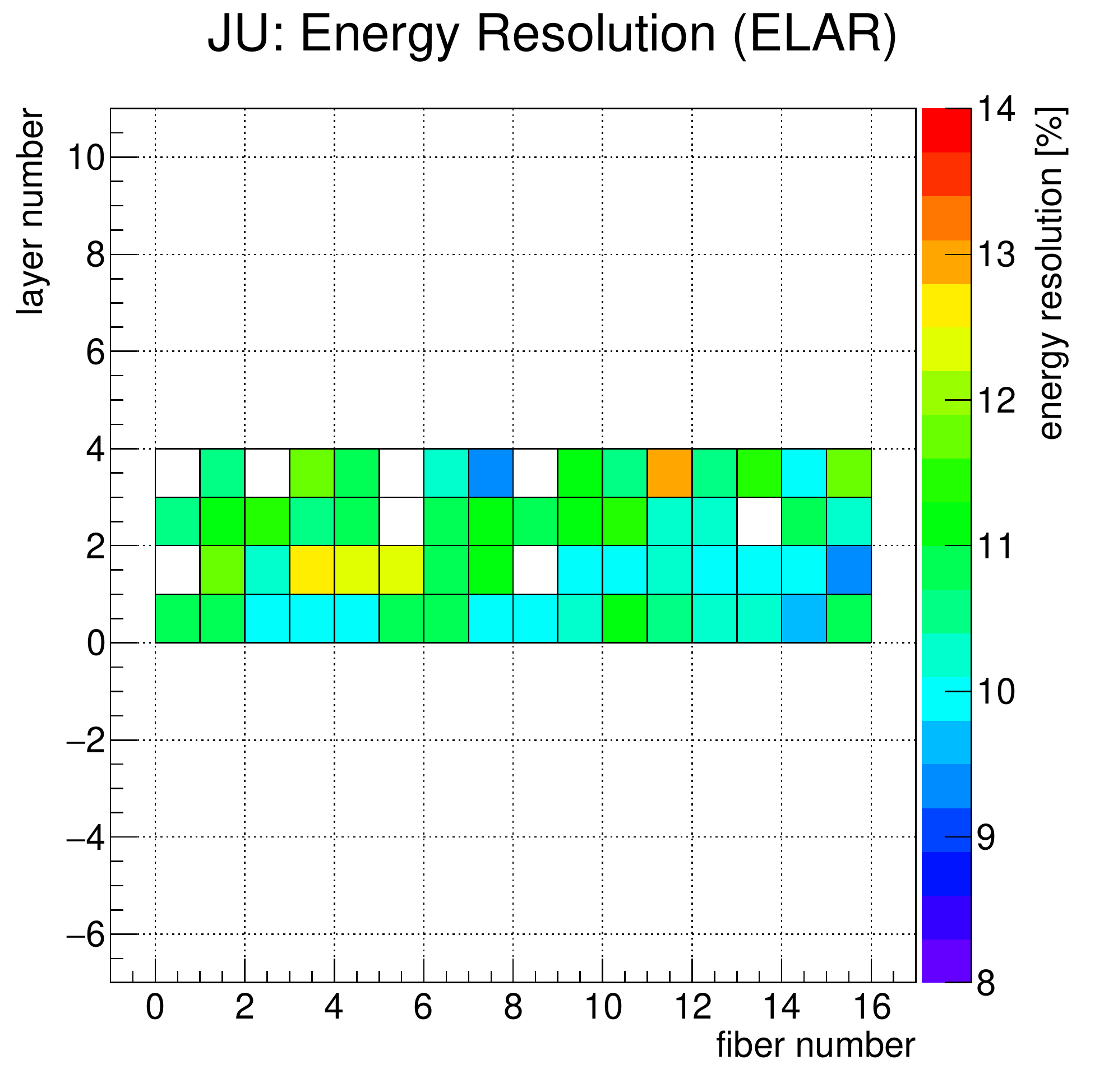}
\includegraphics[width=0.49\textwidth]{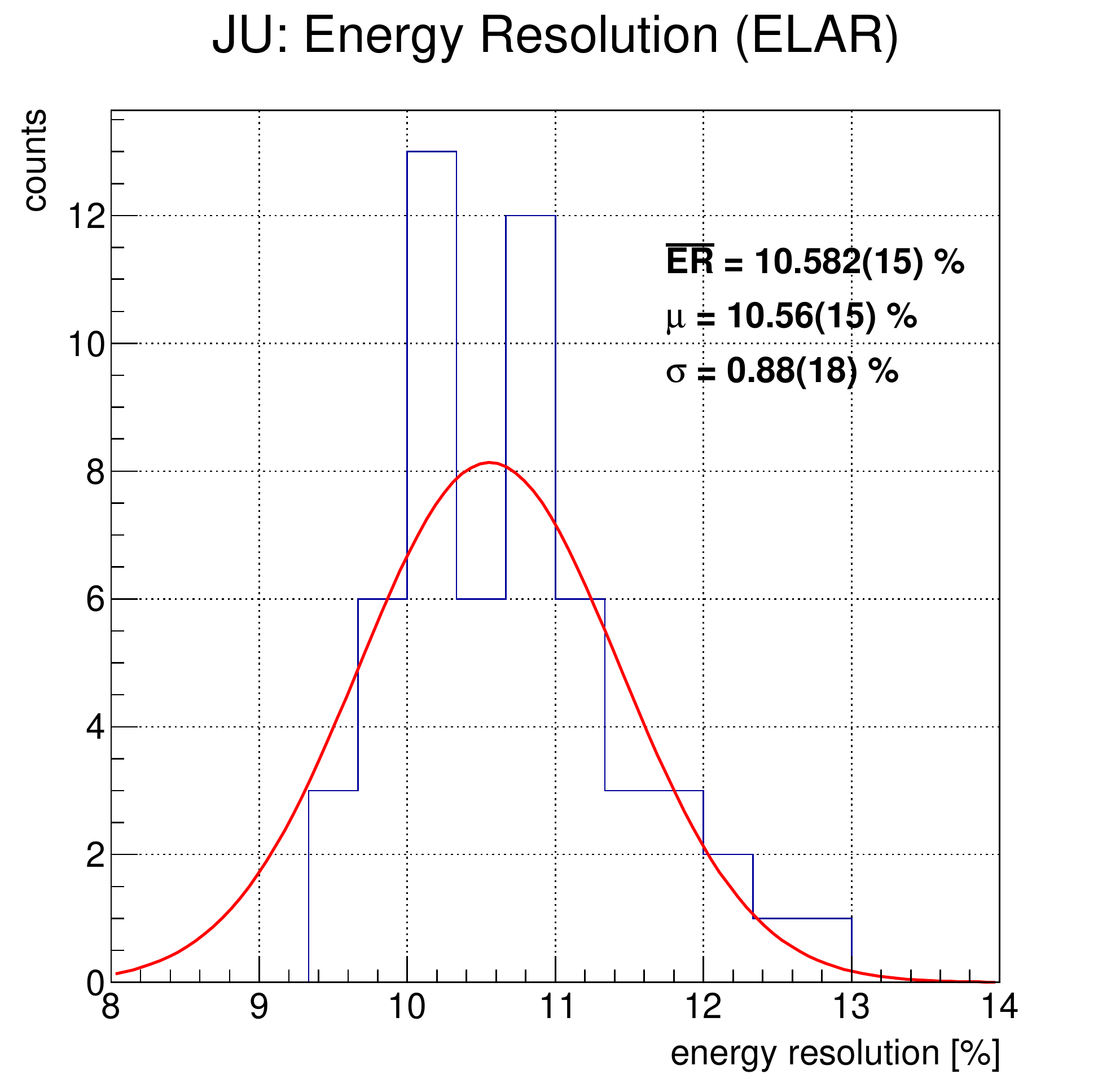}
\caption{Energy resolutions obtained from spectra reconstructed with \gls{gl:Qavg} (top row) and \acrshort{gl:ELAR} (bottom row) methods. Left histograms present the values determined for different fibers in the prototype, while the right plots show their statistical distributions and their parameterizations as Gaussian functions. Parameters of the distributions and the weighted means are listed.}
\label{fig:ju-energyres}
\end{figure}

The obtained results of energy resolution analysis are presented in \cref{fig:ju-energyres}. The distributions of obtained values for both methods are consistent in terms of the mean and standard deviation. Moreover, the 2D histograms show that the spatial distributions of the energy resolutions exhibit the same pattern. 
The average energy resolutions are \SI{10.27}{\percent} and \SI{10.58}{\percent} for the \gls{gl:Qavg} and \acrshort{gl:ELAR}-based methods, respectively. This energy resolution is significantly worse than the one observed for the analogous fiber configuration in the single-fiber study (\SI{8.56}{\percent}, see \cref{tab:diff-wrapping}). This difference can be attributed to different types of \acrshort{gl:SiPM}s used in the two experiments and thus smaller light collection in case of the prototype. 

\subsubsection*{Position reconstruction}

To evaluate the position reconstruction two characteristics were taken into account. The first was the mean of the integrated distribution of position residuals (\gls{gl:Xreco} $-$ \gls{gl:Xreal}). For exact 
position reconstruction, the distribution is expected to be centered at \SI{0}{\milli\meter}. The second assessed characteristics was \acrshort{gl:FWHM} of the residuals distribution, \ie the position resolution. The position reconstruction was carried out using the two previously described methods, namely using the \gls{gl:MLR} quantity and the \gls{gl:MLRstar} quantity calculated with the \acrshort{gl:ELAR} model parameters. 

\begin{figure}[hp]
\centering
\includegraphics[width=0.49\textwidth]{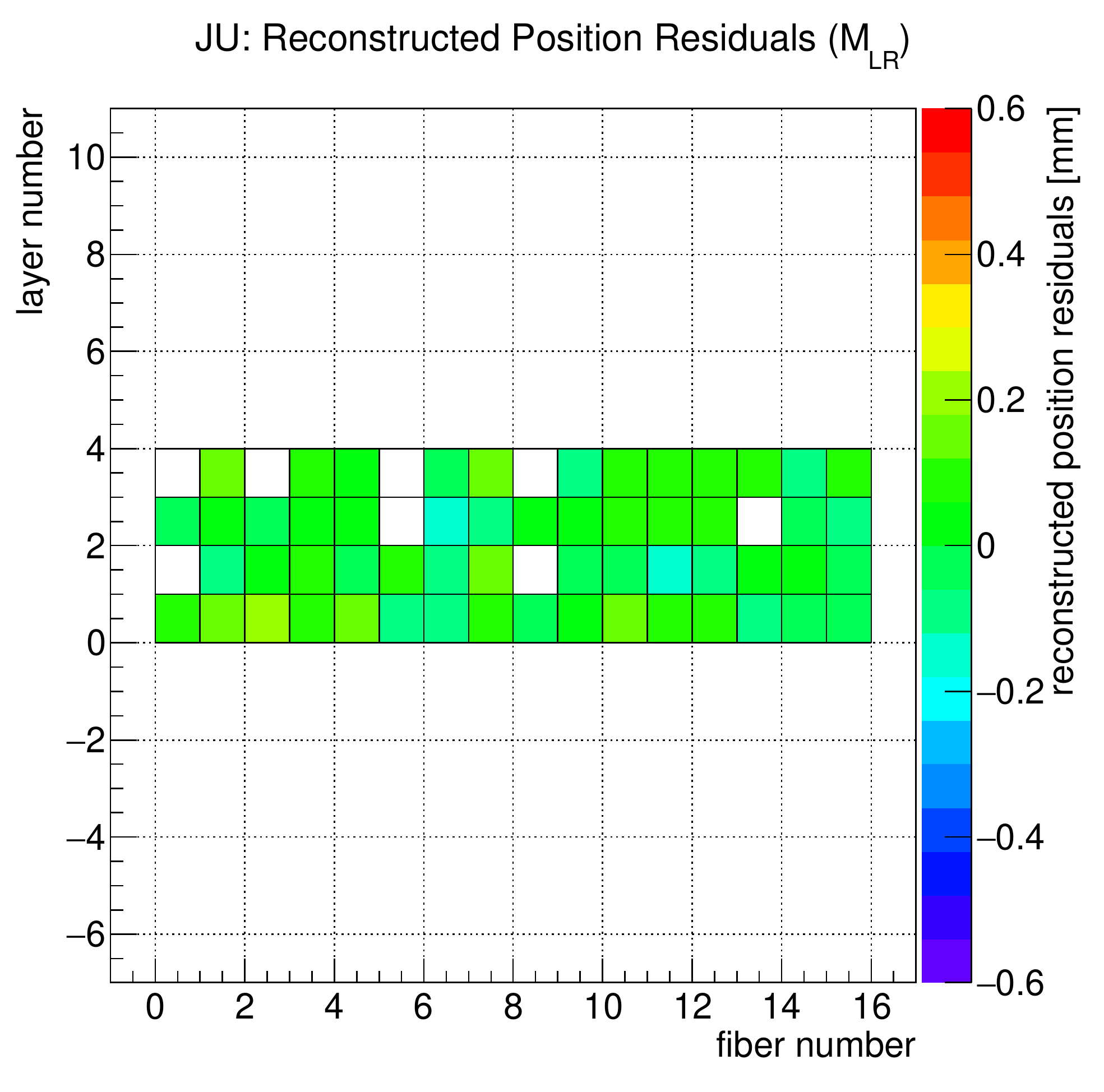}
\includegraphics[width=0.49\textwidth]{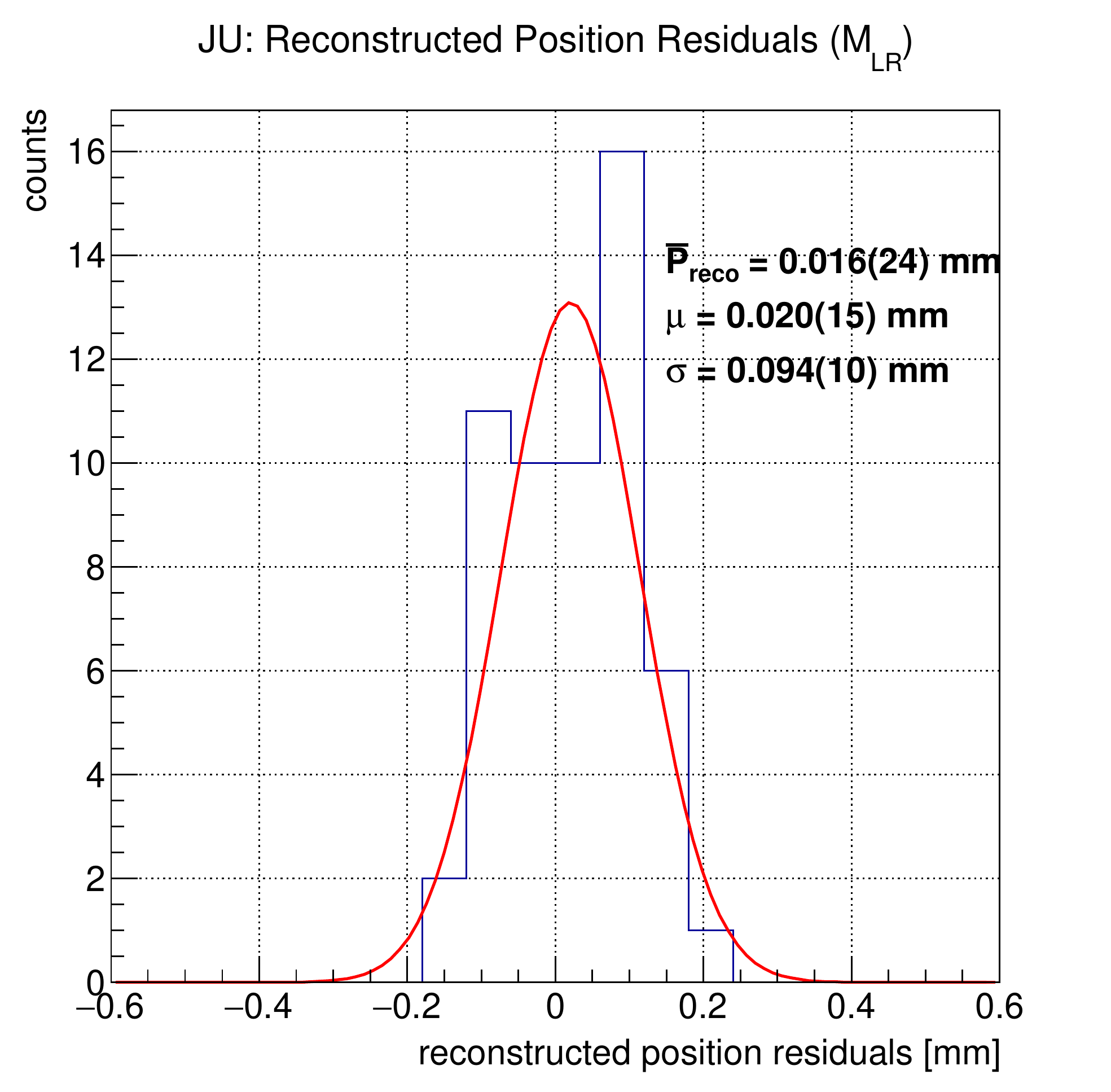}
\includegraphics[width=0.49\textwidth]{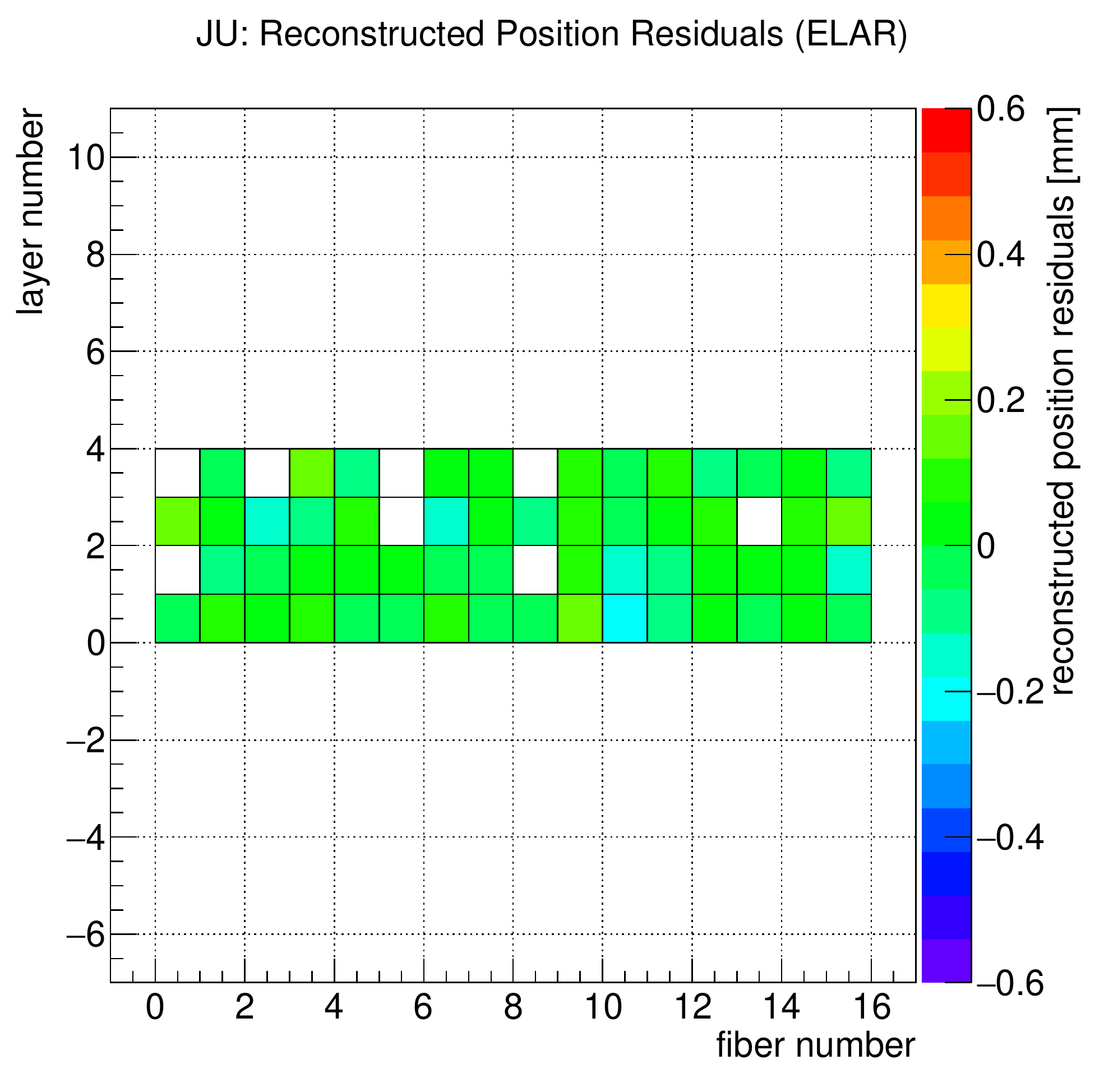}
\includegraphics[width=0.49\textwidth]{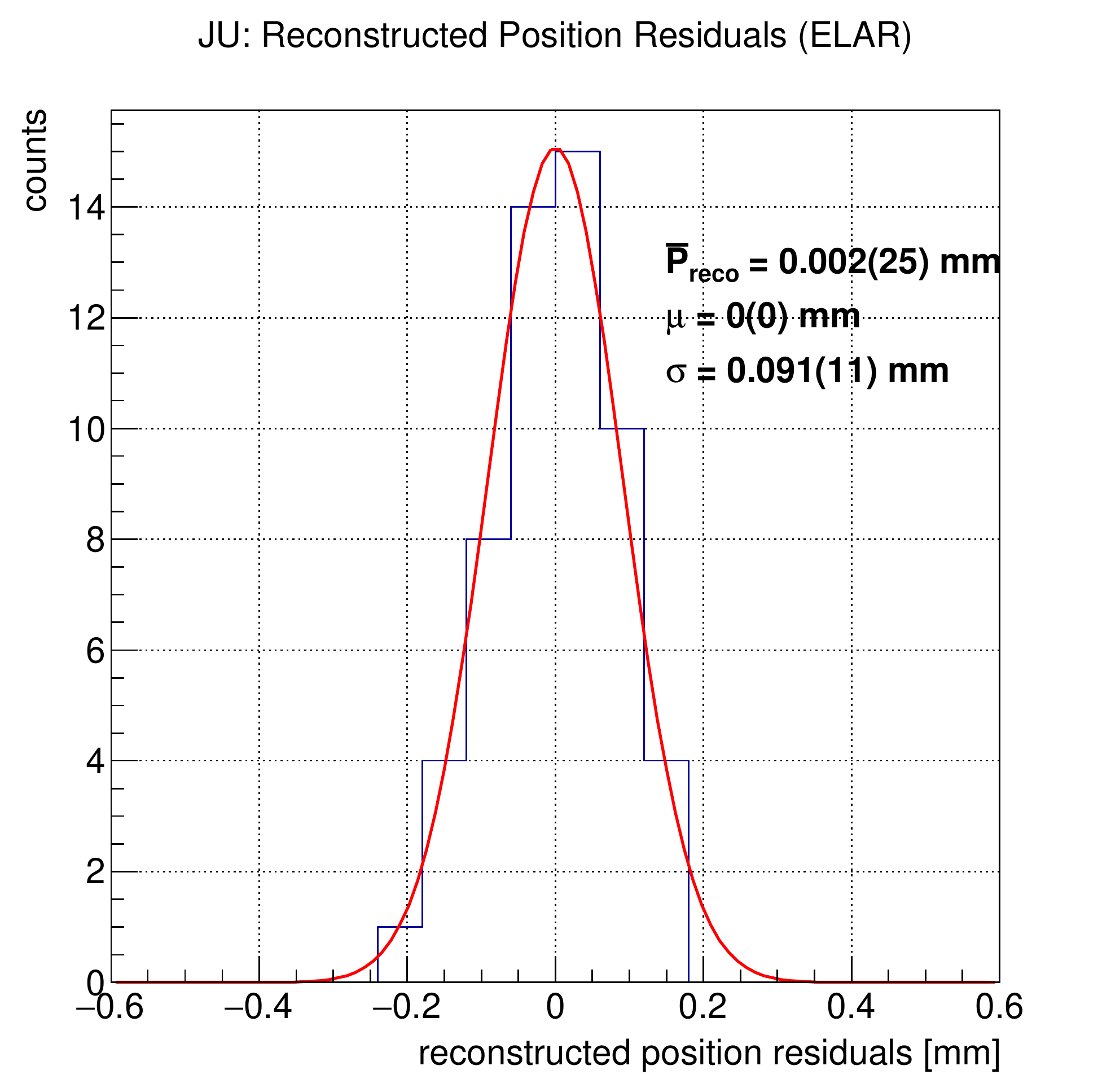}
\caption{Mean position residuals obtained with the two position reconstruction methods: \gls{gl:MLR} (top row) and \acrshort{gl:ELAR} (botton row). Left histograms present the values of the reconstructed position residuals for different fibers in the prototype, while the right plots show their statistical distributions and their parametrizations as Gaussian functions. Fit parameters and weighted means are also listed.}
\label{fig:ju-posreco}
\end{figure}

\Cref{fig:ju-posreco} shows results of the determination of the mean position residuals. For both of the reconstruction methods discussed, the obtained results are consistent. In both cases, the mean position residuals are consistent with \SI{0}{\milli\meter} within uncertainties, as expected. 

\begin{figure}[hp]
\centering
\includegraphics[width=0.49\textwidth]{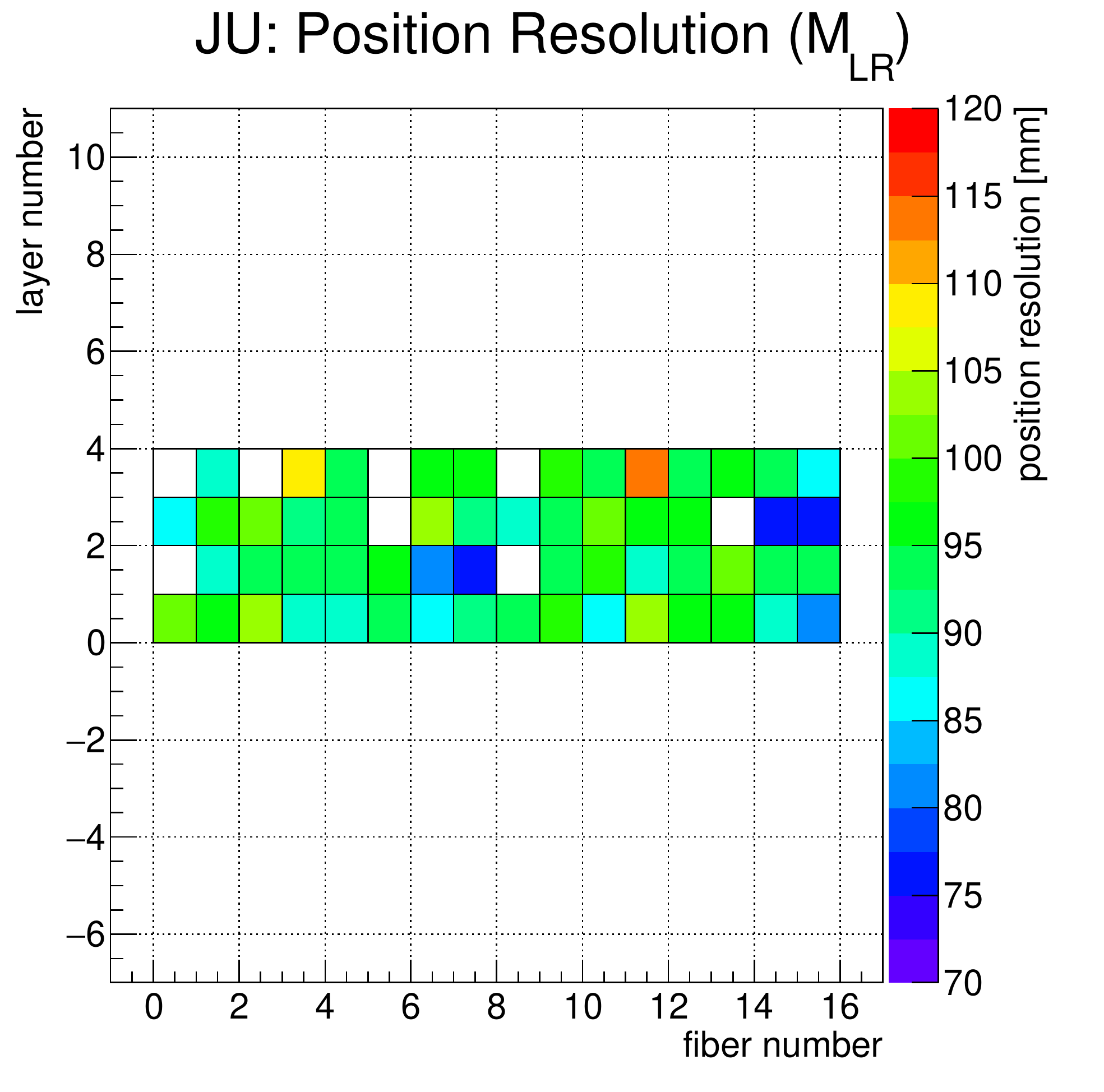}
\includegraphics[width=0.49\textwidth]{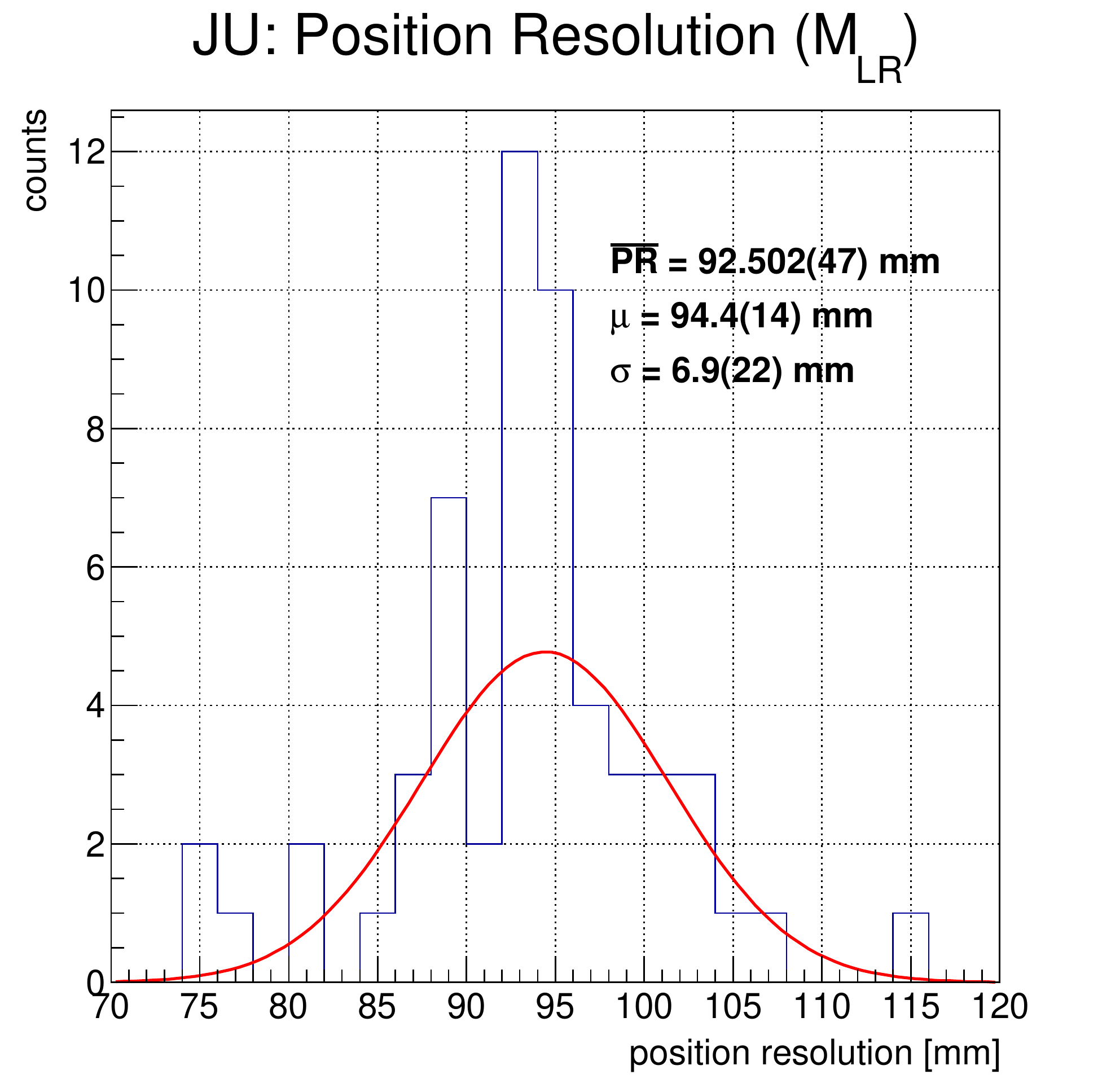}
\includegraphics[width=0.49\textwidth]{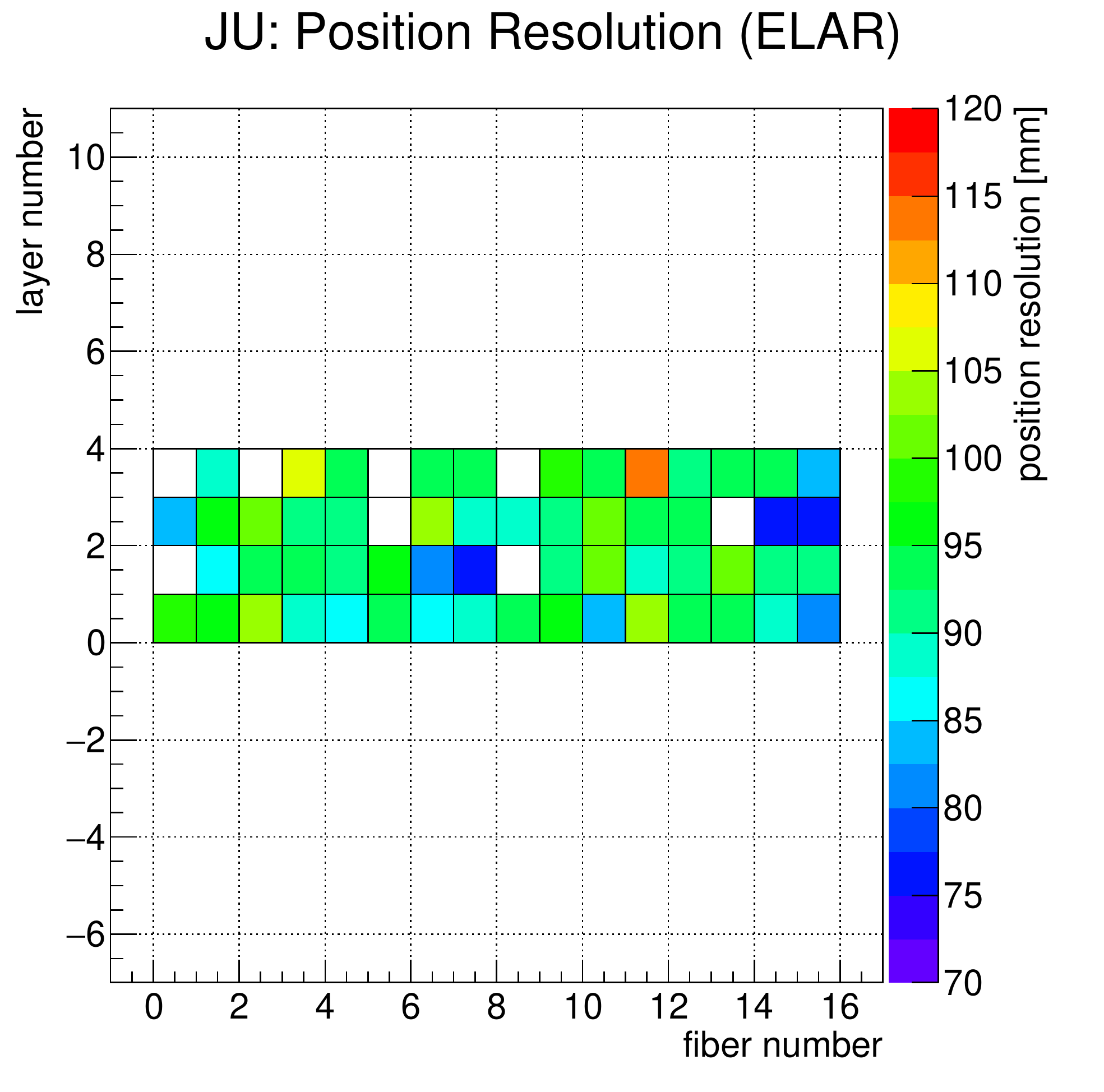}
\includegraphics[width=0.49\textwidth]{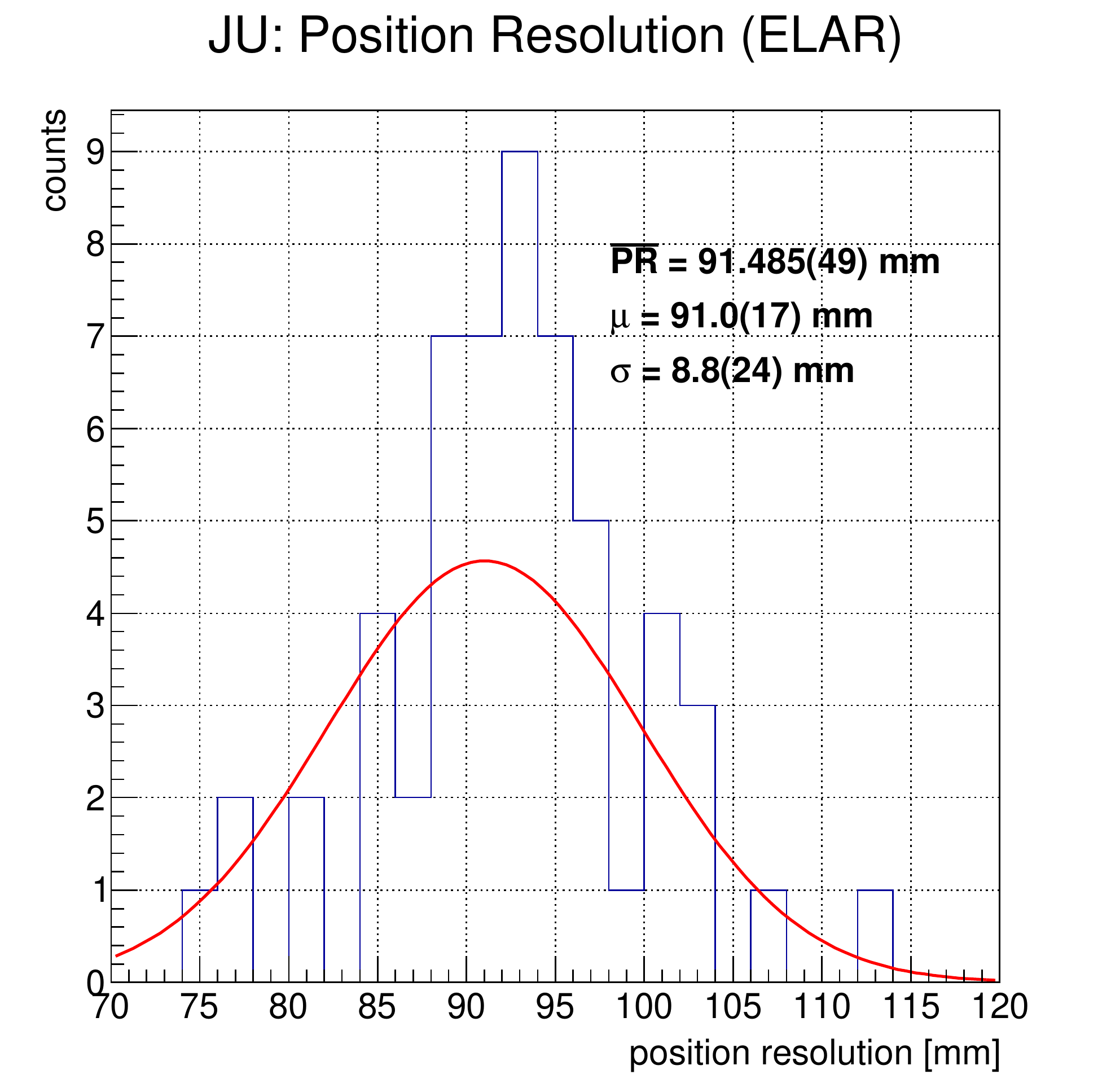}
\caption{Position resolutions obtained with the two position reconstruction methods: \gls{gl:MLR} (top row) and \acrshort{gl:ELAR} (botton row). Left histograms present the spatial distributions of the obtained values within the prototype, while the right plots show their statistical distributions. Histograms were fitted with Gaussian functions. Obtained fit parameters and weighted means are also listed.}
\label{fig:ju-posres}
\end{figure}

The results of the position resolution analysis are presented in \cref{fig:ju-posres}. Here, the \gls{gl:MLR} and \acrshort{gl:ELAR} methods also yielded comparable results, in terms of the mean position resolution as well as the standard deviation of the distributions. In particular, the spatial distributions of the values for both methods show the same pattern.
This proves that, despite the inconsistencies in the physical interpretation of the \acrshort{gl:ELAR} model, it still correctly describes the experimental data and thus allows for position reconstruction. 
 
The achieved average position resolution is \SI{92.50}{\milli\meter} and \SI{91.49}{\milli\meter} for the \gls{gl:MLR} and \acrshort{gl:ELAR} methods, respectively. This result is significantly worse than the one obtained for the analogous fiber configuration in the single-fiber tests (\SI{32}{\milli\meter}, see \cref{tab:diff-wrapping}). Moreover, the position resolutions of the prototype are in the order of the length of \acrshort{gl:LYSO:Ce} fibers, which makes accurate position reconstruction practically impossible. This effect, similarly to the worsening of the energy resolution, can be attributed to the different type of \acrshort{gl:SiPM}s used in \acrshort{gl:JU} prototype measurements and single-fiber measurements. As described in previous section, the small size of KETEK \acrshort{gl:SiPM}s influenced the attenuation of the scintillating light experienced by the system. Furthermore, the smaller amount of light registered by the KETEK \acrshort{gl:SiPM}s negatively affected the position resolution as well.

Performance of the \gls{gl:MLR} and \acrshort{gl:ELAR} methods in position reconstruction is comparable. In the case of the energy reconstruction, performance of the method based on \acrshort{gl:ELAR} parameterization was significantly worse. This suggests that the energy reconstruction is more sensitive to inconsistencies in the fit of the model. It can be caused by the fact that it involves a product of the reconstructed primary signal components (\cref{eq:energy-reco-elar}). On the other hand, during the position reconstruction, the ratio of the reconstructed primary components is calculated (\cref{eq:MLR-corrected}). This means that in the \gls{gl:MLRstar} ratio some components cancel, while in the \gls{gl:Qavgstar} they are enhanced.

\subsubsection*{Light collection}

\begin{figure}[hptb]
\centering
\includegraphics[width=0.49\textwidth]{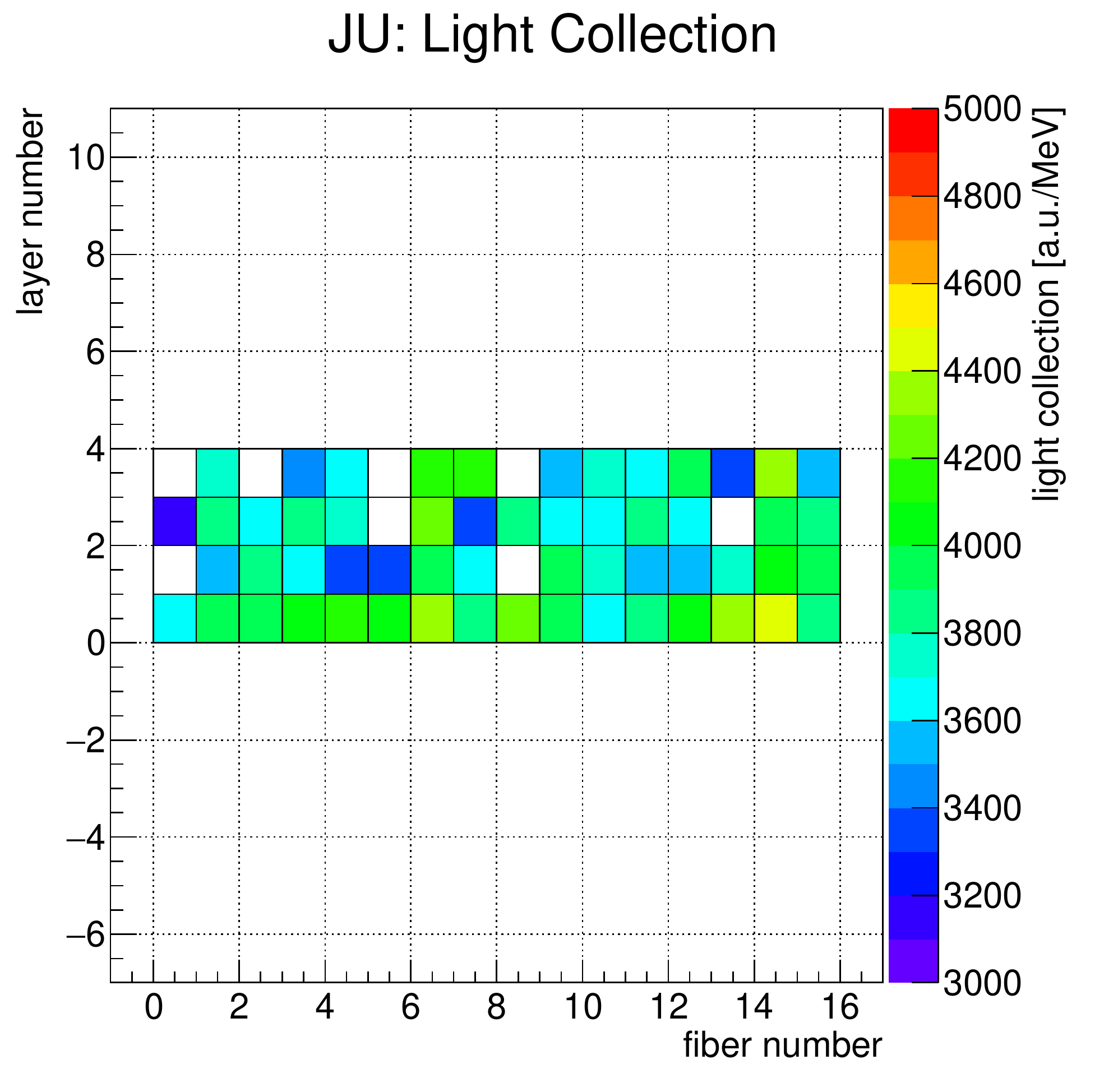}
\includegraphics[width=0.49\textwidth]{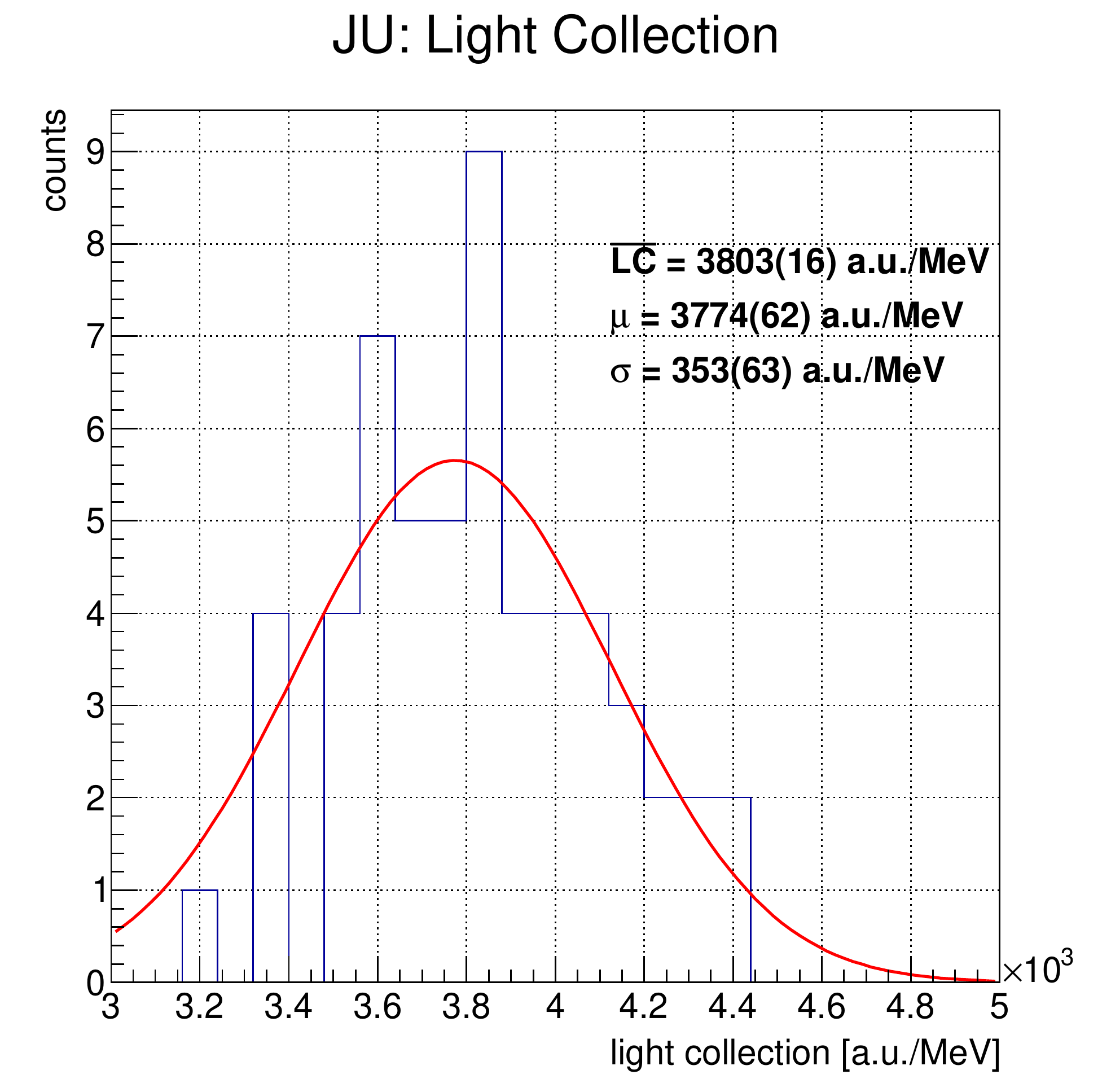}
\caption{Left: histogram showing the light collection values for each investigated fiber in \acrshort{gl:SSP}. Right: Statistical distribution of light collection values fitted with the Gaussian function. The listed values indicate the weighted mean and parameters of the distribution obtained from the fit.}
\label{fig:ju-lcol}
\end{figure}

\Cref{fig:ju-lcol} presents results of the light collection analysis in the \acrshort{gl:SSP}. Since there was no \acrshort{gl:PE} calibration performed for the \acrshort{gl:JU} measurements the charge was expressed in arbitrary units. The lack of calibration makes it impossible to compare obtained results with other \acrshort{gl:DAQ} systems. However, it is sufficient to assess the uniformity of the performance of the \acrshort{gl:LYSO:Ce} crystals in the prototype. The determined light collection values range from \SI{2707}{\au\per\mega\electronvolt} to \SI{4420}{\au\per\mega\electronvolt} with the mean value at \SI{3774}{\au\per\mega\electronvolt} and the standard deviation on the order of \SI{10}{\percent}. The 2D histogram shows that the spatial distribution of light collection is not uniform within the prototype. The differences between scintillating fibers may be a result of surface treatment during the production process, quality of wrapping, and local properties of the coupling.

\subsubsection*{Timing resolution}

\begin{figure}[hptb]
\centering
\includegraphics[width=0.49\textwidth]{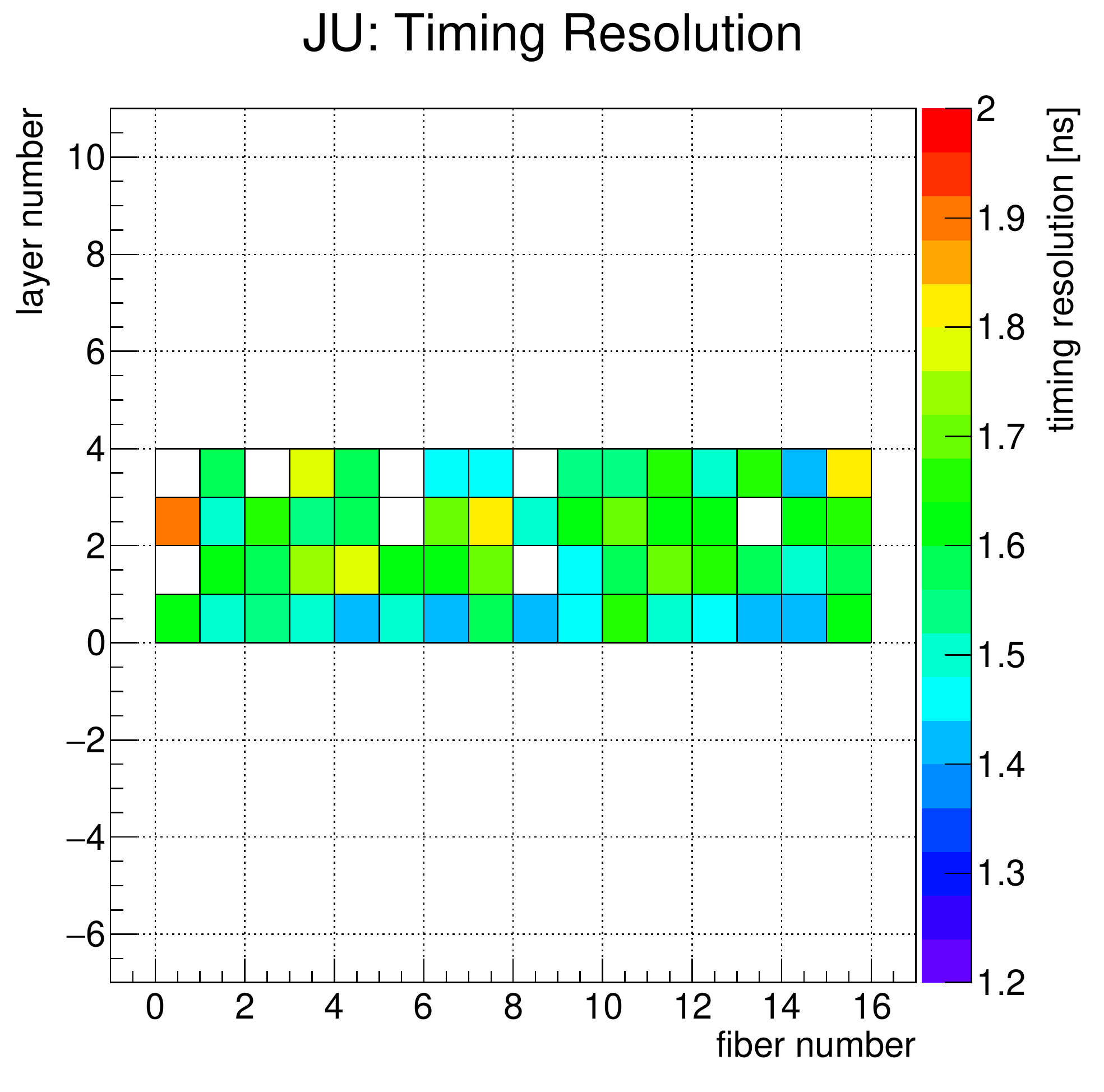}
\includegraphics[width=0.49\textwidth]{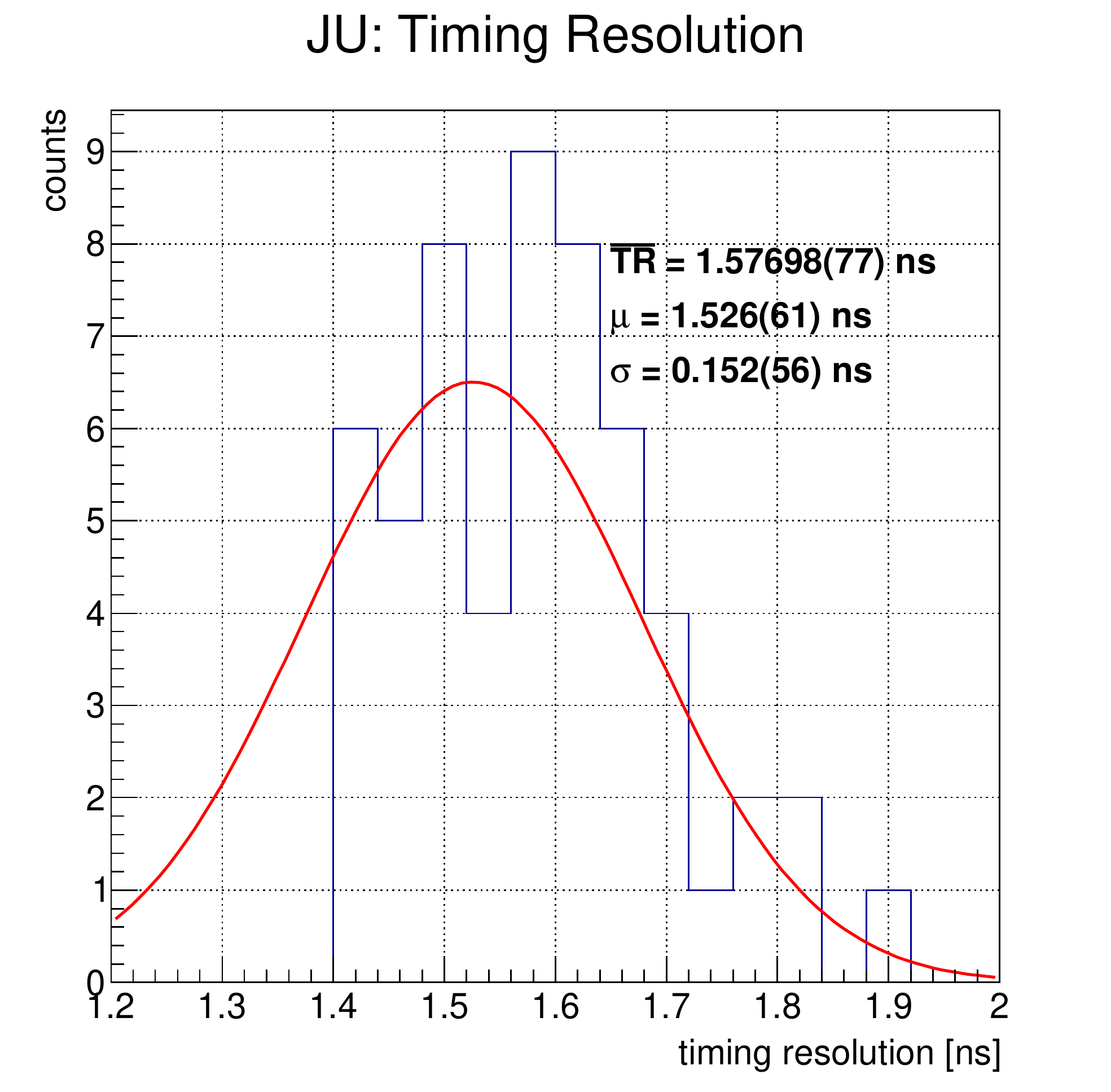}
\caption{Left: histogram showing the timing resolution values for each investigated fiber in \acrshort{gl:SSP}. Right: Statistical distribution of light collection values fitted with the Gaussian function. The listed values indicate the weighted mean and parameters of the distribution obtained from the fit.}
\label{fig:ju-tres}
\end{figure}

Results of the timing properties analysis are presented in \cref{fig:ju-tres}. The timing resolution of the \acrshort{gl:LYSO:Ce} crystals in the prototype ranges from \SI{1.40}{\nano\second} to \SI{1.88}{\nano\second}. The average value (\SI{1.58}{\nano\second}) is higher than the timing resolution obtained for analogous fibers in studies of single fibers (\SI{1.26}{\nano\second}, see \cref{tab:diff-wrapping}). This is caused mainly by different type of \acrshort{gl:SiPM}s and \acrshort{gl:FEE} used in both measurements. Similarly as in the case of single fibers, the achieved timing resolution was not sufficient to perform satisfactory position reconstruction based on the time difference between the two correlated signals. Therefore, the only method used for the determination of hit position was the previously described method based on the charge ratio of the two correlated signals. In the 2D histogram presenting the spatial distribution of timing resolution values within the prototype (\cref{fig:ju-tres} left), it can be observed that some differences occur between single fibers. These differences may result from diversities of the fibers, as well as the \acrshort{gl:SiPM}s.

\subsubsection*{Summary of \acrshort{gl:JU} measurements}


\Cref{tab:prototype-ju-results} summarizes the results of the prototype characterization in \acrshort{gl:JU} setup. Listed uncertainties are purely statistical. The obtained results yield much worse performance in comparison with the single \acrshort{gl:LYSO:Ce} fibers tested in a similar configuration (\cref{sec:sf-results}). All of the determined properties deteriorated significantly, including the key ones, such as the energy- and position resolution, the latter being in the order of the scintillating fibers length. Therefore the reconstruction of the interaction position along the fibers is practically impossible. 

Such drastic deterioration of the position resolution is connected with the different photosensors used in single fibers test and during the prototype characterization. The SensL \acrshort{gl:SiPM}s used in the single-fiber tests had a photosensitive area of \num{3} $\times$ \SI{3}{\cubic\milli\meter}, which was nine times larger than the area of the investigated fibers surface. This allowed the collection of scintillating light which was leaving the fiber at large angles. As explained in \cref{ssec:sf-diff-couplings}, this component of scintillating light is more prone to attenuation since it traverses a longer optical path, and undergoes more reflections, than the light traveling within a narrow cone. In the case of prototype characterization in the \acrshort{gl:JU} setup, the photosensitive area of the KETEK \acrshort{gl:SiPM}s was \num{1} $\times$ \SI{1}{\square\milli\meter}, matching exactly surface area of the LYSO:Ce fibers. Therefore, the light collection angle was limited and thus, the obtained attenuation length was larger.

\begin{table}[!ht]
\centering
\caption{Numerical results of prototype characterization with one-to-one readout.}
\begin{tabularx}{1.0\textwidth}{|X|p{2.7cm}|p{2.7cm}|p{2.7cm}|}
\hline
Property & Mean & Standard \newline deviation & Weighted \newline mean \\ \hline
Attenuation length (\gls{gl:MLR}) [\si{\milli\metre}] & \num{483.7(92)} & \num{53(13)} & \num{465.4(13)} \\
Attenuation length (\acrshort{gl:ELAR}) [\si{\milli\metre}] & \num{161(12)} \newline \num{341(32)} & \num{33(11)} \newline \num{89(46)} & \num{182.8(14)} \\
Light collection [\si{\au\per\mega\electronvolt}] & \num{3774(62)} & \num{353(63)} & \num{3803(16)} \\
Timing resolution [\si{\nano\second}] & \num{1.526(61)} & \num{0.152(56)} & \num{1.57698(77)} \\
Energy resolution (\gls{gl:Qavg}) [\si{\percent}] & \num{10.32(11)} & \num{0.68(10)} & \num{10.272(14)} \\
Reconstructed energy of annihilation \newline peak (\gls{gl:Qavg}) [\si{\kilo\electronvolt}] & \num{514.27(43)} & \num{2.50(61)} & \num{514.306(66)} \\
Energy resolution (\acrshort{gl:ELAR}) [\si{\percent}] & \num{10.56(15)} & \num{0.88(18)} & \num{10.582(15)} \\
Reconstructed energy of annihilation \newline peak (\acrshort{gl:ELAR}) [\si{\kilo\electronvolt}] & \num{512.4(21)} & \num{7.3(34)} & \num{511.949(70)} \\
Position resolution (\gls{gl:MLR}) [\si{\milli\meter}] & \num{94.4(14)} & \num{6.9(22)} & \num{92.502(47)} \\
Reconstructed position residuals \newline (\gls{gl:MLR}) [\si{\milli\meter}] & \num{0.020(15)} & \num{0.094(10)} & \num{0.016(24)} \\
Position resolution (\acrshort{gl:ELAR}) [\si{\milli\meter}] & \num{91.0(17)} & \num{8.8(24)} & \num{91.485(49)} \\
Reconstructed position residuals \newline (\acrshort{gl:ELAR}) [\si{\milli\meter}] & \num{0(0)} & \num{0.091(11)} & \num{0.002(25)} \\
\hline
\end{tabularx}
\label{tab:prototype-ju-results}
\end{table}

Moreover, the \acrshort{gl:ELAR} model did not perform satisfactory in the analysis of the \acrshort{gl:JU} data. For some of the investigated fibers inconsistencies in the fitted curves were observed, such as negative values of $\eta$ parameters, large uncertainties of the reconstructed primary component data points, incorrect reconstruction of the primary light component function. These observations hint, that the \acrshort{gl:ELAR} model does not perform well when weak light attenuation is observed. In such a situation the simple exponential model is sufficient for the description of the scintillating light propagation.

The problem of the increased attenuation length and poor position resolution was addressed in the second experiment with the \acrshort{gl:SSP}, which is described further in this chapter.

\section{Module characterization - light sharing based readout}
\label{sec:pmi-tests}

The \acrshort{gl:SSP} was additionally tested at the Department of Physics of Molecular Imaging Systems at RWTH Aachen University. These measurements will hereinafter be referred to as \acrshort{gl:PMI} measurements. In the measurements, the \acrshort{gl:DAQ} and \acrshort{gl:FEE} developed by the local research group were used. In fact, one of the motivations to repeat the characterization of the prototype at \acrshort{gl:PMI} was a verification of its performance with another photosensor and \acrshort{gl:DAQ} system. 

\subsection{Experimental setup}
\label{subsec:pmi-tets-setup}

The experimental setup used during the measurements at \acrshort{gl:PMI} consisted of several components, as presented in \cref{fig:pmi-setup}. The investigated prototype was readout with the use of photosensors produced by Phillips Digital Photon Counting (\acrshort{gl:PDPC}) called the Power Tiles \cite{PowerTileShort, PowerTileFull, Frach2009, Frach2010}. The coupling medium between the \acrshort{gl:SSP} and the photosensor was an optical interface pad made with Elastosil RT 604 silicone rubber, \ie the same as in the \acrshort{gl:JU} measurements. The size of the pads was \num{8}$\times$\SI{48}{\square\milli\meter} and the thickness was \SI{0.4}{\milli\meter}. In the central part of the experimental system, there was a fan-beam collimator (\acrshort{gl:FBC}) responsible for shaping and positioning of the gamma radiation beam emitted from the \Na radioactive source \cite{Muller2018, Muller2019, Hetzel2020}. Finally, on the opposite side of the \acrshort{gl:FBC} there was a structured \acrshort{gl:PET} crystal, which was serving as a reference detector in the system. Further in this chapter, the listed elements are described in more detail. 

The characterization of the prototype detector, similarly as in previously described experiments, consisted of the series of measurements taken with the radioactive source placed at different positions, in \SI{10}{\milli\meter} steps, along the \acrshort{gl:LYSO:Ce} fibers. In each measurement, the timing information and photon counts were recorded. 

\subsubsection*{Fan-beam collimator (\acrshort{gl:FBC})}

The fan-beam collimator combines the features of the passive and electronic collimator. It consists of the lead block with a slot housing \Na radioactive source in the center. On both sides of the radioactive source, there are slits that shape the $\gamma$ beam emitted from the source in a fan-like shape. On one side of the \acrshort{gl:FBC} the investigated detector was placed, and on the opposite side, there was a reference detector. For this purpose, a \acrshort{gl:PET} crystal detector was used.

The \acrshort{gl:PET} crystal was developed at \acrshort{gl:PMI} as a part of the project aiming to develop \acrshort{gl:PET} inserts operational in high-field, for simultaneous \acrshort{gl:PET}/\acrshort{gl:MRI} \cite{Weissler2015}.
The \acrshort{gl:PET} module consists of three layers of \acrshort{gl:LYSO:Ce} crystal needles. The scintillators were separated with a layer of epoxy mixed with \baso powder. The layers are shifted relative to each other to allow identification of the interaction position in the detector. The module is read out with the Power Tile photosensor \cite{thesis-perez-gonzalez}. 

\begin{figure}[ht]
\centering
\includegraphics[width=0.80\textwidth]{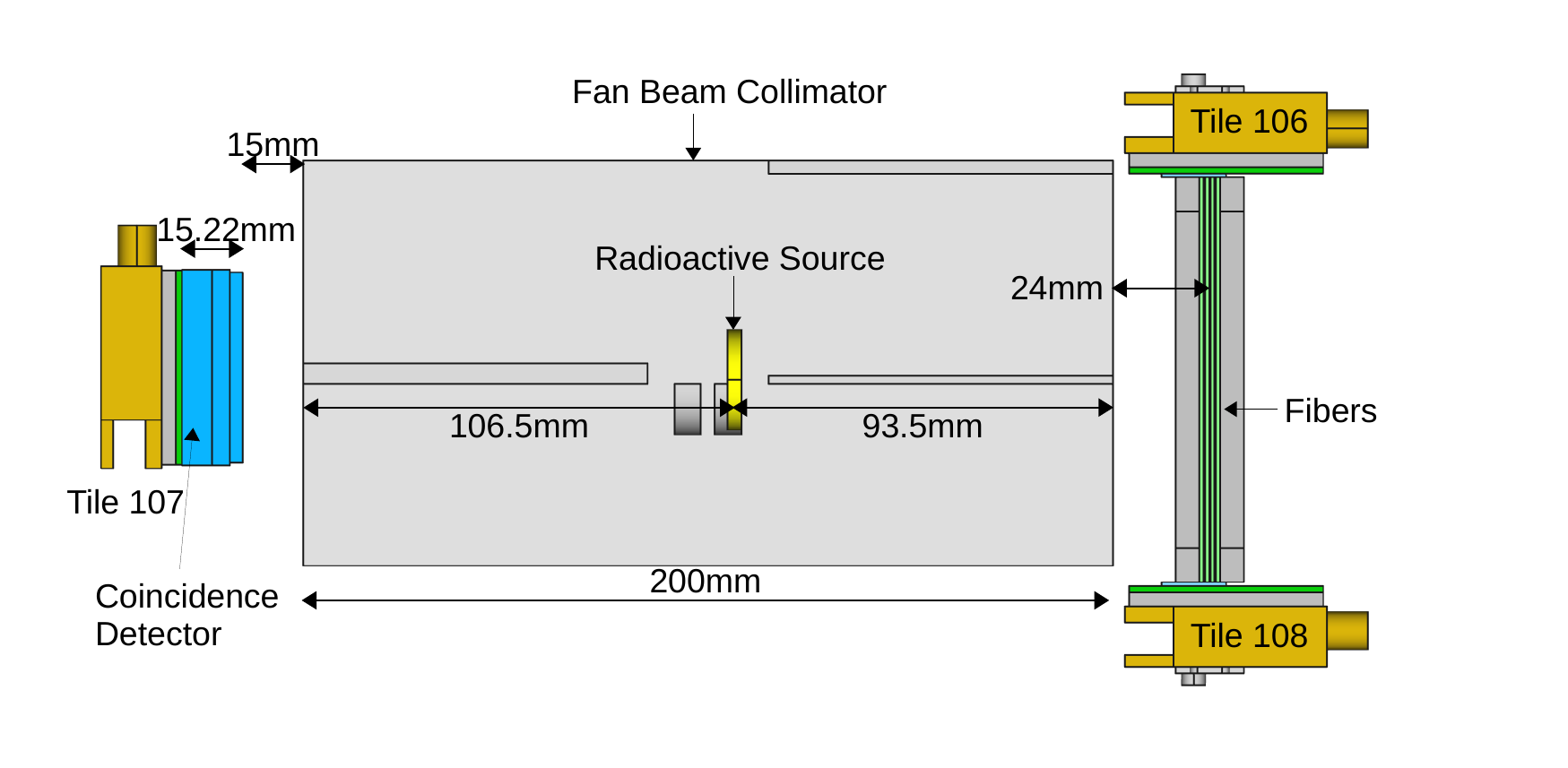} \\ \vspace{0.5cm}
\includegraphics[width=0.49\textwidth]{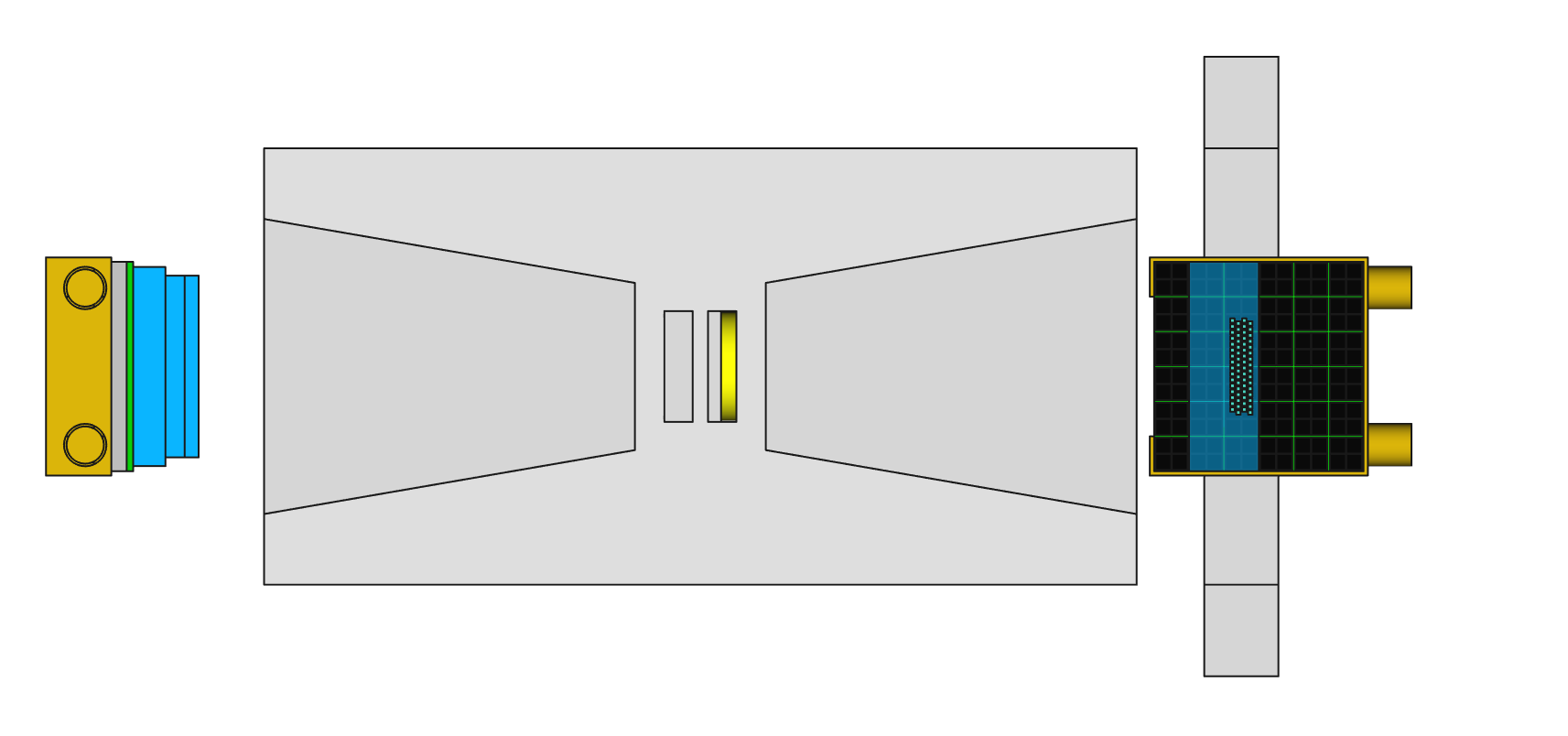}
\includegraphics[width=0.49\textwidth]{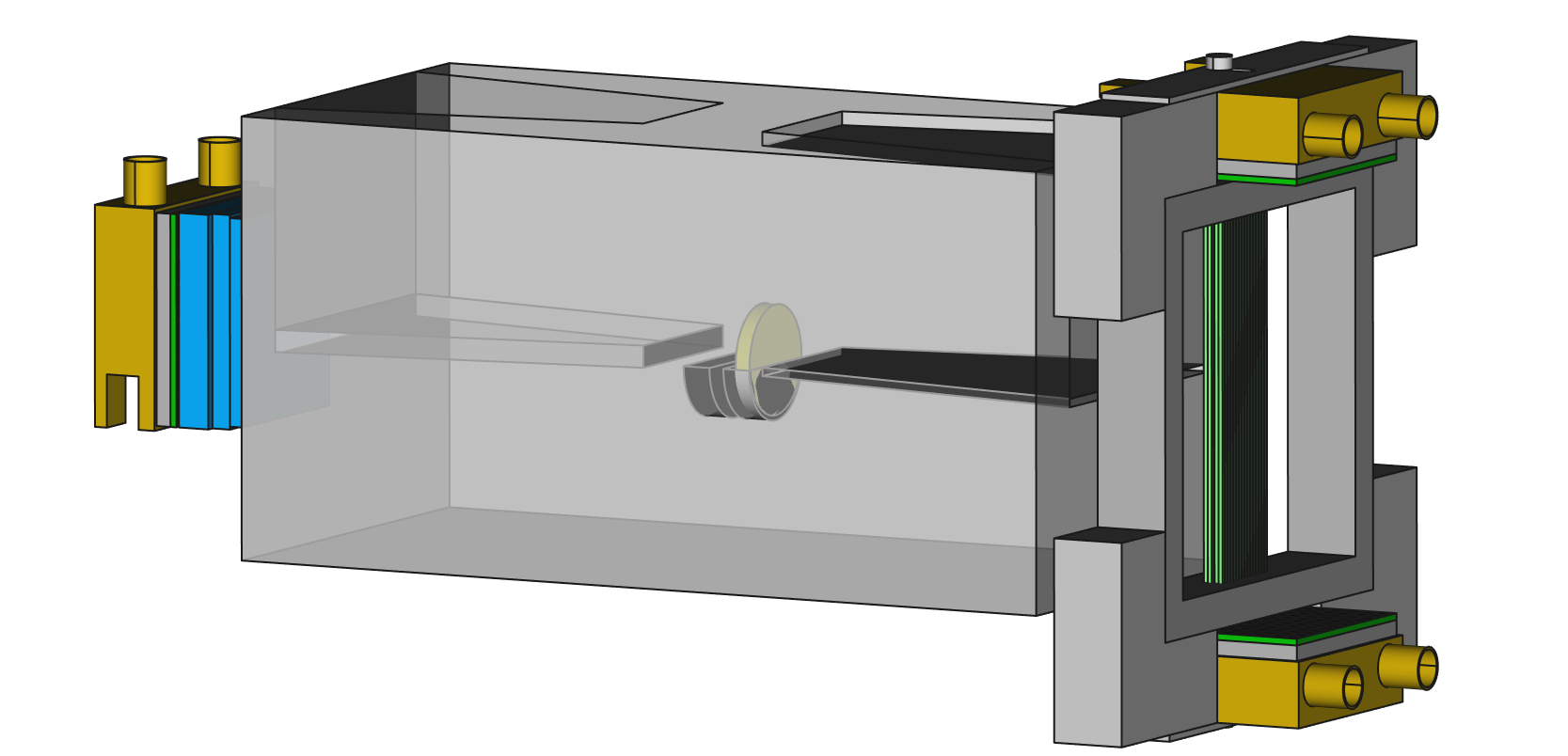}
\caption{Top: side view of the setup in the \acrshort{gl:PMI} measurements, featuring the fan-beam collimator. Tile 107 read out the \acrshort{gl:PET} crystal and Tiles 106 and 108 read out the investigated prototype detector. Bottom left: top view of the experimental setup. Bottom right: perspective view of the experimental setup \cite{Ronja-private}.}
\label{fig:pmi-setup}
\end{figure}

In the presented experimental setup, a single \acrshort{gl:PET} module was used. The data recorded by this detector were utilized to build a coincidence with the events registered in the investigated detector, thus ensuring the operation of the electronic collimation, as described in \cref{susec:krk-tests-setup}. The dimensions of the \acrshort{gl:FBC} and distances in the setup are shown in \cref{fig:pmi-setup}. They defined the irradiated length of \acrshort{gl:LYSO:Ce} fibers in the investigated \acrshort{gl:SSP} to be \SI{2.5}{\milli\meter}.

\subsubsection*{Readout electronics and \acrshort{gl:DAQ}}

\begin{figure}
    \centering
    \includegraphics[width=0.80\textwidth]{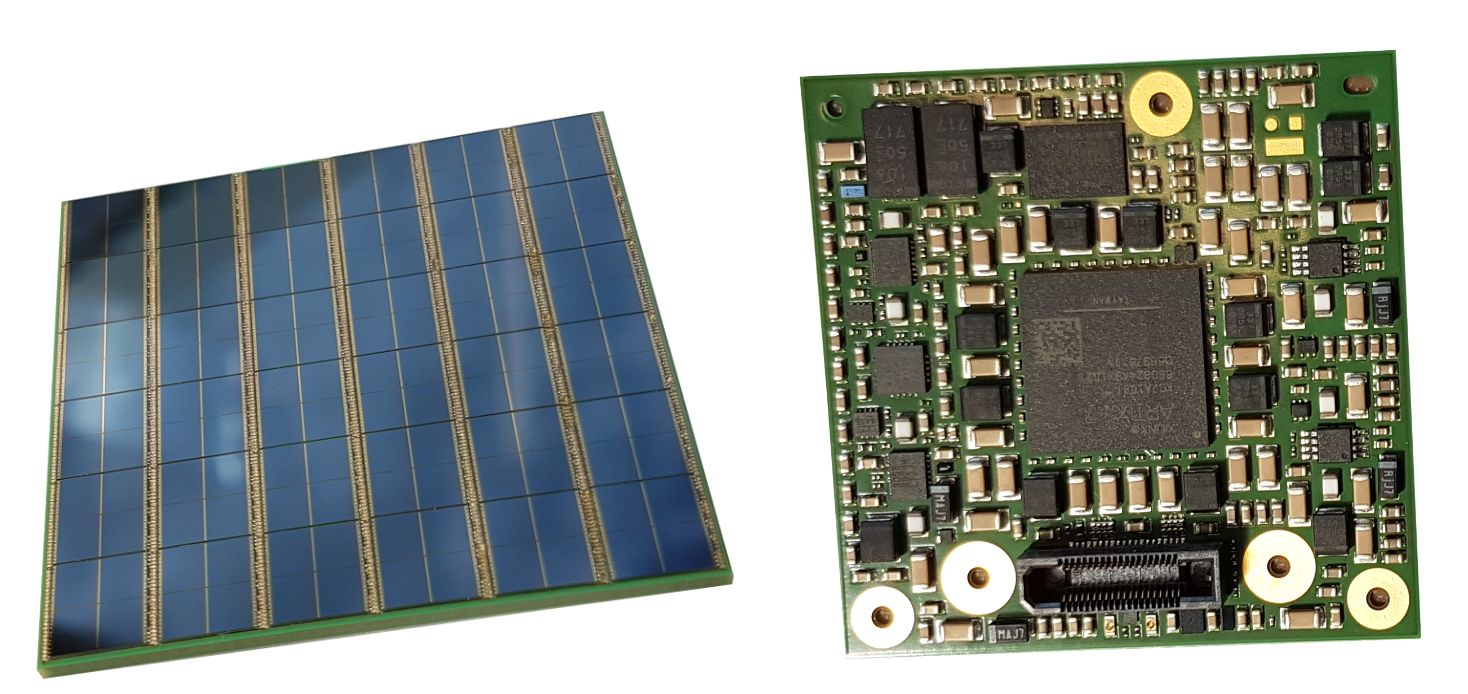}
    \caption{Front (left) and back (right) side of the \acrshort{gl:PDPC} Power Tile sensor \cite{Ronja-private}.}
    \label{fig:power-tile-photo}
\end{figure}

\begin{figure}
    \centering
    \includegraphics[width=0.75\textwidth]{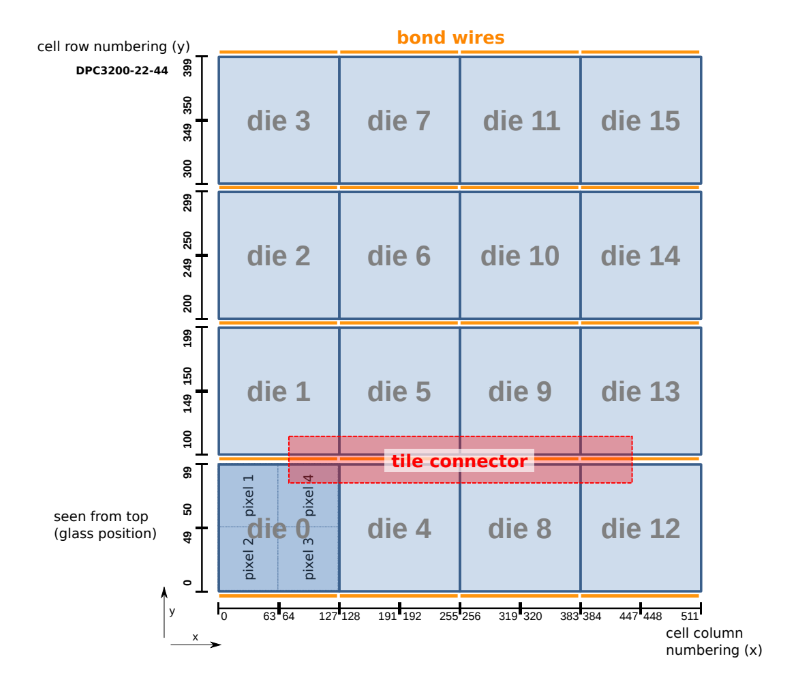}
    \caption{Layout of the Power Tile surface. The picture presents sensor consisting of \num{4} $\times$ \num{4} dies, for which the technical documentation is available \cite{PowerTileFull}. However, the technical details of the sensor are scalable and can be applied to the \num{6} $\times$ \num{6} sensor used in this experiment.}
    \label{fig:power-tile-layout}
\end{figure}

As mentioned in the introduction of this section, \acrshort{gl:PDPC} Power Tiles (model DPC3200-22) were used as photosensors in the experiment presented. The photosensor is presented in \cref{fig:power-tile-photo}. The parameters defining the performance of the \acrshort{gl:PDPC} Power Tiles are listed in \cref{tab:digit-sipms} in \cref{app:photodetectors}. The sensors consist of 36 digital photon counters (\acrshort{gl:DPC}) referred to as dies, arranged in a \num{6} $\times$ \num{6} matrix. Each die is connected to a field-programmable gate array (\acrshort{gl:FPGA}) at the back of the sensor. Total area of the Power Tile is \num{48} $\times$ \SI{48}{\square\milli\meter}. Each die is further divided into 4 pixels with a pitch of \SI{4}{\milli\meter}. Each pixel comprises \num{3200} \acrshort{gl:SPAD}s. The schematic layout of the photosensor surface is presented in \cref{fig:power-tile-layout} \cite{PowerTileFull, thesis-perez-gonzalez}.

\begin{figure}
    \centering
    \includegraphics[width=0.90\textwidth]{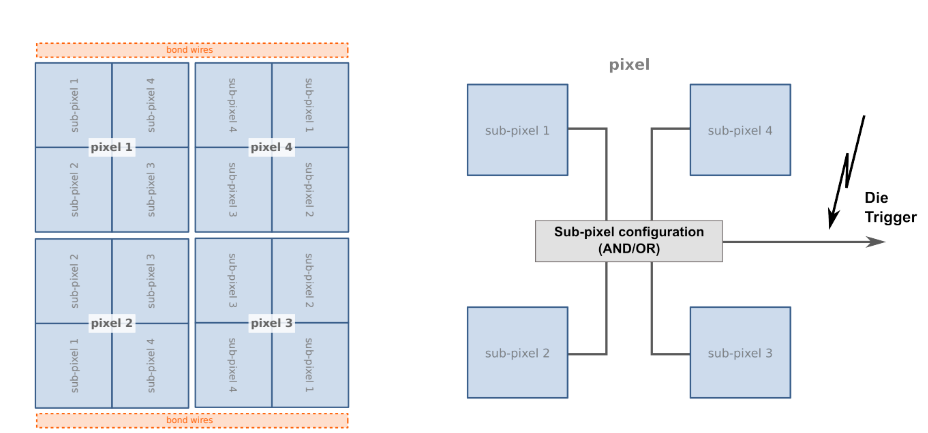}
    \caption{Left: division of pixels into sub-pixels. Right: sub-pixel trigger logic scheme. Pictures reprinted from \cite{PowerTileFull}.}
    \label{fig:power-tile-subpixels}
\end{figure}

Each pixel is further divided into four sub-pixels. They deliver signals used for triggering. 
The user can choose between several triggering schemes, defining how many sub-pixels and in which geometrical configuration presented a logical TRUE signal. The TRUE signal was sent by a sub-pixel, when one of its 800 \acrshort{gl:SPAD}s underwent breakdown. Therefore, the trigger scheme defined the threshold for the acquisition expressed as the average number of breakdowns of \acrshort{gl:SPAD}s within a pixel. The division of the pixels into sub-pixels and the trigger logic scheme are presented in \cref{fig:power-tile-subpixels} \cite{PowerTileFull, thesis-perez-gonzalez}. 

When a trigger occurred, a die proceeded to the validation step, performed for each sub-pixel separately. The sub-pixels were divided into 25 row trigger lines, with 32 \acrshort{gl:SPAD}s in each line. If there were breakdowns of the \acrshort{gl:SPAD}s in the trigger line detected, it confirmed validity of the event by sending the logic TRUE signal. The validation logic scheme, analogous to the trigger logic scheme described above, was used to evaluate the combined output of all trigger lines in a sub-pixel. This validation scheme can also be configured according to the needs of the experiment and further tunes the acquisition threshold. The role of this additional step was to verify whether the trigger was caused by a real event or random noise in photosensitive elements \cite{PowerTileFull, thesis-perez-gonzalez, Schug2015}.  

When the validation condition was met, the die proceeded to the integration stage, in which all breakdowns of \acrshort{gl:SPAD}s were summed in a given time interval. The integration time in the presented measurements was \SI{325}{\nano\second}. The final information from the die consisted of the four values of photon counts (one for each pixel in the die) and a trigger timestamp. Moreover, when one of the dies fulfilled the trigger and validation conditions, the integration was also performed for the neighboring dies, regardless if they also met the required conditions \cite{PowerTileFull, thesis-perez-gonzalez, Schug2015}. This feature allowed for the analysis of light sharing between the dies and a more precise determination of the position of interaction on the sensor, which will be described in detail further in this section. Moreover, the optical pads used for coupling in the described experiment, were purposely larger than the surface of the detector itself, and covered the area of one row of dies, to enable undisturbed light sharing in all dies coupled to the detector . 

Finally, the die entered the recharge phase and the sensor was ready for subsequent events. The recharge phase can also start earlier, if the validation condition was not fulfilled \cite{PowerTileFull, thesis-perez-gonzalez, Schug2015}. 

In the presented experiment, three Power Tiles were used and assigned unique ID numbers. Tile 107 was coupled with the \acrshort{gl:PET} crystal detector and tiles 108 and 106 were coupled with the investigated prototype at its top and bottom, respectively. The operation of the photosensor and the \acrshort{gl:FPGA} which was present at its back surface generated a lot of heat. To avoid overheating and maintain a constant temperature of \SI{15}{\celsius} throughout the measurement, a closed-circuit liquid cooling system was installed. 

The Power Tiles were operated with the custom \acrshort{gl:DAQ} system, which was designed and built by Hyperion Hybrid Imaging Systems and provided hardware, firmware and software solutions for data acquisition \cite{Hyperion, Weissler2015}. Data recorded by Power Tiles were collected by custom electronic boards called singles processing units \acrshort{gl:SPU}. The \acrshort{gl:SPU} was also responsible for the control of the Power Tiles power supply, synchronization of the data and transmission to the data acquisition and processing server (\acrshort{gl:DAPS}). The task of the \acrshort{gl:DAPS} was storing of the measured data and measurement parameters as well as application of the sensor calibration. The calibration of the sensor was conducted prior to the actual measurement and consisted of the adjustment of the power supply and the dark count rate. It allowed to optimize performance of the sensor and minimize negative influence of increased \acrshort{gl:DCR} \cite{thesis-perez-gonzalez}.

\subsubsection*{Data preprocessing}

As described in the previous subsection, the area of the unit of the photosensor (pixel) was larger than that of the single \acrshort{gl:LYSO:Ce} fiber in the prototype (photosensor pixel pitch \SI{4}{\milli\meter}). Therefore, in contrast to the \acrshort{gl:JU} measurements, a straightforward identification of the fiber that recorded an interaction was not possible. Instead, the fiber identification was carried out in several steps based on the recorded photon counts in the pixels and die time stamps.

As already mentioned, the light sharing between the sensor pixels was allowed and desired in the described setup. Furthermore, not only the triggered dies were read out by the \acrshort{gl:DAQ} system, but also the neighboring dies, even if they did not meet the trigger condition. This allowed to use the center-of-gravity (\acrshort{gl:COG}) algorithm for the determination of position of the interaction projected onto the photosensor surface. The \acrshort{gl:COG} algorithms analyzed the collected numbers of photons counted in the pixels and trigger time stamps of the activated dies. In general, the performance of the \acrshort{gl:COG} algorithm is improved when the information from many sensors spread over a large area is considered. In this approach, the contribution of each pixel is weighted with the number of registered photons. However, this method is costly in terms of computational resources and time. In the alternative approach, so called regions of interest (\acrshort{gl:ROI}s) around the main activated die are considered. The main die is identified as the one with the largest number of photon counts and the earliest trigger time stamp. These two characteristics suggest that the \acrshort{gl:LYSO:Ce} fiber emitting scintillating light is directly above the main die. Additionally, \acrshort{gl:ROI}s can contain one or more neighboring dies: horizontal, vertical and diagonal (\acrshort{gl:HVD}). Different combinations of \acrshort{gl:ROI}s can be analyzed, however, it needs to be noted that each type of \acrshort{gl:ROI} yields different hit topology on the sensor surface. Therefore, to improve the performance of the algorithm many different \acrshort{gl:ROI}s should be considered and analyzed. This approach requires more elaborate analysis techniques, but once implemented, it is less costly in terms of computational resources compared to the weighted \acrshort{gl:COG} algorithm.

In this work, the \acrshort{gl:COG} algorithm based on \acrshort{gl:ROI}s was used. Since the investigated photodetector occupied only one row of dies on the photosensor, only two types of \acrshort{gl:ROI}s were significant for the analysis: 
\begin{itemize}
\item \textbf{HVD000NoH} meaning that only main die was triggered, and the neighboring horizontal die was not triggered. \acrshort{gl:COG} was calculated based only on the data recorded by the main die. 
\item \textbf{HVD100} meaning that the main die, as well as neighboring horizontal die were triggered. The \acrshort{gl:COG} was calculated based on the data from both dies. 
\end{itemize}
In the measurements with the \acrshort{gl:SSP}, due to the size of the detector and its placement on the photosensor, almost all of the recorded events belonged to one of the two considered \acrshort{gl:ROI} types. Both of the listed \acrshort{gl:ROI}s yielded different hit topology on the surface of the photosensor and therefore they were analyzed independently of each other.

\begin{figure}[hp]
\centering
\includegraphics[width=0.99\textwidth]{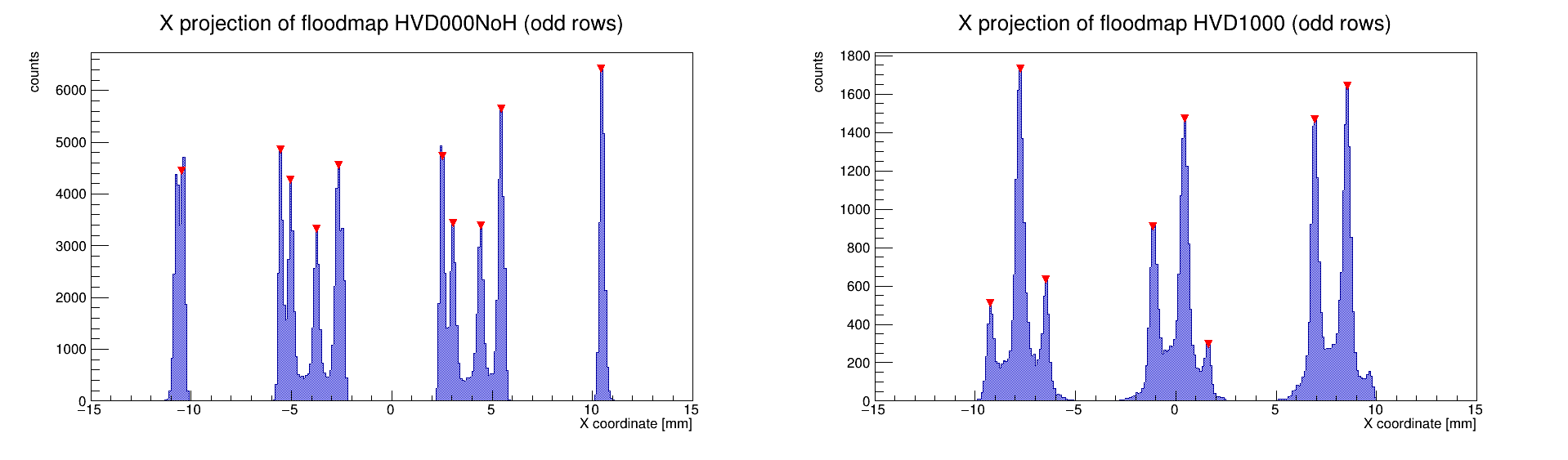}
\includegraphics[width=0.99\textwidth]{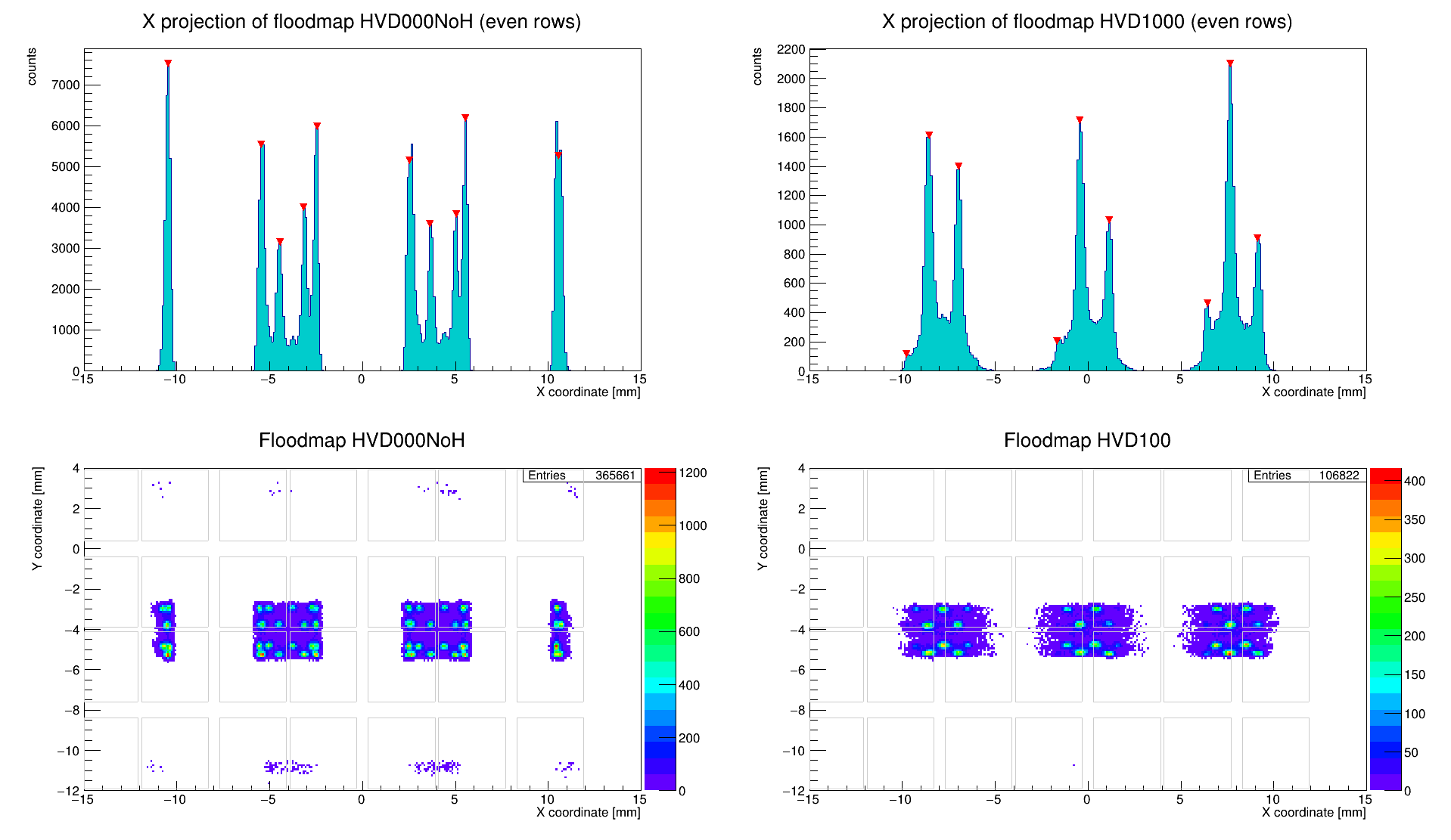}
\caption{Flood maps of tile 108 for two considered \acrshort{gl:ROI}s (bottom panels). Projections on $X$ axis of the flood maps were taken separately for subranges on the $Y$ axis corresponding to even and odd layers of the prototype in order to determine the positions of the peaks associated with hits in individual fibers (two upper rows of panels). Red triangles indicate identified peaks.}
\label{fig:floodmaps_proj}
\end{figure}

Once the \acrshort{gl:COG} was calculated for all events, the flood maps of reconstructed hit positions on the sensor surface were plotted separately for each \acrshort{gl:ROI}. Sample flood maps are presented in the bottom panels of \cref{fig:floodmaps_proj}. It can be observed that there are clear regions with increased number of hits, which correspond to the positions of individual fibers on the photosensor. Another important observation is that the hit maps for individual \acrshort{gl:ROI}s are not complete, meaning that there are areas of the photosensor which did not register any hits, despite being coupled to the scintillating elements. However, when superimposing flood maps for different \acrshort{gl:ROI}s, the whole photosensor area is covered.
This demonstrates that more than one \acrshort{gl:ROI} is necessary for fiber identification, as they are related with different response pattern of the photosensor.

In the next step, the projections of the flood maps on the $Y$ axis were taken. From the projections, the borders between the layers were determined. Subsequently, the projections on the $X$ axis were taken separately for even and odd layers of the prototype. This was necessary to account for the half-pitch shift between neighboring fiber layers. The projections taken from the flood maps of both considered \acrshort{gl:ROI} types are presented in the top and middle panels of \cref{fig:floodmaps_proj}. Clear peaks visible in the projection histograms correspond to events which occurred in the \acrshort{gl:LYSO:Ce} fibers in the prototype. The positions of the peaks were determined using ROOT's TSpectrum tool \cite{TSpectrum}. 

\begin{figure}
\centering
\includegraphics[width=0.99\textwidth]{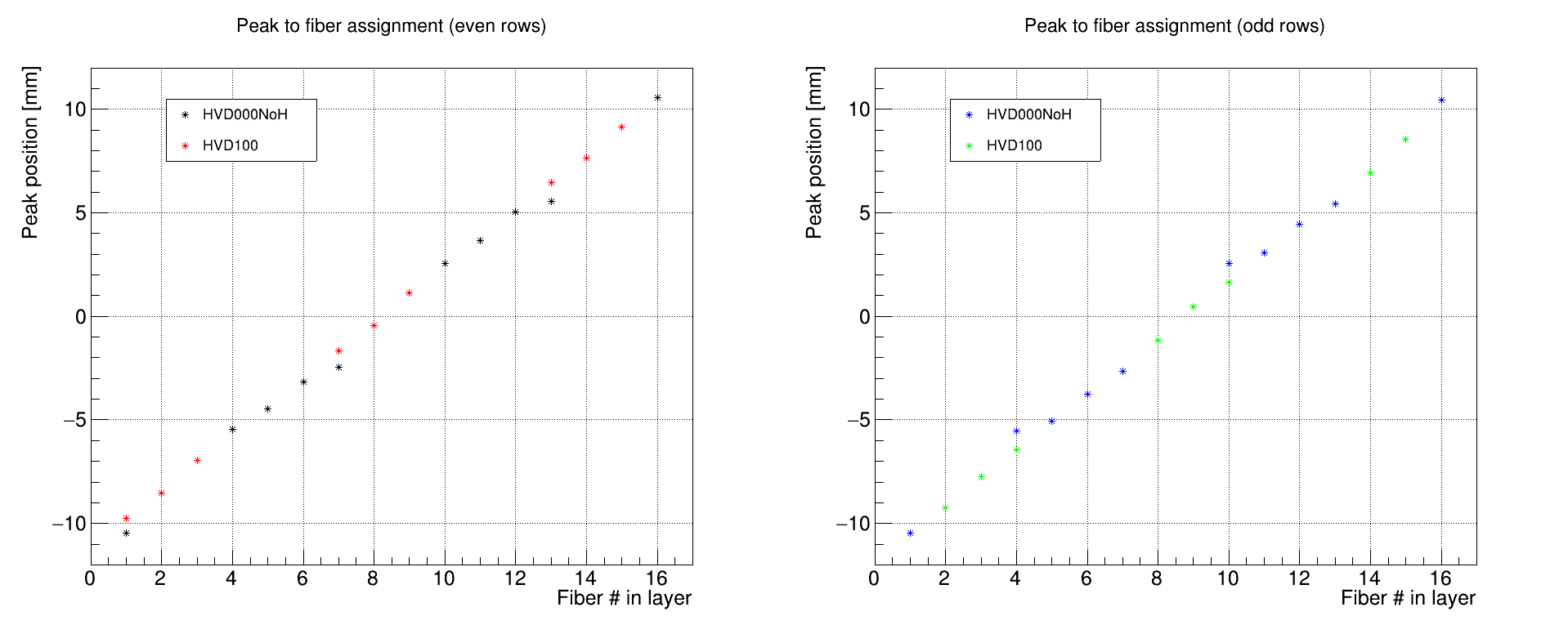}
\caption{Assignment of the peaks identified in the $X$-projections of hit maps to fibers in the prototype. The analysis was performed separately for even and odd layers due to the half-pitch shift of layers in the prototype, and separately for the different \acrshort{gl:ROI}s.}
\label{fig:peak-assignment}
\end{figure}

As a next step, all the determined peak positions were mapped onto the fibers in the prototype. The assignment was done manually, based on the knowledge of the prototype geometry and its placement on the photodetector during the measurement. The mapping is graphically presented in \cref{fig:peak-assignment}.  

\begin{figure}
\centering
\includegraphics[width=0.99\textwidth]{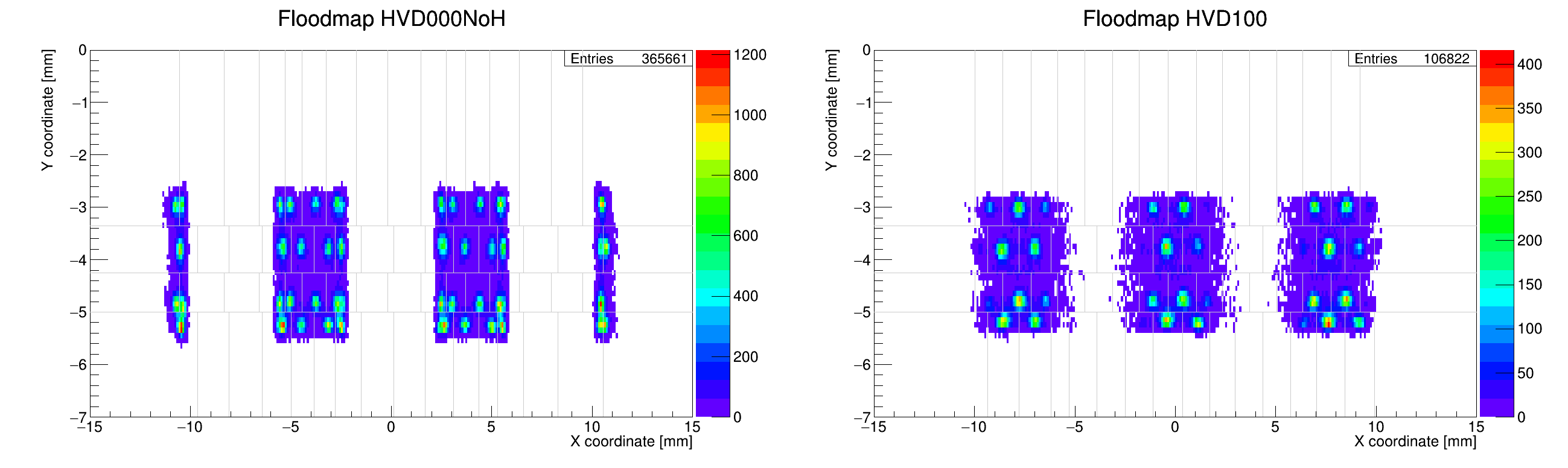}
\caption{Flood maps of tile 108 for two considered \acrshort{gl:ROI}s. Gray lines indicate determined boundaries of response areas for the single fibers in the prototype detector.}
\label{fig:floodmaps_fibers}
\end{figure}

The peak assignment allowed to determine areas on the photosensor which corresponded to the response to single fibers in the prototype. The borders of those areas were determined as half the distance between the neighboring points on the graphs plotted in \cref{fig:peak-assignment}. \Cref{fig:floodmaps_fibers} presents the resulting grids for each of the analyzed \acrshort{gl:ROI}s with the corresponding hit maps overlayed. It needs to be underlined, that the grids do not represent the physical borders of the fibers. They rather represent the boundaries of areas of the photosensor that are activated by the scintillating light emitted by individual fibers. 

\begin{figure}
\centering
\includegraphics[width=0.99\textwidth]{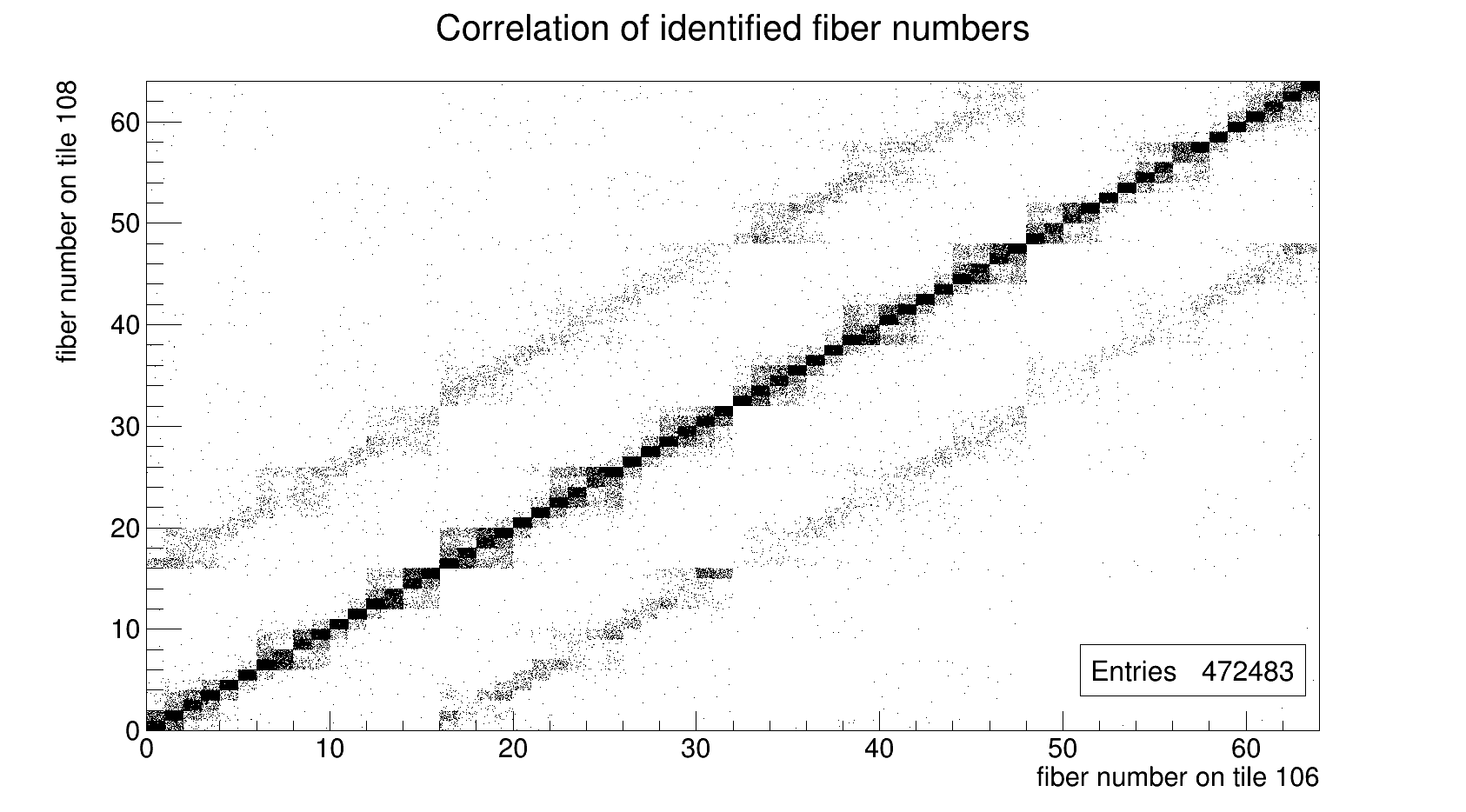}
\caption{The correlation plot of identified fiber numbers on the photosensors reading out two sides of the detector. Correctly identified fiber numbers contribute to the diagonal, events contributing to the bins adjacent to the diagonal correspond to the misidentification in which neighboring fiber number within the same layer was incorrectly assigned. The lines below and above the diagonal correspond to the misidentification in which fibers belong to neighboring layers were assigned.}
\label{fig:fiber-correlation}
\end{figure}

In the next step, the assignment of the recorded events to the fibers in the prototype was performed based on the determined grids. In order to evaluate the quality of fiber identification, a correlation histogram of the identified fiber numbers on both Power Tiles (top and bottom of the detector) was plotted, as presented in \cref{fig:fiber-correlation}. For simplicity, the fibers in the correlation histogram were numbered from \num{0} to \num{63}. Only events in which the same fiber number was identified on both tiles were considered correct. Those events contribute to the diagonal of the correlation histogram. The misidentified events fall into bins adjacent to the diagonal or the lines below or above the diagonal. For the presented case, \SI{75.39}{\percent} of all coincident events were correctly assigned. The remaining misidentified events were rejected from further analysis.

\begin{figure}
\centering
\includegraphics[width=0.99\textwidth]{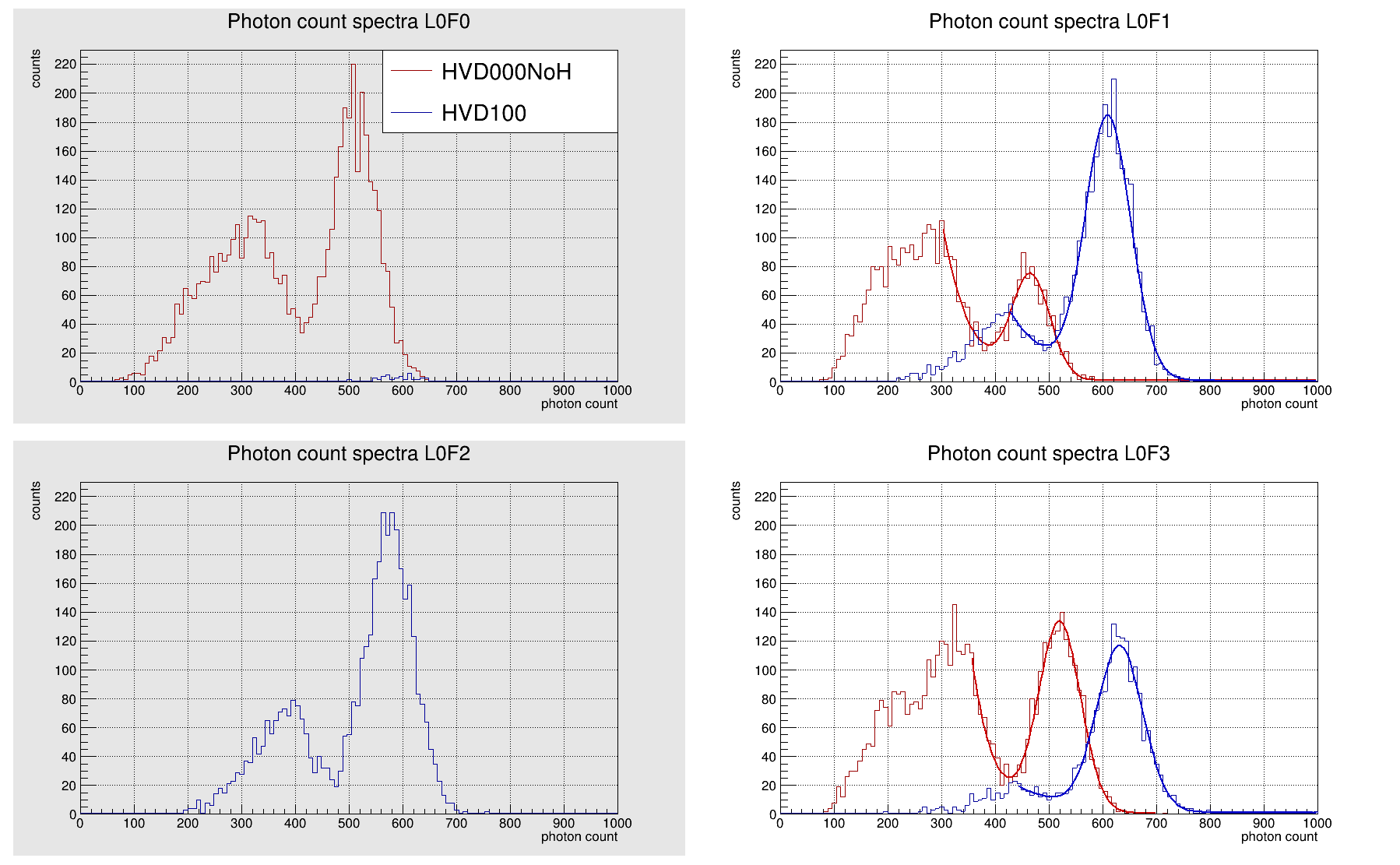}
\caption{Example of photon count spectra for first four fibers in the prototype. Spectra of HVD000NoH and HVD100 contributions were plotted separately. It can be observed that for some fibers both contributions are present (L0F1 and L0F3). In order to align both spectra a scaling of HVD000NoH contribution was performed. Gray background means, that one of the components was not registered or not significant, therefore scaling was not performed (L0F0 and L0F2).}
\label{fig:pmi-spectra-hvd}
\end{figure}

\begin{figure}
\centering
\includegraphics[width=0.70\textwidth]{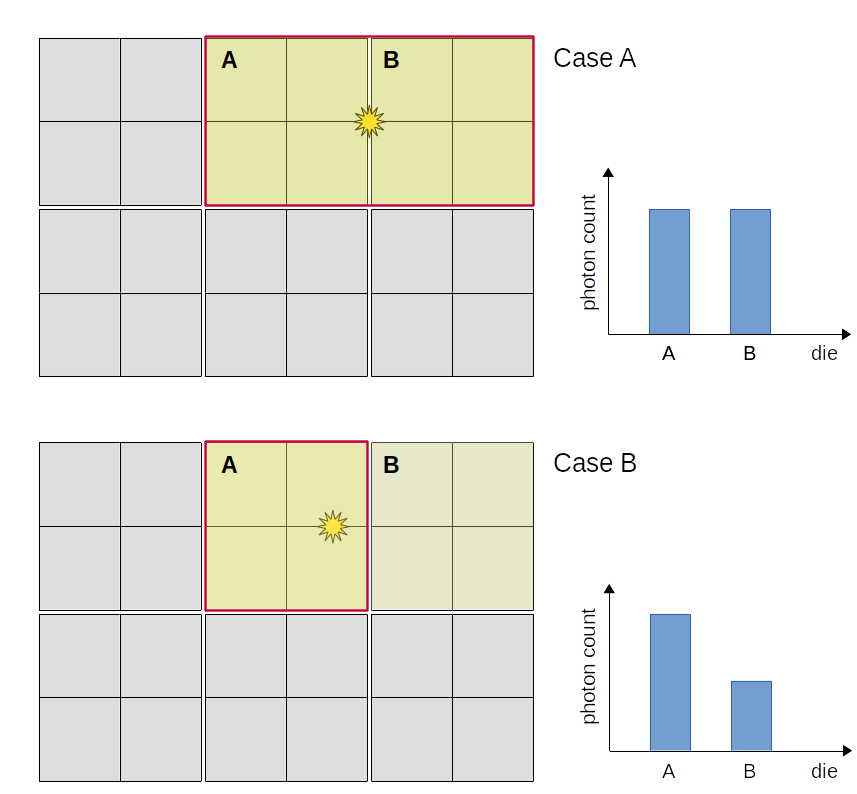}
\includegraphics[width=0.70\textwidth]{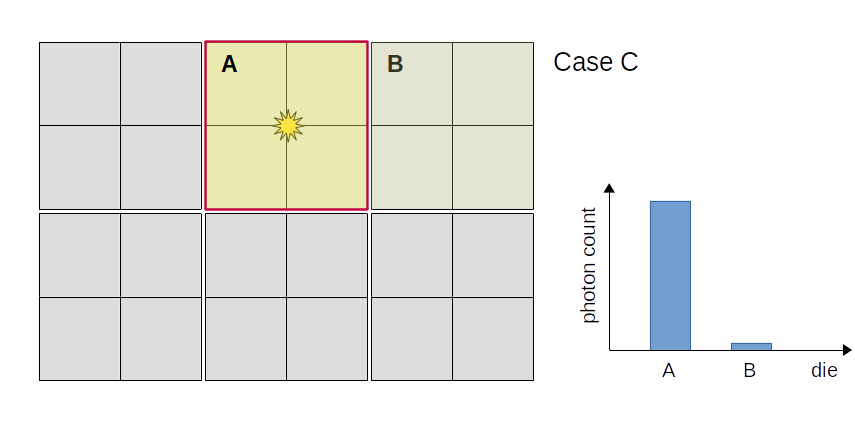}
\caption{Schematic representation of different event topologies registered by \acrshort{gl:PDPC} Power Tiles. Case A: a fiber located at the border of two dies emits scintillating light triggering adjacent dies. Consequently, most of registered events will have the HVD100 signature on that tile. Case B: a fiber located between the die center and border emits light; in some cases only one die is triggered, in others two. Here both event types HVD100 and HVD000NoH will have similar populations. Case C: a fiber located at the center of the die responds, mostly that die is activated, \ie event type HVD000NoH is dominant.}
\label{fig:power-tiles-light-sharing}
\end{figure}

After all events were correctly assigned to the scintillating fibers in the prototype, it was possible to plot their photon count spectra fiber-wise. However, similarly to the previous steps, the spectra were plotted separately for the two considered \acrshort{gl:ROI} types. \Cref{fig:pmi-spectra-hvd} shows example of the photon count spectra for the first four fibers in the prototype, registered in a measurement with a \Na source. It can be observed that for two of the chosen fibers there are contributions from only one \acrshort{gl:ROI}, while for the other two, both \acrshort{gl:ROI}s contribute. Moreover, for the two fibers for which both \acrshort{gl:ROI} spectra are present, the annihilation peaks have different positions. Both of those effects are caused by the placement of the fibers on the sensor tile and the resulting response of the photodetector, which is illustrated in \cref{fig:power-tiles-light-sharing}. Case A presented in the scheme depicts the situation in which the considered scintillating fiber is placed at the border of two dies. Then, the scintillating light is distributed evenly between the dies and in majority of cases both of them are triggered. In this situation the HVD100 event type will be predominant. This corresponds to the spectrum of L0F2 in \cref{fig:pmi-spectra-hvd}. Case B presented in the scheme shows a situation in which the \acrshort{gl:LYSO:Ce} fiber is placed asymmetrically on the die. In that case one of the dies will receive more light than the neighboring one. However, in some cases both dies will be triggered. This means that some events will have HVD000NoH signature and some will have the HVD100 signature. Therefore, it corresponds with the spectra of L0F1 and L0F2 presented in \cref{fig:pmi-spectra-hvd}. Finally, in case C the scintillating fiber is placed at the center of the die. Only a marginal fraction of the scintillating light reaches the adjacent die and therefore it is rarely triggered. In this situation the HVD000NoH event type will be predominant, which corresponds with the L0F0 spectrum. 

\begin{figure}[ht]
\centering
\includegraphics[width=0.99\textwidth]{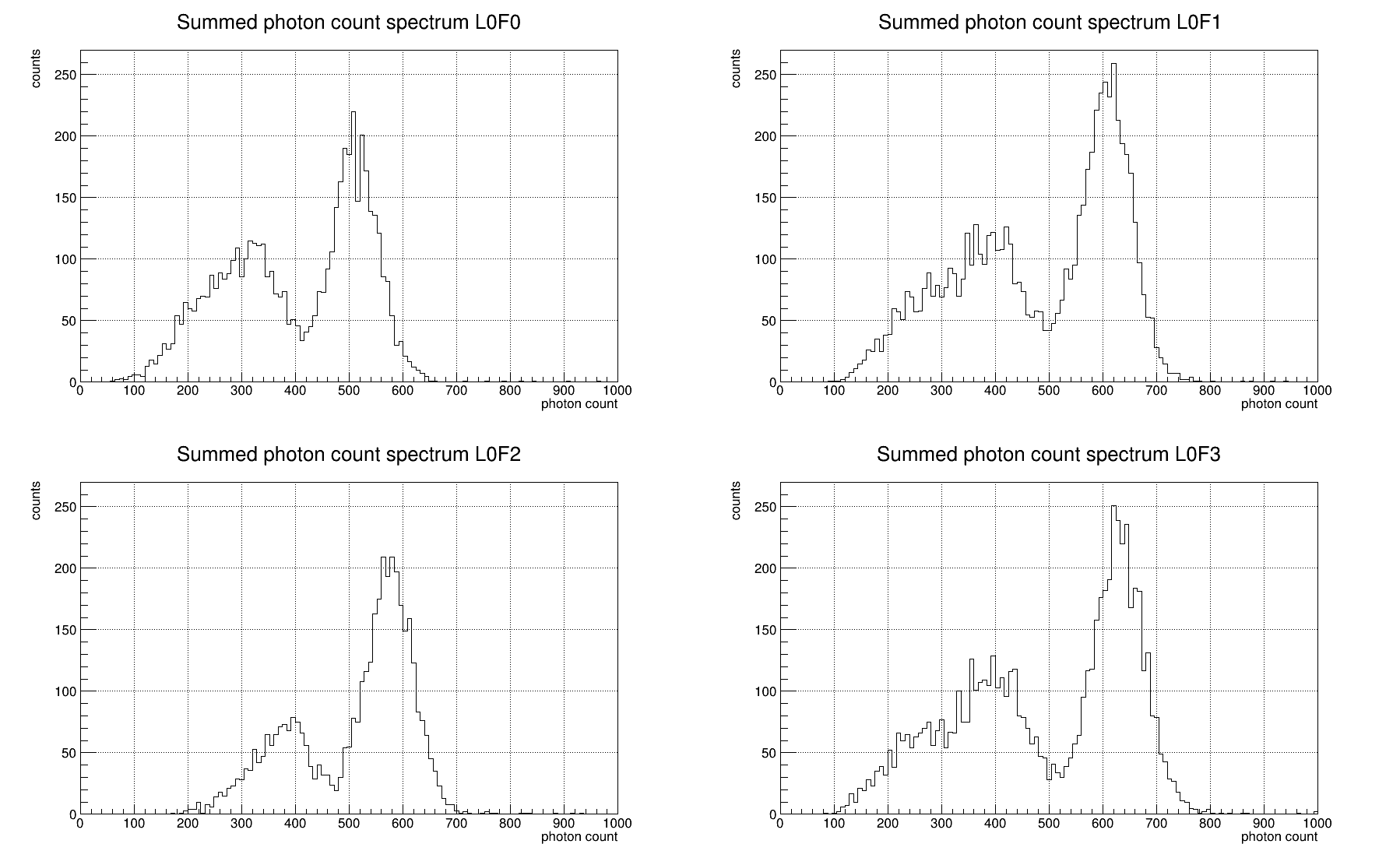}
\caption{Final photon count spectra for the first four fibers in the prototype. After alignment of the annihilation peaks both contributions were added.}
\label{fig:pmi-spectra-sum}
\end{figure}

In cases A and C all or almost all scintillating light emitted is registered. However, in case B some fraction of scintillating light can reach the neighboring die without triggering the data acquisition due to too low number of photons. If we only take into account dies that were triggered, like in this analysis, information about this fraction of scintillating light is lost. This causes artificial compression of the HVD000NoH spectrum, as observed for L0F1 and L0F3. In order to compensate for the light losses occurring due to the event topology, the HVD000NoH spectra were scaled to match the corresponding HVD100 spectra. For that reason spectra pairs were fitted with the function \cref{eq:fitted-function} parameterizing the \anhpeak peak. The scaling factor was calculated as a ratio of the \gls{gl:peakpos} in the two spectra: 
\begin{equation}
    f_{\textrm{HVD}} = \frac{\mu_{\textrm{511~keV}}(\textrm{HVD100})}{\mu_{\textrm{511~keV}}(\textrm{HVD000NoH})}
\end{equation}
After the scaling the resulting HVD000NoH and HVD100 spectra could be accumulated and plotted together. \Cref{fig:pmi-spectra-sum} presents the final result of spectra scaling and accumulation with the same four fibers chosen from the prototype, which were previously shown in \cref{fig:pmi-spectra-hvd}.

The calibration procedure comprising of the determination of the grid representing the response to the scintillating light of individual fibers was done for a single chosen measurement in the series. The determined grid was subsequently used for fiber identification for all measurements in the series. Similarly, the scaling factors were determined only for the calibration measurement and applied later to the remaining measurements in the series\footnote{Determination of the grid of sensor response presented in this thesis was performed by mgr Magdalena Ko\l{}odziej. The author of this thesis performed an analogous analysis for the 2x32 prototype geometry, which is not included in this work. The author performed fiber identification for all measurements, determination of the scaling factors and spectra scaling for all measurements}.

The obtained photon count spectra and timing information from the Power Tiles constituted the input for the further analysis. As previously, the basis for this analysis was the annihilation peak. Therefore, each spectrum was fitted with the \cref{eq:fitted-function} in order to extract the parameters of the annihilation peak. 

\subsection{Characterization results}
\label{pmi-tests-results}

Data recorded in the \acrshort{gl:PMI} measurements were analyzed in the same way as the data collected in the \acrshort{gl:JU} measurements (\cref{subsec:krk-tests-results}). The same set of characteristics was determined, and the same analysis framework was used for that purpose. In \acrshort{gl:PMI} measurements, good quality data were collected for 63 out of 64 \acrshort{gl:LYSO:Ce} fibers in the prototype. Only one fiber was excluded from the analysis due to a threshold set too high during the measurements, which caused annihilation peak to be only partially recorded. In the following part of this section, the obtained results of the prototype characterization are described in detail and illustrated with the 2D spatial distribution maps and the 1D distribution histograms. 

\subsubsection*{Accuracy of the light propagation models}

\begin{figure}[!hp]
\centering
\includegraphics[width=0.49\textwidth]{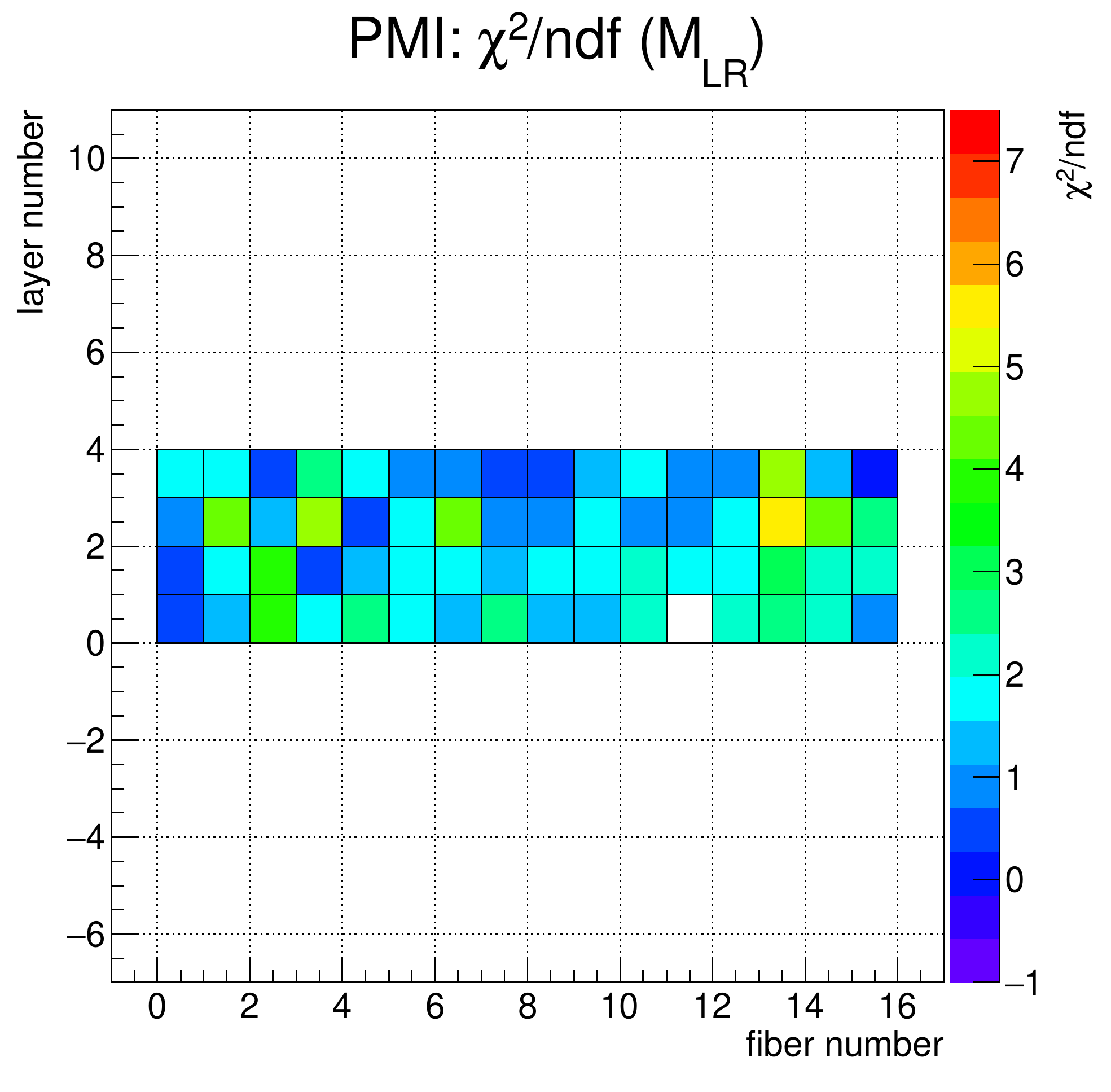}
\includegraphics[width=0.49\textwidth]{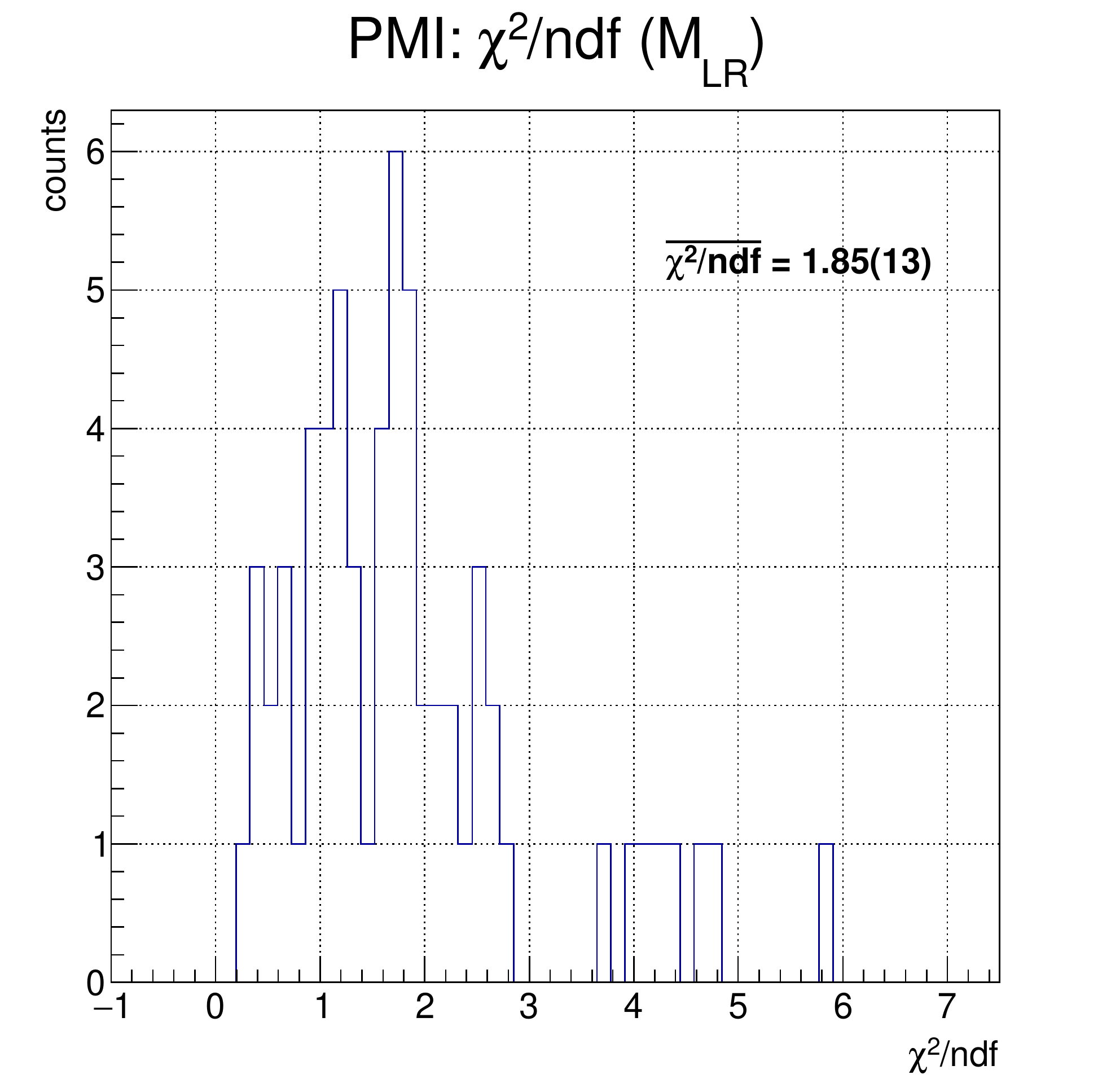}
\includegraphics[width=0.49\textwidth]{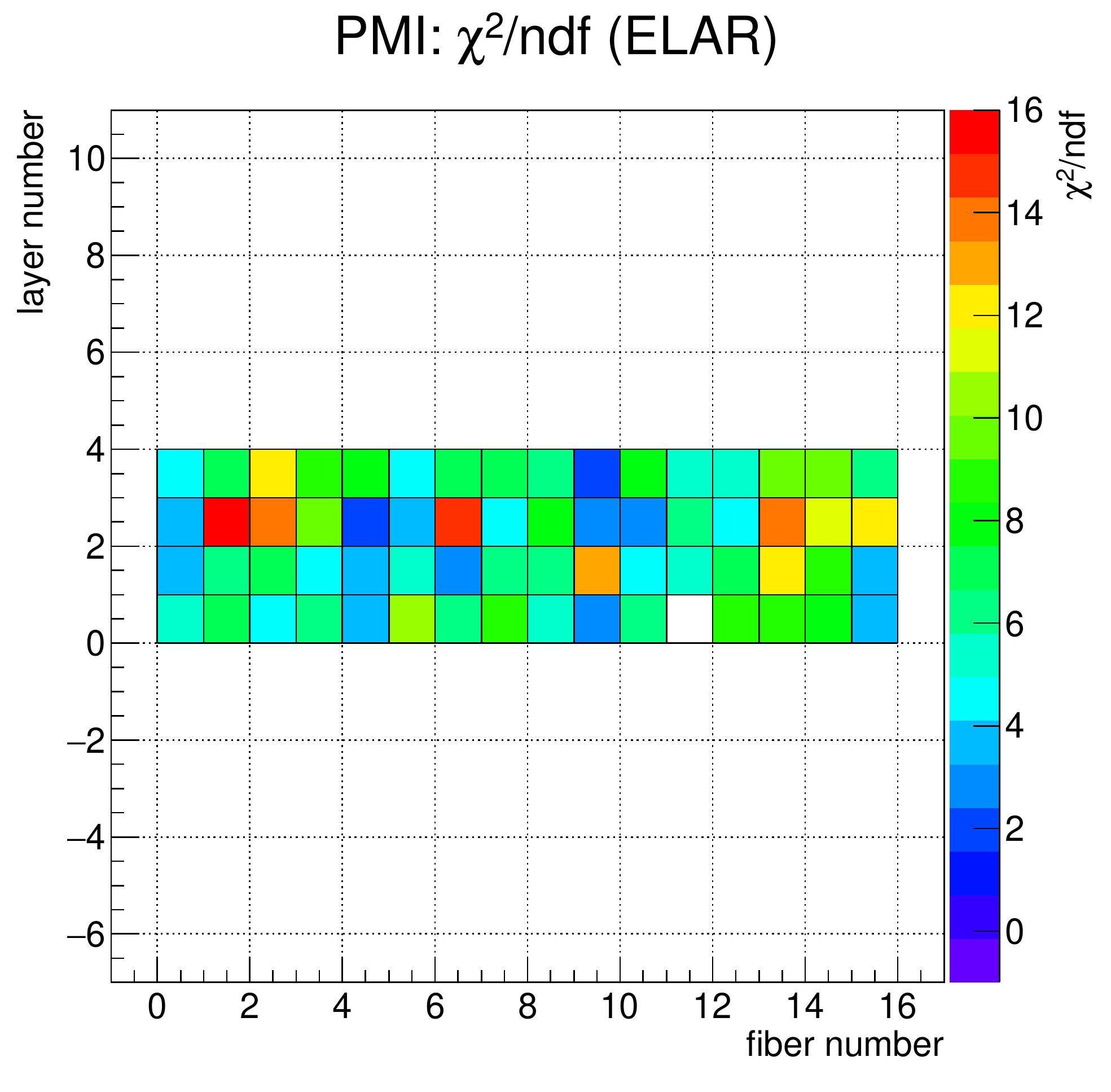}
\includegraphics[width=0.49\textwidth]{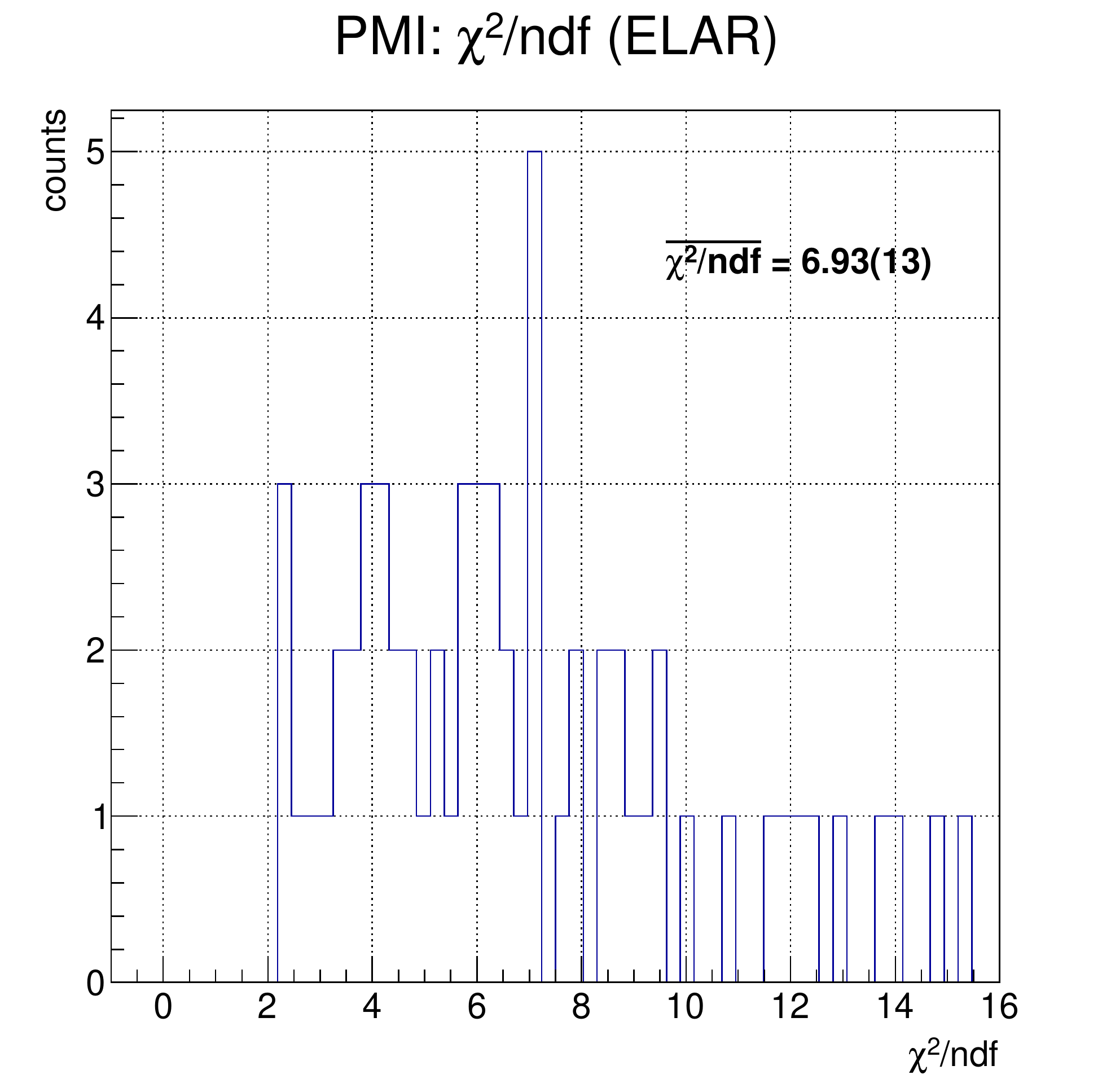}
\caption{Values of \chiNDF obtained in \gls{gl:MLR} (top row) and \acrshort{gl:ELAR} fits (bottom row). Left histograms present spatial distribution of the obtained values within the prototype, while the right plots show their statistical distributions. The weighted mean is listed in both histograms.}
\label{fig:pmi-chindf}
\end{figure}

Similarly as for the \acrshort{gl:JU} measurements, only \gls{gl:MLR} method and \acrshort{gl:ELAR} model were used for description of light propagation in \acrshort{gl:LYSO:Ce} fibers in the prototype. \Cref{fig:pmi-chindf} presents \chiNDF values of light attenuation function fits. In case of the \gls{gl:MLR} method, most of the fits yielded \chiNDF within \num{0.20} -- \num{2.84} range. There were six outliers with slightly larger \chiNDF, below \num{5.77}. For the \acrshort{gl:ELAR} model, the distribution of \chiNDF values is more spread, with the values ranging from \num{2.20} to \num{15.30}. Obtained results show, that the performance of the light attenuation models is slightly worse than in the case of the \acrshort{gl:JU} measurements. 

\subsubsection*{Attenuation length}

\begin{figure}[!hp]
\centering
\includegraphics[width=0.49\textwidth]{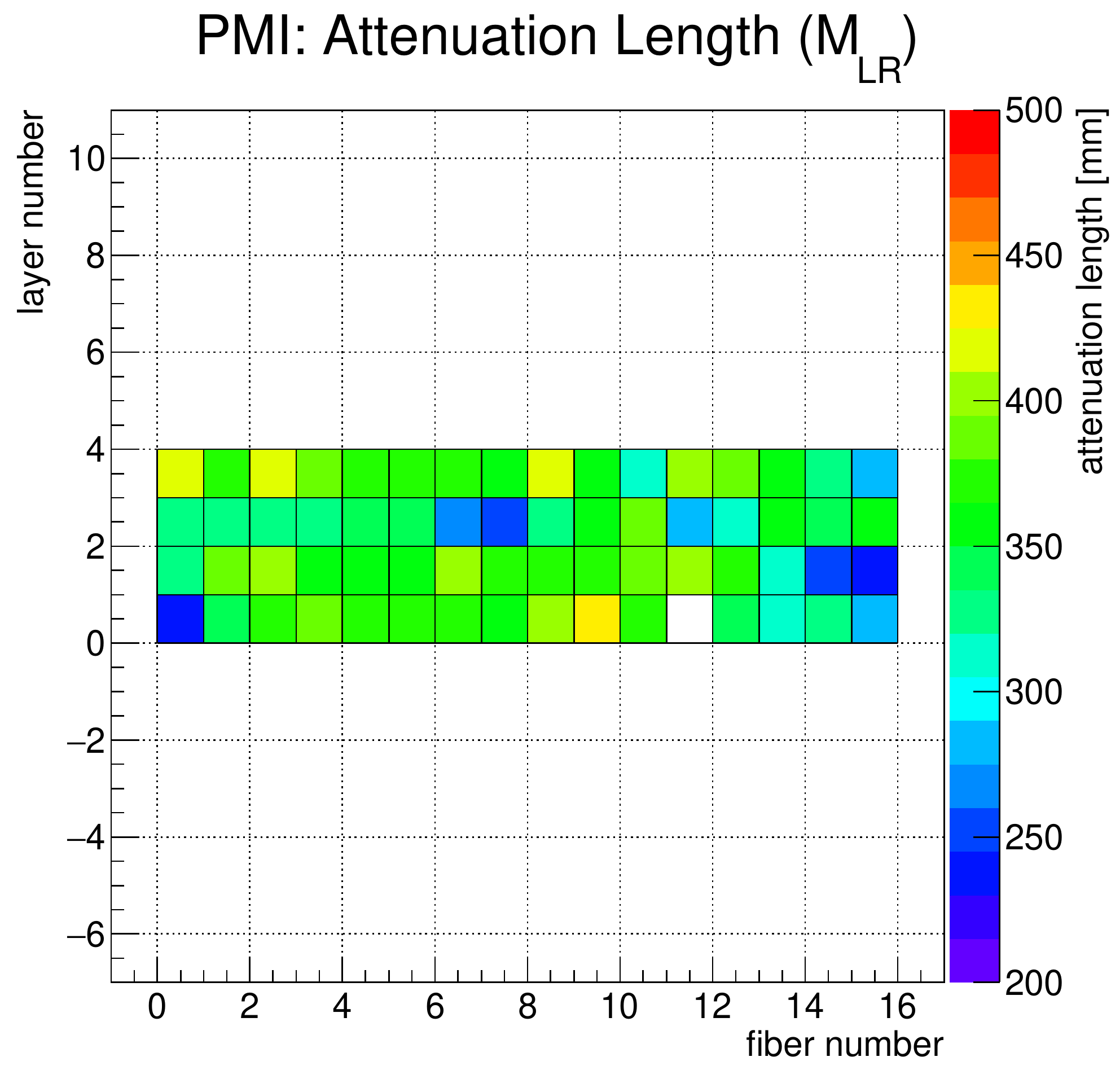}
\includegraphics[width=0.49\textwidth]{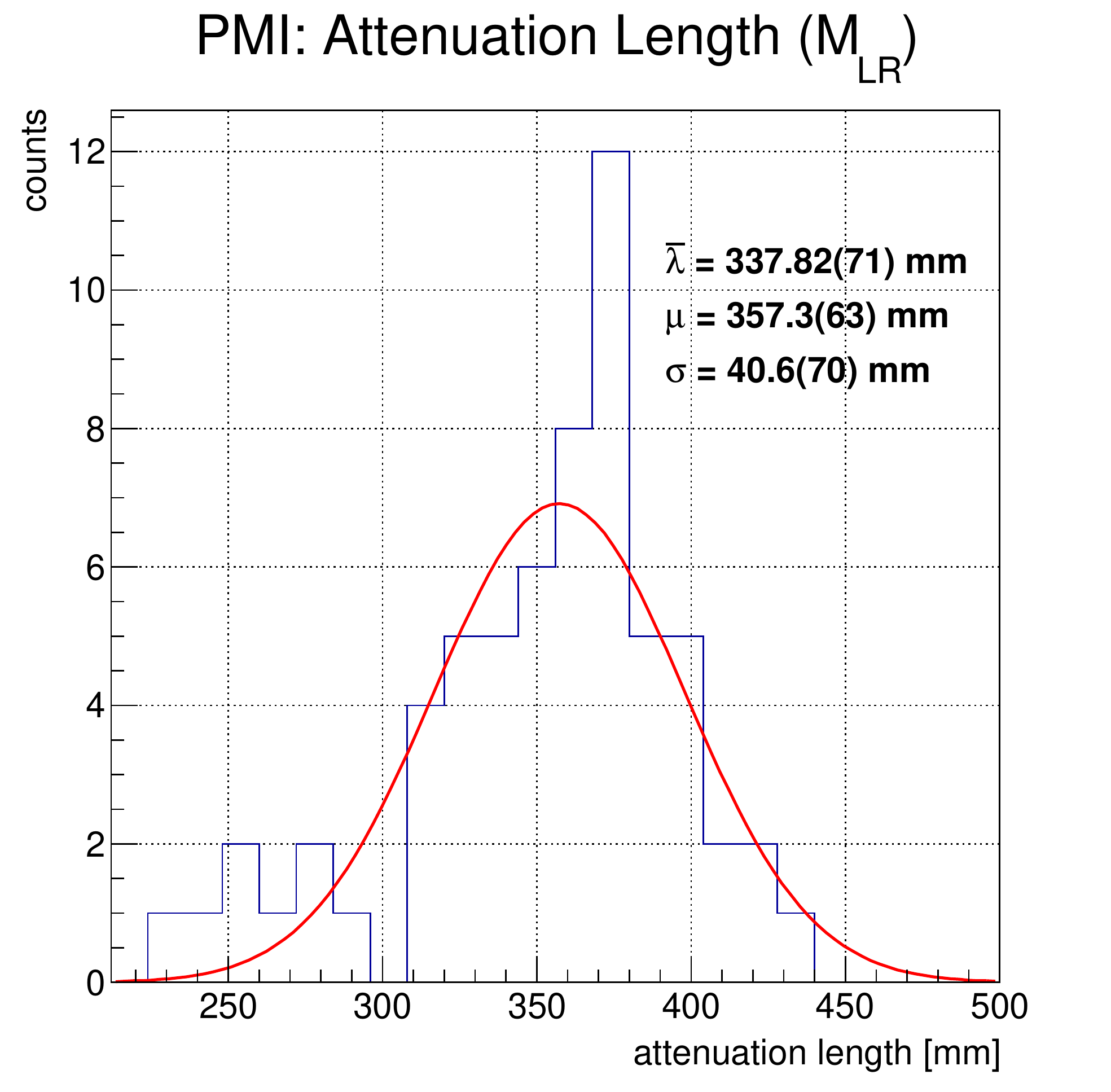}
\includegraphics[width=0.49\textwidth]{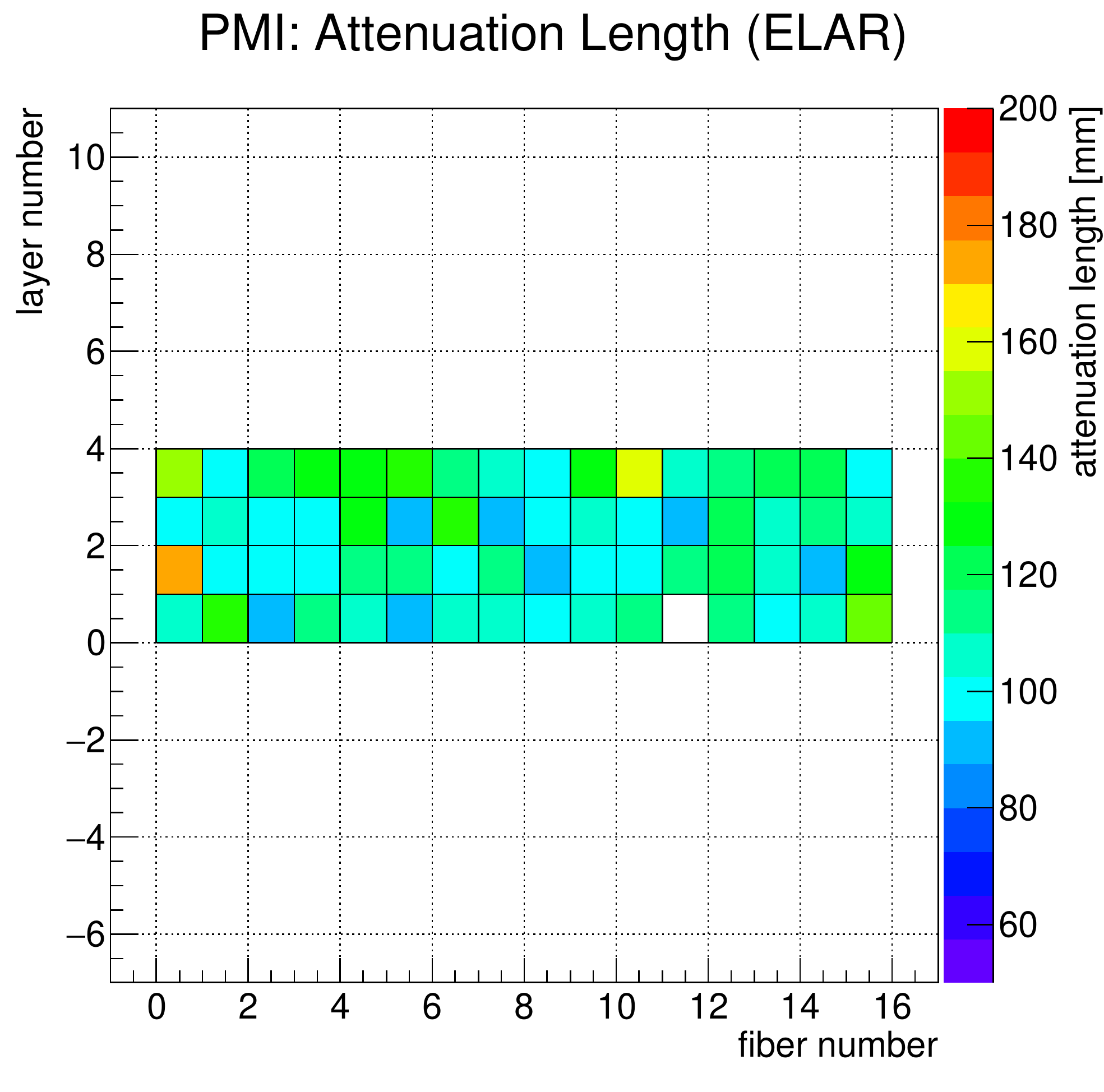}
\includegraphics[width=0.49\textwidth]{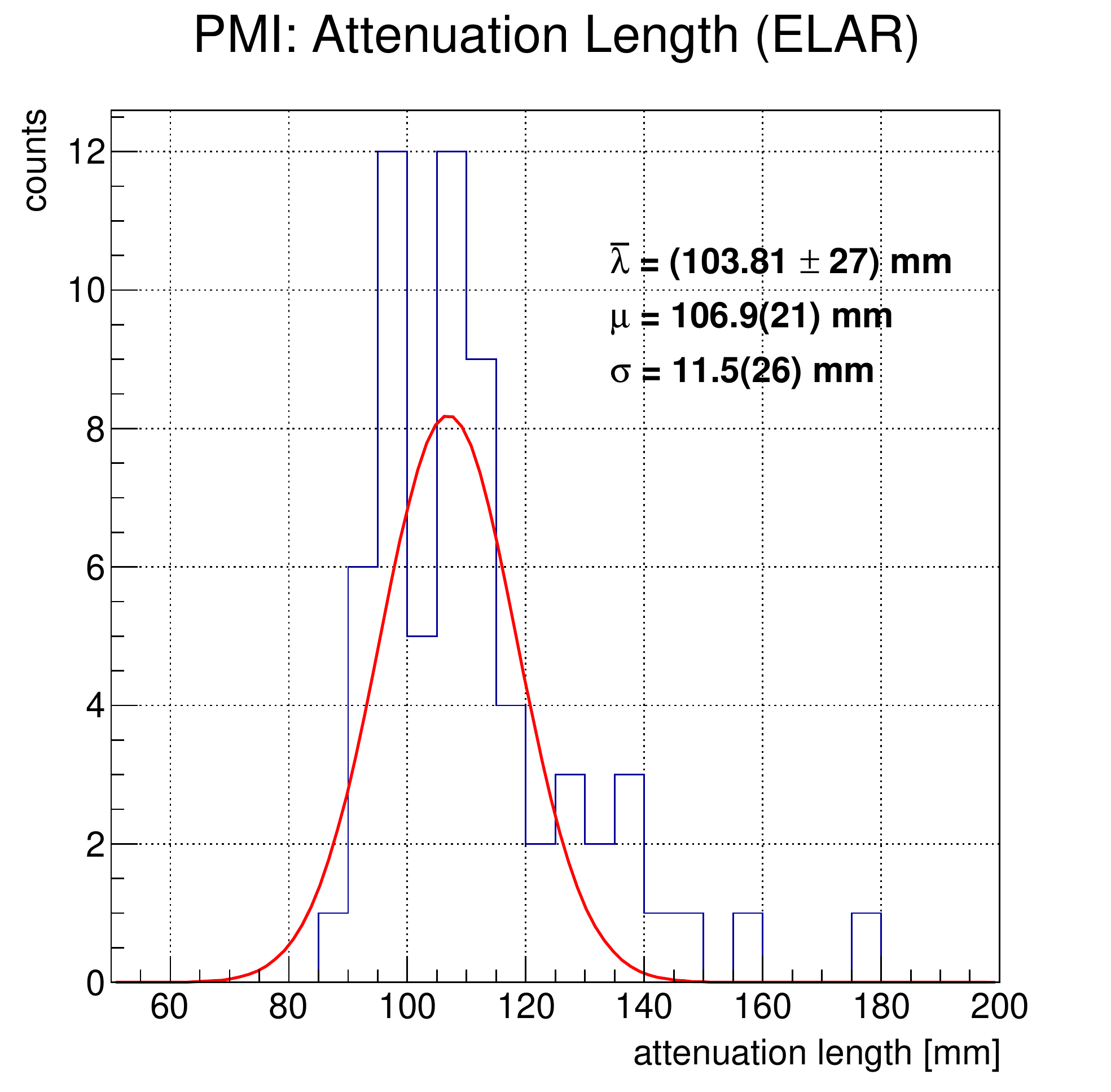}
\caption{Attenuation lengths obtained with \gls{gl:MLR} (top row) and \acrshort{gl:ELAR} methods (bottom row). Left histograms present spatial distribution of the obtained values within the prototype, while the right plots show their statistical distributions. The histograms were additionally fitted with a Gaussian function. Parameters of the fitted functions and weighted means are listed in both histograms.}
\label{fig:pmi-attenuation}
\end{figure}

Obtained results of the scintillating light propagation analysis are presented in \cref{fig:pmi-attenuation}. The upper part shows the results of the \gls{gl:MLR} analysis. The weighted mean of the attenuation length was \SI{338}{\milli\meter}, which is significantly smaller than obtained in \acrshort{gl:JU} data analysis (\SI{465}{\milli\meter}, see \cref{tab:prototype-ju-results}). However, it is still larger than the value determined for the analogous fiber configuration in the single-fiber test (\SI{211}{\milli\meter}, see \cref{tab:diff-wrapping}). The relative standard deviation of the obtained attenuation length distributions in \acrshort{gl:PMI} and \acrshort{gl:JU} measurements are comparable (approximately \SI{11}{\percent} in both cases). 

In the lower part of \cref{fig:pmi-attenuation} the results of the \acrshort{gl:ELAR} analysis are presented. The inconsistencies regarding the model fit that were present in the \acrshort{gl:JU} analysis are much rarer in this experiment, with only 9 cases. The weighted mean of the determined attenuation lengths was \SI{103.81}{\milli\meter}, which is smaller than both the value obtained in \acrshort{gl:JU} analysis (\SI{161}{\milli\meter}, see \cref{tab:prototype-ju-results}) and single-fiber tests for the analogous fiber setup (\SI{113}{\milli\meter}, see \cref{tab:diff-wrapping}). Similarly as above, the standard deviation of the attenuation length distribution is approximately \SI{11}{\percent}. Differences in the obtained light attenuation values in different experiments can be attributed to the different types of used photodetectors as explained in \cref{subsec:krk-tests-results}.

\subsubsection*{Energy reconstruction}

\begin{figure}[htbp]
\centering
\includegraphics[width=0.49\textwidth]{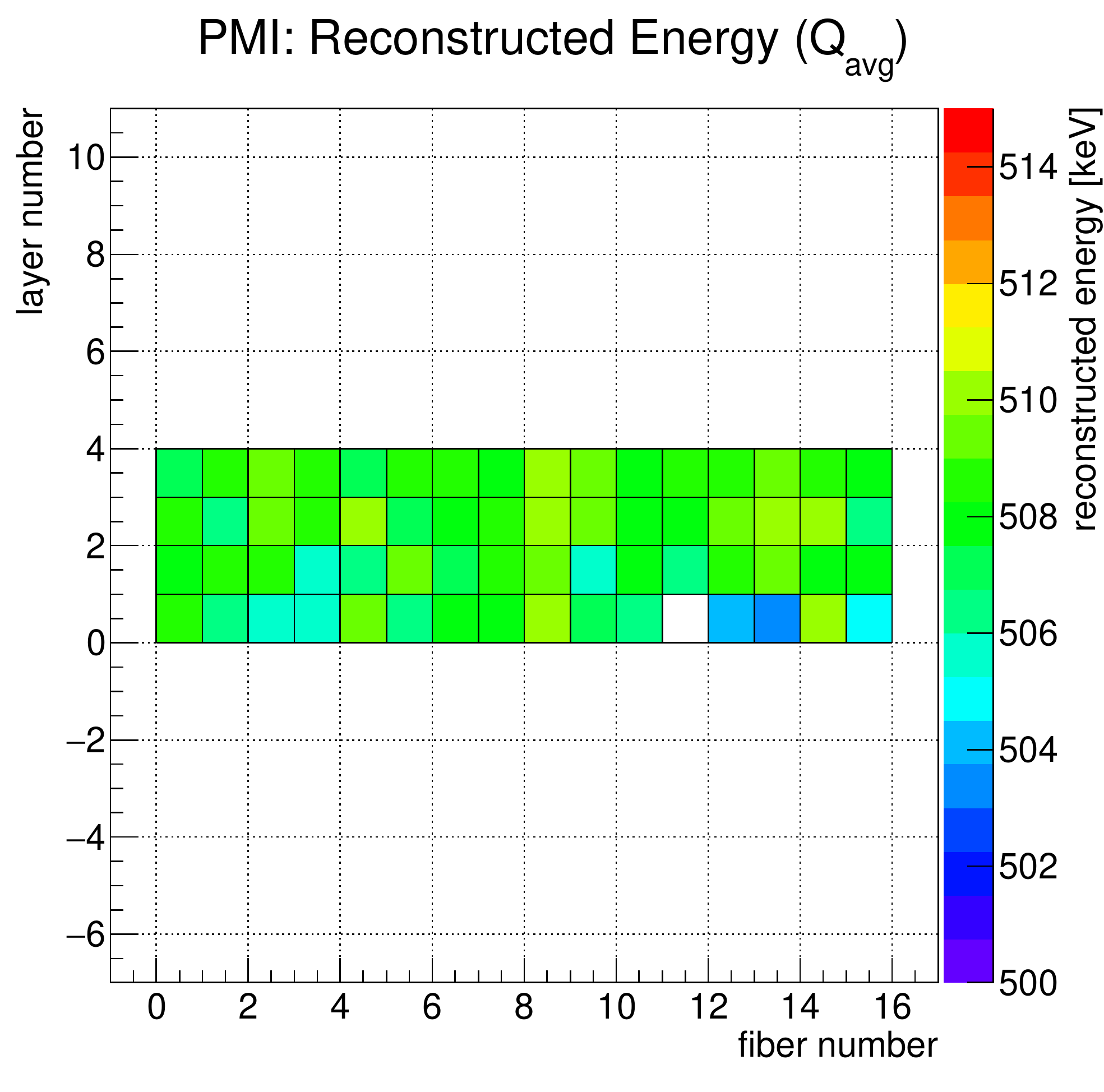}
\includegraphics[width=0.49\textwidth]{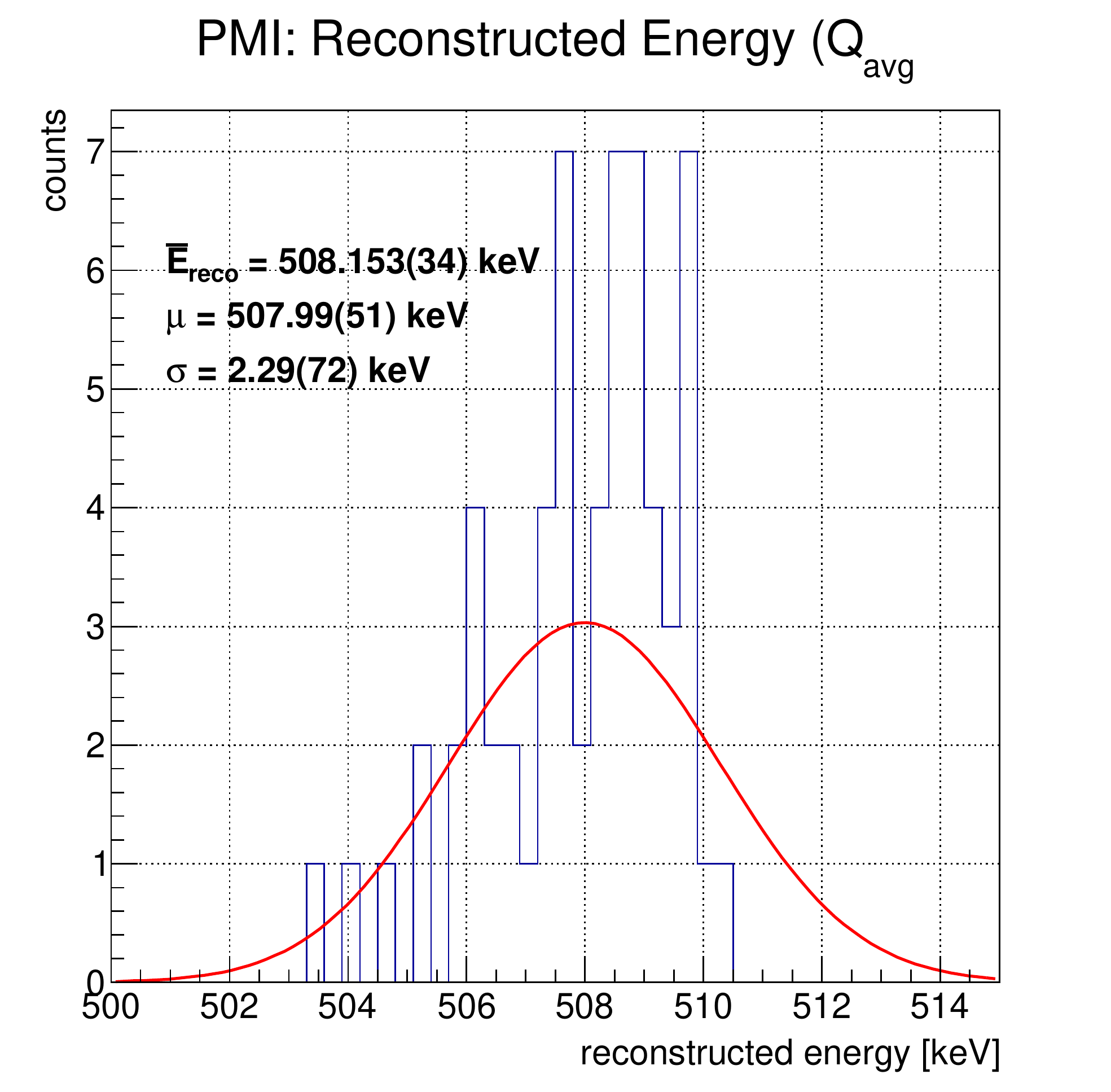}
\includegraphics[width=0.49\textwidth]{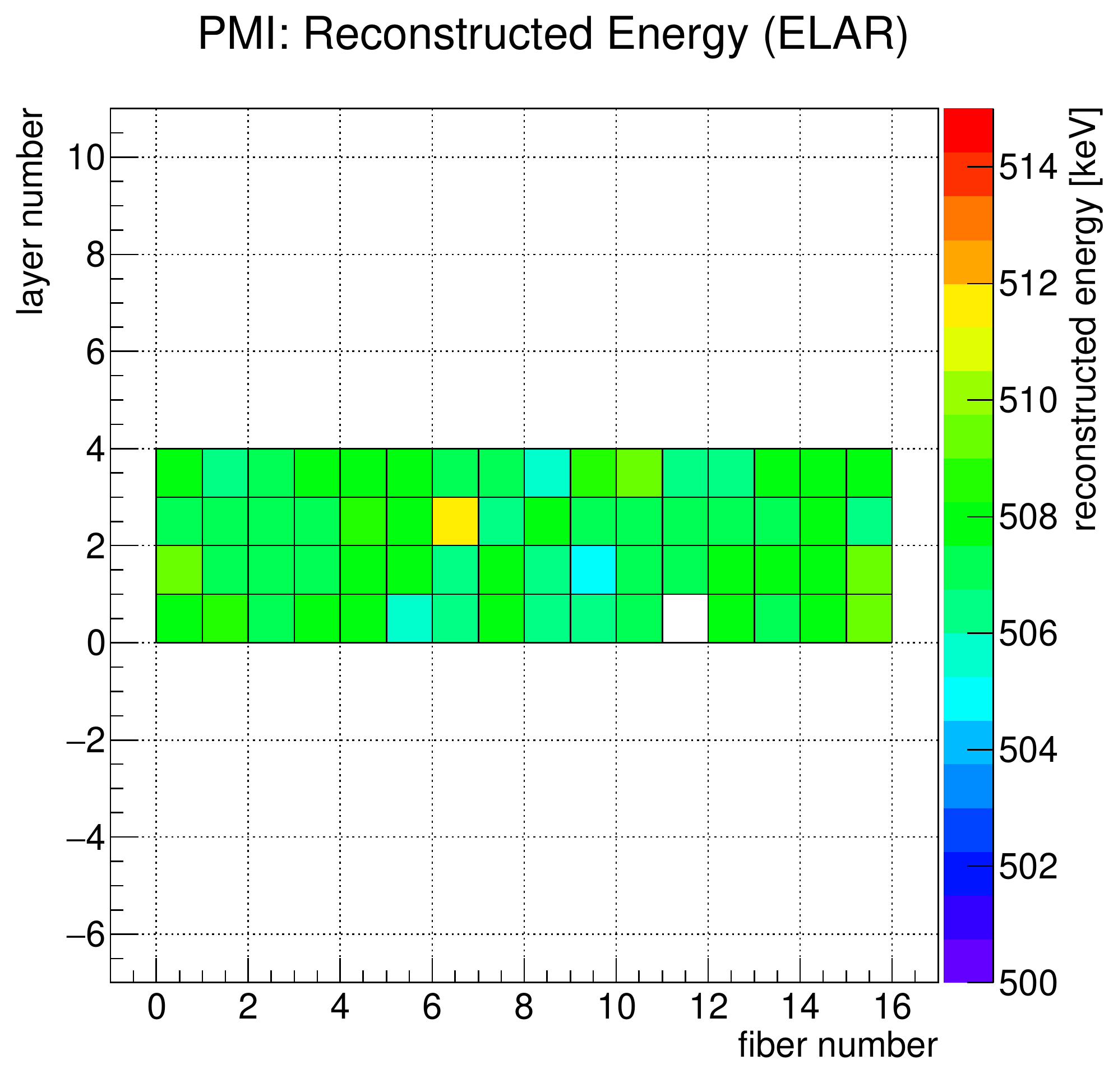}
\includegraphics[width=0.49\textwidth]{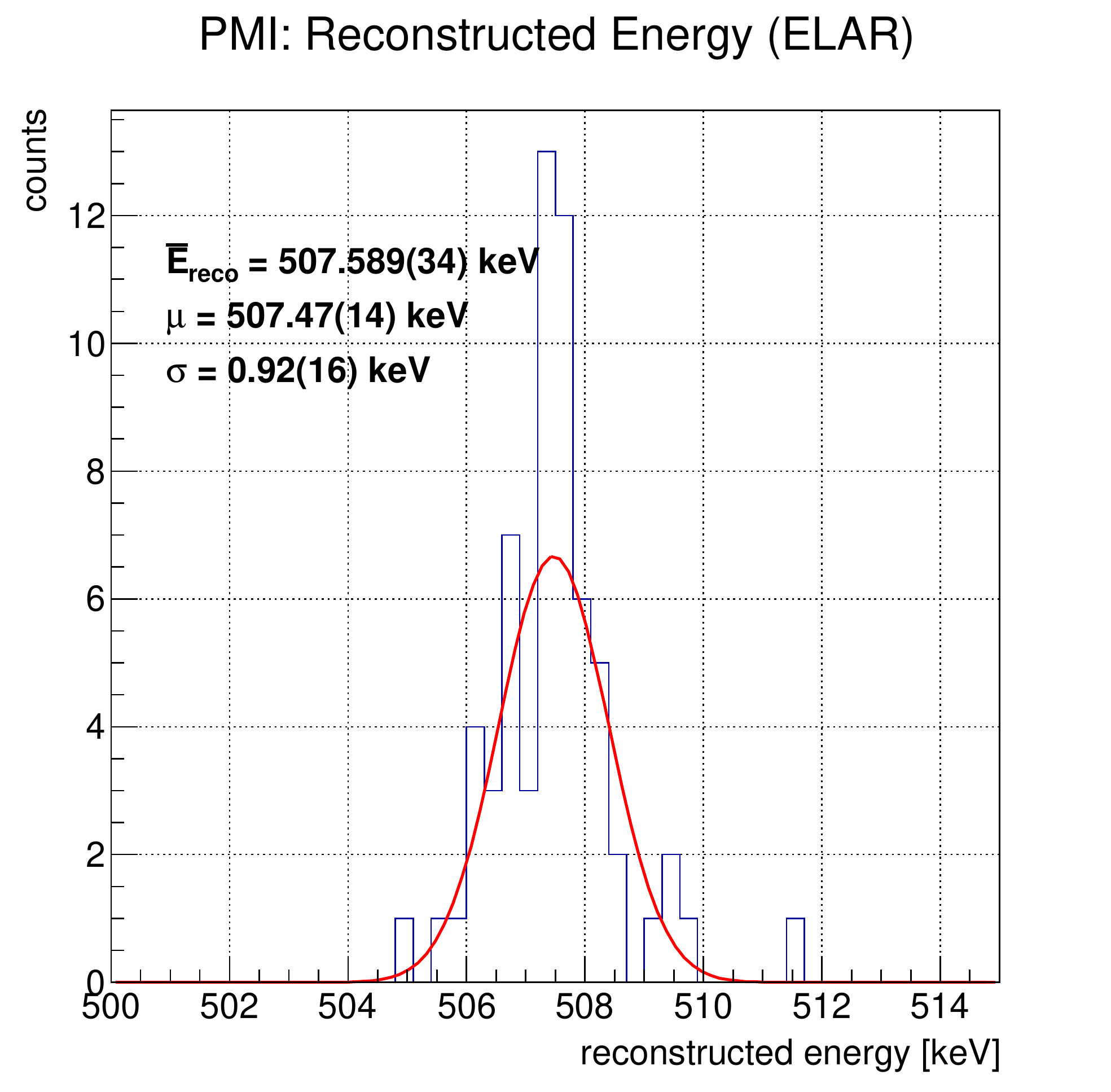}
\caption{Energies of the annihilation peak reconstructed with the two methods: \gls{gl:Qavg} (top row) and \acrshort{gl:ELAR} (bottom row). Left histograms present the values of the reconstructed peak centers for different fibers in the prototype, while the right plots show their statistical distributions and their parametrizations as Gaussian functions (red lines) and the list of their parameters. Fit parameters and weighted means are also listed.}
\label{fig:pmi-energyreco}
\end{figure}

Analogously as for the \acrshort{gl:JU} measurements, the quality of energy reconstruction was evaluated with the accuracy of the reconstruction of the \anhpeak peak in the recorded photon count spectra. The obtained results are presented in \cref{fig:pmi-energyreco}. Both of the used methods yielded similar results, with the average reconstructed energy of the annihilation peak equal to approximately \SI{508}{\kilo\electronvolt}. However, the relative standard deviation of the distribution obtained with the \acrshort{gl:ELAR}-based method is two times smaller than for the \gls{gl:Qavg}-based method. In contrast to the \acrshort{gl:JU} measurements, in this case both energy reconstruction methods resulted in the average reconstructed energy of the annihilation peak slightly smaller than expected. 

\begin{figure}[htbp]
\centering
\includegraphics[width=0.49\textwidth]{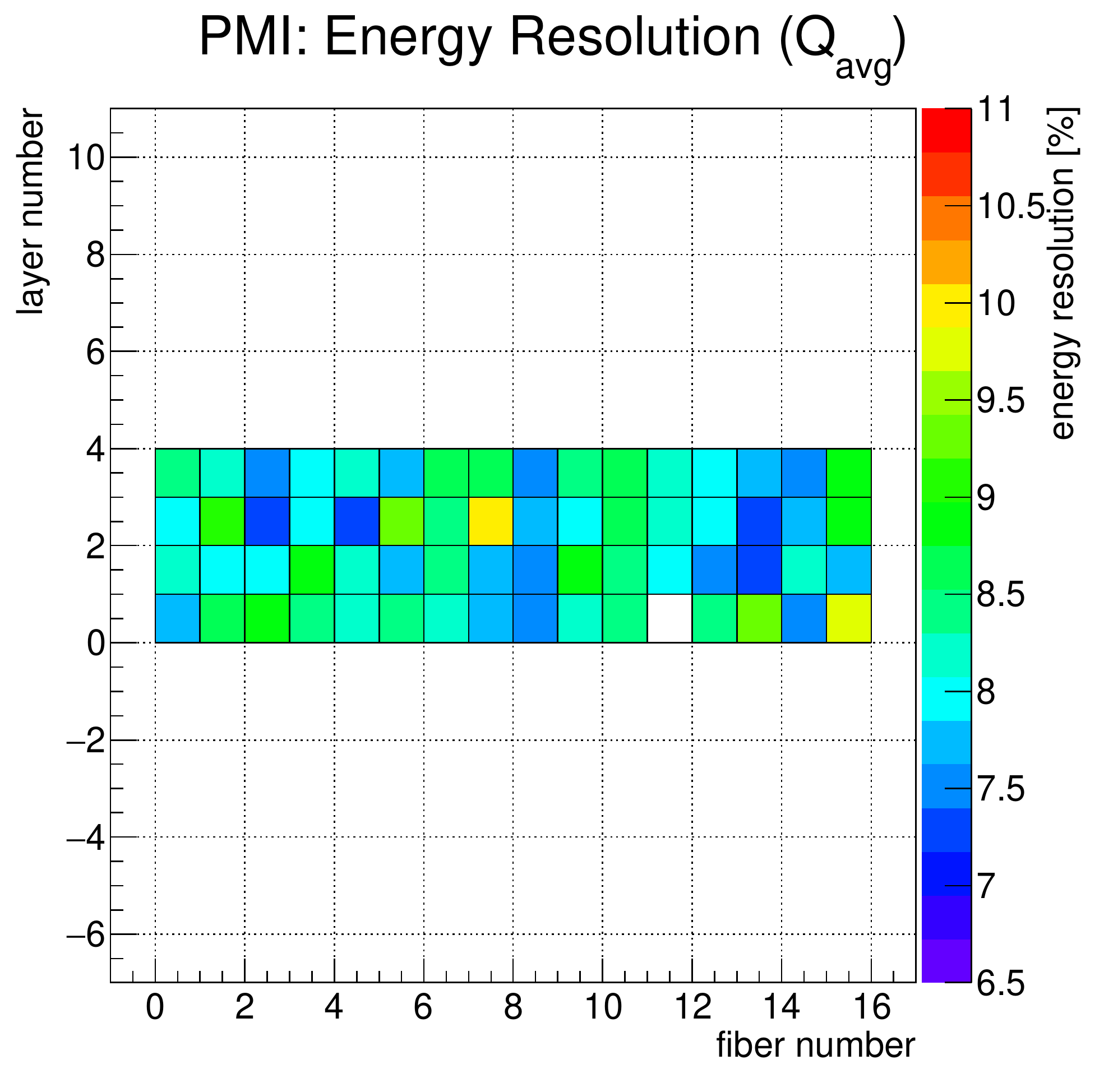}
\includegraphics[width=0.49\textwidth]{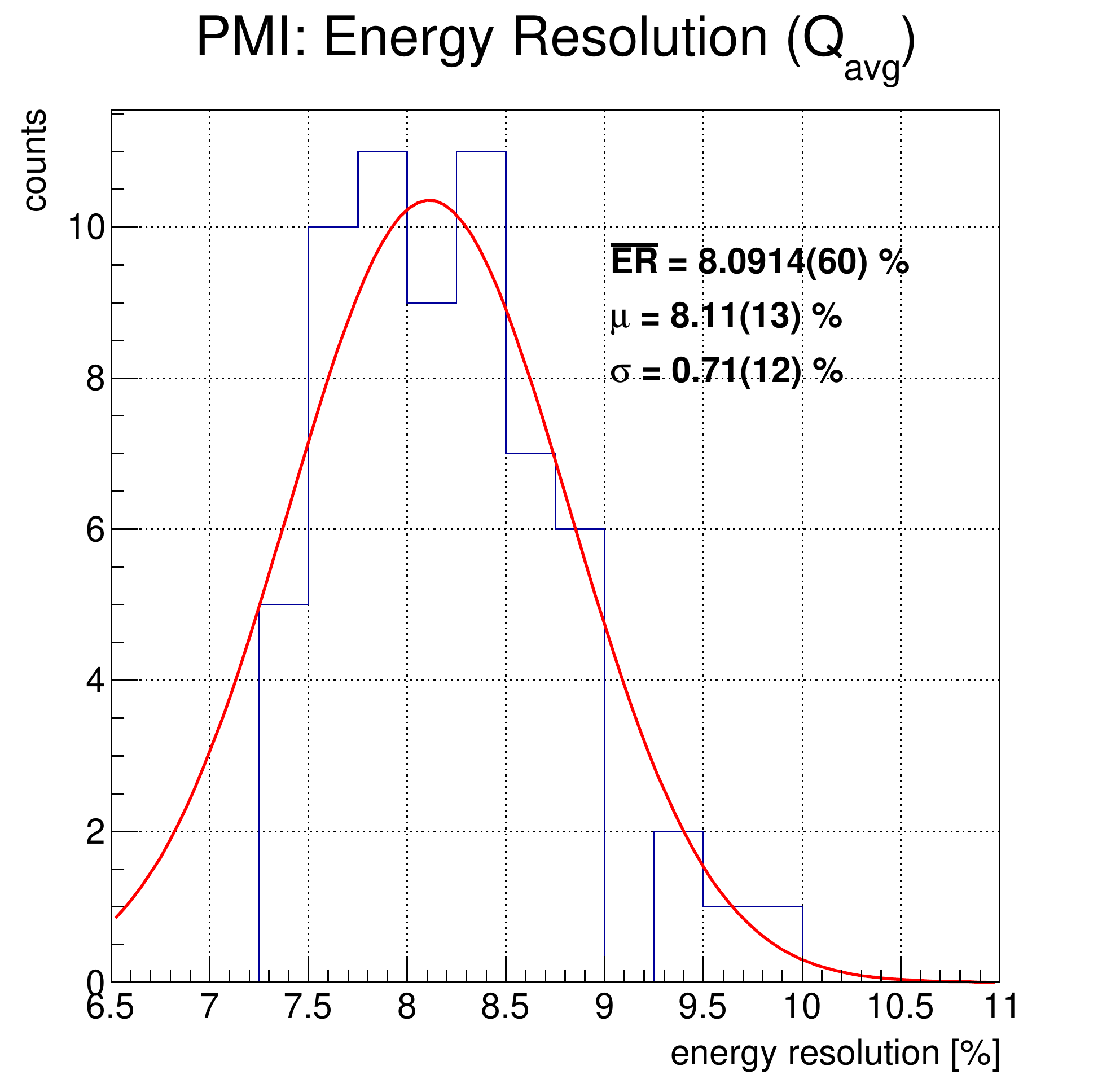}
\includegraphics[width=0.49\textwidth]{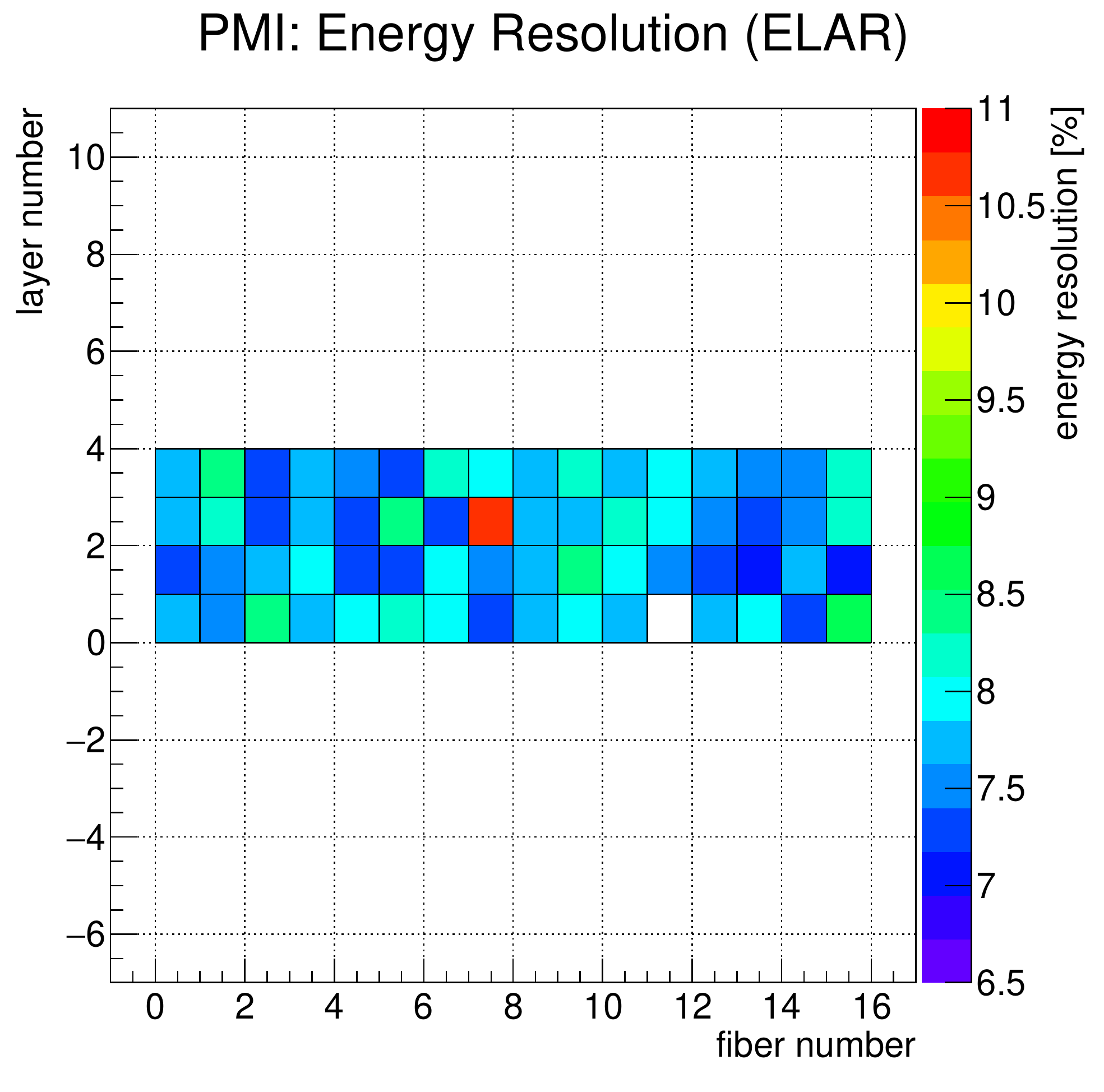}
\includegraphics[width=0.49\textwidth]{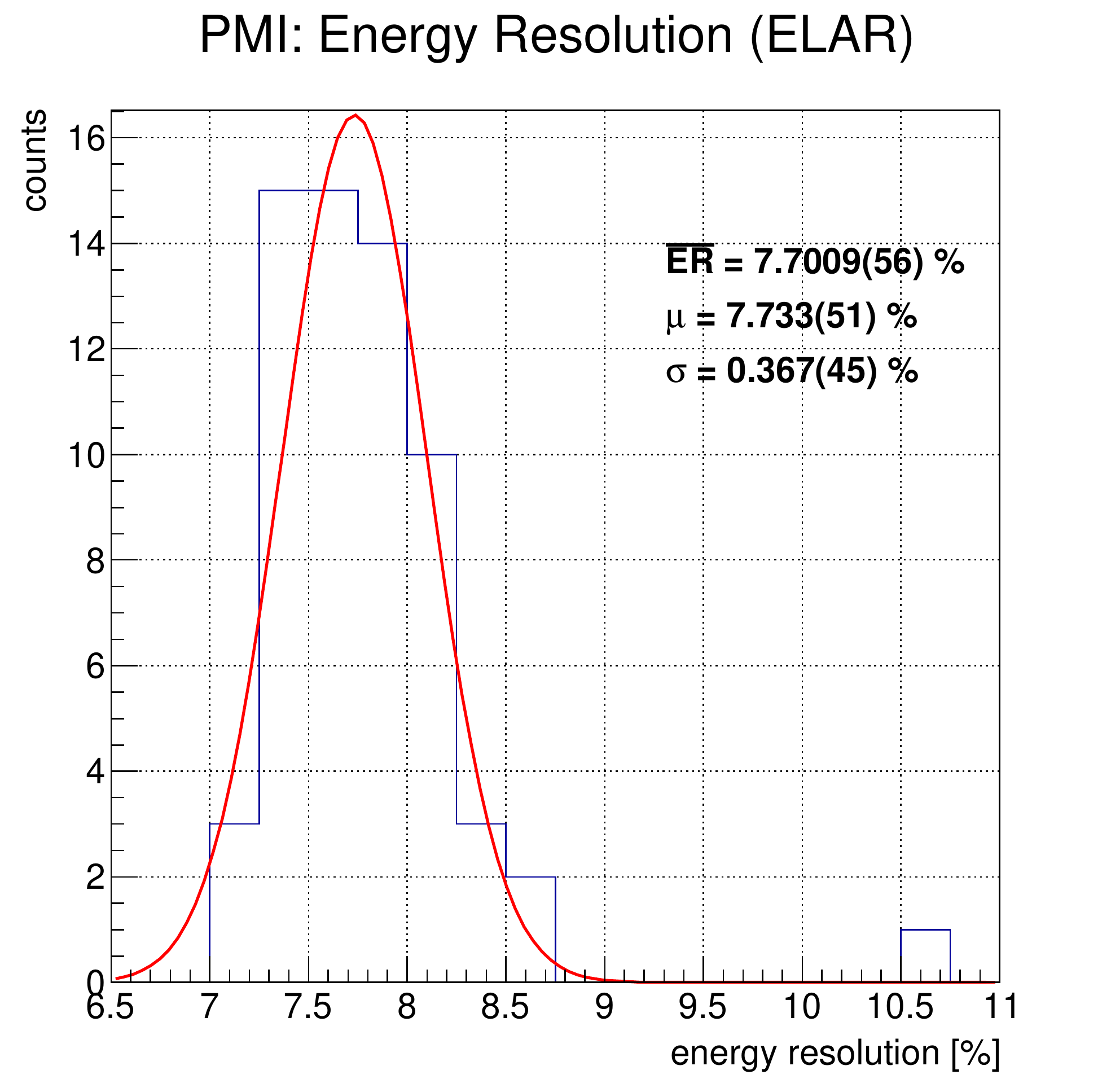}
\caption{Energy resolutions obtained from spectra reconstructed with \gls{gl:Qavg} (top row) and \acrshort{gl:ELAR} (bottom row) methods. Left histograms present the values determined for different fibers in the prototype, while the right plots show their statistical distributions and their parameterizations as Gaussian functions (red lines). Parameters of the distributions and the weighted means are listed.}
\label{fig:pmi-energyres}
\end{figure}

Another important aspect of the evaluation of energy reconstruction was the study of energy resolution. The results are presented in \cref{fig:pmi-energyres}. The average energy resolution obtained from the spectra reconstructed with the \gls{gl:Qavg} method was \SI{8.09}{\percent} which is a significant improvement when compared with the \acrshort{gl:JU} results (\SI{10.27}{\percent}, see \cref{tab:prototype-ju-results}). This result is comparable to that obtained in the single fiber tests for a similar fiber configuration (\SI{8.56}{\percent}, see \cref{tab:diff-wrapping}). The results obtained with \acrshort{gl:ELAR}-based method were even more promising, with the average energy resolution of \SI{7.70}{\percent}. Furthermore, the relative standard deviation of the distribution of obtained energy resolution values is approximately two times smaller for the \acrshort{gl:ELAR}-based reconstruction method. This improvement is expected, since the \acrshort{gl:ELAR} correction compensates the variations of the gain and thus also energy of the annihilation peak, caused by small asymmetries of light propagation in the fiber or differences in coupling. Fiber L2F7 yielded much worse performance than the remaining scintillating fibers in the prototype. The reason for that is a very small contribution of Compton continuum registered for that crystal. This made it impossible to accurately model the background in the spectrum. The overall improvement of the energy resolution in the \acrshort{gl:PMI} measurements can be attributed to the used photodetector. 

\subsubsection*{Position reconstruction}

\begin{figure}[htbp]
\centering
\includegraphics[width=0.49\textwidth]{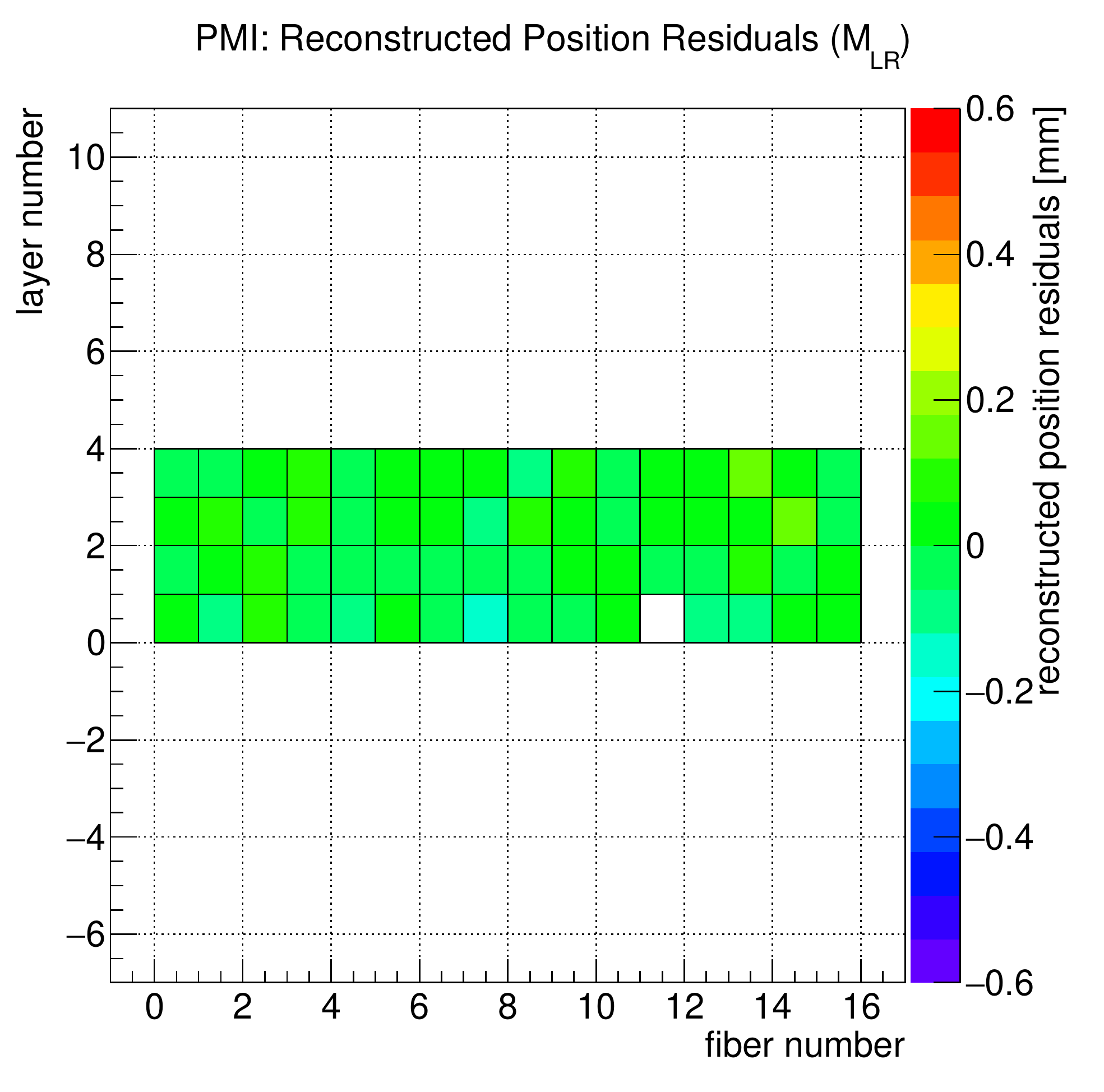}
\includegraphics[width=0.49\textwidth]{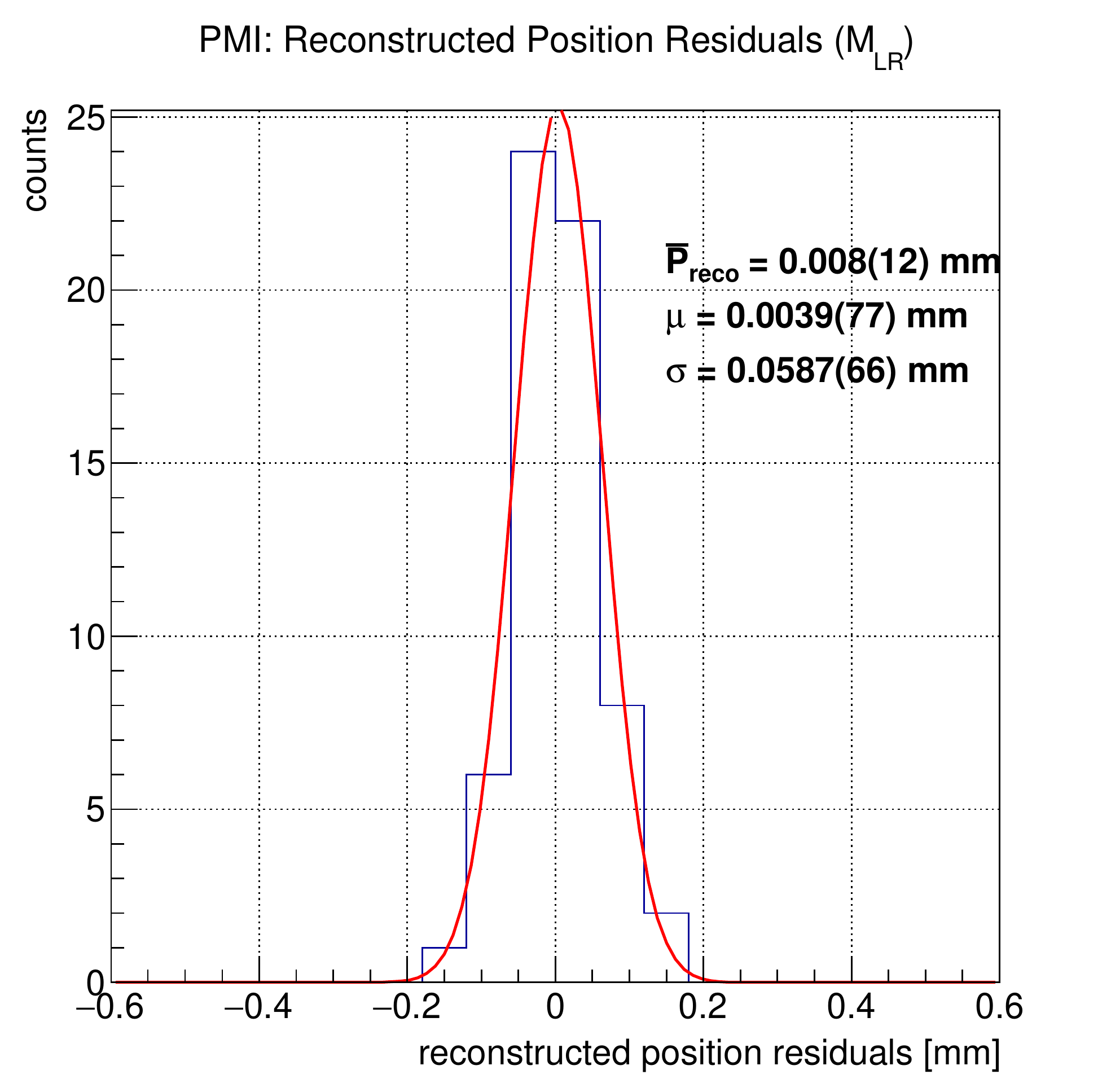}
\includegraphics[width=0.49\textwidth]{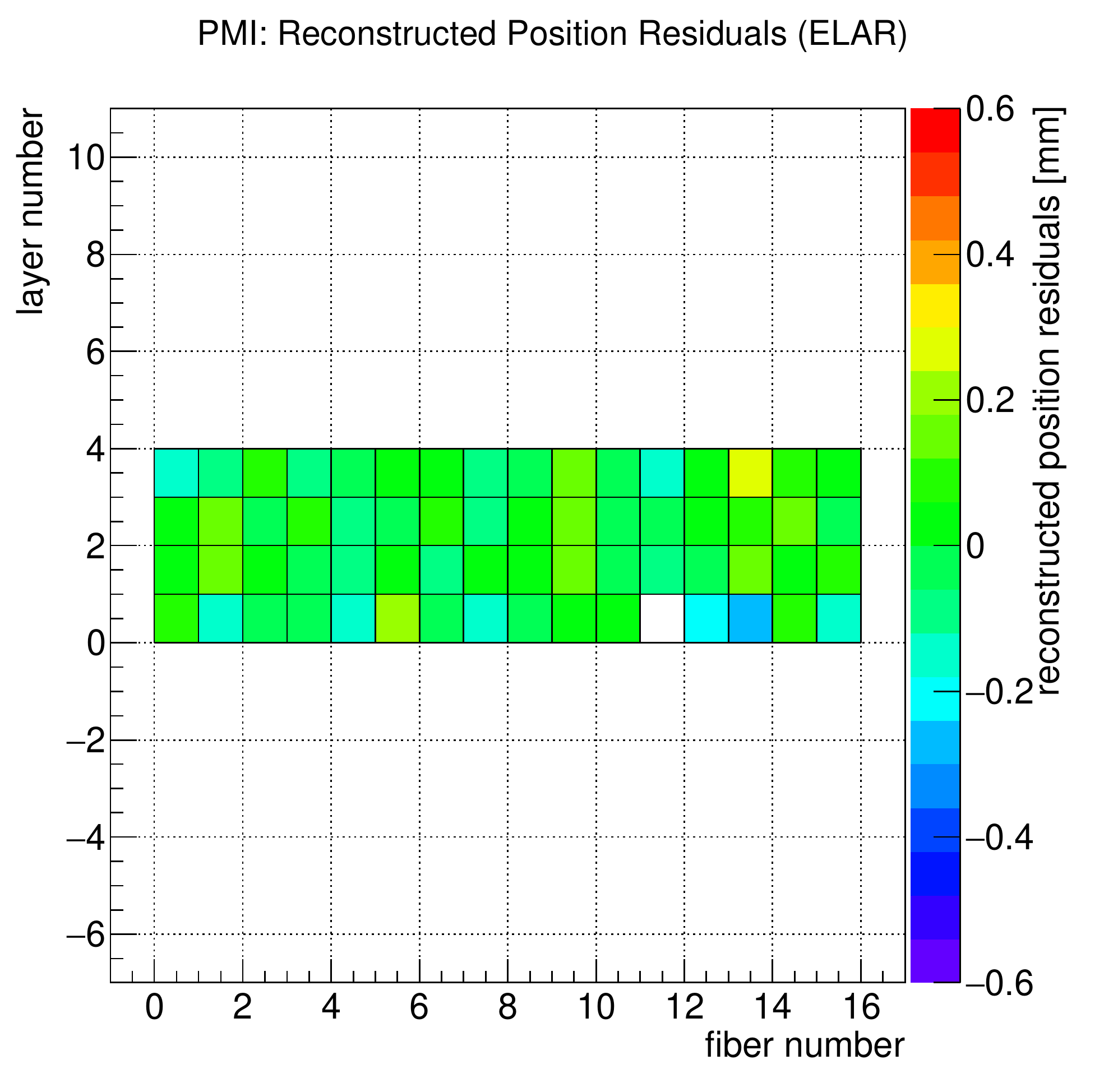}
\includegraphics[width=0.49\textwidth]{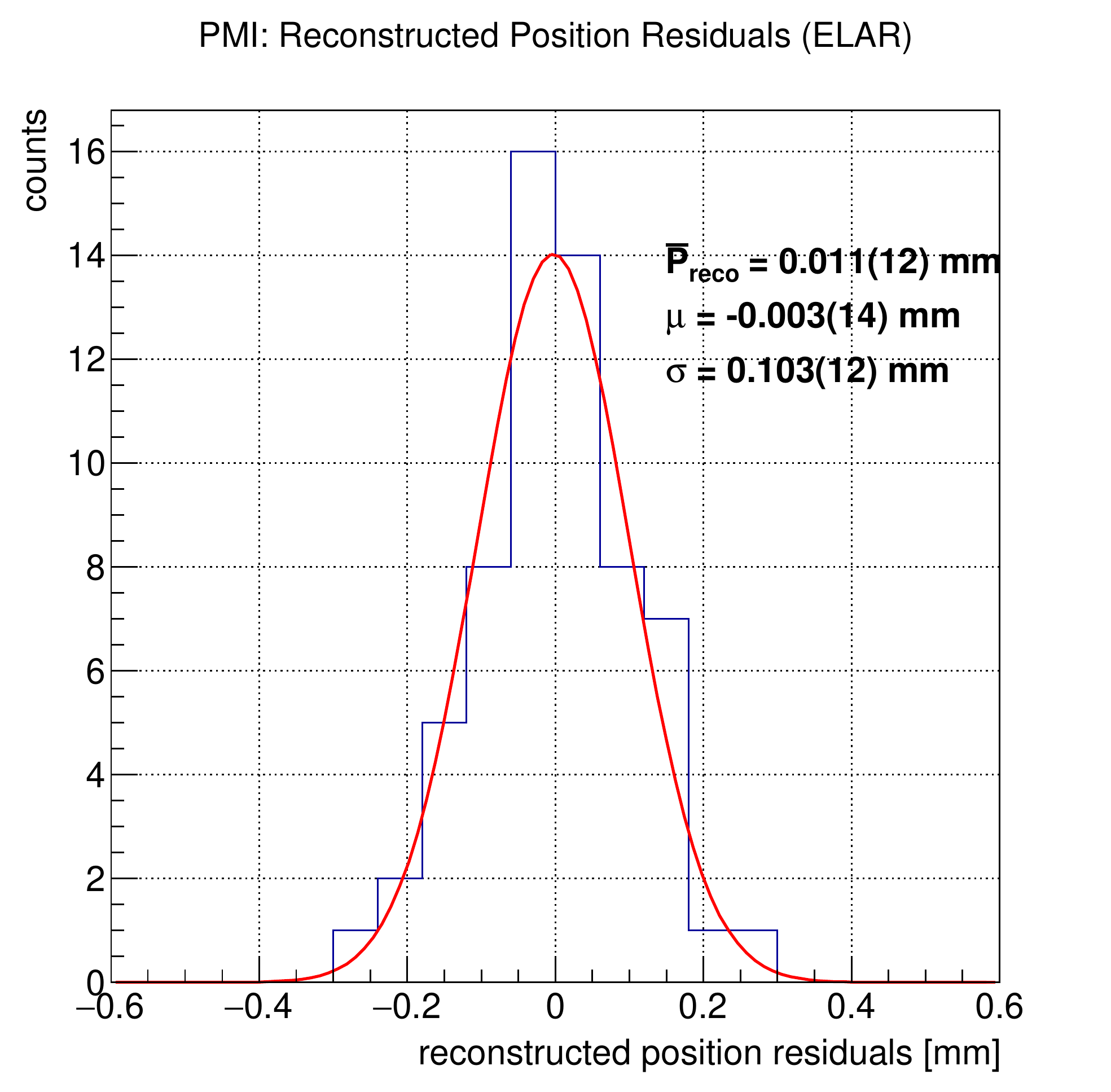}
\caption{Mean position residuals obtained with the two position reconstruction methods: \gls{gl:MLR} (top row) and \acrshort{gl:ELAR} (botton row). Left histograms present the values of the reconstructed position residuals for different fibers in the prototype, while the right plots show their statistical distributions and their parametrizations as Gaussian functions. Fit parameters and weighted means are also listed.}
\label{fig:pmi-posreco}
\end{figure}

The results of position reconstruction, as previously, were expressed with the means of the integrated residual distributions (\gls{gl:Xreco} $-$ \gls{gl:Xreal}). The results obtained for all \acrshort{gl:LYSO:Ce} fibers in the prototype are presented in \cref{fig:pmi-posreco}. Both of the used methods yielded comparable results, with the mean of the obtained distributions consistent with \SI{0}{\milli\meter} within one standard deviation, as expected. This outcome remains in agreement with the results obtained for the \acrshort{gl:JU} measurements. The relative standard deviation of the distribution representing results of the \acrshort{gl:ELAR}-based position reconstruction is two times larger when compared with the distribution of the \gls{gl:MLR} results. As a result, the spatial distribution of the reconstructed values is also more uniform within the prototype for the \gls{gl:MLR} method.

\begin{figure}[htbp]
\centering
\includegraphics[width=0.49\textwidth]{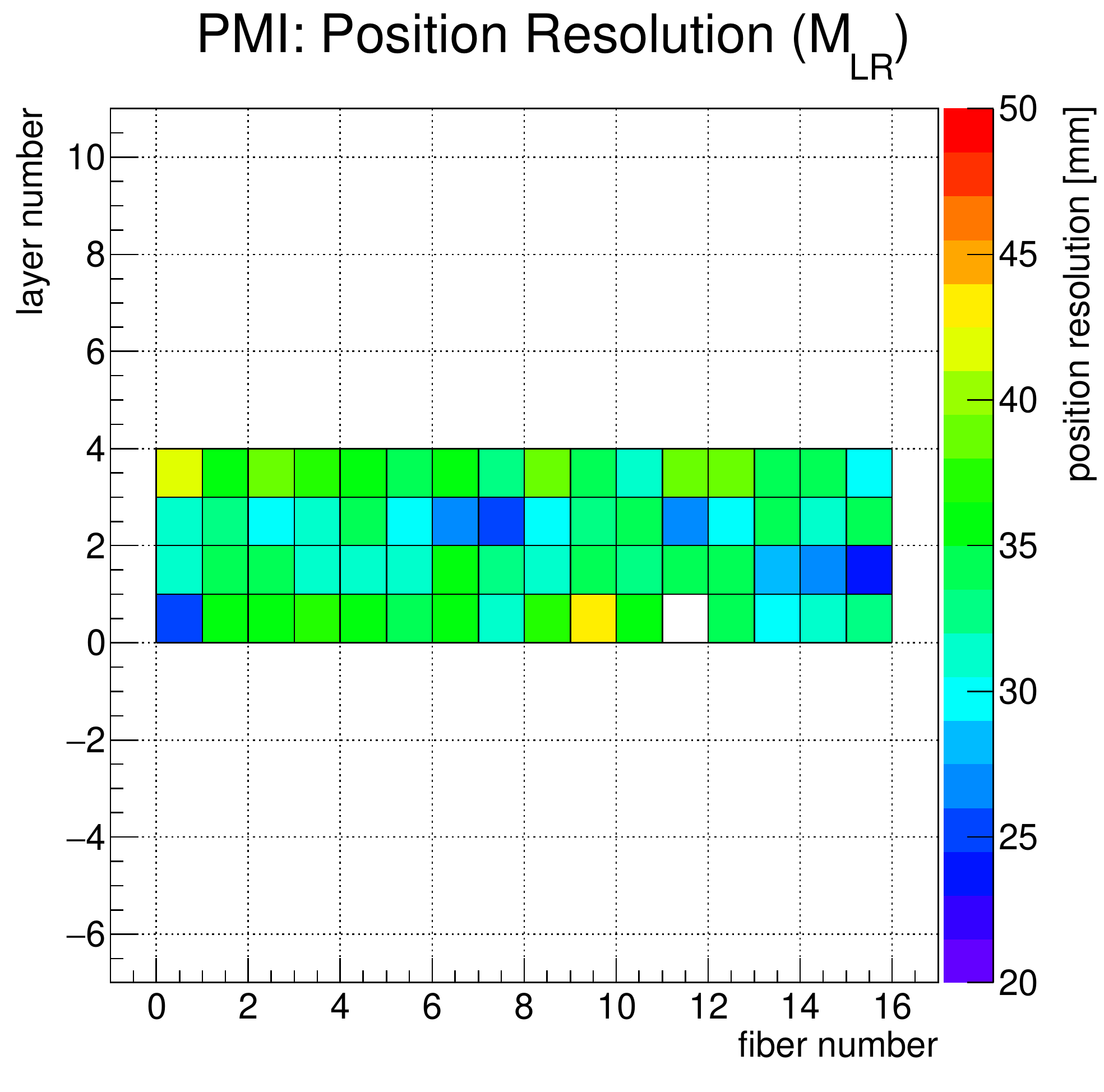}
\includegraphics[width=0.49\textwidth]{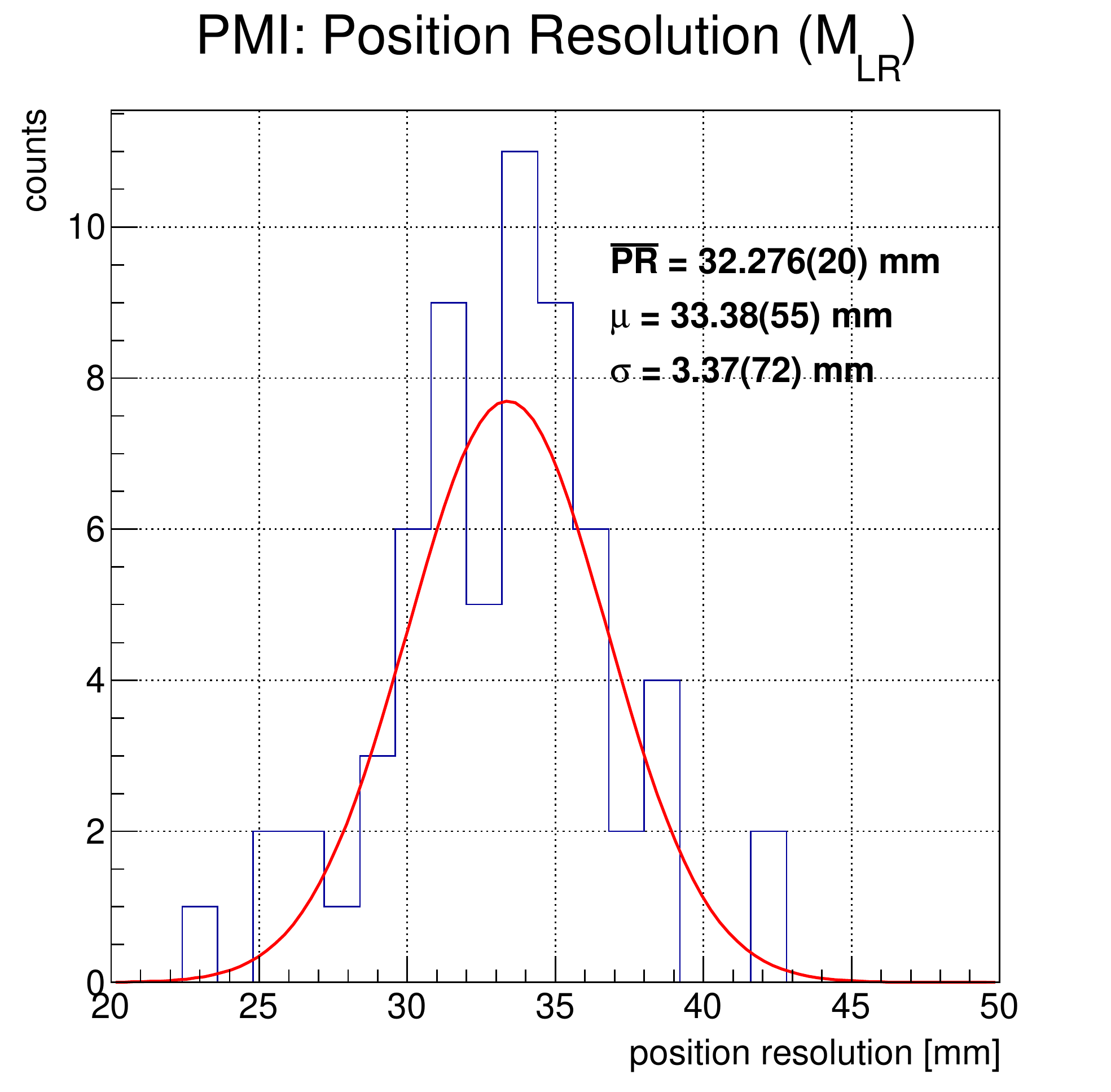}
\includegraphics[width=0.49\textwidth]{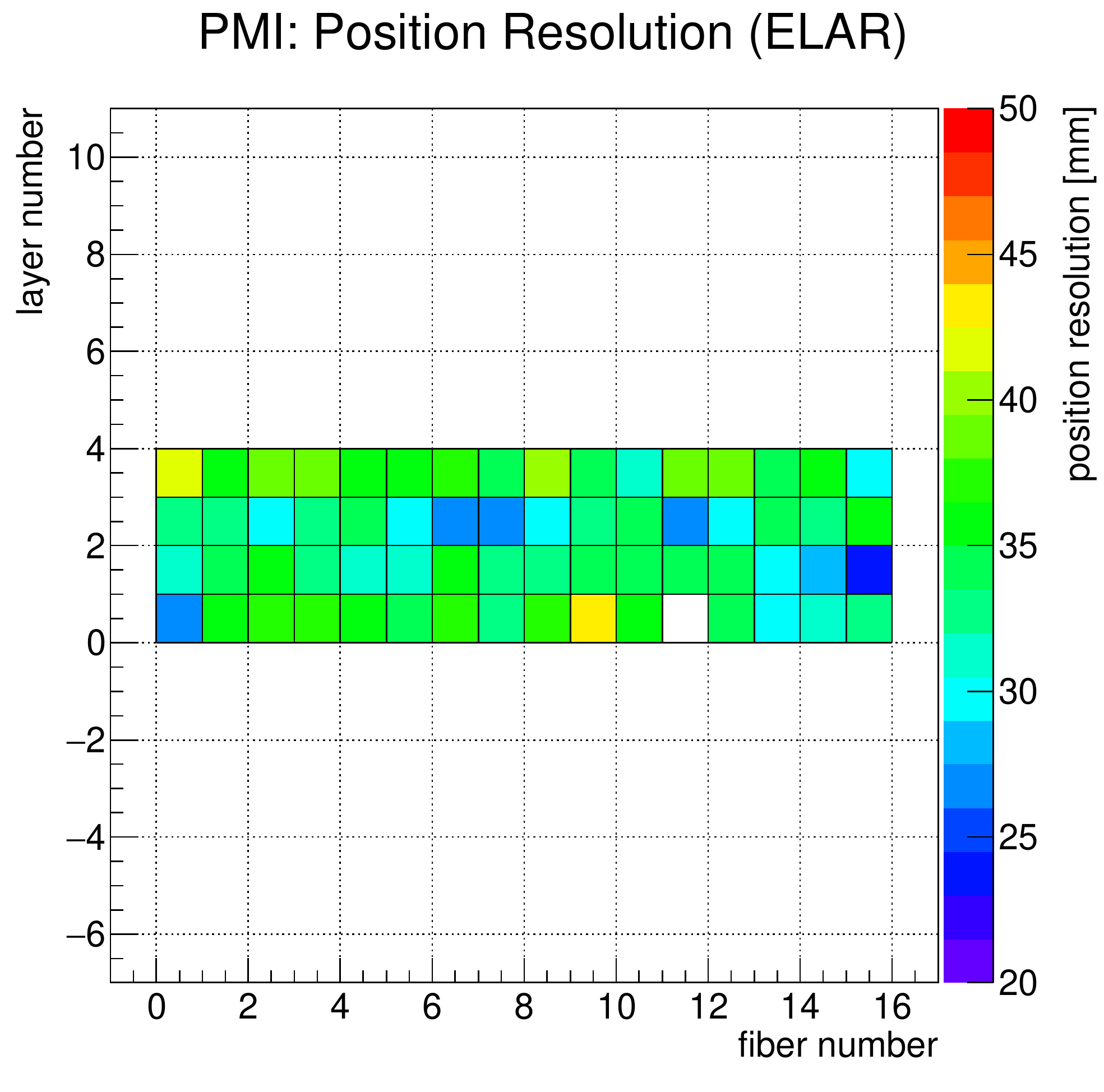}
\includegraphics[width=0.49\textwidth]{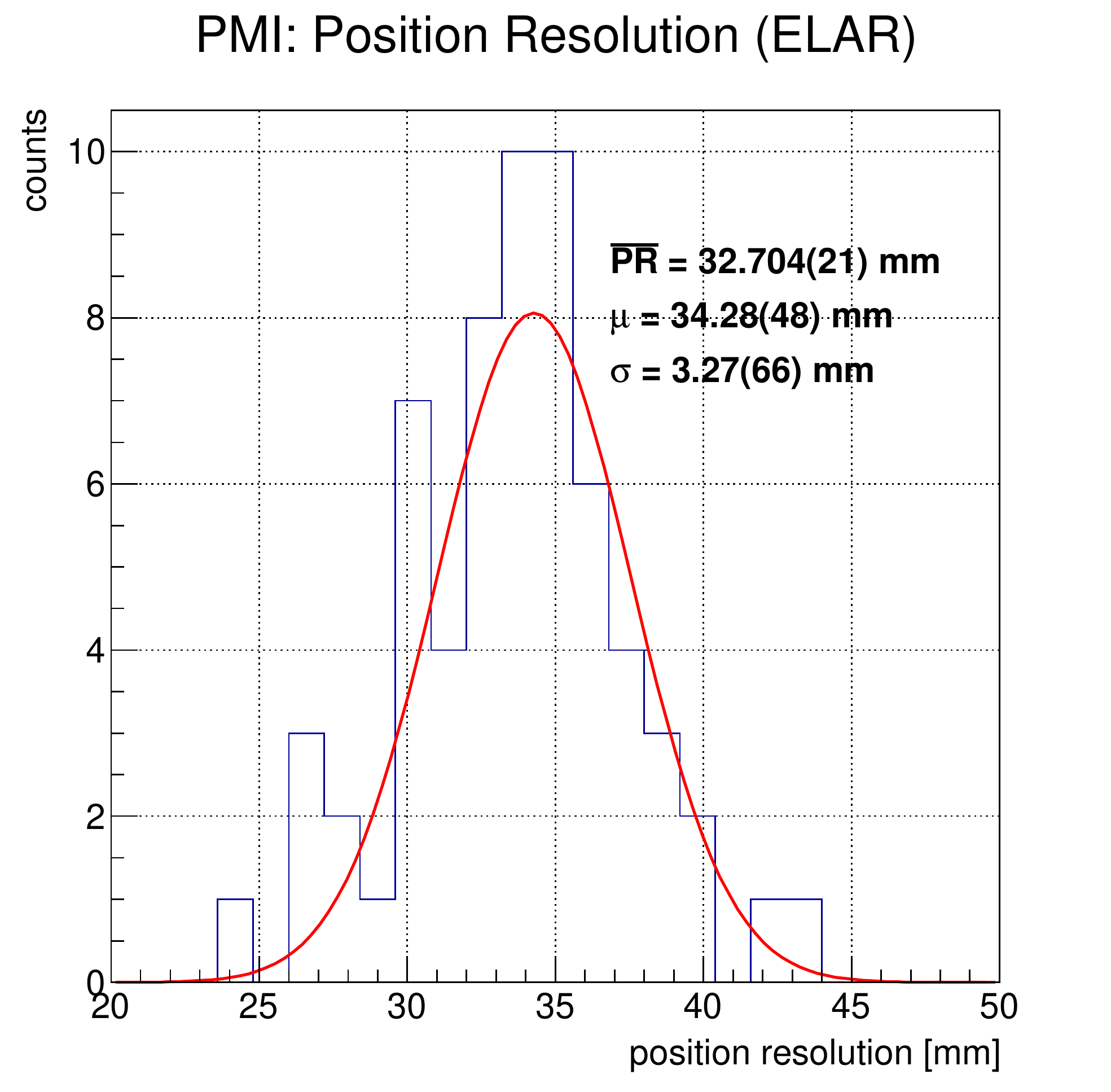}
\caption{Position resolutions obtained with the two position reconstruction methods: \gls{gl:MLR} (top row) and \acrshort{gl:ELAR} (botton row). Left histograms present the spatial distributions of the obtained values within the prototype, while the right plots show their statistical distributions. Histograms were fitted with Gaussian functions. Obtained fit parameters and weighted means are also listed.}
\label{fig:pmi-posres}
\end{figure}

The position resolution was another metrics analyzed for the evaluation of the position reconstruction and the prototype performance. The results are presented in \cref{fig:pmi-posres}. The values of position resolution of \acrshort{gl:LYSO:Ce} fibers obtained with the method using the \gls{gl:MLR} function fit were ranging from \SI{23.32}{\milli\meter} to \SI{42.54}{\milli\meter}, with the weighted mean at \SI{32.28}{\milli\meter}. The results obtained with the \acrshort{gl:ELAR} method are very similar, ranging from \SI{23.67}{\milli\meter} to \SI{43.48}{\milli\meter} with the weighted mean at \SI{32.70}{\milli\meter}. This is a significant improvement compared to the results obtained in the \acrshort{gl:JU} measurements (\SI{94.4}{\milli\meter} and \SI{91.0}{\milli\meter} for \gls{gl:MLR} and \acrshort{gl:ELAR}, respectively, see \cref{tab:prototype-ju-results}). It is caused by the different types of photodetector used, which yielded a larger light collection and allowed for the detection of scintillating light at large angles. At the same time, the position resolution results in \acrshort{gl:PMI} measurements are consistent with those obtained in the single-fiber tests conducted for the analogous fiber setup (\SI{32.0}{\milli\meter}, see \cref{tab:diff-wrapping}).

\subsubsection*{Light collection}

\begin{figure}[htbp]
\centering
\includegraphics[width=0.49\textwidth]{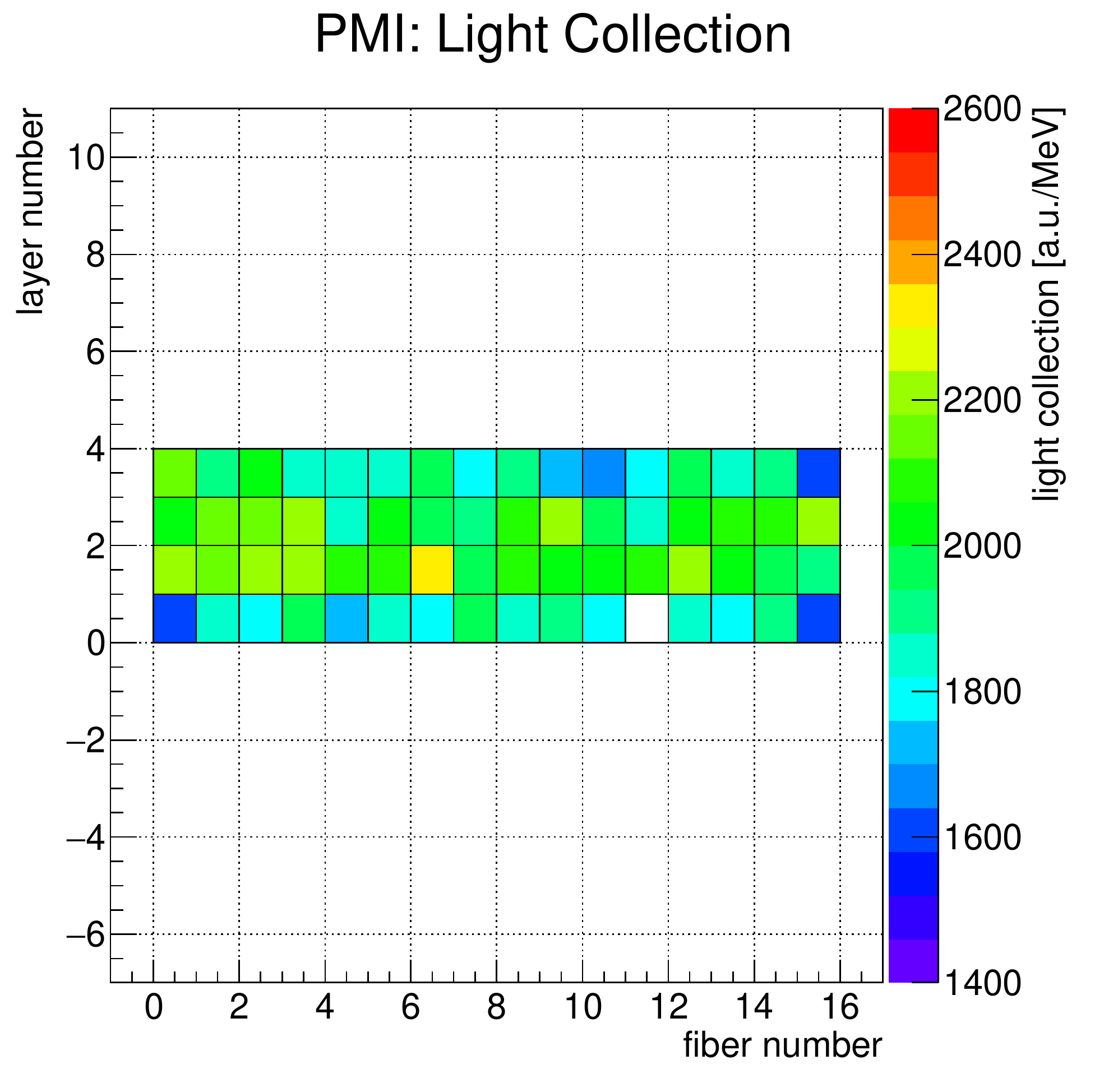}
\includegraphics[width=0.49\textwidth]{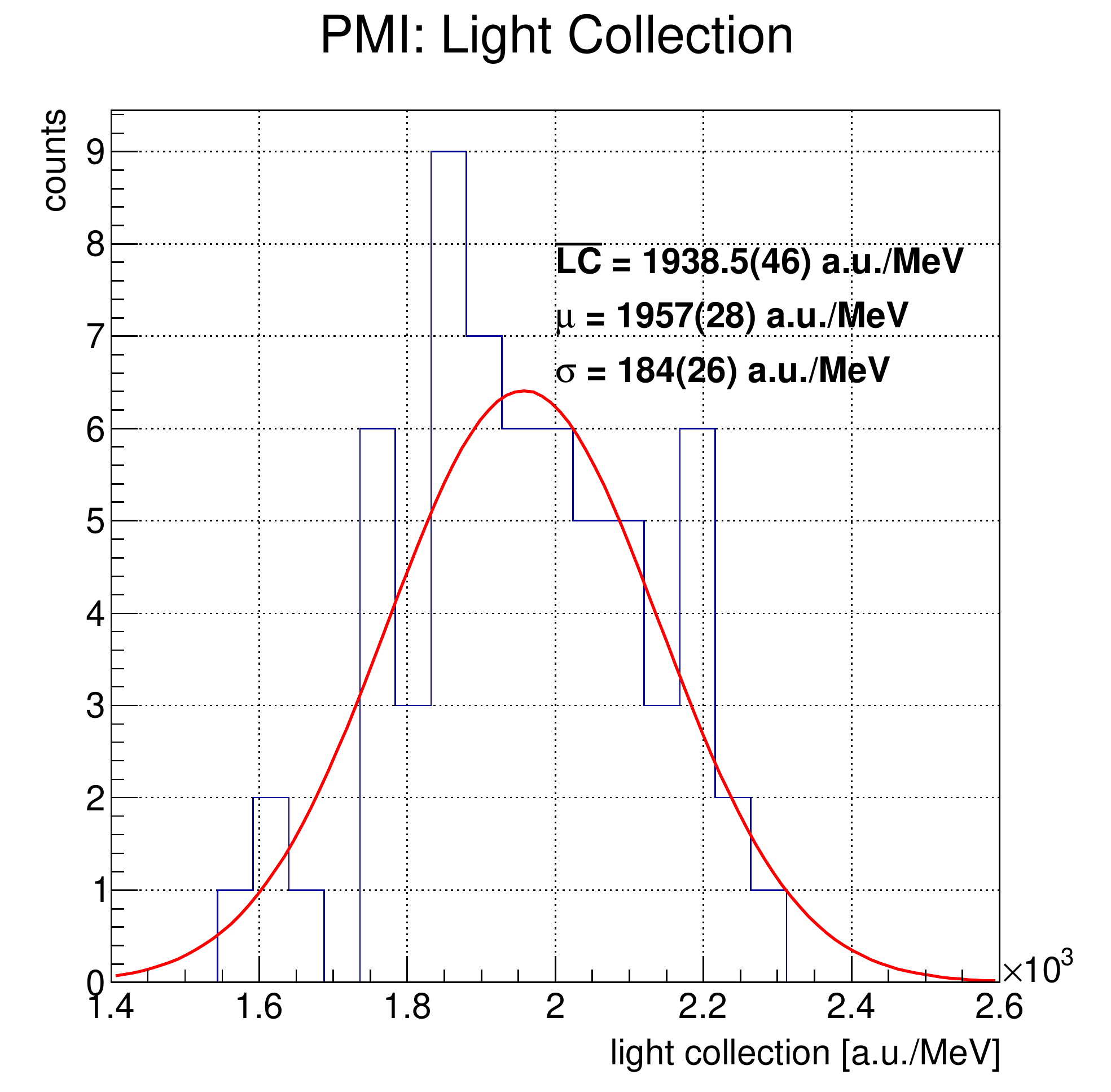}
\caption{Left: histogram showing the light collection values for each investigated fiber in \acrshort{gl:SSP}. Right: Statistical distribution of light collection values fitted with the Gaussian function. The listed values indicate the weighted mean and parameters of the distribution obtained from the fit.}
\label{fig:pmi-lcol}
\end{figure}

The results of the light collection analysis are presented in \cref{fig:pmi-lcol}. The weighted mean of the light collection was \SI{1939}{\au\per\mega\electronvolt}. The values ranged from \SI{1585}{\au\per\mega\electronvolt} to \SI{2308}{\au\per\mega\electronvolt}. The relative standard deviation of the distribution was \SI{9.4}{\percent}, which is in accordance with the results obtained for other fiber characteristics. Comparison with \acrshort{gl:JU} and single-fiber results is not possible, since a universal photoelectron calibration for all measurements was not possible. Relative light collection can only be estimated based on other characteristics, such as energy and position resolution. Both of them are improved when the light collection is increased. 

\subsubsection*{Timing resolution}

\begin{figure}[htbp]
\centering
\includegraphics[width=0.49\textwidth]{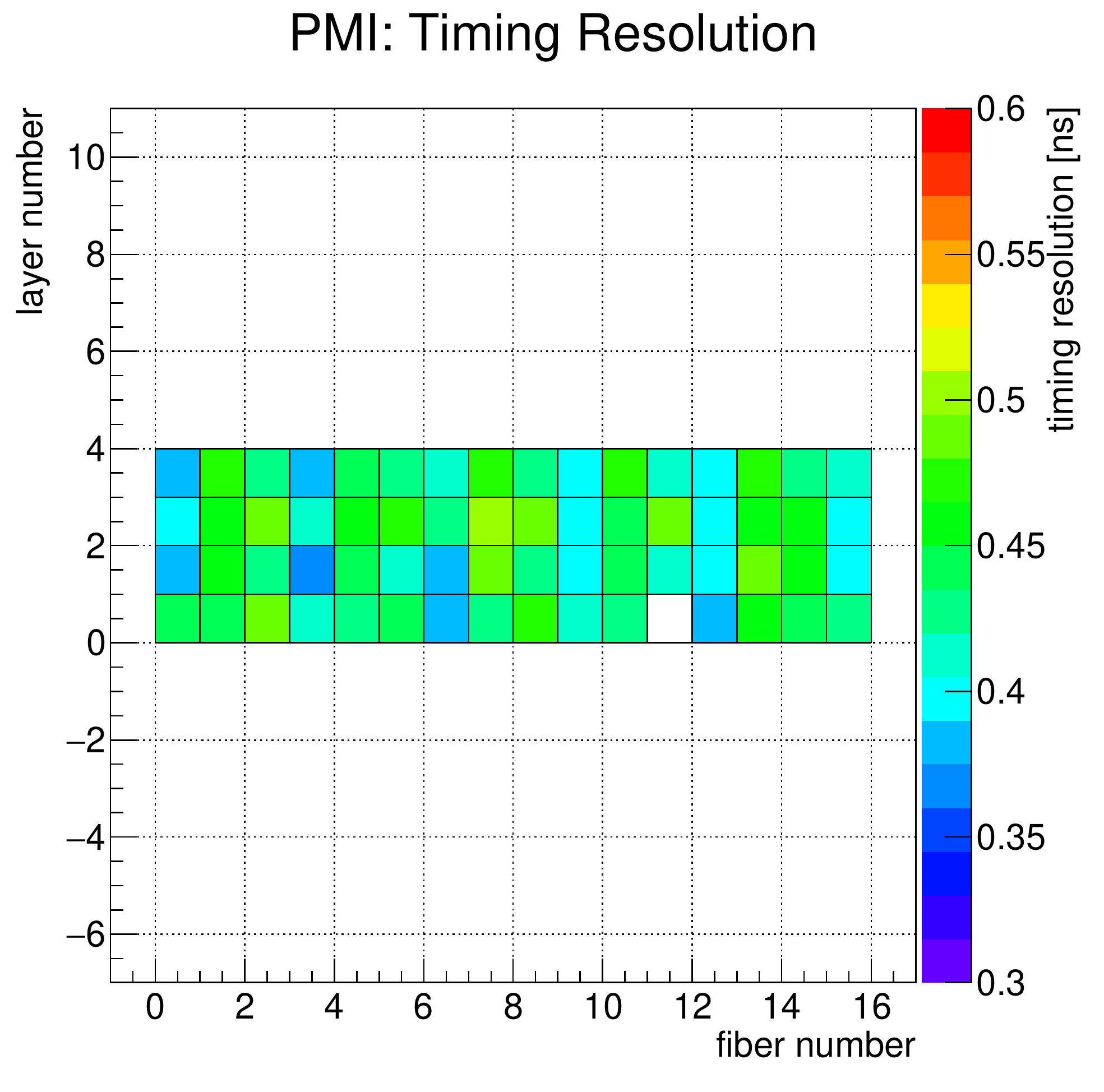}
\includegraphics[width=0.49\textwidth]{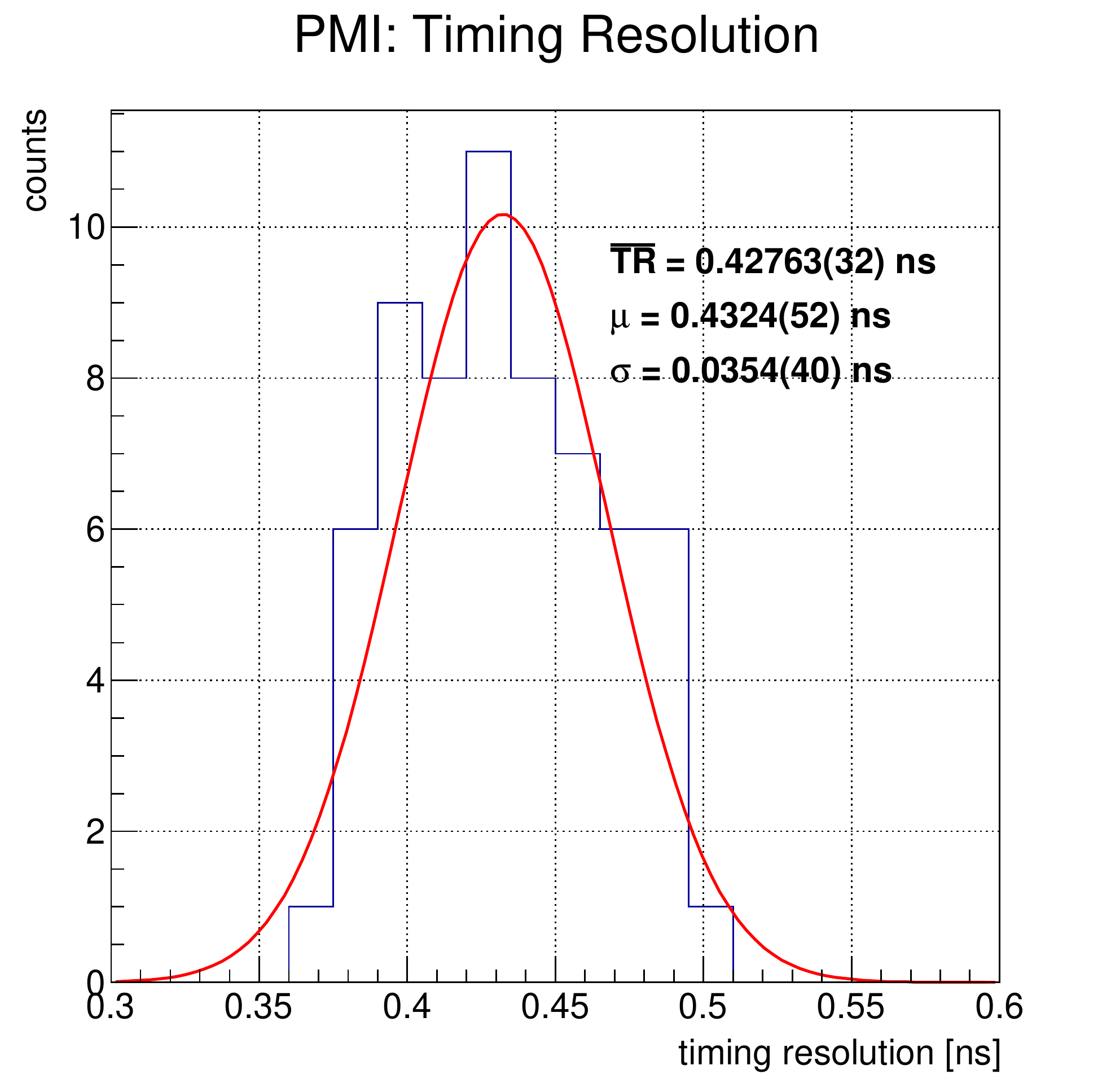}
\caption{Left: histogram showing the timing resolution values for each investigated fiber in \acrshort{gl:SSP}. Right: Statistical distribution of light collection values fitted with the Gaussian function. The listed values indicate the weighted mean and parameters of the distribution obtained from the fit.}
\label{fig:pmi-tres}
\end{figure}

The results of the timing properties analysis are presented in \cref{fig:pmi-tres}. The values of timing resolution ranged from \SI{0.37}{\nano\second} to \SI{0.50}{\nano\second} and the weighted mean of all values was \SI{0.43}{\nano\second}. This result is significantly better than the ones obtained for \acrshort{gl:JU} measurements (\SI{1.58}{\nano\second}, see \cref{tab:prototype-ju-results}) and single-fiber tests (\SI{1.26}{\nano\second}, see \cref{tab:diff-wrapping}). However, it is comparable with the average timing resolution of \acrshort{gl:LYSO:Ce} fibers determined in single-fiber measurements with the Hamamatsu \acrshort{gl:SiPM}s (\SI{0.43}{\nano\second}, see \cref{tab:diff-materials}). The large differences in timing resolutions are a result of different photosensors and their timing capabilities.

\subsubsection*{Summary of \acrshort{gl:PMI} measurements}

Results of the prototype characterization in the \acrshort{gl:PMI} setup are summarized in \Cref{tab:prototype-pmi-results}. Similarly as for the \acrshort{gl:JU} measurements, uncertainties listed in the table are purely statistical. The overall performance of the prototype tested in the \acrshort{gl:PMI} setup improved significantly in comparison with the \acrshort{gl:JU} tests. Obtained key parameters, such as attenuation length, energy resolution and position resolution are comparable with those determined in single fiber tests for analogous fiber configuration. The improved performance of the prototype was a result of the used photodetector, which allowed for acceptance of the scintillating light leaving the \acrshort{gl:LYSO:Ce} at wide angles. As described previously, it resulted in a shortened attenuation length and thus better position resolution. The energy resolution of the detector also improved, which was connected with an increased light collection. Moreover, the superior timing properties of the photosensor and the \acrshort{gl:DAQ} system resulted in improved timing resolution. To summarize, the overall performance of the \acrshort{gl:SSP} in the \acrshort{gl:PMI} tests was satisfactory. 

\begin{table}[!ht]
\centering
\caption{Numerical results of the prototype characterization with the light sharing based readout.}
\begin{tabularx}{1.0\textwidth}{|X|p{2.7cm}|p{2.7cm}|p{2.7cm}|}
\hline
Property & Mean & Standard \newline deviation & Weighted \newline mean \\ \hline
Attenuation length (\gls{gl:MLR}) [\si{\milli\metre}] & \num{357.3(63)} & \num{40.6(70)} & \num{337.82(71)} \\
Attenuation length (\acrshort{gl:ELAR}) [\si{\milli\metre}] & \num{106.9(21)} & \num{11.5(26)} & \num{103.81(27)} \\
Light collection [\si{\au\per\mega\electronvolt}] & \num{1957(28)} & \num{184(26)} & \num{1938.5(46)} \\
Timing resolution [\si{\nano\second}] & \num{0.4324(52)} & \num{0.0354(40)} & \num{0.42763(32)} \\
Energy resolution (\gls{gl:Qavg}) [\si{\percent}] & \num{8.03(10)} & \num{0.59(10)} & \num{8.0915(60)} \\
Reconstructed energy of annihilation \newline peak (\gls{gl:Qavg}) [\si{\kilo\electronvolt}] & \num{508.11(34)} & \num{1.87(47)} & \num{508.153(34)} \\
Energy resolution (\acrshort{gl:ELAR}) [\si{\percent}] & \num{7.733(51)} & \num{0.367(45)} & \num{7.7005(56)} \\
Reconstructed energy of annihilation \newline peak (\acrshort{gl:ELAR}) [\si{\kilo\electronvolt}] & \num{507.28(18)} & \num{0.94(29)} & \num{507.589(34)} \\
Position resolution (\gls{gl:MLR}) [\si{\milli\meter}] & \num{33.38(55)} & \num{3.37(72)} & \num{32.276(20)} \\
Reconstructed position residual \newline (\gls{gl:MLR}) [\si{\milli\meter}] & \num{0.0039(77)} & \num{0.0587(66)} & \num{0.008(12)} \\
Position resolution (\acrshort{gl:ELAR}) [\si{\milli\meter}] & \num{34.28(48)} & \num{3.27(66)} & \num{32.704(21)} \\
Reconstructed position residual \newline (\acrshort{gl:ELAR}) [\si{\milli\meter}] & \num{-0.003(14)} & \num{0.103(12)} & \num{0.011(12)} \\
\hline
\end{tabularx}
\label{tab:prototype-pmi-results}
\end{table}

\chapter{Summary and conclusions}
\label{chap:summary}

The aim of the presented work was design optimization of the \acrshort{gl:SiFi-CC} detector components and subsequent construction and characterization of a small-scale prototype of the future detector setup. The ultimate motivation of the presented research and the goal of the \acrshort{gl:SiFi-CC} project is development of the real-time method to monitor the dose distribution administered to a patient during proton therapy. Such a method would significantly improve the quality of proton therapy. Therefore, intensive research is being conducted in this subject worldwide. A short overview of the current state of research on real-time monitoring of proton therapy is given in \cref{chap:intro}. As stated in that overview, many of the constructed detectors suffer from a too small efficiency or large background. The \acrshort{gl:SiFi-CC} collaboration addresses this issue with its proposed dedicated detection setup. The \acrshort{gl:SiFi-CC} design takes advantage of the techniques and materials well established in particle and high energy physics, such as state-of-the-art photodetectors and heavy scintillating fibers, and applies them in medical imaging. 

In \cref{chap:scintillating-materials}, an overview of topics related to scintillation detectors was given, including: different types of scintillating materials and their scintillation mechanisms, the most commonly used types of photodetectors and their principles of operation, a description of important aspects of the design and assembly of scintillation detectors, and finally the role of scintillating detectors in medical imaging.

Further in that chapter various characteristics of scintillators were described along with the methods of determining them. In particular, the two models of scintillating light propagation were described in detail. The first of the models assumed straightforward exponential light attenuation (\acrshort{gl:ELA}). The alternative implementation of that model was additionally presented using the quantity \gls{gl:MLR}. The second described model assumed that the scintillating light emitted from the interaction point undergoes exponential attenuation, however, when it reaches the fiber end it can be reflected and travels towards the opposite side of the fiber, where it is registered by a photodetector. The model of scintillating light attenuation including  light reflection (\acrshort{gl:ELAR}) provided more rigorous and at the same time more complicated description of the light propagation process in the elongated scintillators. All of the listed models allow to determine light attenuation length, which parameterizes the transparency of scintillator to its own scintillating light.

The two presented formalisms of light propagation description led to the corresponding methods for reconstruction of the energy deposit and the position of the interaction in the scintillator. The two energy reconstruction methods were presented: first based on the \gls{gl:Qavg} quantity and the other, using the \acrshort{gl:ELAR} data parameterization (\gls{gl:Qavgstar}). Having the reconstructed energy spectra, it was also possible to determine the energy resolution of the investigated scintillator. Similarly, the position reconstruction was conducted using the two methods corresponding to light attenuation models. 
The position reconstruction additionally yielded the position resolution of the investigated scintillating fibers. The remaining properties of the scintillating materials presented in that chapter were the light collection and the timing characteristics. Determination and analysis of all the listed properties was crucial for the assessment of the scintillator performance and therefore design optimization of the scintillating detector.

The optimization of the design of the future \acrshort{gl:SiFi-CC} detector consisted of a series of tests of the detector components, which were described in \cref{chap:single-fibers}. An extensive, systematic, comparative study of different scintillating materials, different types of coatings and wrappings as well as different types and sizes of the optical couplings was conducted. The measurements were performed with the use of a dedicated experimental setup, featuring a remotely controlled electronic collimator, independent temperature monitoring and precise fiber positioning systems. Data preprocessing and analysis were done using custom-written software. 

The light propagation analysis showed that out of the three methods of light attenuation determination \gls{gl:MLR} and \gls{gl:ELAR} perform the best, based on the \chiNDF values obtained from the fits. However, since the two methods are based on different assumptions regarding the light propagation pattern in the elongated scintillator, the obtained light attenuation values differ significantly. Therefore, the results of light attenuation analysis should be considered along with the corresponding model formalism. On the other hand, the \acrshort{gl:ELA} model and its alternative implementation based on the \gls{gl:MLR} parameter show very similar results. 

Based on the obtained results, the following conclusions for the future \acrshort{gl:SiFi-CC} detector were made:
\begin{itemize}
\item \acrshort{gl:LYSO:Ce} was chosen as an active part of the detector. Of the three investigated scintillating materials, it offered the best light collection, timing properties, as well as position and energy resolution. Additionally, it is affordable and easily available;
\item the Al foil (the bright side facing towards the fiber) was selected as a fiber wrapping. It allowed to maintain satisfactory light collection and energy resolution. At the same time, a decreased light attenuation length caused improved position resolution;
\item silicone pads were selected as the coupling interface in the detector, as they offer a compromise between the improved system performance and its stability and durability;
\item the size of the interface silicone pads has no influence on the performance of the scintillator in the setup.
\end{itemize}

In \cref{chap:prototype}, further experimental investigations are presented, of a small-scale prototype detector module built exploiting the conclusions of the single-fiber studies. 
The prototype consisted of 64 \acrshort{gl:LYSO:Ce} scintillating fibers wrapped in Al foil (bright side facing the fiber). The fibers were inserted into custom made frames, which allowed to assemble the prototype into desired geometry. This work focused on the geometry, which was earlier determined as optimal in the Monte Carlo simulations, namely four parallel layers, with the neighboring layers shifted by a half-fiber pitch.

The following part of \cref{chap:prototype} describes the prototype tests and the obtained results of characterization. The prototype was tested in two configurations, which featured different readout electronics and \acrshort{gl:DAQ}. In the first configuration (\acrshort{gl:JU}) the readout system consisted of custom designed \acrshort{gl:PCB} boards housing the \acrshort{gl:SiPM}s. The size of the \acrshort{gl:SiPM}s and their layout was matching that of the scintillating fibers arranged in the prototype, allowing for one-to-one readout. To carry out the tests and complete the characterization of the prototype, a dedicated test bench was constructed. Similarly as in the case of single-fiber tests, it featured a remotely controlled electronic collimator. However, this time, the reference detector had a form of an elongated fiber to enable collimation of multiple fibers simultaneously. The \acrshort{gl:DAQ} was the same as used in the single-fiber measurements.

As a result of the \acrshort{gl:JU} measurement 56 out of 64 fibers in the prototype were characterized.
The obtained results were not satisfactory, with the mean values of main parameters characterizing the detector performance: \gls{gl:MLR} attenuation length \SI{483.7}{\milli\meter}, \gls{gl:Qavg} energy resolution \SI{10.32}{\percent}, \gls{gl:MLR} position resolution \SI{94.4}{\milli\meter} and timing resolution \SI{1.526}{\nano\second}. Due to weak attenuation of scintillating light, a part of the results obtained with the \acrshort{gl:ELAR} model of light propagation were disputable and therefore they are not quoted in this summary. This indicates, that the simplified exponential model performs better in situations where the expected light attenuation is strong. It can be observed that all of the analyzed properties were worse than previously determined for the analogous scintillator configuration in single-fiber tests. In particular, the position resolution value was inadequate, as it was comparable with the length of the \acrshort{gl:LYSO:Ce} fibers, thus making the position reconstruction impossible. Such a significant deterioration of the performance in comparison with the single-fiber tests can be attributed to the different photodetectors used in both experiments. SensL \acrshort{gl:SiPM}s used in the single-fiber tests had a photosensitive area of \SI[parse-numbers=false]{3 \times 3}{\square\milli\meter}, which enabled collection of the scintillating light leaving the crystal at large angles. On the contrary, Ketek \acrshort{gl:SiPM}s used in the \acrshort{gl:JU} prototype measurements had a photosensitive area of \SI[parse-numbers=false]{1 \times 1}{\square\milli\meter}. This significantly restrained the angular acceptance of the photodetector. As a result, only light leaving the scintillator at small angles was registered. It should be underlined that the light which leaves the scintillator at large angles is mostly responsible for the position sensitivity, as it experiences larger optical path due to multiple internal reflections and thus is strongly attenuated. Therefore, restricting the collection angle is not favorable for the detector performance.

This issue was addressed in the second experiment with the prototype which was conducted at the \acrshort{gl:PMI} department of the RWTH Aachen University. In those measurements, the prototype was read out with the Phillips Power Tile digital sensors. In contrast to the previously used \acrshort{gl:SiPM}s, light from each single fiber was let to spread and was registered in several pixels of the photosensor (i.e. light-sharing-based readout). This allowed to use the center-of-gravity method to reconstruct the two lateral coordinates of the interaction point. 
The experimental setup featured the fan beam collimator, which combined both passive and electronic collimation. An analogous prototype characterization to that of the \acrshort{gl:JU} measurements was performed. 63 out of 64 fibers in the detector were characterized, yielding the mean results as follows: \gls{gl:MLR} attenuation length \SI{357.3}{\milli\meter}, \acrshort{gl:ELAR} attenuation length \SI{106.9}{\milli\meter}, \gls{gl:Qavg} energy resolution \SI{8.03}{\percent}, \gls{gl:Qavgstar} energy resolution \SI{7.73}{\percent}, \gls{gl:MLR} position resolution \SI{33.38}{\milli\meter}, \gls{gl:MLRstar} position resolution \SI{34.28}{\milli\meter} and timing resolution \SI{0.4324}{\nano\second}. Therefore, a significant improvement of the detector performance was observed in comparison to the \acrshort{gl:JU} measurements. It was caused by the superior timing characteristics of the photosensor and the improved light collection in the system, but most importantly the acceptance of the light leaving the scintillating crystals at wide angles, leading to better position resolution. This shows the crucial role of the photodetector in the designed scintillating detector.

In addition to detector characterization, during the experiment at \acrshort{gl:PMI} additional measurements with the prototype were performed in the 1D coded mask mode and point-like sources. Images were reconstructed using the \acrshort{gl:MLEM} algorithm. The projections of the reconstructed images showed clear peaks that represent radioactive point-like sources without noise or artifacts. The projections yielded promising results with $\sigma$ of the peaks in the order of \SI{1.2}{\milli\meter}. The results are currently under review for publication in the scientific journal.

\chapter{Outlook}
\label{chap:outlook}

\begin{figure}
\centering
\includegraphics[width=0.7\textwidth]{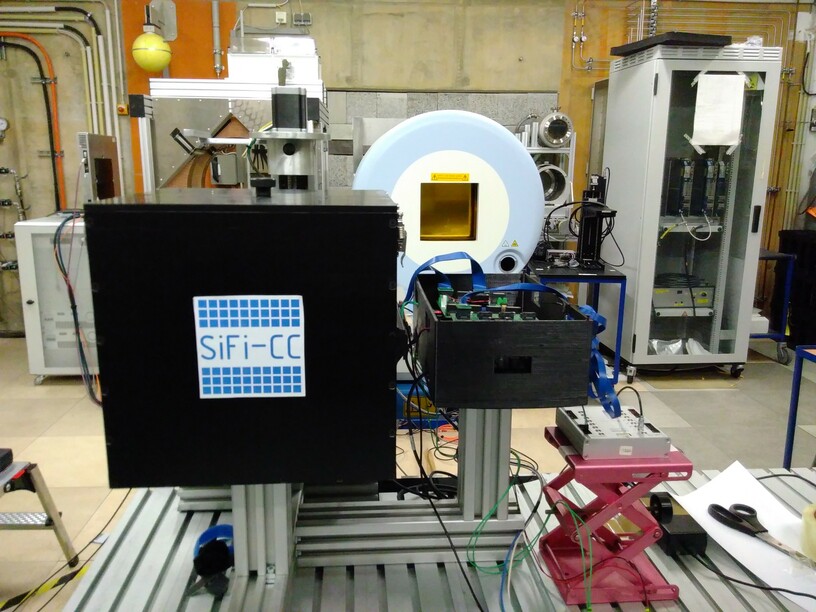} \\
\vspace{0.5cm}
\includegraphics[width=0.7\textwidth]{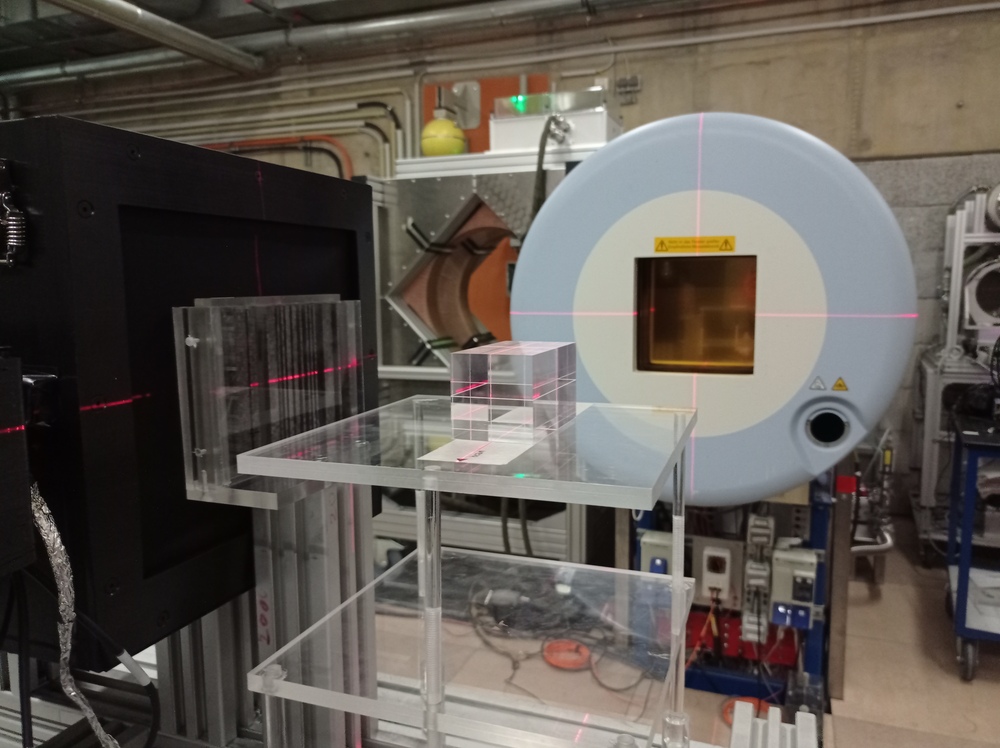}
\caption{Pictures taken during the beam time at \acrshort{gl:HIT} in January 2023. Top: view on the detection setup. The scatterer module was placed in the light-tight box (left). A TOFPET2-based system was used as the \acrshort{gl:DAQ} (right). In the background the beam nozzle is visible. Bottom: setup prepared for measurements with the 1D coded mask. The mask can be seen positioned in front of the light-tight box housing the detector. In the central part of the picture a \acrshort{gl:PMMA} target is visible.}
\label{fig:hit-setup}
\end{figure}

The extensive tests conducted with the single scintillating fibers and subsequently with the small-scale prototype resulted in valuable conclusions for the design of the scatterer - the first module of the final \acrshort{gl:SiFi-CC} setup. Multiple measurements allowed not only to find the optimal combination of a scintillating material, a wrapper and a coupling interface, but also gain experience in the technical aspects of the detector design such as handling of the fragile scintillating fibers and photodetectors. As a result, the full-size scatterer module was designed and constructed. It consists of 55 \acrshort{gl:LYSO:Ce} fibers organized in seven layers. The fiber dimensions are \SI[parse-numbers=false]{1.94 \times 1.94 \times 100}{\cubic\milli\meter}. All fibers were mechanically polished by the producer and glued together in a solid block. The fibers are separated with the layer of Al foil. Tests with the prototype allowed to choose the final geometry for the detector, and therefore the positioning frames were no longer needed. Moreover, the detector in the form of a solid block is significantly easier to handle and less fragile. Additionally, in the final setup, larger \acrshort{gl:SiPM}s are used for readout, with the photosensitive area of \SI[parse-numbers=false]{3.72 \times 3.62}{\square\milli\meter}. This means that each \acrshort{gl:SiPM} is attached to four \acrshort{gl:LYSO:Ce} fibers. The relative placement of the top \acrshort{gl:PCB} board and bottom \acrshort{gl:PCB} board with a one-fiber shift in each direction allows for straightforward fiber identification. At the same time, acceptance of the scintillating light leaving the crystals at large angles is enabled. First tests with the scatterer module were performed in January 2023 at the Heidelberg Ion-Beam Therapy Center (\acrshort{gl:HIT}), Germany (see \cref{fig:hit-setup}, top). The scatterer was also tested in the 1D coded-mask mode with the \acrshort{gl:PMMA} phantom and a proton beam (see \cref{fig:hit-setup}, bottom). The collected data are currently being analyzed. Once the data analysis is finalized, the second module of the planned \acrshort{gl:SiFi-CC} detector, \ie the absorber, will be constructed. Subsequently the full detection setup will be tested with the proton beam, both in Compton camera and coded mask mode. 

\begin{appendices}
\label[app]{appendix}

\chapter{Exponential light attenuation with light reflection (ELAR) - derivation of the model}
\label[app]{app:ELAR}

\begin{table}[ht]
\centering
\caption{List of symbols used in the derivation of the \acrshort{gl:ELAR} model.}
\label{tab:elar-model-symbols}
\begin{tabularx}{1.0\textwidth}{p{2,5cm}X}
Symbol & Meaning \\ \midrule
$P_\mathrm{l}$, $P_\mathrm{r}$ & direct component for left and right side of the scintillator \\ 
$R_\mathrm{l}$, $R_\mathrm{r}$ & reflected component for left and right side of the scintillator \\
$S_\mathrm{l}$, $S_\mathrm{r}$ & total signal for left and right side of the scintillator \\
$x$ & position of the interaction along the scintillator \\
$S_\mathrm{0}$ & signal amplitude at the place of the scintillating light emission \\
$\lambda$ & attenuation length \\
$L$ & scintillating fiber length \\
$\eta_\mathrm{ l}$, $\eta_\mathrm{ r}$ & coefficients related to the light reflection at the left and right side of the scintillator \\
$\xi_\mathrm{l}$, $\xi_\mathrm{r}$ & 
light transmission factors associated with the coupling quality at the corresponding fiber ends \\
$\eta_\mathrm{ l'}$, $\eta_\mathrm{ r'}$ & transformed $\eta_\mathrm{l}$ and $\eta_\mathrm{r}$ coefficients \\
$S_\mathrm{0'}$ & transformed $S_\mathrm{0}$ coefficient \\
$\xi$ & asymmetry coefficient \\
$P_\mathrm{l}^*$, $P_\mathrm{r}^*$ & reconstructed direct component for left and right side of the scintillator \\
$\sigma^2_{\mathrm{f P_l^*}}$, $\sigma^2_{\mathrm{f P_l^*}}$ & variance of $P_\mathrm{l}^*$ and $P_\mathrm{r}^*$ \\
$\sigma_\mathrm{P_l^*}$, $\sigma_\mathrm{P_r^*}$ & uncertainty of the reconstructed direct component for left and right side of the scintillator \\
\end{tabularx}
\end{table}

\newpage
\noindent
The exponential light attenuation model with light reflection (\acrshort{gl:ELAR}) describes the propagation of the scintillating light in the elongated scintillator. In this approach, it is assumed that part of the produced scintillating light is recorded by the photodetector as soon as it reaches one of the scintillator ends. This fraction of scintillating light is called \textbf{direct component} and can be expressed as follows:
\begin{equation}
\label{app:eq:primary-component}
\begin{cases}
  P_\mathrm{l}(x) = S_\mathrm{0} \exp\left(\frac{-x}{\lambda}\right) \\
  P_\mathrm{r}(x) = S_\mathrm{0} \exp\left(\frac{-(L-x)}{\lambda}\right) \ .
\end{cases}
\end{equation}
The remaining light undergoes reflection, travels to the opposite side of the scintillator, and is recorded there. It is called \textbf{reflected component} and can be described as follows:
\begin{equation}
\label{app:eq:reflected-component}
\begin{cases}
  R_\mathrm{l}(x) = \eta_\mathrm{r} P_\mathrm{r}(x) \exp(\frac{-L}{\lambda}) \\
  R_\mathrm{r}(x) = \eta_\mathrm{l} P_\mathrm{l}(x) \exp(\frac{-L}{\lambda}) \ .
\end{cases}
\end{equation}
The \textbf{total signals} recorded at both ends of the scintillator are the sums of the corresponding direct and reflected components. It is assumed that no additional light losses occur in the scintillator, and thus the fractions of the corresponding direct and reflected components sum up to unity. Then, the equations to calculate total signals are the following:
\begin{equation}
\label{app:eq:measured-signal-1}
\begin{cases}
  S_\mathrm{l}(x) = \xi_\mathrm{l} \left[ (1-\eta_\mathrm{l}) P_\mathrm{l}(x) + \eta_\mathrm{r} R_\mathrm{l}(x) \right] \\
  S_\mathrm{r}(x) = \xi_\mathrm{r} \left[ (1-\eta_\mathrm{r}) P_\mathrm{r}(x) + \eta_\mathrm{l} R_\mathrm{r}(x) \right] \ .
\end{cases}
\end{equation}
After substituting $P_\mathrm{l}(x)$, $P_\mathrm{r}(x)$, $R_\mathrm{l}(x)$, and $R_\mathrm{r}(x)$ into \cref{app:eq:measured-signal-1} the following equations are obtained:
\begin{equation}
\label{app:eq:measured-signal-2}
\begin{cases}
  S_\mathrm{l}(x) = \xi_\mathrm{l} S_\mathrm{0} \left[ (1-\eta_\mathrm{l}) \exp{\left( \frac{-x}{\lambda} \right)} + \eta_\mathrm{r} \exp{\left( \frac{-2L+x}{\lambda} \right)} \right] \\
  S_\mathrm{r}(x) = \xi_\mathrm{r} S_\mathrm{0} \left[ (1-\eta_\mathrm{r}) \exp{\left( \frac{-L+x}{\lambda} \right)} + \eta_\mathrm{l} \exp{\left( \frac{-L-x}{\lambda}\right)} \right] \ .
\end{cases}
\end{equation}
In \cref{app:eq:measured-signal-2}, it is not possible to resolve the values of $S_\mathrm{0}$ and $\xi_\mathrm{i}$ based on the fit of the experimental data. Therefore, the following parameterization is introduced to remove the ambiguity and reduce the number of parameters:
\begin{equation}
\label{app:eq:ELAR-transformation}
\begin{cases}
  \eta_\mathrm{r'} = \frac{\eta_\mathrm{r}}{1-\eta_\mathrm{l}} \\
  \eta_\mathrm{l'} = \frac{\eta_\mathrm{l}}{1-\eta_\mathrm{r}} \ ,
\end{cases}
\qquad\qquad
\begin{cases}
  S_\mathrm{0'} = \xi_\mathrm{l} S_\mathrm{0} (1-\eta_\mathrm{l}) \\
  \xi = \frac{\xi_\mathrm{r} (1-\eta_\mathrm{r})}{\xi_\mathrm{l}(1-\eta_\mathrm{l})} \ .
\end{cases}
\end{equation}
After applying the parameterization \cref{app:eq:ELAR-transformation}, the total signal formulas are the following:
\begin{equation}
\label{app:eq:ELAR-final}
\begin{cases}
  S_\mathrm{l}(x) = S_\mathrm{0'} \left[ \exp{\left( \frac{-x}{\lambda} \right)} + \eta_\mathrm{r'} \exp{\left( \frac{-2L+x}{\lambda} \right)} \right] \\
  S_\mathrm{r}(x) = \xi S_\mathrm{0'} \left[ \exp{\left( \frac{-L+x}{\lambda} \right) } + \eta_\mathrm{l'} \exp{\left( \frac{-L-x}{\lambda} \right)} \right] \ .
\end{cases}
\end{equation}

\subsubsection*{Reconstruction of the direct component}

In the calibration measurement, the total signals $S_\mathrm{l}$ and $S_\mathrm{r}$ are recorded for different known source positions $x$ along the fiber. Equations \ref{app:eq:ELAR-final} can be fitted simultaneously to the experimental data obtained for the left and right sides of the scintillating fiber. Having the parameters of the \acrshort{gl:ELAR} model determined in the calibration measurement, it is possible to reconstruct direct components $P_\mathrm{l}$ and $P_\mathrm{r}$ based on measured signals. For that the following equation system has to be solved: 
\begin{equation}
 \begin{cases}
 S_\mathrm{l} = P_\mathrm{l} + \eta_\mathrm{r'} P_\mathrm{r} \exp\left( \frac{-L}{\lambda} \right) \\
 S_\mathrm{r} = \xi \cdot \left[P_\mathrm{r} + \eta_\mathrm{l'} P_\mathrm{l} \exp \left( \frac{-L}{\lambda} \right) \right] \ .
 \end{cases}
\end{equation}
%
%
%
%
%
%
%
%
%
The solution of this system of equations allows to reconstruct direct components $P_\mathrm{l}$ and $P_\mathrm{r}$, having measured values $S_\mathrm{l}$ and $S_\mathrm{r}$ and parameters of the \acrshort{gl:ELAR} model as follows:
\begin{equation}
 \begin{cases}
 P_\mathrm{l}^{*}(S_\mathrm{l}, S_\mathrm{r}) = \frac{e^{L/\lambda} \left( e^{L/\lambda} \xi S_\mathrm{l} - S_\mathrm{r} \eta_\mathrm{r'} \right)}{\xi \left( e^{2L/\lambda} - \eta_\mathrm{l'} \eta_\mathrm{r'} \right)} \\ \\
 P_\mathrm{r}^{*}(S_\mathrm{l}, S_\mathrm{r}) = - \frac{e^{L/\lambda} \left( -e^{L/\lambda} S_\mathrm{r} + \xi S_\mathrm{l} \eta_l' \right)}{\xi \left( e^{2L/\lambda} - \eta_\mathrm{l'} \eta_\mathrm{r'} \right)} \ .
 \end{cases}
\end{equation}

\subsubsection*{Uncertainty calculation}

Uncertainties of the reconstructed direct components are calculated according to the matrix expression for error propagation:
\begin{equation}
\label{eq:error-matrix}
\sigma^2_\mathrm{f} = \mathbf{g}^T \mathbf{V} \mathbf{g} \ ,
\end{equation}
where $\sigma^2_f$ is the variance of the function $f$ with a set of parameters $\beta$, $\mathbf{V}$ is the variance-covariance matrix and $\mathbf{g}$ is a vector in which $i$-th element is defined as the partial derivative~$\frac{\partial f}{\partial \beta_i}$ 
\cite{Nowak}. \\ \\
The partial derivatives of $P_\mathrm{l}^{*}(S_\mathrm{l}, S_\mathrm{r})$ are the following: \\ 
\begin{equation}
\frac{\partial P_\mathrm{l}^{*}(S_\mathrm{l}, S_\mathrm{r})}{\partial \lambda} = - \frac{e^{L/\lambda} L \eta_\mathrm{r'} \left( -2 e^{L/\lambda} \xi S_\mathrm{l} \eta_\mathrm{l'} + S_\mathrm{r} \left( e^{2L/\lambda} + \eta_\mathrm{l'} \eta_\mathrm{r'} \right) \right)}{\lambda^2 \xi \left( e^{2L/\lambda} - \eta_\mathrm{l'} \eta_\mathrm{r'} \right)^2} \ ,
\end{equation}
\begin{equation}
\frac{\partial P_\mathrm{l}^{*}(S_\mathrm{l}, S_\mathrm{r})}{\partial \xi} = \frac{e^{L/\lambda} S_\mathrm{r} \eta_\mathrm{r'}}{\xi^2 \left( e^{2L/\lambda} - \eta_\mathrm{l'} \eta_\mathrm{r'} \right)} \ ,
\end{equation}
\begin{equation}
\frac{\partial P_\mathrm{l}^{*}(S_\mathrm{l}, S_\mathrm{r})}{\partial \eta_\mathrm{l'}} = \frac{e^{L/\lambda} \eta_\mathrm{r'} (e^{L/\lambda} \xi S_\mathrm{l} - S_\mathrm{r} \eta_\mathrm{r'})}{\xi (e^{2L/\lambda} - \eta_\mathrm{l'} \eta_\mathrm{r'})^2 } \ ,
\end{equation}
\begin{equation}
\frac{\partial P_\mathrm{l}^{*}(S_\mathrm{l}, S_\mathrm{r})}{\partial \eta_\mathrm{r'}} = \frac{e^{2L/\lambda} \left( -e^{L/\lambda} S_\mathrm{r} + \xi S_\mathrm{l} \eta_\mathrm{l'} \right)}{\xi \left( e^{2L/\lambda} - \eta_\mathrm{l'} \eta_\mathrm{r'} \right)^2} \ .
\end{equation}
Experimental uncertainties of $S_l$ and $S_r$ also contribute to overall uncertainty, so they need to be taken into account:
\begin{equation}
\label{app:eq:uncert-pl-star}
\sigma_\mathrm{P_l^{*}} = \sqrt{\sigma_\mathrm{f P_l^{*}}^2 + \left( \frac{\partial P_\mathrm{l}^{*}(S_\mathrm{l}, S_\mathrm{r})}{\partial S_\mathrm{l}} \sigma_\mathrm{S_l} \right)^2 + \left( \frac{\partial P_\mathrm{l}^{*}(S_\mathrm{l}, S_\mathrm{r})}{\partial S_\mathrm{r}} \sigma_\mathrm{S_r} \right)^2} \ ,
\end{equation}
where:
\begin{equation}
\frac{\partial P_\mathrm{l}^{*} (S_\mathrm{l}, S_\mathrm{r})}{\partial S_\mathrm{l}} = \frac{e^{2L/\lambda}}{e^{2L/\lambda} - \eta_\mathrm{l'} \eta_\mathrm{r'}} \ ,
\end{equation}
\begin{equation}
\frac{\partial P_\mathrm{l}^{*}(S_\mathrm{l}, S_\mathrm{r})}{\partial S_\mathrm{r}} = - \frac{e^{L/\lambda} \eta_\mathrm{r'}}{\xi \left( e^{2L/\lambda} - \eta_\mathrm{l'} \eta_\mathrm{r'} \right)} \ .
\end{equation}
Similar derivation for $\sigma_\mathrm{P_r^{*}}$ leads to the formula:
\begin{equation}
\label{app:eq:uncert-pr-star}
\sigma_\mathrm{P_r^{*}} = \sqrt{\sigma_\mathrm{f P_r^{*}}^2 + \left( \frac{\partial P_\mathrm{r}^{*}(S_\mathrm{l}, S_\mathrm{r})}{\partial S_\mathrm{l}} \sigma_\mathrm{S_l} \right)^2 + \left( \frac{\partial P_\mathrm{r}^{*}(S_\mathrm{l}, S_\mathrm{r})}{\partial S_\mathrm{r}} \sigma_\mathrm{S_r} \right)^2} \ ,
\end{equation}
where $\sigma_\mathrm{f P_r^{*}}^2$ is calculated according to \cref{eq:error-matrix} and $\frac{\partial P_\mathrm{r}^{*}(S_\mathrm{l}, S_\mathrm{r})}{\partial S_\mathrm{l}}$ and $\frac{\partial P_\mathrm{r}^{*}(S_\mathrm{l}, S_\mathrm{r})}{\partial S_\mathrm{r}}$ are corresponding contributions of the experimental uncertainties of $S_\mathrm{l}$ and $S_\mathrm{r}$.

Equations \cref{app:eq:uncert-pl-star} and \cref{app:eq:uncert-pr-star} allow to calculate uncertainties of the reconstructed direct scintillating light components $P_\mathrm{l}^{*}$ and $P_\mathrm{r}^{*}$.

\chapter{Energy and position reconstruction}
\label[app]{app:energy-position-reconstruction}

In the following appendix the examples of energy and position reconstruction are presented.

\subsubsection*{Energy reconstruction}

\begin{sidewaysfigure}[htbp]
\centering
\includegraphics[width=.99\textwidth]{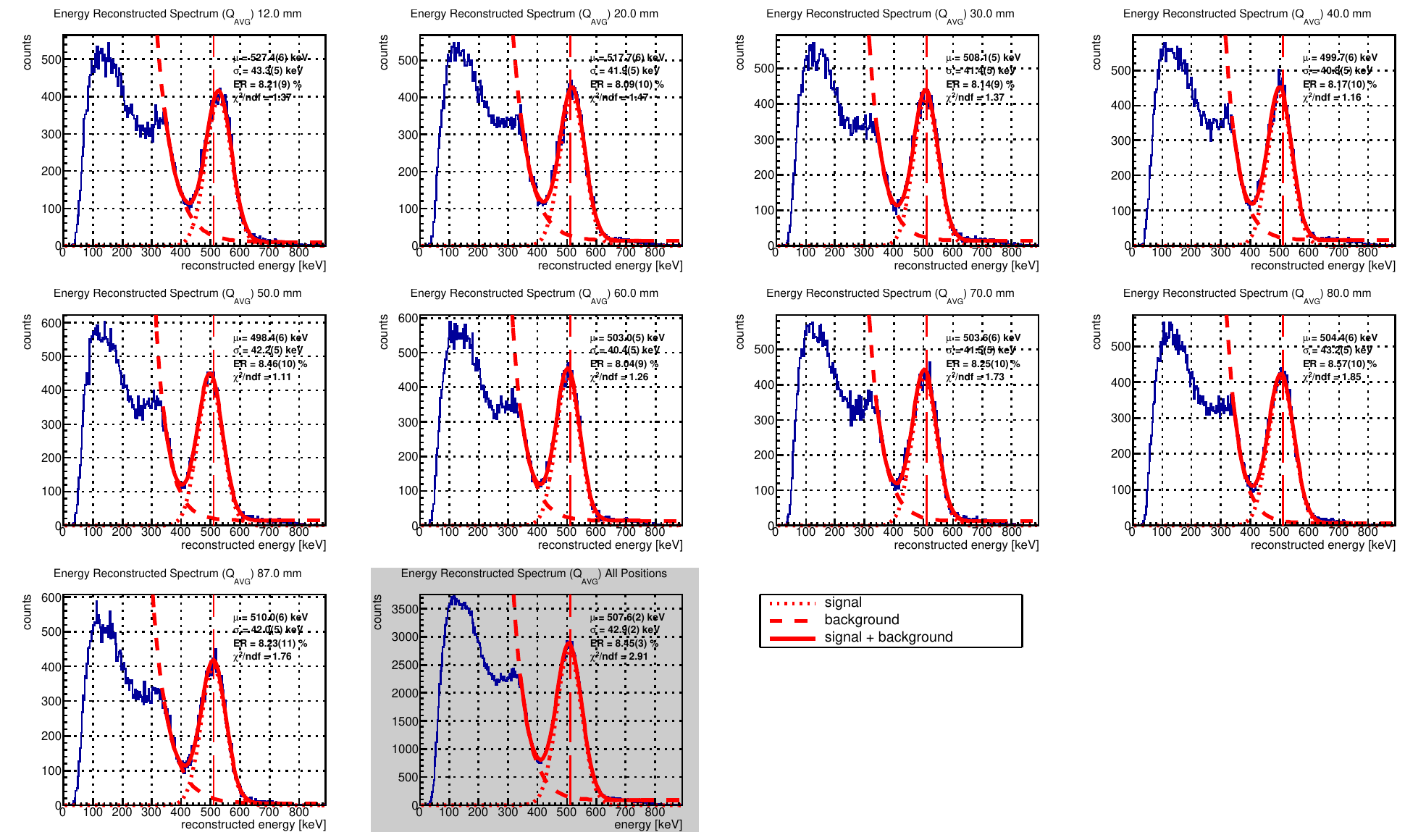}
\caption{Energy spectra reconstructed with the use of the \gls{gl:Qavg} method. The vertical line indicates the expected position of the annihilation peak. The highlighted panel shows the cumulative energy spectrum for all measurements in the series. Presented data come from the measurement of a single \acrshort{gl:LYSO:Ce} fiber (S109).}
\label{fig:energy_reconstruction_all_spectra_qavg}
\end{sidewaysfigure}

\begin{sidewaysfigure}[htbp]
\centering
\includegraphics[width=.99\textwidth]{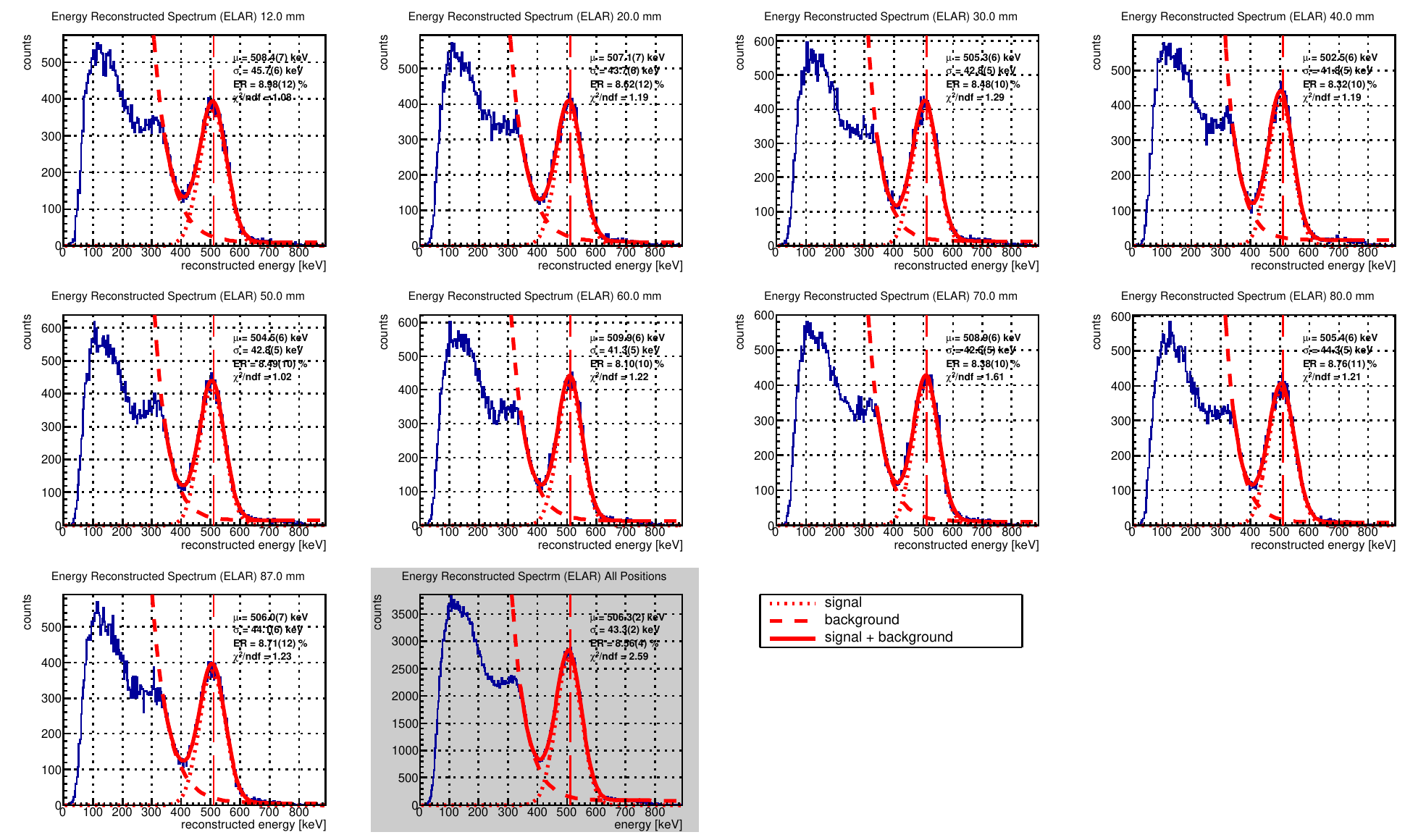}
\caption{Energy spectra reconstructed with the use of the \acrshort{gl:ELAR} method. The vertical line indicates the expected position of the annihilation peak. The highlighted panel shows the cumulative energy spectrum for all measurements in the series. Presented data come from the measurement of a single \acrshort{gl:LYSO:Ce} fiber (S109).}
\label{fig:energy_reconstruction_all_spectra_elar}
\end{sidewaysfigure}

The energy reconstruction was performed using \gls{gl:Qavg} and \acrshort{gl:ELAR} methods, as shown in \cref{fig:energy_reconstruction_all_spectra_qavg} and \cref{fig:energy_reconstruction_all_spectra_elar}. In both cases the energy reconstruction was done for all nine measurements taken at different positions of the radioactive source along the scintillating fiber. In addition, a cumulative energy spectrum was reconstructed (highlighted). In all spectra, the annihilation peak was fitted to calculate energy resolution.  

\subsubsection*{Position reconstruction}

\begin{sidewaysfigure}[htbp]
\centering
\includegraphics[width=.99\textwidth]{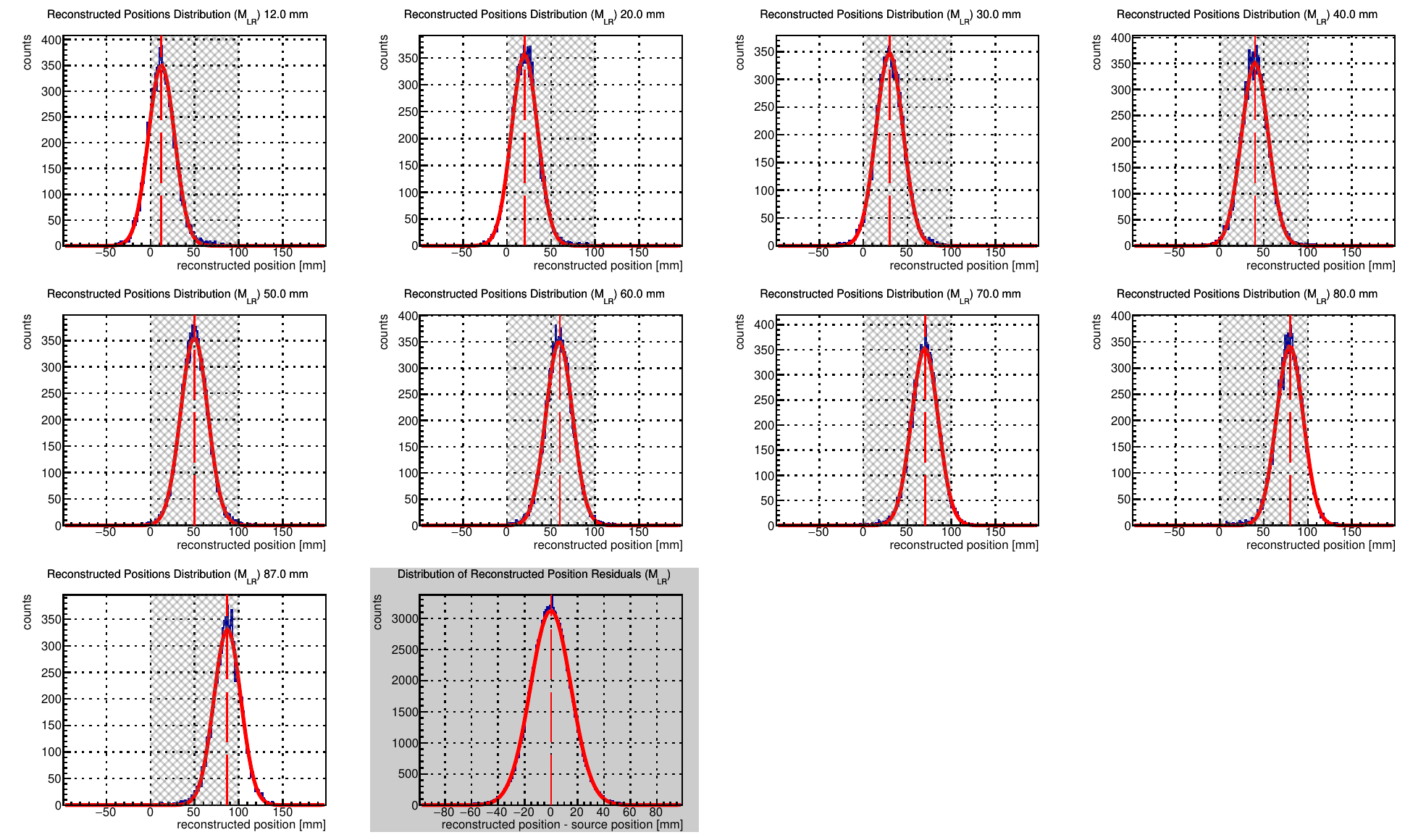}
\caption{Distributions of the reconstructed interaction positions in the scintillating fiber. Position reconstruction was performed using the \gls{gl:MLR} method. The rectangle indicates the size of the scintillating fiber. The highlighted panel shows distribution of the reconstructed position residuals for all measurements in the series. Presented data come from the measurement of a single \acrshort{gl:LYSO:Ce} fiber (S109).}
\label{fig:position_reconstruction_all_mlr}
\end{sidewaysfigure}

\begin{sidewaysfigure}[htbp]
\centering
\includegraphics[width=.99\textwidth]{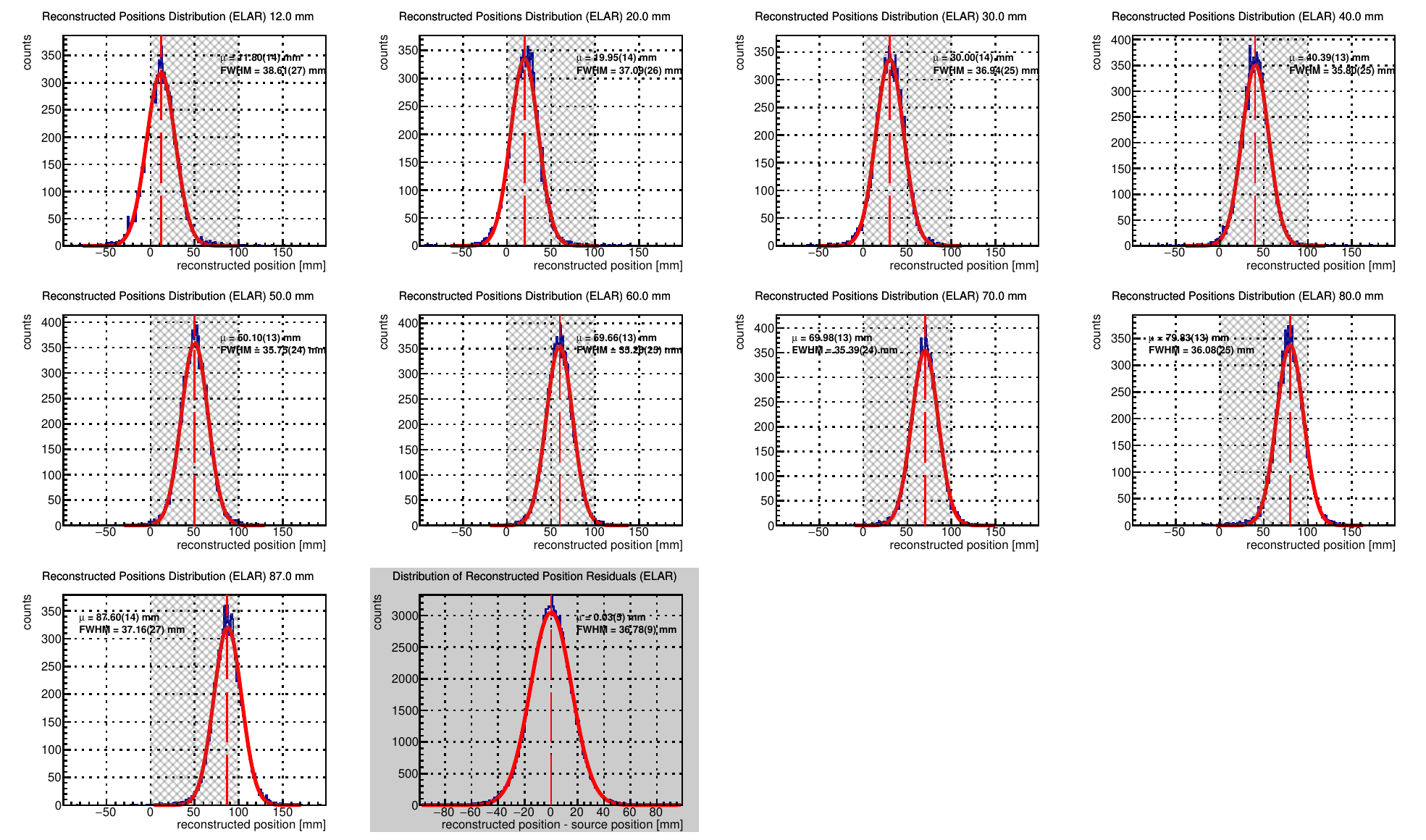}
\caption{Distributions of the reconstructed interaction positions in the scintillating fiber. Position reconstruction was performed using the \acrshort{gl:ELAR} method. The rectangle indicates the size of the scintillating fiber. The highlighted panel shows distribution of the reconstructed position residuals for all measurements in the series. Presented data come from the measurement of a single \acrshort{gl:LYSO:Ce} fiber (S109).}
\label{fig:position_reconstruction_all_elar}
\end{sidewaysfigure}

Reconstruction of the interaction position in the scintiallting fiber was performed using \gls{gl:MLR} and \acrshort{gl:ELAR} methods, as shown in \cref{fig:position_reconstruction_all_mlr} and \cref{fig:position_reconstruction_all_elar}. Similarly as to what was described above, the position reconstruction was done for all nine measurements in the experimental series. Additionally, a cumulative distribution of position residuals was reconstructed (highlighted). All distributions were fitted with the Gaussian function in order to obtain the reconstructed position (mean of the distribution) and the position resolution (\acrshort{gl:FWHM} of the distribution).

\chapter{Characteristics of used photodetectors}
\label[app]{app:photodetectors}

\begin{table}[ht]
\caption{Comparison of the analog \acrshort{gl:SiPM}s used in the single-fiber measurements (Hamamatsu and SensL) and \acrshort{gl:JU} prototype measurements (Ketek).Values with $*$ symbol apply to the PM1125-WB-C0 model of Ketek \acrshort{gl:SiPM}s.}
\label{tab:sipms}
\centering
\begin{threeparttable}[ht]
\begin{tabularx}{\textwidth}{X p{3.0cm} p{2.5cm} p{3.2cm}}
\toprule
{} & \textbf{Hamamatsu} \newline S13360-3050VE & \textbf{SensL} \newline C30020 & \textbf{Ketek} \newline PM1125-WB-B0 \newline PM1125-WB-C0$^{*}$ \\ \midrule
Effective photosensitive \newline area [\si{\milli\meter}] & $3 \times 3$ & $3 \times 3$ & $1 \times 1$ \\
Number of microcells & \num{3584} & \num{10998} & \num {1600} \\
Fill factor [\si{\percent}] & \num{74} & \num{48} & no data \\
Pixel size [\si{\micro\meter}] & \num{50} & \num{20}
& \num{25} \\
Breakdown voltage [\si{\volt}] & 48.0 -- 58.0 & 24.2 -- 24.7 & 25.1 -- 25.7 \newline 23.5 -- 25.5${}^{*}$ \\
Recommended operating \newline overvoltage $V_\textrm{ov}$ [\si{\volt}] & \num{3} & 1 -- 5 & 2 -- 5 \\
Spectral response range [\si{\nano\meter}] & 320 -- 900 & 300 -- 950 & 300 -- 900 \\
Peak sensitivity wavelength $\lambda_\textrm{p}$ [\si{\nano\meter}] & \num{450} & \num{420} & \num{430} \\
Photon detection efficiency at \newline max. $V_\textrm{ov}$ and $\lambda_\textrm{p}$ [\si{\percent}] & \num{40} & \num{31} &  \num{43} \newline \num{45}${}^{*}$ \\
Typical dark count at max. $V_\textrm{ov}$ & \num{0.5} \si{\mega\cps} & \SI{50}{\nano\ampere}\tnote{1} & \SI{125}{\kilo\cps\per\milli\meter\squared} \\
Source & \cite{hamamatsu} & \cite{sensl} & \cite{ketek-b0, ketek-c0} \\
\bottomrule
\end{tabularx}
\begin{tablenotes}
\item[1] At $V_\textrm{ov} = \SI{2.5}{\volt}$.
\end{tablenotes}
\end{threeparttable}
\end{table}

\begin{table}[ht]
\caption{Characteristics of the DPC3200-22 photosensor manufactured by Philips Digital Photon Counting.}
\label{tab:digit-sipms}
\centering
\begin{tabularx}{0.70\textwidth}{Xp{3.0cm}}
\toprule
Physical characteristics & Value  \\ \midrule
Dimensions & $48 \times 48$ \si{\milli\meter\squared} \\
Number of dies & \num{36} \\
Pixel active area & $3.8 \times 3.2$ \si{\milli\meter\squared} \\
Pixel pitch & \SI{4.0}{\milli\meter} \\
Cells per pixel & \num{3200} \\
Cells per subpixel & \num{800} \\
Cell size & $59.4 \times 64$ \si{\micro\meter\squared} \\
Pixel fill factor & \SI{74}{\percent} \\
Tile fill factor & \SI{55}{\percent} \\
Recommended operating voltage & $27 \pm 0.5$ \si{\volt} \\
Spectral response range & 380 -- 700 \si{\nano\meter} \\
Peak sensitivity wavelength $\lambda_\mathrm{p}$ & \SI{420}{\nano\meter} \\
Photon detection efficiency at $\lambda_\mathrm{p}$ & \SI{40}{\percent} \\
Dark count rate & \SI{140}{\kilo\cps\per\milli\meter\squared} \\
Sources & \cite{PowerTileShort, PowerTileFull, thesis-perez-gonzalez} \\
\bottomrule
\end{tabularx}
\end{table}


\chapter{Signal processing chain in single-fiber measurements}
\label[app]{app:signal-processing}

\begin{figure}[ht]
\centering
\includegraphics[width=.85\textwidth]{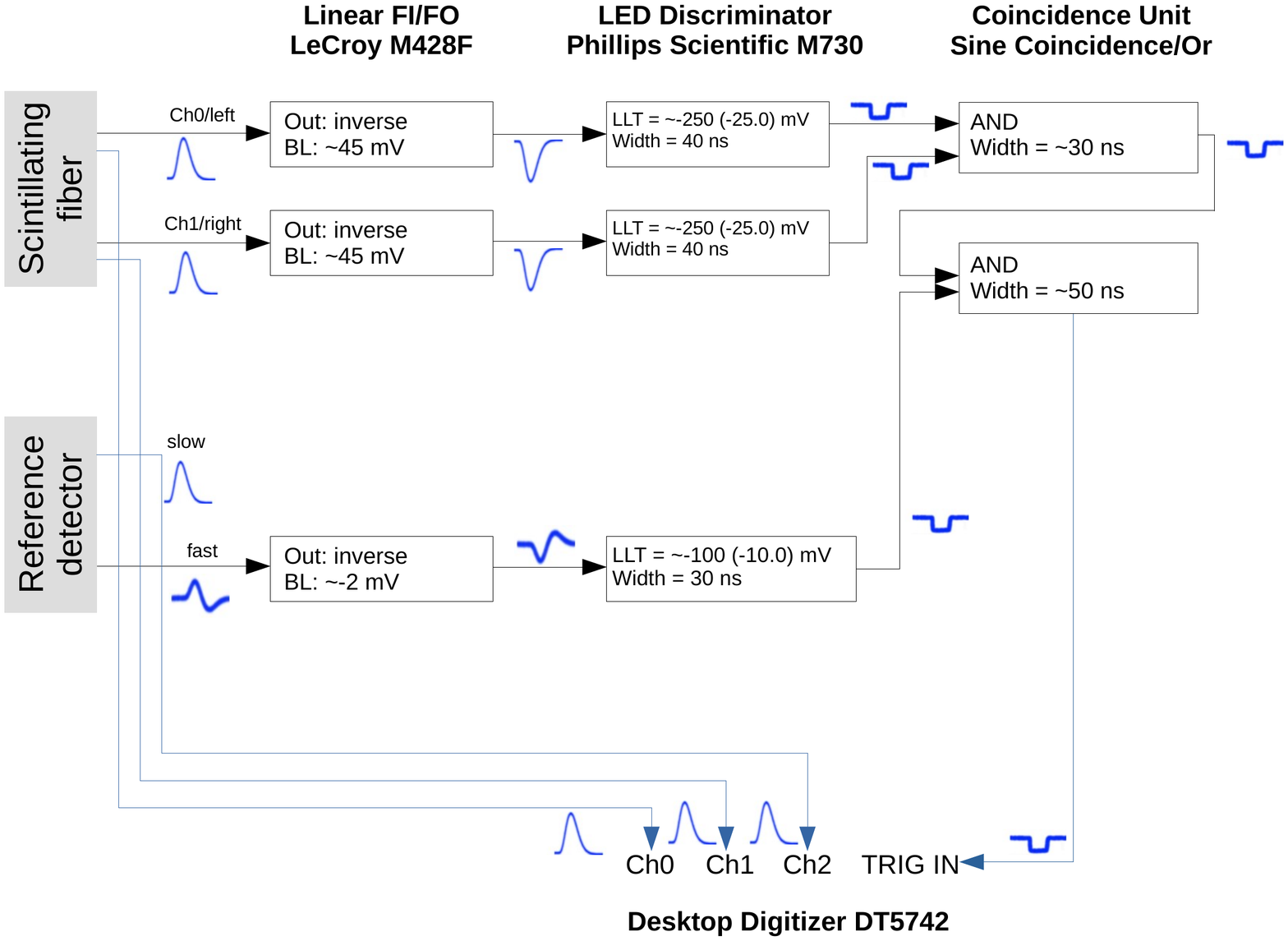}
\caption{Analog signal processing chain used in measurement series 30 -- 62 (see \cref{tab:measurements}. The two values of the low-level threshold (\acrshort{gl:LLT}) listed for the leading edge discriminator (\acrshort{gl:LED}) denote the value set in the module and the effective value applied in the signal processing chain, respectively.}
\label{fig:electronics-hamamatsu-1}
\end{figure}

\begin{figure}[ht]
\centering
\includegraphics[width=.85\textwidth]{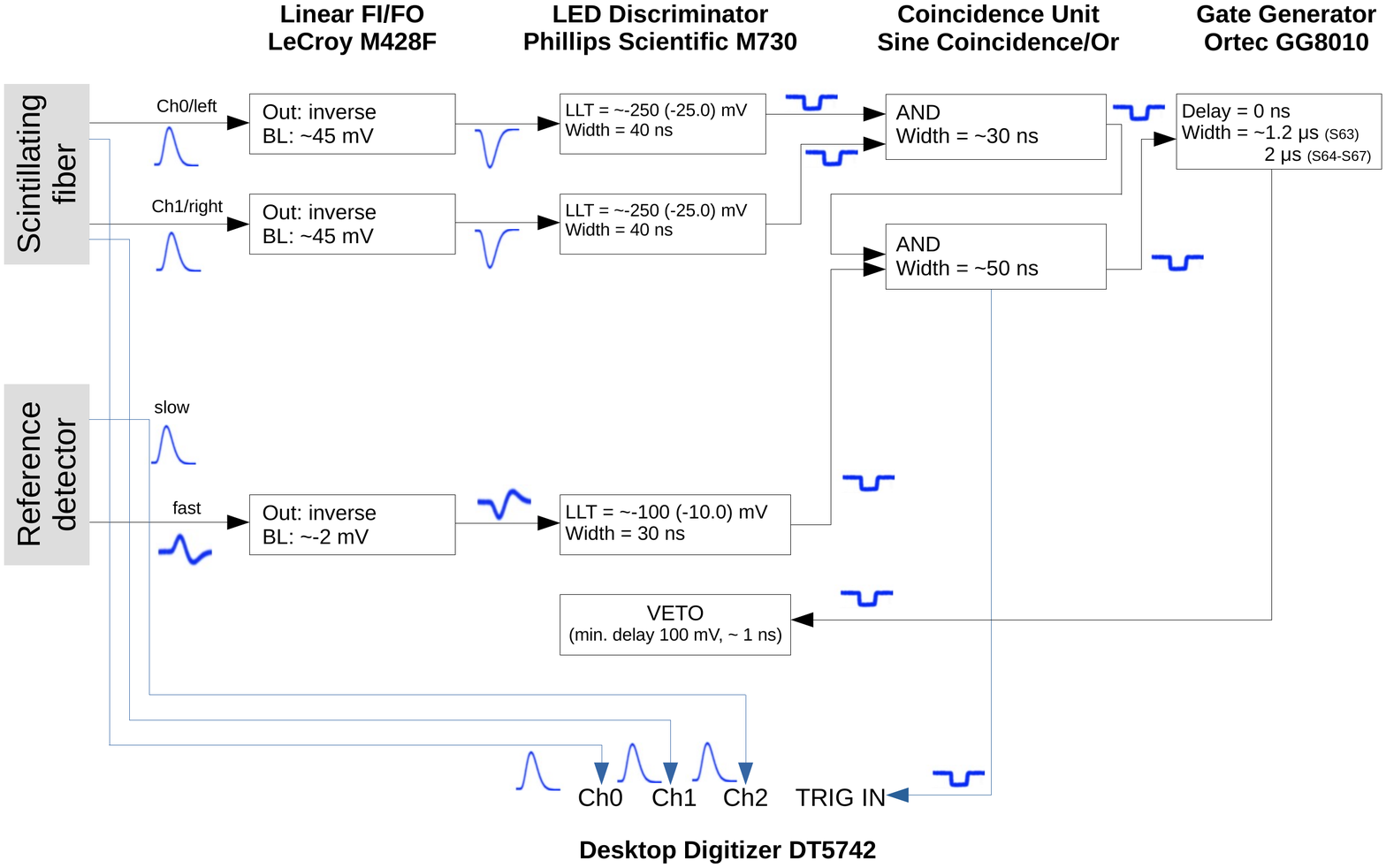}
\caption{Analog signal processing chain used in measurement series 63 -- 67 (see \cref{tab:measurements}). Relatively long \acrshort{gl:LuAG:Ce} signals often caused incorrect triggering, when the tail of an earlier signal triggered registration of new event. For that reason the gate generator was introduced in the chain. The width of the veto gate was increased to further suppress the number of incorrectly triggered events.  
}
\label{fig:electronics-hamamatsu-2}
\end{figure}

\begin{figure}[ht]
\centering\includegraphics[width=.85\textwidth]{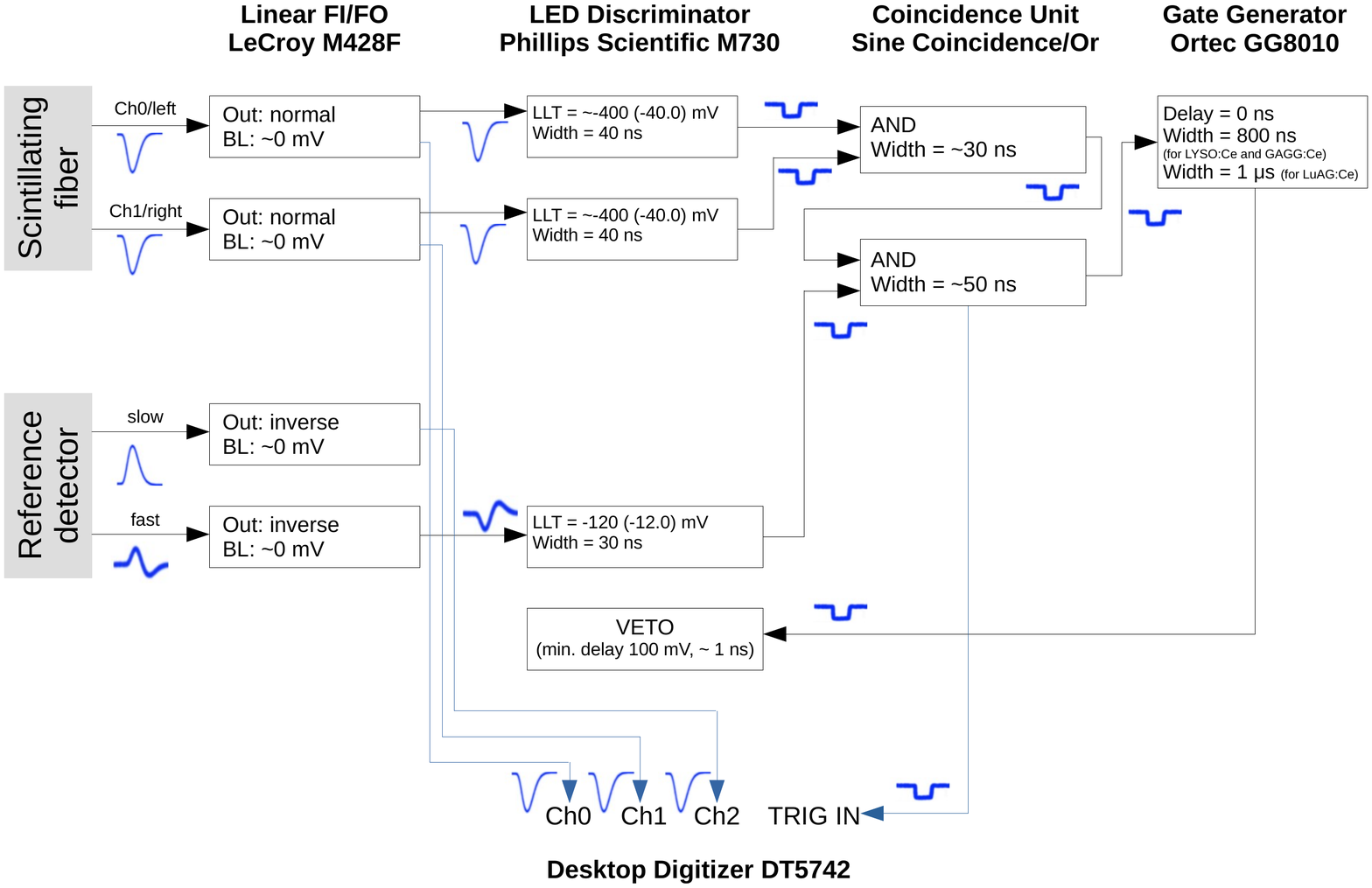}
\caption{Analog signal processing chain used in measurement series 98 -- 242 (see \cref{tab:measurements}). The length of the veto signal generated by the gate generator was changed depending on the decay constant of the investigated scintillating material to suppress the number of signals that incorrectly triggered the \acrshort{gl:DAQ}.}
\label{fig:electronics-sensl-1}
\end{figure}

\begin{figure}[ht]
\centering
\includegraphics[width=.85\textwidth]{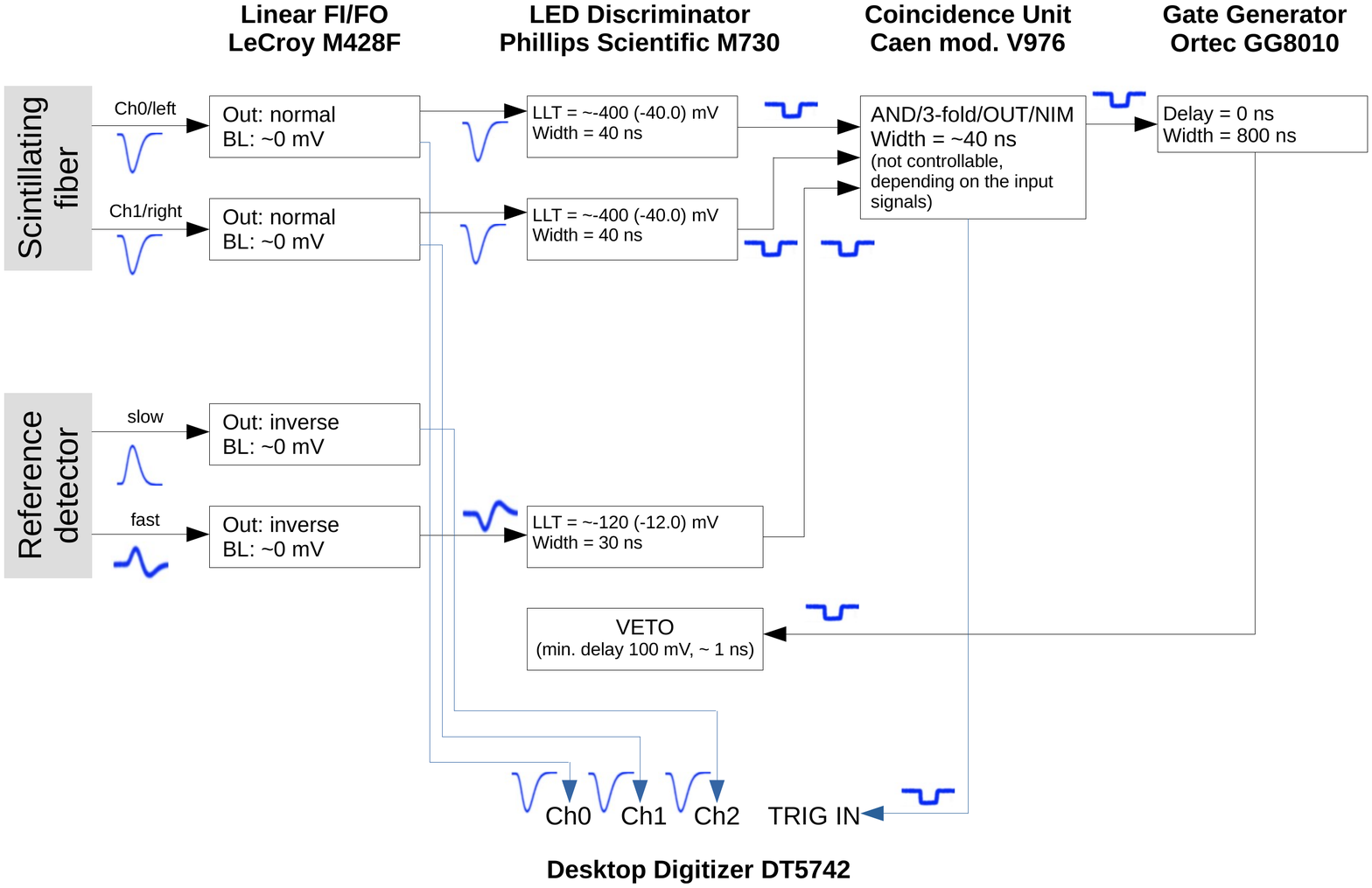}
\caption{Analog signal processing chain used in measurement series 243 -- 262 (see \cref{tab:measurements}).}
\label{fig:electronics-sensl-2}
\end{figure}

In the following appendix the diagrams of analog signal processing chain in single-fiber measurements are presented (see \cref{chap:single-fibers}). In particular, the diagrams show the three-fold coincidence constructed for the electronic collimation system. The signal processing chain was modified throughout the experiment in order to adjust to changing \acrshort{gl:SiPM}s and different scintillating materials, as depicted in \cref{fig:electronics-hamamatsu-1}, \cref{fig:electronics-hamamatsu-2}, \cref{fig:electronics-sensl-1} and \cref{fig:electronics-sensl-2}.

\end{appendices}
\printglossary[title=List of Symbols,
               toctitle=List of Symbols,
               type=main,
               nonumberlist]
\printglossary[title=List of Abbreviations,
               toctitle=List of Abbreviations,
               type=acronym,
               nonumberlist]
\thispagestyle{empty}
\thispagestyle{empty}

\chapter*{Acknowledgments}
\addcontentsline{toc}{chapter}{Acknowledgments}

This work was supported by the Polish National Science Centre (grants No. 2017/26/E/ST2/ 00618 and 2019/33/N/ST2/02780). The exchange of staff and students between Poland and Germany was financed by the Polish National Agency of Academic Exchange (NAWA) as well as German Academic Exchange Service (DAAD) - project ID 57562042. In this context, the project on which part of this thesis is based was funded by the German Federal Ministry of Education and Research (BMBF).

\vspace{2cm}

\noindent
I would like to express sincere gratitude to people which contributed to the success of the presented research: supervisors prof. dr hab. Andrzej Magiera and dr Aleksandra Wrońska, colleagues from the \acrshort{gl:SiFi-CC} group, especially Magda, Mark, Ronja and Jonas. To my family: parents Alina and Andrzej and sister Anna. Finally, to Rafał.
\clearpage
\thispagestyle{empty}

\addcontentsline{toc}{chapter}{Bibliography}
\printbibliography

\end{document}